\documentclass[a4paper,12pt]{report}

\usepackage{graphicx,amssymb,amsmath,epic,amsfonts,cite,bbm,enumerate,subfigure}
\usepackage{caption}
\usepackage{oldgerm}
\usepackage{epsfig}
\usepackage{color}
\DeclareMathOperator{\id}{id}

\DeclareMathOperator{\im}{Im}

\DeclareMathOperator{\tr}{tr}
\DeclareMathOperator{\Tr}{Tr}
\DeclareMathOperator{\ch}{ch}
\DeclareMathOperator{\aut}{aut}
\DeclareMathOperator{\Lc}{L}
\DeclareMathOperator{\Td}{Td}
\DeclareMathOperator{\diag}{diag}
\renewcommand{\d}{\text{d}}

\newcommand{\I}{\text{i}}

\def\clap#1{\hbox to 0pt{\hss#1\hss}}
\def\mathclap{\mathpalette\mathclapinternal}
\def\mathrlap{\mathpalette\mathrlapinternal}

\def\mathclapinternal#1#2{\clap{$\mathsurround=0pt#1{#2}$}}
\def\mathrlapinternal#1#2{\rlap{$\mathsurround=0pt#1{#2}$}}

\newcommand{\gt}{\ensuremath{\tilde{g}}}
\newcommand{\pt}{\ensuremath{\tilde{p}}}
\newcommand{\qt}{\ensuremath{\tilde{q}}}
\newcommand{\et}{\ensuremath{\tilde{e}}}
\newcommand{\vt}{\ensuremath{\tilde{v}}}
\newcommand{\ut}{\ensuremath{\tilde{u}}}
\newcommand{\wt}{\ensuremath{\tilde{w}}}

\newcommand{\rhot}{\ensuremath{\tilde{\rho}}}

\newcommand{\omegat}{\ensuremath{\widetilde{\omega}}}

\newcommand{\Mt}{\ensuremath{\widetilde{M}}}
\newcommand{\Pt}{\ensuremath{\widetilde{P}}}
\newcommand{\Rt}{\ensuremath{\widetilde{R}}}
\newcommand{\Sigmat}{{\ensuremath{\widetilde{\Sigma}}}}
\newcommand{\Kt}{{\widetilde{K3}}}
\newcommand{\Vt}{\widetilde{V}}
\newcommand{\Ct}{\widetilde{C}}
\newcommand{\nut}{\widetilde{\nu}}
\newcommand{\jt}{\widetilde{J}}
\newcommand{\etat}{\widetilde{\eta}}
\newcommand{\Vv}{\mathcal{V}}
\newcommand{\Nn}{\mathcal{N}}

\newcommand{\Pp}{\mathbb{P}}
\newcommand{\Ppt}{\widetilde{\mathbb{P}}}

\def\aut{\operatorname{aut}}
\def\Aut{\operatorname{Aut}}

\newcommand{\be}{\begin{equation}}  
\newcommand{\ee}{\end{equation}}

\newcommand{\nn}{\nonumber}
\newcommand\e{\mathrm{e}}
\newcommand\iu{\operatorname{i}}
\def\Hirz[#1]{\mathbbm{F}_{#1}}
\def\ord{\mbox{ord}}
\def\C{\mathbb{C}}
\def\Z{\mathbb{Z}}
\def\R{\mathbb{R}}
\def\P{\mathbb{P}}
\def\nbo{N_{O\setminus B}}
\def\ord{\operatorname{ord}}
\def\PD{\operatorname{PD}}
\def\Pic{\operatorname{Pic}}
\def\Hirz[#1]{\mathbbm{F}_{#1}}
\def\o[#1]{\overline{#1}}
\def\fib[#1, #2, #3]{
\begin{array}{ccc}
#1 & \hookrightarrow & #2 \\
&& \downarrow \\ \vspace{1cm}
&& #3
\end{array}
}

\begin{document}

\thispagestyle{empty}
\begin{center}{\bf
Dissertation \\
submitted to the \\
Combined Faculties of the Natural Sciences and Mathematics \\
of the Ruperto-Carola-University of Heidelberg. Germany \\
for the degree of \\ 
Doctor of Natural Sciences \\
\vspace{15cm}

Put forward by \\
Andreas Braun \\
born in Koblenz \\
Oral examination: February 5th, 2010 }\\
\end{center}

\newpage
\thispagestyle{empty}
\mbox{}

\newpage
\thispagestyle{empty}
\begin{center}
 {\Large\bf F-Theory and the Landscape of \vspace{1ex} \\ Intersecting D7-Branes}
\end{center}
\vspace{17cm}
\begin{align}
 \mbox{\bf Referees:}\hspace{6cm}  & \mbox{\bf Prof. Dr. Arthur Hebecker} \nn\\
& \mbox{\bf Prof. Dr. Michael G. Schmidt}\nn
\end{align}

\newpage
\thispagestyle{empty}
\mbox{}
\newpage

\thispagestyle{empty}

\section*{Abstract}

In this work, the moduli of D7-branes in type IIB orientifold compactifications and their stabilization by fluxes
is studied from the perspective of F-theory. In F-theory, the moduli of the D7-branes and the moduli of the orientifold
are unified in the moduli space of an elliptic Calabi-Yau manifold. This makes it possible to study the flux stabilization 
of D7-branes in an elegant manner. To answer phenomenological questions, one has to translate the deformations of the elliptic Calabi-Yau 
manifold of F-theory back to the positions and the shape of the D7-branes. We address this problem by constructing the 
homology cycles that are relevant for the deformations of the elliptic Calabi-Yau manifold. We show the viability of our 
approach for the case of elliptic two- and three-folds. Furthermore, we discuss consistency conditions related to the 
intersections between D7-branes and orientifold planes which are automatically fulfilled in F-theory. Finally, we use our 
results to study the flux stabilization of D7-branes on the orientifold $K3\times T^2/\Z_2$ using F-theory on $K3\times K3$.
In this context, we derive conditions on the fluxes to stabilize a given configuration of D7-branes. This thesis furthermore contains
an introduction to F-theory and a brief review of some mathematical background.

\tableofcontents

\chapter{Introduction}

Unification has always been among the central themes of theoretical physics. String theory is one of
the most ambitious projects in this development: it is aimed at nothing less than unifying all 
fundamental forces, including gravity, into a single theory. As all fundamental forces apart from gravity 
are quantum theories, this involves in particular giving a consistent
quantum theory of gravitation. String theory is arguably the best-developed scheme for a theory with which this
might be achieved.

In the last decade it was found that the five consistent superstring theories, all defined in ten 
space-time dimensions, are linked by an astonishing network of dualities. This lead to the conclusion 
that they are all limits of a single, more fundamental theory: M-theory (see e.g.\cite{Polchinski:1998rq}). In many cases, 
dualities open up completely new points of view and methods of computation. One of the most striking examples is given by the 
AdS/CFT correspondence \cite{Maldacena:1997re}. On the mathematical side, dualities have inspired many far-reaching conjectures, the 
most famous example being homological mirror symmetry \cite{kontsevich-1994}. 

A fundamental challenge, which becomes even more pressing in the age of the LHC, is to connect string theory to 
existing models of particle physics and cosmology. Even though we seem to be holding a unique theory in our hands, 
this theory exists only in ten space-time dimensions. To remedy this
problem one has to ``compactify'' string theory, so that only four dimensions, the ones we perceive, remain large. 
As superstring theory naturally incorporates supersymmetry (SUSY), one is hence tempted to aim for a supersymmetric
version of the standard model, or some extension thereof. This choice has also a technical side: compactifications
of string theory in which SUSY is broken at a low scale are under much better control than compactifications in which
SUSY is broken at a high scale. 

Compactifications of string theory involve the choice of a six-dimensional manifold. Even though the requirement of 
low-energy SUSY limits the choices that can be made to so-called Calabi-Yau manifolds, a huge freedom still remains.
This freedom is further enhanced by the inclusion of so-called D-branes. D-branes are higher-dimensional 
objects which lead to gauge theories in the low-energy theory, see \cite{Johnson:2000ch} for a review. As we are aiming for the gauge 
group of the standard model (or some GUT group), we are interested in studying compactifications with 
D-branes\footnote{This is not the case in heterotic string theory, which has an $E_8\times E_8$ 
or $SO(32)$ gauge theory already in ten dimensions.
We discuss the dualities that connect compactifications of heterotic strings with compactifications of string theory
with D-branes in various parts of this thesis.}. Intersecting brane models are a promising candidate for
constructing standard-like models \cite{Lust:2004ks}.

String theory contains gravity in ten dimensions, which means that the ten-dimensional space-time is dynamical. Hence the 
manifold we have compactified string theory on will be dynamical as well. Without further input, the deformations of 
this manifold have no potential and give rise to massless scalar fields in four dimensions, called moduli. As D-branes
are dynamical objects, they contribute further moduli, the study of which is one of the main topics of this work.
Moduli are a phenomenological disaster because they mediate interactions that are comparable to gravity in strength 
and thus spoil the equivalence principle. We will refer to this as the moduli problem of string compactifications.

Fortunately, string theories contain further ingredients which solve the moduli problem in a natural way. 
The crucial ingredient are so-called fluxes, background values of $p$-form fields along the compact directions. 
These fluxes have to obey quantization conditions similar to Dirac monopoles, and can only exist along non-trivial 
homology cycles of the compactification manifold. There has been a large amount of work devoted to the study of compactifications 
with fluxes in recent years, see \cite{Grana:2005jc, Blumenhagen:2006ci, Douglas:2006es, Denef:2007pq} for reviews. In particular, it has 
been realized that the potential which is generated by the energy density of the fluxes is capable of fixing geometric as well 
as D-brane moduli, hence solving the moduli problem. In this work, we use a framework called F-theory to study the flux 
stabilization of D7-branes\footnote{A D$p$-brane extends in $p$ spatial directions, so that it is a $p+1$-dimensional object 
in space-time.} in compactifications of type IIB string theory in detail.

From the perspective of the underlying physics, the choice of fluxes and the inclusion of D-branes 
is very similar to the choice of a compactification manifold. On the one hand, all three involve a choice for the 
background value of some degrees of freedom. If we allow ourselves to choose a background value for the space-time metric 
such that six dimensions are compact, we might as well allow background values for other degrees of freedom. In fact, it 
would seem unnatural not to. On the other hand, the actual choices we have to make to specify the compactification 
manifold, fluxes and D-branes are all discrete. Once a manifold with D-branes and fluxes is chosen, the flux potential, together
with a contribution from non-perturbative effects, will fix the geometric moduli and the positions and shapes of the D-branes. 

\section{The string theory landscape}

The topology of the compactification manifold, D-branes and the fluxes allow, in principle, to compute the complete 
low-energy spectrum including all coupling constants. The effective four-dimensional theories one
obtains are hence parameterized by the values of a set of integers. The number of possible effective
theories can be obscenely large, with rough estimates leading to numbers such as $10^{500}$ \cite{Denef:2004dm}.
All of these compactifications correspond to valid solutions of string theory.

Compactifications of string theory on different manifolds are not as distinct as it may seem at first.
In fact, there can be transitions between topologically different Calabi-Yau manifolds. Geometrically,
these transitions involve deforming a Calabi-Yau manifold until a singularity develops and subsequently
resolving this singularity such that one ends up with a topologically different manifold. It turns out that this
process is, however, perfectly well-behaved in string theory, see e.g. \cite{Strominger:1995cz, Aspinwall:1993nu}. 
Even though there is no general proof of the conjecture that all Calabi-Yau spaces are connected by
such singular transitions\footnote{This idea is goes back to the mathematician Miles
Reid and is sometimes called ``Reid's phantasy'' \cite{Reid}.}, all examples that are known so far are part of
a gigantic web \cite{Avram:1995pu, Kreuzer:2000xy}. Similar insights have been obtained regarding D-branes and 
fluxes \cite{Aspinwall:2005qw}. Hence different flux compactifications with D-branes should be properly viewed 
as different vacua in the ``landscape'' of all string compactifications.

As we have discussed, the uniqueness of string theory in ten dimensions is lost upon compactification. Instead, 
there is a (possibly finite) discretuum of string vacua in four dimensions. This result can be interpreted in a number 
of ways. One approach is to postpone worries about uniqueness until one has constructed a string compactification which 
is consistent with experiments. This is a difficult task and much of the work in string phenomenology has been following this path. 
Another conclusion one might draw is that we have not properly understood string theory yet. A logical possibility is that 
we are missing a fundamental principle which can tell us how to single out one specific compactification, which then, 
hopefully, reproduces the Standard Model of particle physics as its low-energy description. 

The point of view which is the basis for this work, is to accept the existence of the multitude of string vacua as a
fundamental feature of string theory, see \cite{Lerche:1986cx} for an early proposal of this idea.
Conceptually, a good analogy for the state of affairs is given by chemistry \cite{Douglas:2006za}. Even though there 
are only three building blocks, neutrons, protons and electrons, there are many stable (or metastable) vacua: atoms. A 
crucial difference is, of course, that we can observe many different atoms, including excited states, in nature, whereas 
other string vacua seem practically inaccessible to us.

The general idea to handle such a landscape of string vacua is hence to look for correlations between aspects of the emerging 
low-energy theories and compare the findings to established experimental facts. Investigations of this sort were pioneered in 
\cite{Ashok:2003gk,Douglas:2004kp,Denef:2004ze}, see also \cite{Douglas:2006es,Denef:2008wq} for reviews. In the
context of intersecting D-brane models on toroidal orientifolds, a statistical study has been performed in 
\cite{Blumenhagen:2004xx,Gmeiner:2005vz}. 

The picture of the landscape of string vacua we have drawn has an interesting interplay with cosmology. As the
different vacua can be connected by tunneling, or decay into each other, the landscape can become
populated statistically in the evolution of the early universe. This offers solutions to several cosmological problems, 
see e.g. \cite{Gasperini:2007zz} and \cite{Erdmenger}. One of the most prominent cosmological problems
that can be addressed in flux compactifications is the cosmological constant problem.
As has been proposed in \cite{Bousso:2000xa}, the cosmological constant, being equal to the energy density of the 
vacuum, can be small due to cancellations among the many possible fluxes. 

The vacuum energy for supersymmetric compactifications, however, is always non-positive. Hence a positive
cosmological constant can only be obtained after SUSY is broken. It has been shown in \cite{Kachru:2003aw} 
(see also \cite{Becker:2007zj} for a summary) that this can be achieved in a controlled way in the setting of 
flux compactifications of type IIB string theory, to be discussed below. The SUSY-breaking metastable minimum 
one obtains have a tunably small positive cosmological constant.

The complicated effective potential that arises in the string landscape is also an ideal playground for implementing models of inflation, 
see \cite{Baumann:2009ni} for a review. There is growing evidence that the number of meta-stable string vacua which are compatible 
with very general properties of the Standard Model (minimal mass for lightest Kaluza-Klein excitations, small positive cosmological 
constant, not too large extra-dimensions) is finite~\cite{Acharya:2006zw}. 

Flux compactifications on Calabi-Yau manifolds generically contain so-called throat geometries which describe strongly warped 
regions~\cite{Klebanov:2000hb}. The magnitude of the warping, which gives rise to large hierarchies of 
mass-scales from the 4-dimensional perspective, depends on the ratio of certain flux numbers~\cite{gkp01}. (This can be mapped 
fairly explicitly~\cite{Verlinde:1999fy,Chan:2000ms,Brummer:2005sh} to the famous 5-dimensional model of Randall and 
Sundrum~\cite{Randall:1999ee}, which explains the hierarchy between the electroweak and the Plank scale.) A statistical analysis 
shows that throats are generically expected in compact spaces with many cycles~\cite{Hebecker:2006bn}.

\section{Type IIB orientifold compactifications and F-theory}\label{ibm}

One of the best-understood branch of the string theory landscape is provided by Calabi-Yau orientifold compactifications of type 
IIB string theory with D3 and D7 branes. This is also the framework in which most of the ideas presented above were implemented. In this section, we 
only give some details related to this work, see \cite{Blumenhagen:2006ci, Douglas:2006es, Denef:2008wq} for a thorough introduction to type IIB 
orientifolds. 

At low energies, type IIB string theory is well described by the ten-dimensional type IIB supergravity \cite{Polchinski:1998rq}. 
The bosonic field content of type IIB supergravity splits into the Neuveu-Schwarz (NS-NS) and the Ramond-Ramond (R-R) sector.
In the Neuveu-Schwarz sector there is the metric $g_{MN}$, the dilaton $\phi$, a 2-form potential $B_2$. 
The 2-form $B_2$ gives rise to a 3-form field-strength $H_3=\d B_2$. The Ramond-Ramond sector contains the $p$-form 
potentials $C_p$, $i=0,2,4$. The exterior derivatives of the forms $C_p$ yield the field strengths $F_{p+1}=\d C_p$.
The five form $F_5$ furthemore has to fulfill the self-duality condition $F_5=\ast F_5$, where $\ast$ denotes the 
Hodge-$\ast$ operation. For convenience we introduce the fields
\begin{align}
\tau&=C_0+ \iu \e^{-\phi}\nn \\
G_3&=F_3-\tau H_3  \nn \\
\tilde{F}_5&=F_5-\frac{1}{2}C_2\wedge H_3+\frac{1}{2}B_2\wedge F_3 \ . 
\end{align}

The bosonic equations of motion of type IIB supergravity can be derived from the action
\begin{align}
S_{IIB}=&2\pi\int d^{10}x\sqrt{-g}R-\frac{1}{2(\operatorname{Im}\tau)^2}\d\tau\wedge\ast \d\overline{\tau} \nn \\
& +\frac{1}{\operatorname{Im}\tau}G_3\wedge\ast\overline{G}_3+\frac{1}{2}\tilde{F}_5\wedge\ast \tilde{F}_5+C_4\wedge H_3\wedge F_3 \ ,
\end{align}
supplemented by $\tilde{F}_5=\ast \tilde{F}_5$. As the self-duality  makes the kinetic term of $\tilde{F}_5$ vanish, it can only
be enforced at the level of the equations of motion. We have used units in which the string length, which is related to the string 
tension $\alpha'$ by $l_s=2\pi\sqrt{\alpha'}$, is set to unity.

Type IIB string theory contains D-branes of all even dimensions. Their effective action consists of two pieces, one 
of which, the so-called Dirac-Born-Infeld action, describes their dynamics and the gauge theory living on their worldvolume. 
The remaining piece, the Chern-Simons action, describes the coupling of a D-brane to the various Ramond-Ramond
fields of string theory \cite{Johnson:2000ch}:
\begin{equation}\label{diibcs}
 S_{D, cs}=2\pi\int_{{\cal D}} C\wedge \e^{-B}\wedge\ch(F)\wedge\sqrt{\frac{\hat{A}(T{\cal D})}{\hat{A}(N{\cal D})}} \ .
\end{equation}
In the above equation we have abbreviated $C=C_0+C_2+C_4+C_6+C_8$, where $\d C_8=\ast \d C_0$ and $\d C_6=\ast \d C_2$. 
Furthermore, $\ch(F)$ denotes the Chern class and $\hat{A}$ is the $A$-roof genus, see Appendix~\ref{sumchernclass}.
The 2-form of the gauge theory living on the worldvolume ${\cal D}$ of the D-brane is denoted by $F$. $T{\cal D}$
and $N{\cal D}$ denote the tangent and normal bundle of ${\cal D}$, respectively. The integrand in the expression above 
only makes sense when expanded in terms of the various differential forms that appear in it. The integral then picks
out those forms matching the dimension of the D-brane. Note that the Chern-Simons action of a D$p$-brane contains 
a term
\begin{equation}\label{diibcs2}
2\pi\int_{{\cal D}} C_{p+1} \ .
\end{equation}
Hence a D$p$-brane sources the field $F_{p+2}=\d C_{p+1}$.  In this sense, the Chern-Simons action is a generalization 
of the coupling of a point-particle to the electromagnetic potential.

Taking the quotient of type IIB string theory on a Calabi-Yau manifold by 
\be
\Z_{2, O}=\Z_{2, g}P(-1)^{F_L} \ .
\ee
produces a so-called orientifold compactification. Here $P$ denotes the parity transformation on the string world-sheet, $F_L$ is the 
left-moving fermion number on the world-sheet and $\Z_{2, g}$ is a holomorphic involution of the Calabi-Yau manifold which projects
out the holomorphic 3-form $\Omega^{3,0}$. The geometric quotient of the Calabi-Yau by $\Z_{2, g}$ is hence no longer Calabi-Yau. 
Compactifications of type IIB string theory on Calabi-Yau manifolds have $N=2$ SUSY in four dimensions. Orientifolding reduces
the supersymmetry further to $N=1$.  The action of $P(-1)^{F_L}$ on the fields of type IIB supergravity is 
\begin{align}
 C_0\rightarrow C_0 &\hspace{1cm}  C_4\rightarrow C_4 \nn\\
 C_2\rightarrow -C_2&\hspace{1cm}  B_2\rightarrow -B_2 \nn\\
g_{MN}\rightarrow g_{MN} & \ .
\end{align}

String theory compactifications must obey certain consistency conditions, which are known as the tadpole cancellation conditions. 
They originate from Gauss' law for the charges (which correspond to various differential forms) induced
by D-branes and fluxes, applied in a compact geometry. From the point of view of string theory, the tadpole cancellation conditions 
guarantee the absence of anomalies. To cancel the charge of D7-branes, compactifications of type IIB string theory must contain 
orientifold 7-planes (O7-planes). O7-planes are given by the fixed locus of the involution $\Z_{2, g}$. They are lie at the
fixed-point locus of a holomorphic involution, so that they are locally given by the vanishing of holomorphic function.

The coupling of orientifold planes to the fields of type IIB supergravity can also be described by an action
similar to \eqref{diibcs}. Denoting the worldvolume of an O$p$-plane by $O$, it is given by \cite{Johnson:2000ch}
\be\label{oiibcs}
S_O=-2\pi \cdot 2^{p-4}\int_O C\wedge\sqrt{\frac{L(\frac{1}{4}TO)}{L(\frac{1}{4}NO)}} \ .
\ee 
In the above equation, $L$ denotes the Hirzebruch $L$-genus, see Appendix~\ref{sumchernclass}.
The coupling of an O$p$-plane to $C_{p+1}$ is hence
\be\label{oiibcs2}
-2\pi \cdot 2^{p-4}\int_O C_{p+1} \ .
\ee
This is the same as the coupling of a D$p$-brane except of a factor of $2^{p-4}$ and a minus sign. This sign is
crucial for the cancellation of the charges between D-branes by O-planes. For O7-planes and D7-branes, the
relative factor is $2^3=8$. In the double cover, we hence need to have eight times as many D7-branes as
O7-planes to cancel their charges. After orientifolding, only half of the D7-branes are present. Denoting 
the homology class of all the O7-planes in the quotient by $\Gamma$, the homology class of all D7-branes must hence 
be $4\Gamma$ to cancel the D7-brane tadpole. Hence we cannot simply add D7-branes to a compact Calabi-Yau 
compactification as they can only be consistently included in type IIB orientifolds\footnote{It is possible to include 
D7-branes even in the absence of O7-planes in non-compact models}. Note that this means in particular that the total 
homology class of all D7-branes is fixed.

In order not to spoil Lorentz invariance in the effective four-dimensional theory, we need all D-branes to 
fill out all four non-compact dimensions. This means that D3-branes are points in the six compact directions, 
whereas the D7-branes extend in four compact directions. As we need the right O-planes to cancel the charges
of the D7-branes, a consistent model should contain O7-planes which also fill out the four non-compact directions
and have the real codimension two in the six compact dimensions. The involution $\Z_{2, g}$ should hence have a fixed
point locus which has codimension two in the compact geometry. As the involution is furthermore
required to be holomorphic, the O7-planes are given by divisors of the Calabi-Yau manifold.

The simplest way to cancel the D7-brane tadpole is to place four D7-branes on top of the O-plane. When one
moves the D7-branes off the O7-plane, supersymmetry requires them to be given by the vanishing
of holomorphic functions on $B$. Hence both the O7-planes and the D7-branes are described by divisors.

Let us briefly review some aspects of the moduli of such compactifications, see also \cite{Grimm:2004uq, Jockers:2004yj}. 
The moduli of Calabi-Yau threefolds split into K\"ahler and complex structure moduli, see Appendix~\ref{cyapp}.
These can be thought of as variations of the K\"ahler form $J$ and the holomorphic 3-form $\Omega^{3,0}$, 
respectively, see Appendix~\ref{cyapp}. In particular, complex structure moduli can be expressed in terms of
the periods
\be\label{periodsintro}
z_i=\int_{A_i}\Omega^{3,0} \ ,
\ee
where the $A_i$ are elements of the third homology group of the Calabi-Yau manifold.

As $\Omega^{3,0}$ is odd and $J$ is even under $Z_{2, g}$, we find that Calabi-Yau orientifolds have $h^{1,1}_+$ K\"ahler 
and $h^{2,1}_-$ complex structure deformations. Whereas each complex structure deformation corresponds 
to a complex degree of freedom, the bosonic field content of a chiral multiplet, there is only one real degree of freedom 
associated with each K\"ahler deformation. In the four-dimensional effective theory, the K\"ahler moduli pair up with the 
zero modes of the two form $C_4$ to form the bosonic components of chiral multiplets\footnote{Although these moduli can be
naturally included in linear multiplets, one usually dualizes them to chiral multiplets, see \cite{Grimm:2004uq}}. 

As D7-branes are submanifolds of $B$, one expects their deformations to be given by sections of their normal 
bundle \cite{Jockers:2004yj}. As we discuss in Section~\ref{Obstructions}, D7-branes are not completely 
generic hypersurfaces, but have obstructed deformations due to the presence of the O7-plane. They are forced to take a 
shape such that all intersections between the D7-branes and the O7-planes are double intersections. We discuss how this 
arises by using the weak coupling limit of F-theory in Section~\ref{d7locus}. As has been discussed in \cite{Collinucci:2008pf}, 
this can also be seen in perturbative type IIB string theory. 

Due to the curvature terms in \eqref{diibcs} and \eqref{oiibcs}, D7-branes and O7-planes contribute
to the D3-brane tadpole. This contribution can be cancelled by either introducing D3-branes or
3-form fluxes $G_3=F_3-\tau H_3$, which contribute to the D3-brane tadpole due to the coupling
$C_4\wedge F_3\wedge H_3$ in the type IIB supergravity action. 

\begin{figure}
\begin{center}
\includegraphics[height=8cm, angle=0]{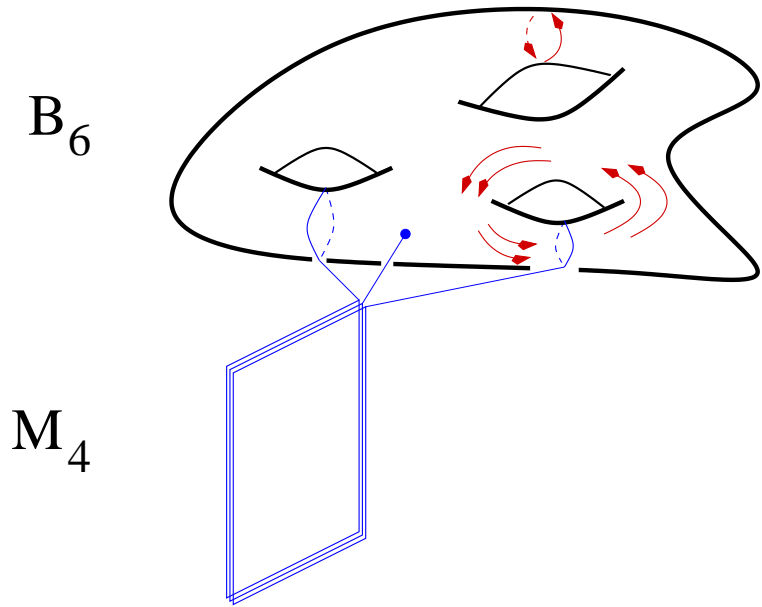}
\end{center}
\caption{\textsl{A cartoon depicting a type IIB orientifold compactification with D7- and D3-branes. The ten-dimensional spacetime 
factors into four non-compact directions $M_4$ and a six-dimensional compact space $B_6$. The D7-branes and O7-planes fill out the four 
non-compact direction and wrap a 4-cycle of the compact geometry. In order not to overload the picture we have not drawn 
intersections. The compactification also contains quantized fluxes that can thread various cycles. D3-branes, which may also be 
part of these compactifications, fill out the four non-compact directions and are points on the compact extra dimensions.}}
\label{fluxcomp}
\end{figure}

It has been shown in \cite{gkp01}, that one can switch on 3-form fluxes in type IIB orientifolds preserving
$N=1$ SUSY. The geometry of these models is, up to warping, still described by a Calabi-Yau orientifold. These fluxes 
are capable of fixing all the complex structure moduli. This is described by the Gukov-Vafa-Witten superpotential:
\begin{equation}\label{gvwpotintro}
 W=\int G_3\wedge\Omega^{3,0}=\int \left(F_3-\tau H_3\right)\wedge\Omega^{3,0} \ .
\end{equation}
This superpotential can be easily translated into an effective potential for the periods, \eqref{periodsintro}.
We have drawn a cartoon of a flux compactification in Figure~\ref{fluxcomp}

The remaining K\"ahler moduli can be fixed by 2-form fluxes on the D-branes and non-perturbative effects, such as 
gaugino condensates. As has been argued by \cite{Kachru:2003aw}, these models can have metastable de Sitter vacua in 
which SUSY is broken. Furthermore, one can show that the moduli of D-branes are fixed by fluxes as well, 
see e.g. \cite{Gomis:2005wc, Douglas:2006xy}. An explicit picture of this, however, is lacking except for simple
examples \cite{Dbranestorus1,Dbranestorus2,Dbranestorus3}.

D7-branes and O-planes couple to the form $C_8$ via \eqref{diibcs2} and \eqref{oiibcs2}. As $dC_0=\ast d C_8$, they
magnetically source $C_0$. This means that compactifications in which we displace D-branes 
from the O-planes necessarily have a non-constant axiodilaton. The equations of motion force $\tau=C_0+i\e^{-\phi}$ to
be a holomorphic function of $B$. This means that $\tau$ undergoes $\tau\rightarrow \tau +1$ upon 
encircling a D-brane and $\tau\rightarrow \tau -4$ upon encircling an O-brane. Because type IIB has a $SL(2,\Z)$ self-duality, 
there can be consistent solutions in which the D7-brane charge is not cancelled locally. This $SL(2,\Z)$ maps
\be \label{sl2zintro}
\tau\rightarrow\frac{a\tau+b}{c\tau+d} \ ,   \ a,b,c,d\in \Z \ ,
\ee
so that in particular $\tau$ and $\tau +n, n\in \Z $ are identified. The corresponding solutions
are the similar to the backgrounds one obtains for cosmic strings \cite{Greene:1989ya}, which also
are codimension-two objects (in four dimensions).

\begin{figure}
\begin{center}
\includegraphics[height=5cm, angle=0]{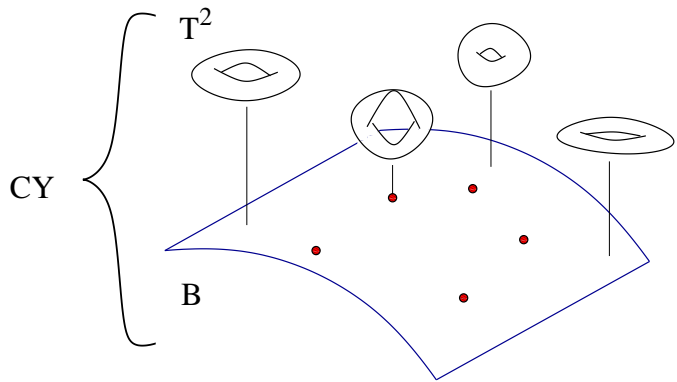}
\end{center}
\caption{\textsl{In F-theory, the value of $\tau$ over every point of $B$ is represented by a torus. Consistency requires
the whole space of this fibration to be a Calabi-Yau manifold. We have only drawn the two directions of $B$ which are orthogonal
to the D7-branes, which are thus shown as points. The fibre torus degenerates at the positions of the D7-branes.}}
\label{ellcartoon}
\end{figure}

Backgrounds with varying dilaton are most naturally described by F-theory.
The idea leading to F-theory is to interpret $\tau$ as the complex structure modulus of a torus \cite{Vafa:1996xn}.
This is a natural identification, because two tori that have complex structure moduli connected by
an $SL(2,\Z)$ transformation are diffeomorphic as complex manifolds \cite{riemannsurfaces, Chandrasekharan:1985}.
One way to see this is the construction of a torus as a quotient of $\C$. One identifies $z\sim z +n +m\tau, \ n,m\in \Z$, so 
that any two tori that come from equivalent lattices are also equivalent. Two such lattices are isomorphic if they are related 
by an $SL(2,\Z)$ transformation. Interpreting $\tau$ as the modular parameter of an extra torus geometrizes the $SL(2,\Z)$ 
duality group of type IIB string theory. As $\tau$ is a holomorphic function of the coordinates of the Calabi-Yau orientifold $B$, 
there is a torus over every point of $B$ which varies holomorphically. This is called an elliptic fibration. The way this torus 
is fibred over the base space $B$ encodes the positions of the D7-branes and O7-planes. In particular, the fibration is non-trivial 
due to the fact that the fibre torus degenerates over the positions of the branes, see Figure~\ref{ellcartoon} There is a map which 
relates degenerations of the fibre torus to the gauge symmetry of the corresponding branes. We give an introduction to elliptic 
fibrations in Section~\ref{ellfibsection}.

The fact that D7-branes are (real) codimension-two objects makes them backreact strongly on
the geometry, i.e. there is no probe approximation and their presence can be inferred also from great distance 
\cite{Greene:1989ya, Gibbons:1995vg}. This effect is automatically taken into account by requiring the total space of the 
fibration of the torus over $B$ to be a Calabi-Yau manifold $X$. 

The moduli space of this Calabi-Yau manifold $X$ encodes the moduli space of $B$ plus
the moduli of the fibration of the $T^2$ over $B$. As the details of the fibration are determined by the positions 
of the D7-branes and O7-planes in $B$, the moduli space of this Calabi-Yau manifold unifies the moduli space
of a Calabi-Yau orientifold with the moduli space of the D7-branes wrapped on its 4-cycles. 
Another advantage of this description is that constraints such as double intersections between O-planes and D-branes are 
naturally incorporated in this framework. Recombinations of D-branes, processes in which the intersection between two 
branes is smoothed out such that a single brane results, can also be easily described within this framework. In particular, 
we show that recombinations are related to shrinking cycles of the elliptic Calabi-Yau manifold of F-theory in 
Section~\ref{localConstruction}.

A large part of this thesis is devoted to the question of how complex structure moduli of elliptic Calabi-Yau manifolds 
encode the positions and the shape of D7-branes and O7-planes. As complex structure moduli of Calabi-Yau manifolds come 
from cycles of the middle homology (see Appendix~\ref{cyapp}), our strategy is to analyse how these cycles arise from
the topology of the D7-branes and O7-planes. Using the details of the elliptic fibration over $B$, we
show how this can be achieved for elliptic $K3$ surfaces and Calabi-Yau threefolds, i.e. for a complex
one- and two-dimensional base space $B$. Related work using different techniques recently appeared in 
\cite{openperiods}.

Even though the fibre torus may seem like an auxiliary concept, it becomes quite real in the dual 
M-theory description: type IIB orientifolds on $B$ are dual to M-theory on $X$ in the limit in which the size of the fibre 
torus goes to zero \cite{Witten:1995ex}. This is called the F-theory limit.

From our perspective, the real power of the F-theory description is 
that we can describe flux compactifications from the perspective of M-theory. As 11-dimensional supergravity contains a 3-form
potential, one can switch on 4-form flux $G_4$ in M-theory backgrounds. Under the duality to type IIB,
the $G_4$ flux is mapped to both $G_3$ flux and gauge flux on the branes. 
Compactifications of M-theory on elliptic fourfolds \cite{bb96} with 4-form flux $G_4$ share many of the good properties of 
type IIB flux compactifications: The geometry is, up to warping, still Calabi-Yau in the presence of fluxes \cite{Becker:2001pm} 
and the potential generated by the fluxes \cite{gvw99,hl01} is capable of fixing all the complex structure moduli. Similar 
to type IIB flux compactifications, moduli stabilization is best described by writing the flux potential in terms of
the periods. As the periods of F-theory compactifications on elliptic fourfolds determine the position
and shape of D7-branes and O7-planes, this approach allows us to study flux stabilization of 
D-branes \cite{drs99,Gorlich:2004qm}. An understanding of the map between periods of elliptic Calabi-Yau manifolds and the 
D-brane configurations they describe hence allows a very explicit study of D-brane stabilization by fluxes. This is our 
main motivation for working out how the cycles of elliptic Calabi-Yau manifolds are related to deformations of D7-branes and O7-planes. 
We give an example of this approach in Chapter~\ref{fluxonk3xk3}: we show how to use the map between cycles and
D7-brane positions to work out flux stabilization of D-branes on the orientifold $K3\times T^2/\Z_2$ which is described by 
$K3\times K3$ in F-theory.

F-theory can also be thought of as a non-perturbative description of type IIB backgrounds. One way to see this is to note that the
$SL(2,\Z)$ self-duality group which is used in F-theory contains strong-weak coupling transitions. Hence there can
be compactifications of F-theory which inevitably contain regions where the coupling is strong. Correspondingly, F-theory
vacua can contain branes and couplings which are absent in perturbative type IIB string theory.
In agreement with the interpretation of singularities of the elliptic fibration as gauge groups of the corresponding branes, 
those objects can support gauge theories with exceptional gauge groups on their worldvolume. In the context of GUT model-building, 
this has recently attracted a lot of attention \cite{Beasley:2008dc, Beasley:2008kw, Donagi:2008ca, Donagi:2008kj, Donagi:2009ra}. 
In perturbative type IIB compactifications with D-branes, one needs instanton effects to generate the right Yukawa couplings,
e.g. the $ \ {\bf 10\ 10 \ 5} \ $ coupling in $SU(5)$ GUTs \cite{Blumenhagen:2007zk}. In $SU(5)$ GUTs based on F-theory, 
these couplings can come from singularities that are of $E_6$ type. Other advantages include a natural mechanisms for the
breaking of the GUT gauge group down to the standard model and for the generation of the hierarchy between th GUT and
the Planck scale. Furthermore, one can engineer the absence of higherdimensional operators that lead to Proton decay
by choosing the geometry appropriately. For a study of SUSY-breaking in these models, see 
\cite{Heckman:2008qt, Heckman:2008es, Marsano:2008jq}. Until recently, most of the work on F-theory GUTs has focussed on the
study of desirable effects in local models. Studies of global compactifications include 
\cite{Marsano:2009ym, Marsano:2009gv, Blumenhagen:2009yv}. For the F-theory versions of perturbative type IIB GUT models
see \cite{Blumenhagen:2008zz,Blumenhagen:2009up}.

\section{Overview}

In Chapter~\ref{chapter1} we give an introduction to F-theory and elliptic fibrations. We
introduce the Weierstrass model description and discuss how various stacks of branes are encoded in the
elliptic fibration. We mainly focus on the weak coupling limit, in which F-theory describes perturbative IIB 
orientifold compactifications.

After the first chapter we start with the main part of this thesis, which contains our own work. 
In Chapter~\ref{chapterk3}, which is based on \cite{Braun:2008ua, Braun:2009wh}, we treat F-theory on elliptic $K3$ surfaces in detail. 
$K3$ is the only non-trivial compact two-dimensional Calabi-Yau manifold, so that it represents the simplest non-trivial F-theory 
compactification. We start this section with a review of the properties of elliptic $K3$ surfaces, which have
a two-sphere as their base space.
The enhancement of gauge symmetry which occurs for coincident D-branes can be beautifully mapped to singularities of $K3$. 
This is also evident from the duality to the heterotic string, which we discuss in detail. We then show how to construct the 
second homology group of $K3$ from the presence of D7-branes and O7-planes. In other words, we explicitly understand the motion 
of the 16 D7-branes the background geometry in terms of shrinking or growing cycles stretched
between the branes or the orientifold planes. We then discuss the relation of a $K3$ described by a 
Weierstrass model at the orientifold point to $T^4/\Z_2$. Even though these spaces are different, they describe the
same F-theory background. We discuss an intuitive way to deform these spaces into each other and show that they
approach the same space in the F-theory limit. We also provide an example, given by the Enriques involution, which shows that 
there can be symmetries which are only present in the F-theory limit in the Weierstrass model description.

In Chapter~\ref{chapter3}, we study the moduli D7-branes and O7-planes of F-theory compactifications on elliptic three-folds 
from the perspective of complex structure deformations. This chapter is based on \cite{c3fold}. Working in the weak coupling 
limit, we construct 3-cycles from the topology of the D-branes and O-planes.
We achieve complete agreement between the degrees of freedom encoded in the
Weierstrass model and the complex structure deformations of the elliptic
Calabi-Yau threefold. All relevant quantities can be expressed in terms of 
topological quantities of the base space, allowing us to formulate our results 
for general base spaces. Furthermore, we discuss in detail how the degrees of freedom of D7-branes are
restricted as compared to generic hypersurfaces. 

In Chapter~\ref{fluxonk3xk3}, we consider the stabilization of M-theory on $K3\times K3$ in the
presence of fluxes. Given a certain flux which is consistent with
the F-theory limit, we can explicitly derive the positions at which D7~branes
or stacks of D7~branes are stabilized. Our analysis is based on a parameterization
of the moduli space of type IIB string theory on $T^2/\mathbb{Z}_2$ (including
D7-brane positions) in terms of the periods of integral cycles of M-theory on
$K3$, obtained in Chapter~\ref{chapterk3}. This allows us, in particular, to select a specific 
desired gauge group by the choice of flux numbers. This chapter is based on \cite{Braun:2008pz}.

The appendix contains an introduction to some of the mathematics used throughout this work.

\chapter{Aspects of F-theory and elliptic fibrations}\label{chapter1}

In this chapter we review F-theory and explain the geometrical constructions that are needed.
Some of the underlying mathematics is presented in the appendices. Our exposition mostly focuses on topics and 
methods that are directly relevant to the subsequent chapters, see \cite{Denef:2008wq,Blumenhagen:2010at} for 
recent reviews of similar material.

\section[Type IIB and $SL(2,\Z)$]{Type IIB and \boldmath$SL(2,\Z)$}\label{sl2z}

The type IIB string theory has an $SL(2,\mathbb{Z})$ self-duality. In the 
ten-dimensional IIB supergravity action this shows up as an $SL(2,\mathbb{R})$
which acts on the complexified string coupling $\tau=C_{0}+i\e^{-\phi}=C_{0}+ig_s^{-1}$, also known as the axiodilaton, 
and the NS-NS and R-R 2-forms as 
\be
\tau\rightarrow\frac{a\tau+b}{c\tau+d}\ , \hspace{.5cm}
\left(\begin{array}{c}B_{2}\\C_{2}\end{array}\right)\rightarrow
\left(\begin{array}{cc} 
	d & c \\
	b & a 
	\end{array}\right)\
\left(\begin{array}{c}B_{2}\\C_{2}\end{array}\right)\ ,
\ee
and leaves $\tilde{F}_5$ invariant.
On the quantum level, quantization of the field-strengths associated to the 
2-forms degrades the symmetry to $SL(2,\mathbb{Z})$. Its nonperturbative nature
is apparent from the fact that it exchanges the field strength coupled to
the fundamental string with the field strength of a D1 brane, a non-perturbative
object. This goes along with an exchange of strong and weak coupling, 
$\phi\rightarrow -\phi$ and is generated by the $SL(2,\mathbb{Z})$ element (for $C_0=0$)
\be
S=\left(\begin{array}{cc} 
	0 & 1 \\
	-1 & 0 
	\end{array}\right)\ .
\ee

F-theory \cite{Vafa:1996xn} describes type IIB backgrounds which are patched together
using this $SL(2,\mathbb{Z})$ self-duality. This construction is
inevitable in the presence of D7-branes and orientifold 7-planes (O7-planes) as
they are magnetically charged under the real part of $\tau$. 
Upon encircling\footnote{As these objects have real codimension 2 
in a 10-dimensional spacetime, there is a well-defined notion of a winding number.} 
a D7-brane we have that $\tau\rightarrow\tau+1$. The corresponding element 
of $SL(2,\mathbb{Z})$ is commonly denoted by\footnote{The two 
elements $S,T$ actually generate the whole group $SL(2,\mathbb{Z})$. 
Given their physical interpretation this means that all type IIB vacua with varying coupling are built 
out of D-branes and strong $\leftrightarrow$ weak coupling transitions. This observation will be made 
more precise later on.}
 
\be
T=\left(\begin{array}{cc} 
	1 & 1 \\
	0 & 1 
	\end{array}\right)\ .
\ee

For O7-planes, the action on $\tau$ is given by $\tau\rightarrow\tau-4$.
The corresponding $SL(2,\mathbb{Z})$ element is however not simply $T^{-4}$, but $-T^{-4}$.
Altough the extra minus sign acts trivially on $\tau$, its presence can be inferred
from the orientifold action on the 2-form fields.

\begin{figure}
\begin{center}
\includegraphics[height=6cm]{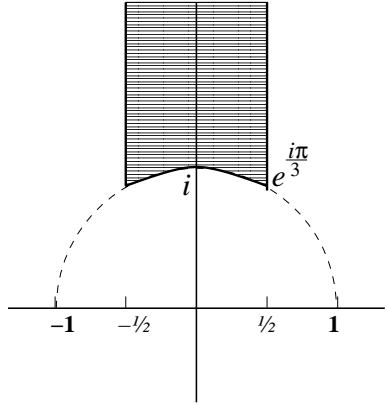}
\caption{\textsl{The fundamental domain of $SL(2,\Z)$ can be displayed as the shaded region drawn above.}}
\label{fundreg}
\end{center}
\end{figure}

Supersymmetry requires $\tau$ to be a holomorphic function of the base. 
From this it follows that the torus degenerates such that $\tau\rightarrow \iu\infty$
at the positions of the D-branes. This is best described by using the modular function 
$j(\tau)$ instead of $\tau$ itself. The function $j$ is a holomorphic bijection from the fundamental 
domain of $SL(2,\mathbb{Z})$ (see Figure~(\ref{fundreg})) onto the Riemann sphere and is invariant under $SL(2,\mathbb{Z})$ 
transformations of $\tau$ (details can be found in~\cite{Chandrasekharan:1985}). 

The fundamental domain contains three singular points which are fixed points under 
a subgroup of $SL(2,\mathbb{Z})$. These points and the generators of this subgroup are\footnote{We have chosen
a normalization of Klein's $j$-function that agrees with the convention used in the physics literature \cite{Sen:1997gv}.}
\begin{table}
\begin{center}
\begin{tabular}[h]{c|c|c}
$\tau$ & $j(\tau)$ & invariant under \\  \hline
$\tau = \e^{\,2\pi \iu/3}$& 0 &$ST$\\
$\tau = \iu$ & $24^3$ &$S$ \\ 
$\tau \to \iu \infty$&  $\sim\e^{-2 \pi \iu \tau}$  &$T$.
\end{tabular} 
\end{center}
\label{singfundregionmono}
\caption{\textsl{The three singular points of fundamental domain of $SL(2,\Z)$ and the corresponding monodromies.}}
\end{table}
Whenever $j(\tau)$ encircles one of these three points, $\tau$ undergoes the corresponding
transformation.

A very elegant way to describe how type IIB backgrounds with varying string coupling
are patched together is to interpret $\tau$ as the complex structure modulus of an 
extra torus. This torus does not describe further dimensions of space-time, but rather gives a geometrization of backgrounds
with varying axiodilaton $\tau$. This is the basic idea of F-theory. In this picture, the $SL(2,\mathbb{Z})$ is the 
monodromy group acting on the 1-cycles, and thus on the complex structure modulus, of the fibre torus. 
The tori can be patched together because two tori which have a complex structure modulus related by an
$SL(2,\mathbb{Z})$ transformation can be mapped holomorphically to one another. As the complex
structure modulus of the extra torus is used to describe the axiodilaton, its volume is of
no physical importance.

From this point of view, F-theory is a description of type IIB backgrounds with D7-branes which is
obtained by holomorphically fibering a torus over spacetime. The monodromies acting on this torus encode 
the positions (and type) of the D-branes. The spaces one obtains in this way are called elliptically fibred 
manifolds, and we will review their properties and construction in the next section. We will finally be interested in 
elliptically fibred Calabi-Yau manifolds \cite{Sen:1997gv}.

\section{Elliptic fibrations}\label{ellfibsection}

An elliptic fibration is a space $X$ together with a holomorphic map $\pi$ (the projection) 
to a space $B$ (the base), such that the generic fibre, $\pi^{-1}(b),\hspace{.2cm}b\in B$, 
is a non-singular elliptic curve, also known as a torus.
\begin{center}
\includegraphics[height=2cm]{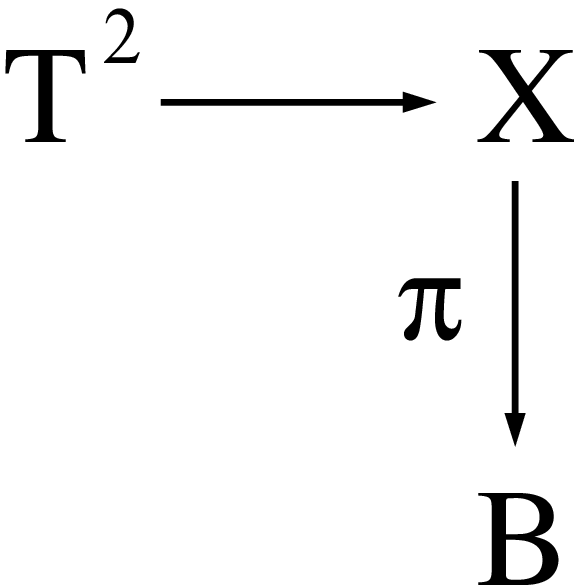}
\end{center}
Elliptic fibrations may have a section $\sigma$, which is a map
from the base into the total space such that $\sigma(B)$ intersects each fibre once.
\begin{center}
\includegraphics[height=2cm]{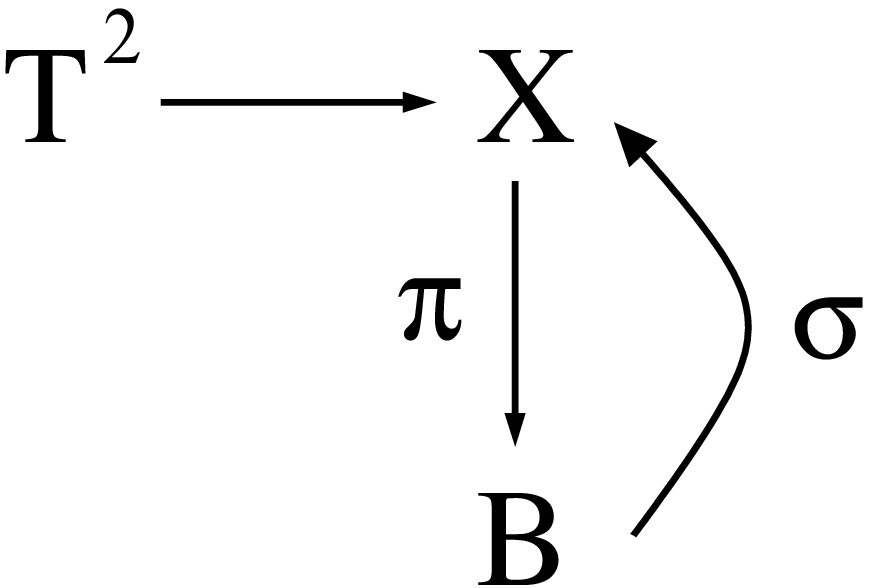}
\end{center}
We can also have n-sections, in which case the section intersects each fibre n-times.

Let us introduce the so-called Weierstrass model description:
\be
y^2=x^3+f(b)x+g(b).\label{urweier}
\ee
There is a very intuitive way to understand this equation. For every point of the base
it describes a double cover of the Riemann sphere, branched over four points:
the three roots of the right hand side plus the point at infinity. Hence it describes a torus, 
which varies its shape when we change the point $b$ on the base.
\begin{center}
\includegraphics[height=3cm]{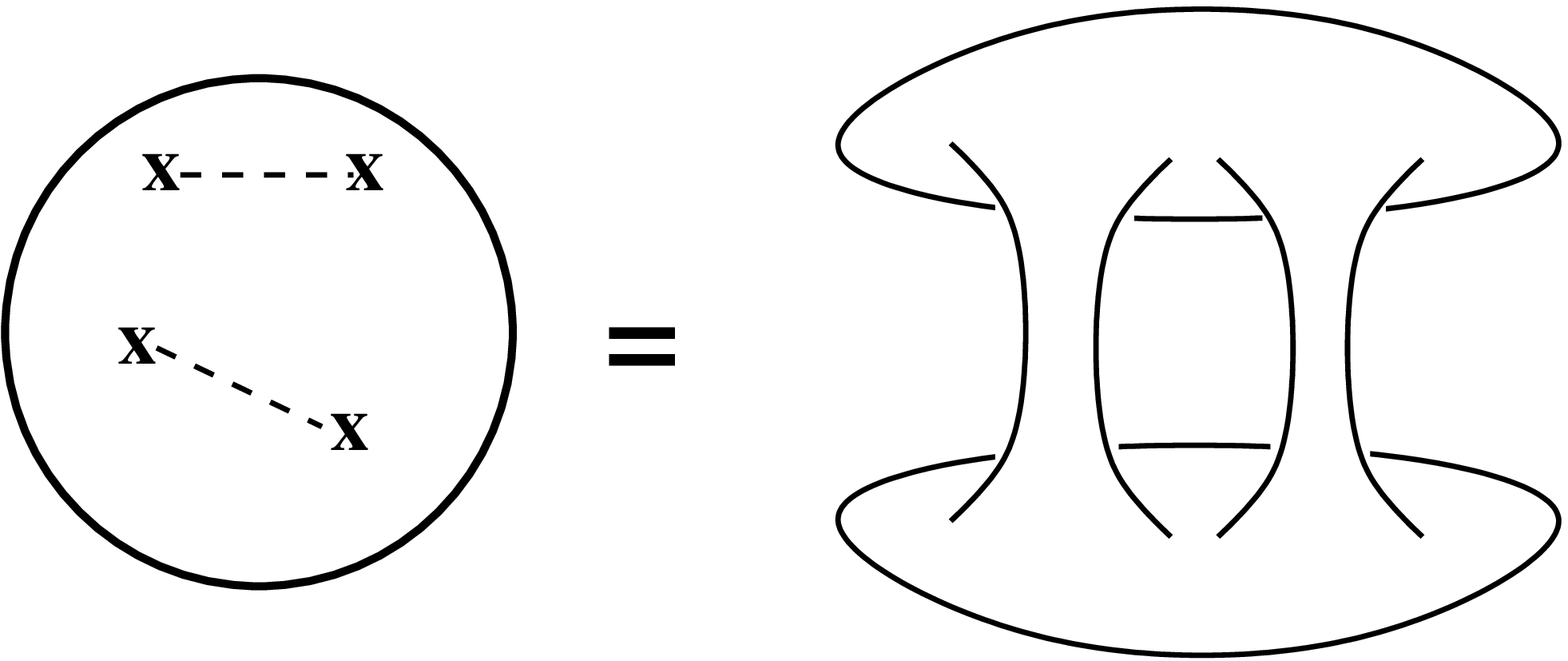}
\end{center}

For our purposes, it will be of advantage to slightly rewrite the Weierstrass equation so that it describes
the elliptic fibre as hypersurface in weighted projective space.
\be
y^2=x^3+f(b)xz^4+g(b)z^6.\label{weier}
\ee

We take $(y,x,z)$ to be the homogeneous coordinates of a $\P_{1,2,3}$ bundle\footnote{One can also use other
spaces for embedding the fibre and in this way finds elliptic fibrations which allow for more than one section
\cite{Klemm:1996ts}. These fibrations do not have the full $SL(2,\Z)$ monodromy group \cite{Berglund:1998va},
allowing a non-zero $B$-field invariant under S-duality to be switched on. This can be exploited to construct 
F-theory duals to the CHL string \cite{Bershadsky:1998vn}.} over $B$.
The $\P_{1,2,3}$ bundle over the base is such that $y$ is a section of $L^{\oplus 3}$ and $x$ is a section of $L^{\oplus 2}$,
for an appropriate line bundle $L$. It follows that $f$ and $g$ are sections of $L^{\oplus 4}$ and $L^{\oplus 6}$,
respectively.  If we take e.g. the projective space $\P^n$ as the base, all this boils down to letting $y$ and $x$ transform
under the scaling of $\P^n$ and making $f$ and $g$ homogeneous polynomials such that \eqref{weier} is a homogeneous
equation with respect to a $\P_{1,2,3}$ bundle over $\P^n$. A similar construction applies if $B$ is any toric variety.
An elliptic fibration described by \eqref{weier} hence is given by a hypersurface in a $\P_{1,2,3}$ bundle over the base $B$
if $B$ is a toric variety. Thus one can use all the standard techniques to analyse it, see e.g. Appendix~\ref{appendix}.
This can be generalized to situations in which the base itself is a hypersurface or complete intersection, see
\cite{Collinucci:2008zs} for the details.

Let us fix some point in the base. In this case \eqref{weier} is a homogeneous equation of
degree 6 in $\P_{1,2,3}$. The singularities of $\P_{1,2,3}$ are located at $x=z=0$ and $y=z=0$. 
As $(y,x,z)=(0,0,0)$ is not part of $\P_{1,2,3}$, the space $E$ described to \eqref{weier} always misses 
these singularities. Bertini's theorem \cite{Hubsch:1992nu}, see also Appendix~\ref{appendix}, 
then ensures that $E$ is generically a smooth space if $f$ and $g$ are sufficiently generic.

The first chern class of $E$ is readily computed. Denoting the hyperplane divisor of $\P_{1,2,3}$, which is
given by $z=0$ by $D_z$ we find that
\be
c(E)=\frac{(1+3D_z)(1+2D_z)(1+D_z)}{1+6D_z}\ .
\ee
Hence
\be
c_1(E)=3D_z+2D_z+D_z-6D_z=0 \ .
\ee
Thus for fixed point in the base, \eqref{weier} describes a one-dimensional Calabi-Yau space,
also known as a torus.
The projection is now given by simply forgetting about the coordinates $y,x,z$. The section of this
fibration is found by mapping any $b\in B$ to the point $(y,x,z,b)=(1,1,0,b)$. Using the scaling
of $\P_{1,2,3}$ one can check that the equation $y^2=x^3$ has actually only one solution. This single
point is the intersection between the section and every fibre. The obvious trick here is that $z=0$ 
eliminates any dependence on the base.

In application of elliptic fibrations to F-theory one is usually interested in a fibration
such that the surface described by \eqref{weier} is Calabi-Yau. This requirement can be
seen from a number of perspectives that will be explained when we come to the physics of
F-theory on elliptic fibrations. Let us denote the divisor associated to the bundle $L$ by
$D_L=[L]$. Using the second adjunction formula, \eqref{adj2gen}, the Chern class of $X$ is 
\be
c(X)=\frac{c(B)(1+3D_z+3D_L)(1+2D_z+2D_L)(1+D_z)}{1+6D_z+6D_L},
\ee
so that the condition for the first Chern class of $X$ to vanish is
\be
c_1(B)-D_L=0.\label{CycondX}
\ee
Thus $L$ has to be equal to the anticanonical bundle $L=[-K_B]$ for
\eqref{weier} to describe a Calabi-Yau manifold.

Choosing $B=\P^1$ is the easiest example. As $c_1(\P^1)=2H$, where $H$ is the 
hyperplane divisor, the anticanonical bundle must vanish on two points of $\P^1$. Thus its
sections are given by homogeneous polynomials of degree two. To get a Calabi-Yau space,
we thus need to choose $f$ and $g$ to be homogeneous polynomials on $\P^1$ of degree
$8$ and $12$, respectively:
\begin{equation}\label{k3}
y^2=x^3+f_{8}(a_i)xz^4+g_{12}(a_i)z^6 \ .
\end{equation}
The weights of the projective coordinates are given by \\
\begin{center}
\begin{tabular}[h]{cccccc}
$y$ & $x$& $z$ & $a_1$ &  $a_2$& \\  
3 & 2 & 1& 0 &  0& \\  
6 & 4& 0 & 1 &  1 & \ .
\end{tabular} 
\end{center}
As there is only one non-trivial Calabi-Yau manifold in two complex dimensions, \eqref{k3} describes
an elliptic $K3$ surface, an object that will reappear frequently throughout this work.

As a second example, consider $B=\P^1\times\P^2$. Now we find that $c_1(B)=2H_a+3H_b$, where
$H_a$ is the hyperplane divisor of $\P^1$ and $H_b$ is the hyperplane divisor of $\P^2$. The sections
of $[-K_B]$ are thus given by homogeneous polynomials that have the bidegree $(2,3)$. We can write
our Weierstrass model,
\be\label{p1p2}
y^2=x^3+f_{8,12}(a_i,b_j)xz^4+g_{12,18}(a_i,b_j)z^6,
\ee
where the weights of the projective coordinates are given by\\
\begin{center}
\begin{tabular}[h]{ccccccccc}
$y$ & $x$& $z$ & $a_1$ &  $a_2$ & $b_1$  & $b_3$  & $b_3$ \\  
3 & 2& 1 & 0 &  0 & 0  & 0  & 0 &\\  
6 & 4& 0 & 1 &  1 & 0  & 0  & 0 &\\  
9 & 6& 0 & 0 &  0 & 1  & 1  & 1 & \ ,
\end{tabular} 
\end{center}
and the degrees of homogeneity of $f$ and $g$ are as indicated. Using the methods explained in the
appendix, one can check that the two examples are indeed Calabi-Yau. For every equivalence relation, the sum
of the weights of all projective coordinates must equal the degree of the equation 
defining the hypersurface.

In our general discussion we have assumed that $L$ actually has holomorphic sections and \eqref{weier} 
describes a smooth space. This can seen to be true without much work for the two examples. 
It is intuitively clear that the construction presented demands the bundle $L$
to be ``positive'' in some sense. A more precise rephrasing of this criterion is the concept of 
ampleness, as described in the appendix. Spaces with an ample anticanonical bundle are called Fano 
varieties and have been classified in dimensions up to three \cite{fano}. The toric
fano threefolds have recently been discussed in the physics literature in \cite{Hanany:2009vx}.

\subsection{Topological invariants}

Given a base $B$, there is only one choice for the bundle $L$ such that \eqref{weier} yields an elliptic Calabi-Yau manifold:
$c_1(B)-D_L$. This allows to derive expressions for its characteristic classes in terms of the 
characteristic classes of the base, which are often called Sethi-Vafa-Witten formulae \cite{Sethi:1996es}.

Elliptic Calabi-Yau twofolds are always elliptic $K3$ surfaces. Hence their Euler characteristic is always given by $\chi=24$.

For the case of elliptic threefolds, we use the trick of \cite{Sethi:1996es}. As the whole structure
of the elliptic threefold is fixed once the Chern classes of the base are known, we expect that we can
write the Euler characteristic of an elliptic threefold $X$ over $B$ as an integral of the Chern classes
of $B$. Furthermore, the elliptic fibration is trivial if and only if $c_1(B)=0$, so that the expression
for the Euler characteristic of $X$ must be proportional to $c_1(B)$. Hence we make the Ansatz
\begin{equation}
 \chi(X)=\int_B \alpha c_1(B)\wedge c_1(B) \ .
\end{equation}
To determine the constant $\alpha$ we consider an example, $B=\P^2$. As $c_1(\P^2)=3H$, where $H$ denotes
the hyperplane divisor, we find that $\int_B \alpha c_1(B)\wedge c_1(B)=9\alpha$. Furthermore, we
can compute the Euler characteristic of $X$ by considering a Calabi-Yau hypersurface in the toric variety
given by the weights
\begin{center}
\begin{tabular}[h]{ccccccc}
$y$ & $x$& $z$ & $a_1$ &  $a_2$ & $a_3$ & \\  
3 & 2& 1 & 0 &  0 & 0  &\\  
9 & 6& 0 & 1 &  1 & 1  & \ ,
\end{tabular} 
\end{center}
where $a_i$ are the homogeneous coordinates on $\P^2$. The Euler characteristic
of this hypersurface Calabi-Yau can be computed by integrating the top Chern class
of $X$, see Appendix~\ref{cyapp} for similar examples. In the present case, the result is
$\chi(X)=-540$. Hence we find $\alpha=-60$, so that we have found
\be
 \chi(X)=-60 \int_B c_1(B)^2 \ .
\ee

We can use the same strategy to compute the Euler characteristic of an elliptic fourfold
by intergrating the Chern classes of the base. The result is \cite{Sethi:1996es}
\be
\chi(X)=\int_B 12c_1(B)c_2(B)+360c_1(B)^3 \ .
\ee

\section{Monodromy and singularities}\label{singfibmono}

The $j$-function of the torus fibres is given in terms of $f$ and $g$ as
\be
j(\tau(b))=\frac{4 (24f)^3}{4f^3+27g^2}.\label{tauofb}
\ee
As $\tau\rightarrow \iu\infty$ for $j\rightarrow\infty$ it follows that the fibre degenerates
over those points of the base, where the discriminant $\Delta=4f^3+27g^2$ vanishes. These are
the locations of the 7-branes of F-theory.

We can also understand this in a second way by directly checking when the torus becomes singular. Writing, 
for a fixed point in the base, 
\be
y^2=(x-\rho_1)(x-\rho_2)(x-\rho_3),\label{weierrho}
\ee
one can check that 
\be
\Delta \sim (\rho_1-\rho_2)^2(\rho_2-\rho_3)^2(\rho_3-\rho_1)^2,\label{deltarho}
\ee
so that the fibre torus degenerates if and only if $\Delta=0$. The difference between the
roots on the right hand side are nothing but the positions of the branch points, if one thinks of the torus as a
double cover over $\P^1$. As the distances between the branch points give
the sizes of the 1-cycles of the torus, the singularities that arise when the branch points coincide
develop because a 1-cycle $\gamma$ of the torus is pinched. In the mathematics literature, $\gamma$ is 
referred to as a \emph{vanishing cycle}.

\begin{figure}\label{torus}
 \begin{center}
  \includegraphics[height=3cm]{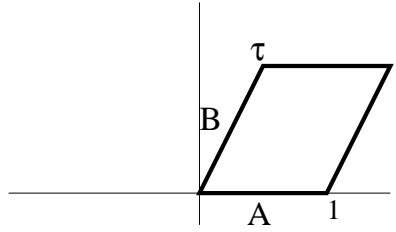}\caption{\textsl{The two 1-cycles of a torus are labelled by $A$ and $B$.}}
 \end{center}
\end{figure}

The relation between the vanishing cycle $\gamma$ and the $SL(2,\Z)$ action on 
a 1-cycles $\eta$ of the torus upon encircling the $\Delta=0$ locus is given by the 
Picard-Lefschetz monodromy formula \cite{peters, Dimca},
\be\label{piclef}
\eta\mapsto\eta-(\eta\cdot\gamma)\gamma.
\ee
Here $\eta\cdot\gamma$ denotes the intersection form. Let us choose a basis for the 1-cycles 
of a torus and call the horizontal 1-cycle $A$ and the vertical 1-cycle $B$, see Figure~\ref{torus}.
We choose the orientation such that $A\cdot B=1$. If the cycle $pA+qB$ vanishes, an arbitrary cycle $aA+bB$ 
undergoes the monodromy
\begin{equation}\label{monpq}
\left(\begin{array}{c}a\\b\end{array}\right)\mapsto
\left(\begin{array}{c}a-(aq-bp)p\\b-(aq-bp)q\end{array}\right)
=
\left(\begin{array}{cc} 
	 1-pq  & p^2 \\
	 -q^2 & 1+pq
	\end{array}\right)\
\left(\begin{array}{c}a\\b\end{array}\right) 
\end{equation}
when it is transported around the singularity. As expected, the transformation is
given by an $SL(2,\Z)$ matrix. A consequence of the fact that monodromies are linked to
vanishing cycles is that monodromies are only defined locally. 
Let $(p,q)$ be the cycle that vanishes at the point $\zeta$ and $M_{\zeta}$ the corresponding
monodromy. If we now encircle another point $\xi$, at which another cycle $(p',q')$ vanishes, the 
cycle $(p,q)$ is transformed to
\be
\left(\begin{array}{c}p\\q\end{array}\right)\mapsto\left(\begin{array}{c}p\\q\end{array}\right)-
(pq'-qp')\left(\begin{array}{c}p'\\q'\end{array}\right).\label{pp'trafo}
\ee
%%%%%% check that the conjugation is the right one
Correspondingly, the monodromy matrix $M_{\zeta}$ is changed to $M_{\xi}M_{\zeta}M_{\xi}^{-1}$, where
$M_{\xi}$ is the monodromy matrix associated with the vanishing of the cycle $(p',q')$. Thus  
monodromies are only defined up to conjugation globally. If two vanishing cycles fulfill the 
relation 
\be\label{mutloc}
pq'-qp'=0, 
\ee
they are said to be mutually local. If this condition is met, the map 
\eqref{pp'trafo} acts as the identity, so that the two monodromies are, in a sense, independent.

\begin{figure}
\begin{center}
\includegraphics[height=5cm]{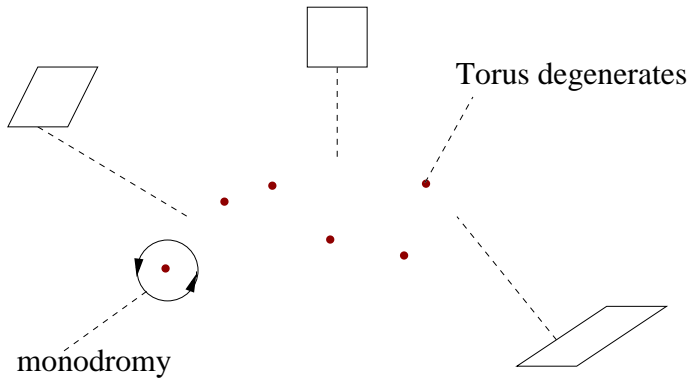}\caption{\textsl{By slicing orthogonally to the $\Delta=0$ locus
through the base, the elliptic fibration looks like a plane that has different tori sitting over different
points. The torus degenerates at the points where $\Delta=0$. Correspondingly, there is an $SL(2,\Z)$ monodromy acting
on it upon encircling these loci.}}\label{ellfibpic}
\end{center}
\end{figure}

Note that the functional dependence of the $j$-function on the base, \eqref{tauofb}, is such that all
monodromies locally are some power of $T$ or $-\id$ \cite{Bergshoeff:2006jj}. By Table~\ref{singfundregionmono}, monodromies 
that are different from $T$ can occur whenever $j(\tau)$ is $0$ or $24^3$. As $j(\tau)\sim f^3$ near $j(\tau)=0$, the monodromy is not
$ST$ but $ST^3=\id$. Similarly, $j(\tau)-24^3$ in the vicinity of $j(\tau)=24^3$ is given by
\begin{equation}
 \frac{4(24f)^3}{4f^3+27g^2}-24^3=\frac{4(24f)^3-24^3(4f^3+27g^2)}{4f^3+27g^2}=\frac{-24^3 27g^2}{4f^3+27g^2} \ ,
\end{equation}
so that the monodromy around $j(\tau)=24^3$ is $S^2=-\id$.

\subsection{Elliptic Surfaces}\label{ellsurf}
Let us first discuss the case of elliptic surfaces as it contains
(almost) all the salient ingredients. For F-theory compactifications we are 
thus talking about elliptic $K3$ surfaces, as described by \eqref{k3}. 
The following analysis will not depend on this, however, as only local properties are 
relevant.

The types of singular fibres that can occur for elliptic surfaces have been classified 
by Kodaira \cite{Kodaira2,peters}. The type of singular fibre only depends on the vanishing 
degree of $f$, $g$ and $\Delta$. The fibre singularities can be directly translated to 
singularities of the complex surface, as is indicated in Table~\ref{kodairaclass}. 

\begin{table}
\begin{center}
\begin{tabular}[h]{|c|c|c|c|c|c|c|}
\hline
 $\ord(f)$ & $\ord(g)$ & $\ord(\Delta)$ & fibre type & singularity type & monodromy \\ \hline \hline
 $\geq 0$ & $\geq 0$ & $0$ & smooth & none &  $\left(\begin{array}{cc} 1 & 0 \\ 0 & 1 \end{array}\right)$ \\ \hline
$0$&$0$&$n$&$I_n$&$A_{n-1}$ & $\left(\begin{array}{cc} 1 & n \\ 0 & 1 \end{array}\right)$   \\ \hline
$2$ & $\geq 3$ & $n+6$ &  $I_n ^*$ & $D_{n+4}$ & $-\left(\begin{array}{cc} 1 & n \\ 0 & 1 \end{array}\right)$ \\ \hline
$\geq 2$ & $3$ & $n+6$ &  $I_n ^*$ & $D_{n+4}$ & $-\left(\begin{array}{cc} 1 & n \\ 0 & 1 \end{array}\right)$\\ \hline
$\geq 1$ & $1$ & $2$& $II$ & none &$\left(\begin{array}{cc} 1 & 1 \\ -1 & 0 \end{array}\right)$\\ \hline
$\geq 4$ & $5$ & $10$ & $II^*$  & $E_8$ & $\left(\begin{array}{cc} 0 & -1 \\ 1 & 1 \end{array}\right)$\\ \hline
$1$ & $\geq 2$ & $3$ &  $III$ & $A_1$ &$\left(\begin{array}{cc} 0 & 1 \\ -1 & 0 \end{array}\right)$\\ \hline
$3$ & $\geq 5$ & $9$ &  $III^*$  & $E_7$ &$\left(\begin{array}{cc} 0 & -1 \\ 1 & 0 \end{array}\right)$\\ \hline
$\geq 2$ & $2$ & $4$ &   $IV$  & $A_2$ &$\left(\begin{array}{cc} 0 & 1 \\ -1 & -1 \end{array}\right)$\\ \hline
$\geq 3$ & $4$ & $8$ &  $IV^*$ & $E_6$&$\left(\begin{array}{cc} -1 & -1 \\ 1 & 0 \end{array}\right)$\\ 
\hline

\end{tabular}
\end{center}
\caption{\label{kodairaclass}\textsl{The classification of bad fibres in terms of the vanishing 
degree of $f$, $g$ and $\Delta$. Also given is the corresponding monodromy and the type of surface singularity.}}
\end{table}

Note that the monodromies are local, i.e. only determined up to $SL(2,\Z)$-equivalence.

As an example we discuss the case of an $I_2$ fibre in some detail.
The according monodromy is $\tau\mapsto\tau+2$, so this configuration describes two coincident 
D7-branes (see Section~\ref{sl2z}).
At a point where the fibre degenerates, two of the $\rho_i$  coincide so that, 
adjusting the normalization for convenience, the Weierstrass model locally reads
\be
y^{2}=(x-x_{0})^{2}\,.
\ee
To see what happens to the whole space we have to keep the dependence on the base coordinates. 
Let us deform away from the degenerate point by shifting 
$x_{0}\rightarrow x_{\pm}=x_{0}\pm\delta$. This means that now
\be
y^{2}=(x-x_{+})(x-x_{-})=(x-x_{0})^{2}+\delta^{2} \ .\label{y2ofdelta}
\ee
The quadratic difference of the now indegenerate roots is given by
\begin{equation}
(x_{+}-x_{-})^{2}=4\delta^{2} \ .
\end{equation}
Comparing this with~\eqref{deltarho} and ignoring the slowly 
varying factor associated with the distant third root, we have 
\begin{equation}
\delta^2\sim\Delta\,.
\end{equation}
Since we also want to see what happens to the full space, let us reintroduce 
the dependence on the base coordinates, $\Delta=\Delta(a,b)$, and write~\eqref{y2ofdelta} as
\be
y^{2}=(x-x_{0})^{2}+\Delta(a,b) \ .
\ee

Without loss of generality we assume $a\neq 0$ and use $a$ as an 
affine coordinate. Near the $I_2$ fibre we can write $\Delta=(a-a_{0})^{2}$, so that
the Weierstrass model reads
\be
y^{2}=(x-x_{0})^{2}+(a-a_{0})^{2}\ .
\label{singularity structure}
\ee
By computing the gradient, one quickly sees that the above equation describes a singular hypersurface. This analysis 
can easily be modified to see that simple roots of $\Delta$ do not lead to any singularity of the whole space, whereas 
higher roots lead to singularities 
in accordance with Kodaira's classification.

We can resolve this singularity in two ways, both of which will make a two-sphere emerge.

\subsubsection{Deformation}
Let us first do the obvious and deform it to
\be
y^2=x^2+a^2+\epsilon, \label{defa1}
\ee
where $\epsilon$ is some complex parameter.

As the situation is somewhat analogous to the conifold case 
\cite{Candelas:1989ug}, we will essentially repeat the analysis that is done there:
We first note that $\epsilon$ can always be chosen to be real by redefining 
the coordinates. Next, we collect $x,y,a$ in a complex vector with real part 
$\xi$ and imaginary part $\eta$. The hypersurface~\eqref{defa1} 
may then be described by the two real equations 
\begin{equation}
\xi^2-\eta^2=\epsilon, \hspace{.5cm}
\xi\cdot\eta=0 \ .
\end{equation}
We can understand the topology of X by considering its intersection with 
a set of 5-spheres in $\mathbb{R}^6$ given by $\xi^2+\eta^2=t$, 
$t>\epsilon$ :
\begin{equation}
\xi^2=\frac{t}{2}+\frac{\epsilon}{2}, \hspace{.5cm}
\eta^2=\frac{t}{2}-\frac{\epsilon}{2}, \hspace{.5cm}
\xi\cdot\eta=0 \ . \label{s2s1}
\end{equation}
If we assume for a moment that $\epsilon=0$, the
equations above describe two $S^2$s of equal size for
every $t$ that are subject to an extra constraint. 
If we take the first $S^2$ to be unconstrained, 
the second and third equation describe the intersection of 
another $S^2$ with a hyperplane. Thus we have an $S^1$ 
bundle over $S^2$ for every finite $t$. This bundle shrinks
to zero size when $t$ approaches zero so that we reach the 
tip of the cone. Furthermore, the bundle is clearly non-trivial since 
the hyperplane intersecting the second $S^2$ rotates as one moves along the 
first $S^2$. Let us now allow for a non-zero $\epsilon$, so that $X$ is no 
longer singular. The fibre $S^1$ still shrinks to zero size at 
$t=\epsilon$, but the base $S^2$ remains at a finite
size. This is the 2-cycle that emerged when resolving the singularity.

\subsubsection{Blow-up}

The blow up of the singularity 
\be
y^2=x^2+a^2.
\ee
is performed in detail in Appendix~\ref{blowup}. It is also done from the perspective of toric geometry 
in Appendix~\ref{blowupt}. The result is the same as the result of the deformation: the resolution contains
a $\P^1$. Shrinking this $\P^1$ leads back to the original space.

Even though blow-ups and deformations may yield very different spaces in the case of e.g.
conifold singularities on Calabi-Yau threefolds, they produce the same space (although with a
different complex structure) for $K3$ surfaces. When discussing the moduli space
of $K3$ surfaces in more general terms in Section~\ref{k3mod}, this result will emerge in a more general 
context.

Although the deformation seems like the more obvious thing to do, the blow-up is technically
much more easy when it comes to more complicated singularities. The result is that the
resolution of an ADE singularity produces a set of two-spheres that intersect according
to the Dynkin diagram of a simply-laced Lie algebra. Of course these singularities will eventually be
connected to gauge enhancements. This will be discussed in all generality in Section~\ref{gaugeenh}. 
We have collected the algebraic equations describing the ADE singularities in Table~\ref{ADEtable}.
The ADE singularities can also be described as quotient singularities, see e.g.~\cite{Aspinwall:1996mn}. \\

\begin{table}
\begin{tabular}[h]{cccc}
Singularity & Lie Algebra & Equation & Dynkin diagram \\
\hline
$A_n$ & $SU(n+1)$  &$y^2+x^2+z^{n-1}=0$ &\includegraphics[height=1.7cm]{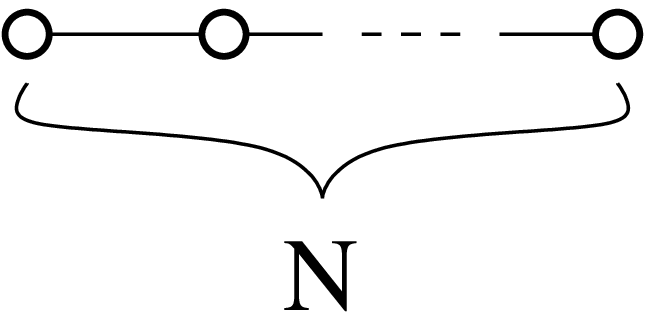} \\
$D_n$ & $SO(2n)$  & $y^2+x^2z+z^{n-1}=0$ & \includegraphics[height=1.7cm]{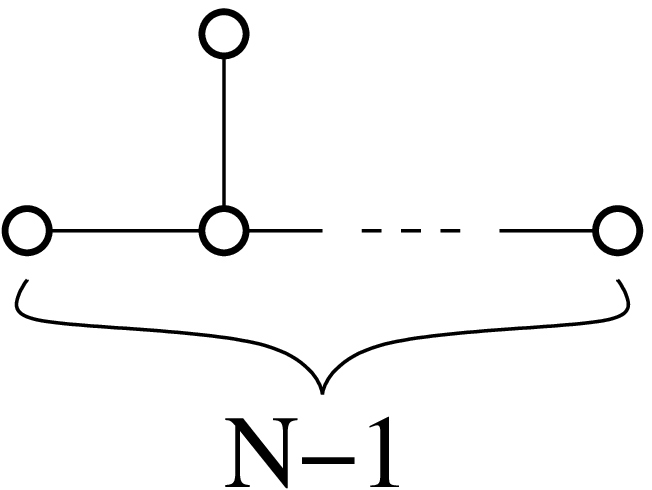} \\
$E_6$ & $E_6$ &$y^2+x^3+z^4=0$ & \includegraphics[height=1cm]{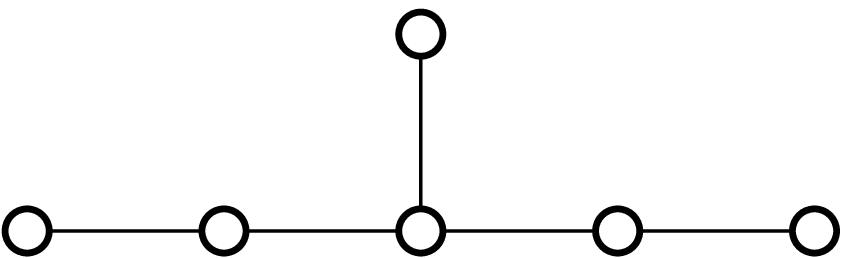} \\
$E_7$ & $E_7$  &$y^2+x^3+xz^3=0$ &  \includegraphics[height=1cm]{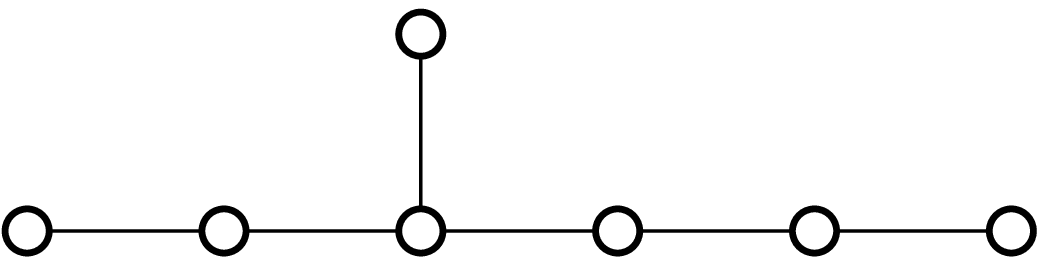} \\
$E_8$ & $E_8$  &$y^2+x^3+z^5=0$ &\includegraphics[height=1cm]{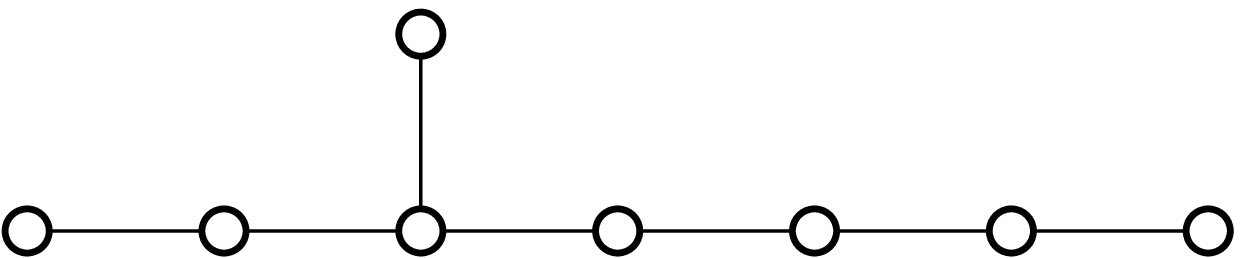} 
\end{tabular}
\caption{\textsl{The ADE singularities.\label{ADEtable}}} 
\end{table}

\subsection{General case}\label{sectionnonsimplaced}

\begin{figure}
\begin{center}
\includegraphics[height=7cm]{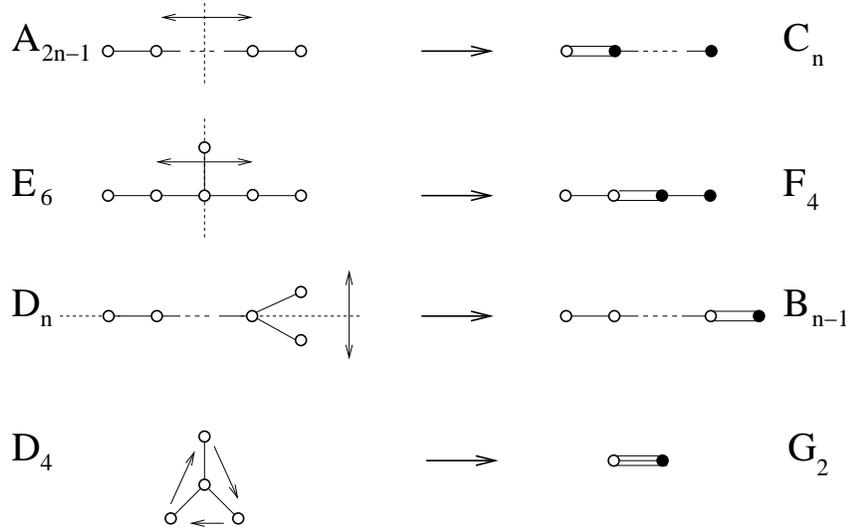}\caption{\textsl{Monodromies can fold the Dynkin diagrams that displays the intersection
pattern of the collapsing two-spheres such that the structure of non-simply laced Lie algebras emerges.}}\label{foldDyn}
\end{center}
\end{figure}

The picture sketched in the last section essentially persists in higher dimensions. 
When the fibre degenerates over some divisor $D$ in the base, there is a corresponding 
singularity over the same locus. If this singularity is resolved, one finds a number of $\P^1$s whose
intersection matrix characterizes the type of singularity. The main difference is that these
two-spheres might be interchanged when moving on $D$  \cite{Aspinwall:1996nk,Perevalov:1997vw,Aspinwall:2000kf,Szendroi:2002rf}.
This corresponds to an outer automorphism of the associated Lie Algebra. The way these outer 
automorphisms fold the Dynkin diagrams of the simply laced Lie Algebras is depicted in Figure~\ref{foldDyn}.

There is an algorithm \cite{Tate}, introduced into the physics literature 
in \cite{Bershadsky:1996nh}, which allows to deduce the type of fibre in the general case. It
is generally called Tate's algorithm. The distinction arises between 
so-called `split' and `non-split' configurations. In the non-split case, monodromies like the ones displayed in Figure~\ref{foldDyn} 
occur, whereas they are absent in the split case. 

\begin{figure}
\begin{center}
\includegraphics[height=6cm]{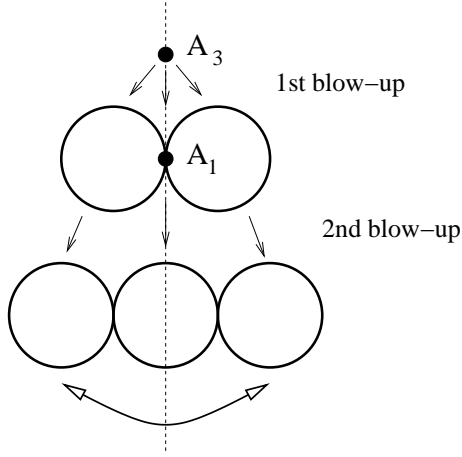}\caption{\textsl{Blowing-up an $A_3$ singularity. In the first blow-up, two $\P^1$s intersecting in
a $A_1$ singularity emerge. After the second blow-up, these spheres do not intersect anymore, because another sphere grows out of
the $A_1$ singularity. If \eqref{blowupA3} does not factorize, the $\P^1$s are permuted when moving in the transverse space.}}
\label{blowa3}
\end{center}
\end{figure}

Let us demonstrate this distinction in a simple example:
Consider the case of an $A_3$ singularity, fibred over another space $T$. Let this space be locally given in terms 
of affine coordinates $x,y,z,t_i$, such that the singularity occurs over $x=y=z=0$. We can then write:
\be
y^2\alpha+x^2\beta+z^4\gamma=0. 
\ee
Here $\alpha, \beta$ and $\gamma$ are functions of the remaining coordinates $t_i$, or, more generally,
sections of some line bundle on $T$.
Blowing up the point $x=y=z=0$ yields an exceptional set given by an equation of the type
\be
\xi_y^2\alpha+\xi_x^2\beta=0.\label{blowupA3}
\ee
inside $\P^2$.
Fixing a point on $T$, $\alpha, \beta$ and $\gamma$ become constants, so that we can factorize \eqref{blowupA3}.
Thus the blow-up produces two $\P^1$s that intersect in a $A_1$ singularity over every point of $T$. 
Globally, however, \eqref{blowupA3} does not factorize in general. Thus \eqref{blowupA3} describes just
one surface, which means that the two $\P^1$s are permuted when moving in $T$. Thus they cannot be considered 
to be independent but are locked together. A further blow-up can be used to resolve the remaining $A_1$ singularity, so
that we find the Dynkin diagram of $A_3$, see Figure~\ref{blowa3}. As the two `outer' $\P^1$s are permuted, 
the Dynkin diagram of $A_3$ is folded to that of $C_2$.

In the example just presented, a factorization condition has to be met in order
for the singularity to be split. This carries over to the general case: generically the 
singularities will be of the non-split type, so that we end up with Lie groups that are not
simply laced. The singularity is only of the split type if certain factorization conditions
are fulfilled. To formulate those conditions, it is convenient to write the Weierstrass model 
in a slightly different shape:
\be
y^2+a_1xyz+a_3yz^3=x^3+a_2x^2z^2+a_4xz^4+a_6z^6.\label{weiertate}
\ee
The $a_i$ denote sections of appropriate line bundle of the base, i.e. homogeneous polynomials
of the projective coordinates. 
Defining the variables:
\begin{align}
b_2&=a_1^2+4a_2  \nn\\
b_4&=a_1a_3+2a_4 \nn\\
b_6&=a_3^2+4a_6 \nn\\
b_8&=b_2a_6-a_1a_3a_4+a_2a_3^2-a_4^2, \label{tate1}
\end{align}
we can write the discriminant as:
\be
\Delta=-b_2^2b_8-8b_4^3-27b_6^2+9b_2b_4b_6.\label{tate2}
\ee
Completing the square and the cube in \eqref{weiertate} we find the relation to the ordinary Weierstrass form:
\begin{align}
f&=-\frac{1}{48}\left(b_2^2-24b_4\right) \nn\\
g&=\frac{1}{864}\left(-b_2^3+36b_2b_4-216b_6\right).\label{tate3}
\end{align}
The classification of singularities and the distinction between the split/non-split case
uses the order of vanishing of the various $a_i$ \cite{Bershadsky:1996nh}, see Table~\ref{tatetable}.
Only the distinction between the groups $SO(4k+4)$ and $SO(4k+3)$ is a bit more subtle, we will come back to 
this point in Section~\ref{tateweak}.
\begin{table}
\begin{tabular}{cccccccc}
 type & group & $ a_1$ &$a_2$ & $a_3$ &$ a_4 $& $ a_6$ &$\Delta$ \\ $I_0 $ & --- &$ 0 $ &$ 0
$ &$ 0 $ &$ 0 $ &$ 0$ &$0$ \\ $I_1 $ & --- &$0 $ &$ 0 $ &$ 1 $ &$ 1
$ &$ 1 $ &$1$ \\ $I_2 $ &$SU(2)$ &$ 0 $ &$ 0 $ &$ 1 $ &$ 1 $ &$2$ &$
2 $ \\ $I_{3}^{ns} $ & unconven. &$0$ &$0$ &$2$ &$2$ &$3$ &$3$ \\
$I_{3}^{s}$ &unconven. &$0$ &$1$ &$1$ &$2$ &$3$ &$3$ \\
$I_{2k}^{ns}$ &$ Sp(k)$ &$0$ &$0$ &$k$ &$k$ &$2k$ &$2k$ \\
$I_{2k}^{s}$ &$SU(2k)$ &$0$ &$1$ &$k$ &$k$ &$2k$ &$2k$ \\
$I_{2k+1}^{ns}$ &unconven. & $0$ &$0$ &$k+1$ &$k+1$ &$2k+1$ &$2k+1$ \\ 
$I_{2k+1}^s$ &$SU(2k+1)$ &$0$ &$1$ &$k$ &$k+1$ &$2k+1$ &$2k+1$ \\ 
$II$ & --- &$1$ &$1$ &$1$ &$1$ &$1$ &$2$ \\ 
$III$ &$SU(2)$ &$1$&$1$ &$1$ &$1$ &$2$ &$3$ \\ 
$IV^{ns} $ &unconven. &$1$ &$1$ &$1$&$2$ &$2$ &$4$ \\ 
$IV^{s}$ &$SU(3)$ &$1$ &$1$ &$1$ &$2$ &$3$ &$4$\\ 
$I_0^{*\,ns} $ &$G_2$ &$1$ &$1$ &$2$ &$2$ &$3$ &$6$ \\
$I_{2k-3}^{*\,ns}$ &$SO(4k+1)$ &$1$ &$1$ &$k$ &$k+1$&$2k$ &$2k+3$ \\ 
$I_{2k-3}^{*\,s}$ &$SO(4k+2)$ &$1$ &$1$ &$k$ &$k+1$
&$2k+1$ &$2k+3$ \\ $I_{2k-2}^{*\,ns}$ &$SO(4k+3)$ &$1$ &$1$ &$k+1$
&$k+1$ &$2k+1$ &$2k+4$ \\ $I_{2k-2}^{*\,s}$ &$SO(4k+4)^*$ &$1$ &$1$
&$k+1$ &$k+1$ &$2k+1$ 
&$2k+4$ \\ $IV^{*\,ns}$ &$F_4 $ &$1$ &$2$ &$2$ &$3$ &$4$
&$8$\\ $IV^{*\,s} $ &$E_6$ &$1$ &$2$ &$2$ &$3$ &$5$ & $8$\\
$III^{*} $ &$E_7$ &$1$ &$2$ &$3$ &$3$ &$5$ & $9$\\ $II^{*} $
&$E_8\,$ &$1$ &$2$ &$3$ &$4$ &$5$ & $10$ \\
 non-min & --- &$ 1$ &$2$ &$3$ &$4$ &$6$ &$12$
\end{tabular}\caption{\label{tatetable}\textsl{Tate's Algorithm gives the following map between
the singularity type and the order of vanishing of the $a_i$ appearing in \eqref{weiertate}.}}
\end{table}
It is important to note that one has to impose an extra condition to have non-simply laced groups
in the case of a base of more than one complex dimension, so that the folding of Dynkin diagrams
happens generically. We discuss these conditions in detail for configurations with a type IIB
dual in Section~\ref{tateweak}.

\section{Sen's weak coupling limit}\label{senweak}

Let us now start to discuss some of the physics that is associated with elliptic fibrations.
From the approach advocated in Section~\ref{sl2z} we interpret the complex
structure modulus of the fibre torus, implicitly given by
\begin{equation}
j(\tau(b))=\frac{4 (24f)^3}{4f^3+27g^2}\label{jatau}
\end{equation}
as the axiodilaton of a type IIB compactification on $B$. D7-branes and O7-planes
couple to the axiodilaton, so that there are monodromies in $SL(2,\Z)$ that act on $\tau$ 
upon circling them. As these monodromies occur precisely upon encircling a locus in the base
over which the elliptic fibre degenerates, 7-branes in F-theory are located at the discriminant locus
\begin{equation}
\Delta=4f^3+27g^2=0.
\end{equation}
As expressed in \eqref{monpq}, monodromies are linked to vanishing
cycles of the fibre torus. The locations in the base over which the cycle $(p,q)$ vanishes 
are the locations of a so-called $(p,q)$ 7-brane.
Note that we can always choose an $SL(2,\Z)$-frame such that any given $(p,q)$-brane
has the charge of a D7-brane: $(1,0)$. As soon as we have different $(p,q)$-branes
which are not mutually local in the sense of \eqref{mutloc}, they cannot both be
a D7-brane. This signals the appearance of strong coupling effects:
Remember that the elliptic fibration of F-theory is patched together 
using the $SL(2,\Z)$ S-duality of type IIB string theory from a physical perspective. Thus we might 
start at weak coupling at one point in the base, but are inevitably mapped to strong coupling after encircling a 
brane with the right monodromy. This is also reflected in the fact that the string coupling will generically be large over 
some regions of the base. Hence our theory becomes intrinsically non-perturbative.
Merging branes that are not mutually local leads to the wealth of different monodromies that occur
in Table~\ref{kodairaclass}.

To connect F-theory to type IIB orientifold compactifications, we thus need to find configurations 
such that the imaginary part of $\tau$ can be large (almost) everywhere, while finding a global 
$SL(2,\Z)$ frame in which  all monodromies are the ones of D-branes ($T$) and O-planes ($-T^{-4}$). 
From the discussion of the last paragraph it should be clear that these two requirements are actually one and the same.

The problem of finding the weak coupling limit of F-theory compactified on elliptic 
fibrations of the form \eqref{weier} has been solved by Sen \cite{Sen:1996vd, Sen:1997kw, Sen:1997gv},
see also \cite{Aluffi:2009tm}.

One rewrites
\begin{equation}
\label{def f}
f = C \eta - 3 h^2
\end{equation}
and
\begin{equation}
\label{def g}
g = h (C \eta - 2 h^2) + C^2 \chi \ ,
\end{equation}
where $C$ is a constant and $\eta$, $h$ and $\chi$ are sections of appropriate line bundles:
$h\in\Gamma([-2K_B])$,  $\eta\in\Gamma([-4K_B])$ and  $\chi\in\Gamma([-6K_B])$. 
Note that $f$ and $g$ are still in the most general form if we parameterize them as above. 
The weak coupling limit is to take $C \to 0$. To see this, consider
the modular function $j(\tau)$ in this limit:
\begin{equation}
j(\tau)=\frac{4 (24f)^3}{4f^3+27g^2}=\frac{4 (24)^3 (C\eta-3h^2)^3}{\Delta} \ .
\label{j weak coupling}
\end{equation}
The discriminant in the weak coupling limit is given by
\begin{equation}\label{deltaweakcoupl}
\Delta=C^2(-9h^2)(\eta^2+12h\chi)+...
\end{equation}
plus terms of the order $C^3$ or higher. We observe that for $C \to 0$ we have 
$|j| \to \infty$ everywhere away from the zeros of $h$, where the numerator of
the right hand side of \eqref{jatau} vanishes. The monodromy around
the points at which $h=0$ is precisely $-T^{-4}$, so that these are the positions
of the O-planes. 

Note that the location of the O-planes corresponds to a double zero in the discriminant $\Delta$.
This is a hint that the O-plane actually is a bound state of two branes. Going beyond
leading order in $C$ reveals that the locus of the O-plane is split into two branches, the splitting
being proportional to  $C$ \cite{Sen:1997gv}. The two branches have the $(p,q)$ charges $(1,1)$ and $(3,1)$, so 
that they are mutually non-local branes and produce the monodromy $-T^{-4}$ when taken together. Thus an 
orientifold plane is nothing but two aligned $(p,q)$ branes in F-theory.

The remaining branes are described by the equation
\begin{equation}
\eta^2 + 12 h \chi = 0 \ ,
\label{brane obstructions}
\end{equation}
and there is an $SL(2,\Z)$ frame in which all of them have monodromy $T$, so
that \eqref{brane obstructions} describes the positions of D-branes.

Having established that $h=0$ describes the position of O-planes and $\eta^2 + 12 h\chi=0$
describes the position of D-branes, we can now analyse how the gauge theories living on the
D-branes are encoded in the geometry. The classification of singularities discussed in
Section~\ref{singfibmono} clearly hints at a direct identification between the two. It
is easy to show this fact for configurations that can be described within type IIB.
We will return to the general case in Section~\ref{gaugeenh}. 

Let us first consider the case of an $A_n$ singularity. The corresponding
monodromy is $T^{n+1}$ (see e.g. Table~\ref{kodairaclass}] so that there are $n+1$ D7-branes 
located at the position of the singularity over the base. As there is a $U(n+1)$ gauge theory
on the worldvolume of a D7-brane, it is clear that the singularity type can only give the
non-Abelian part of the gauge theory. This has a natural explanation from the M-theory perspective,
discussed in Section~\ref{gaugeenh}.

We can also have singularities of the type $D_{n+4}$ in the weak coupling
limit. The corresponding monodromy suggests that this describes an O-plane with $4+n$
coincident D7-branes. This gives rise to a $SO(8+2n)$ gauge theory, further supporting the
identification between singularities and gauge enhancement. Note that an O-plane with
less than four coincident D-branes gives rise to a gauge theory of $SU$ type, i.e. there
is no $D_n$ singularity with $n\leq 3$.

Exceptional gauge groups never occur in perturbative type IIB compactifications. One thus expects 
that the corresponding exceptional singularities do not arise in the weak coupling limit. Indeed, one can 
show that exceptional singularities force $C$ to be of order one by translating Table~\ref{tatetable} to 
\eqref{def f} and \eqref{def g}. Hence all exceptional singularities are destroyed when taking the limit 
$C\rightarrow 0$.

\subsection{Tate's algorithm in the weak coupling limit}\label{tateweak}

In the following, the singularities and corresponding gauge enhancements that
appear in the weak coupling limit are systematically analyzed using Tate's algorithm.
In particular, the distinction between the split and the non-split case is formulated
in terms of D-branes and O-planes.

As only singularities of $SU$, $SO$ and $Sp$ type arise in the weak coupling limit, we
focus on the following entries of Table~\ref{tatetable}:
\begin{center}
\begin{tabular}{c|c|c|c|c|c|c}
 & $a_1$ & $a_2$ &$a_3$ &$a_4$&$a_6$&$\Delta$  \\ \hline
 $SU(2N)$ & $0$ & $1$ &$N$ &$N$&$2N$&$2N$ \\ \hline
 $Sp(N)$ & $0$ & $0$ &$N$ &$N$&$2N$&$2N$ \\ \hline
 $SO(4N+1)$ & $1$ & $1$ &$N$ &$N+1$&$2N$&$2N+3$ \\ \hline
 $SO(4N+2)$ & $1$ & $1$ &$N$ &$N+1$&$2N+1$&$2N+3$ \\ \hline
 $SO(4N+3)$ & $1$ & $1$ &$N+1$ &$N+1$&$2N+1$&$2N+4$ \\ \hline
 $SO(4N+4)$ & $1$ & $1$ &$N+1$ &$N+1$&$2N+1$&$2N+4$ 
\end{tabular}
\end{center}
First note that the discriminant in the weak coupling limit, \eqref{deltaweakcoupl}, is proportional to
\begin{equation}
C^2b_2^2(-b_4^2+b_2b_6)\ .
\end{equation}
Here the definitions given in \eqref{tate1} were used.
The location of the O7-planes is given by $b_2=0$, whereas the D7-branes
are located at $-b_4^2+b_2b_6=0$. Up to normalization, the parameterization
of $f$ and $g$ in terms of $h$, $\eta$ and $\chi$ is the same as the parameterization
in terms of $b_2$, $b_4$ and $b_6$.

Remember that $n$ coincident D7-branes carry $U(n)$ as their worldvolume gauge group and $n$ D7-branes
coincident with an O7-plane have an $SO(2n)$ gauge group on their worldvolume. If a stack of $n$ D7-branes
intersects an O7-plane, the gauge group is not $SU(n)$ but $Sp(n/2)$.

Let us now consider the gauge enhancement along some curve $\sigma$ in the base. If one of the $a_i$ 
contains (some power $m$ of) $\sigma$ as a factor, $\hat{a}_i$ is defined by $a_i=\sigma^m\hat{a}_i$,
so that the $\hat{a}_i$ do not contain $\sigma$ as a factor.

\subsection*{$\mathbf{SU(2n)}$ vs. $\mathbf{Sp(n)}$}
For gauge enhancement $SU(2n)$ along $\sigma$, D-brane and O-plane are described by the vanishing of
\begin{align}
O7:\hspace{.5cm} &\hat{a}_1^2+4\sigma\hat{a}_2\ ,  \nn\\
D7:\hspace{.5cm} &\sigma^{2n}\left((\hat{a}_1\hat{a}_3+2\hat{a}_4)^2+O7(\hat{a}_3^2+4\hat{a}_6) \right)\ .
\end{align}
Note that the intersection between the D7-brane stack at $\sigma=0$ and the O-plane is 
special: at $\sigma=0$ the equation describing the O7-plane is a square, so that two branches 
of the O-plane meet at every intersection with the D7-brane stack. The monodromy action that
corresponds to the O-plane locus contains an involution. As the O-plane is a square at every
intersection point with the D-brane, no path on the D-brane will experience any involutions. 
Once we allow the $SU(2n)$ singularity to be folded, the O-plane can intersect the D-brane in
an arbitrary way. In the $Sp(n)$ case, D-brane and O-plane are given by the vanishing of
\begin{align}
O7:\hspace{.5cm} &\hat{a}_1^2+4\hat{a}_2 \ ,  \nn\\
D7:\hspace{.5cm} &\sigma^{2n}\left((\hat{a}_1\hat{a}_3+2\hat{a}_4)^2+O7(\hat{a}_3^2+4\hat{a}_6) \right)\ .
\end{align}
Note that, contrary to the first case, the intersection between the D-brane stack and
the O-plane locus is now completely generic.

\subsection*{$\mathbf{SO(4n+2)}$ vs. $\mathbf{SO(4n+1)}$}
In the split case, in which the gauge enhancement is given by $SO(4N+2)$, D7-brane and O7-plane
are given by the vanishing of:
\begin{align}
O7:\hspace{.5cm} & \sigma(\sigma\hat{a}_1^2+4\hat{a}_2)  \ ,\nn\\ 
D7:\hspace{.5cm} & \sigma^{2n+1}\left(\sigma(\hat{a}_1\hat{a}_3+2\hat{a}_4)^2+(\sigma\hat{a}_1^2+4\hat{a}_2)(\hat{a}_3^2+4\sigma\hat{a}_6) \right) \ .
\end{align}
There is a stack of $2n+1$ D7-branes and one O7-plane along $\sigma$. The locus of the remaining D-branes is such
that they intersect this stack either together with another O-plane or in pairs.
This restriction is absent in the corresponding non-split case, in which $SO(4n+2)$ is folded to $SO(4n+1)$.
In this case the locus of D7-brane and O7-plane is described by:
\begin{align}
O7:\hspace{.5cm}& \sigma(\sigma\hat{a}_1^2+4\hat{a}_2) \ , \nn \\ 
D7: \hspace{.5cm}&\sigma^{2n+1}\left(\sigma(\hat{a}_1\hat{a}_3+2\hat{a}_4)^2+(\sigma\hat{a}_1^2+4\hat{a}_2)(\hat{a}_3^2+4\hat{a}_6) \right)\ .
\end{align}

\subsection*{$\mathbf{SO(4n+3)}$ vs. $\mathbf{SO(4n+4)}$}
Finally, let us turn to the case $SO(4n+3)$ and $SO(4n+4)$. For both,
the locus of D7-brane and O7-plane is described by:
\begin{align}
O7:\hspace{.5cm} &\sigma(\sigma \hat{a}_1^2+4\hat{a}_2) \nonumber\ , \\
D7:\hspace{.5cm} &\sigma^{2n+2}\left(-(\sigma+2\hat{a}_{4})^2+(\sigma \hat{a}_1^2+4\hat{a}_2)(\sigma \hat{a}_{3}+4\hat{a}_6)\right)\ .
\end{align}
There is a stack of $2n+2$ D7-branes and one O7-plane along $\sigma$, which intersects all other degeneration 
loci in a generic fashion. Again, the singularity is generically of the non-split type, implying $SO(4n+3)$.
For $SO(4n+4)$ to occur, we additionally need the polynomial $\hat{a}_2X^2+\hat{a}4X+\hat{a}_6$ to 
factor \cite{Bershadsky:1996nh}. This means that the discriminant of this equation, which is proportional to 
$-4\hat{a}_4^2+16\hat{a}_2\hat{a}_6$, needs to be a perfect square. But this means that the intersection of the stack at 
$\sigma=0$ with the remaining D7-branes,
\begin{equation}
0= -4\hat{a}_4^2+16\hat{a}_2\hat{a}_6 \ ,
\end{equation}
is a perfect square. Hence the remaining D7-branes are forced to intersect the stack at $\sigma=0$ in pairs if
we want to have $SO(4n+4)$ as the gauge enhancement.

In this section it has been explicitly demonstrated that ADE singularities are generically folded in the
weak coupling limit. Even though the monodromies that can fold simply laced Dynkin diagrams to non-simply laced 
Dynkin diagrams seem like a ``global'' issue, they were seen to arise locally: the distinction between the two cases
can be formulated in terms of how the intersections with the remaining D-branes and O-planes occur.

These facts are not surprising from the perspective of D-branes in orientifolds, see e.g. \cite{Johnson:2000ch}.
It is well-known that a stack of $n$ D-branes that is transversely intersected by an orientifold plane carries an $Sp(n)$ 
instead of a $U(2n)$ gauge group. The situation is more subtle for a stack of D-branes that coincides with an orientifold
plane and furthermore is intersected by another, transverse, O-plane. In the classic example of the Bianchi-Sagnotti-Gimon-Polchinski
orientifold \cite{Bianchi:1989du, Gimon:1996rq}, such a configuration is seen to have the gauge group of $SU(4)$. The
F-theory analysis, however, gives a singularity of $SO(7)$ in the same situation \cite{Sen:1996xx}.
The resolution of this apparent discrepancy is that there is a non-zero fractional $B_2$-flux in the conformal field theory 
which further breaks the gauge symmetry \cite{Aspinwall:1995zi}. This flux is absent in the F-theory description

\subsection{A note on the axiodilaton}

Whereas there is a clear distinction between open string moduli and the overall normalization of the 
axiodilaton for orientifolds, this is not true in general for F-theory compactifications. Another feature 
of the weak coupling limit is that it disentangles the axiodilaton from all other moduli.

Let us first show that it is not possible to change the normalization of $\tau$ in generic
configurations. To see this, we rewrite $f=\alpha \hat{f}$ and $g=\beta \hat{g}$. The axiodilaton is 
then implicitly given by
\begin{equation}\label{jfghat}
j(\tau)=\frac{4\alpha^3(24\hat{f})^3}{4\alpha^3\hat{f}^3+27\beta^2\hat{g}^2}\ ,
\end{equation}
whereas the discriminant locus. i.e. the positions of the 7-branes, is at
\begin{equation}\label{deltafghat}
\Delta=4\alpha^3\hat{f}^3+27\beta^2\hat{g}^2=0.
\end{equation}
Changing the normalization of $j$ is equivalent to changing the normalization of $\tau$.
If we perturb $\hat{f}$ or $\hat{g}$, this will alter the solutions to \eqref{deltafghat}, thereby
deforming some of the 7-branes. We can, however, write
\begin{equation}\label{deltafghat2}
\Delta=\alpha^3\left(4\hat{f}^3+27\frac{\beta^2}{\alpha^3}\hat{g}^2\right)=0\ .
\end{equation}
so that changing $\alpha$ while keeping $\beta^2 / \alpha^3$ will not change the discriminant locus.
This will, however, have no effect on the overall scaling of the axiodilaton because the nominator
of \eqref{jfghat} also contains a factor of $\alpha^3$. Thus we cannot alter the normalization
of $\tau$ without deforming the 7-branes.

The axiodilaton in the weak coupling limit $C\rightarrow 0$ is given by
\begin{equation}
j(\tau)=\frac{4 (24)^3 (C\eta-3h^2)^3}{C^2(-9h^2)(\eta^2+12h\chi)}\ .
\end{equation}
Thus, keeping $C$ small but fixed, we can change $\tau$ without altering the discriminant locus $\Delta$
by rescaling $\eta\rightarrow \alpha \eta$ and $\chi\rightarrow \alpha^2 \chi$. Hence
$\tau$ is separated from the other moduli in the weak coupling limit, so that it can be taken to be small without 
changing the locations of the 7-branes.

The fact that $\tau$ is separated from the 7-brane moduli in the weak coupling limit can also be seen
by comparing the degrees of freedom contained in $\Delta$ in general situations with those that arise 
in the weak coupling limit. As already discussed, the number of brane moduli contained in $f$ and $g$ is given by
\begin{equation}
N_{\Delta,\mbox{gen}}=\#(\underbrace{[-4K_B]}_{f})+\#(\underbrace{[-6K_B]}_{g})-\dim \aut_B-1\label{gensectfg}\ ,
\end{equation}
in a general situation. Here we denote the number of independent sections of a bundle $L$ by $\#(L)$ 
and the dimension of the automorphism group of $B$ by $\dim \aut_B$.

In the weak coupling limit, we have to count the number of sections that can be used for $h$, $\eta$ 
and $\chi$. As the discriminant locus is given by $-9h^2(\eta^2+12h\chi)$ in the weak coupling limit, 
the description in terms of $h$, $\eta$ and $\chi$ is redundant, as terms of the form $h^2 p^2$, with 
$p$ a section of $[-2K_B]$, can originate either from the term $h\chi$ or $\eta^2$ in $\Delta$. 
Furthermore, we can pull out a constant from $h$ and $(\eta^2+12h\chi)$ without changing $\Delta$. 
Hence we find
\begin{align}
N_{\Delta,\mbox{wcl}} =&\#(\underbrace{[-2K_B]}_{h})+\#(\underbrace{[-4K_B]}_{\eta})+\#(\underbrace{[-6K_B]}_{\chi}) \nn\\ 
&  -\#(\underbrace{[-2K_B]}_{p})-\dim \aut_B\underbrace{-1-1}_{\mbox{rescaling}}=N_{gen}-1\label{wcsect}\ .
\end{align}
degrees of freedom. The single degree of freedom that is lost in $\Delta$ is precisely the overall normalization
of $\tau$ which becomes independent in the weak coupling limit. In a sense, the weak coupling limit separates
the axiodilaton from the 7-brane moduli while keeping the total number of degrees of freedom fixed.

\subsection{The double cover Calabi-Yau}\label{double cover}

Having found the locus of the orientifold plane in the weak coupling limit of F-theory, we can 
reconstruct the type IIB orientifold. The usual strategy in type IIB is to start with a Calabi-Yau 
manifold $M$ that allows an involution which maps the holomorphic top form $\Omega^{n,0}\mapsto -\Omega^{n,0}$. 
The fixed point locus of this involution is the locus of an O-plane, so that we find an O7-plane 
whenever the fixed-point locus has complex codimension one. To cancel the RR tadpole associated to the D7-branes, one
has to put D7-branes respecting the involution into this background. The sum of their homology classes has
to equal four times the homology class of the O-plane (with everything being counted in the quotient of the Calabi-Yau
manifold $M$ by the involution: $B$).

As the O-plane is the fixed point locus of an involution, we can reconstruct the Calabi-Yau manifold $M$ by
building a double cover of the base which is branched over the locus of the O-plane. This can be described as
a hypersurface in a projectivized bundle $L$ over the base space $B$:
\begin{equation}
 \xi^2=h(b),\label{doublecover}
\end{equation}
where $\xi$ is a coordinate in $L$, which is chosen such that $[h]=[2D_L]=L^{\oplus 2}$. Note that $L$ is the same
bundle introduced in \eqref{weier}. The Chern class of $M$ is 
\begin{equation}
 c(M)=\frac{c(B)(1+D_L)}{1+2D_L},
\end{equation}
so that the Calabi-Yau condition for $M$ reads
\begin{equation}
 c_1(B)=-K_B=c_1(L).
\end{equation}
Note that this is precisely the condition for the elliptic manifold $X$ that is used in the corresponding
F-theory description to be Calabi-Yau, see \eqref{CycondX}. Thus we have shown that the Calabi-Yau
condition of F-theory is a consequence of the Calabi-Yau condition of type IIB before orientifolding.

The simplest example of this has (implicitly) already been discussed. Consider $B=\P^1$. We can
build a double cover of $\P^1$ that is Calabi-Yau (a torus in this case) by branching over four points, 
see the argument and figure below \eqref{urweier}. This is described by the equation
\be
\xi^2=h(z_1,z_2),
\ee
in $\P^2_{1,1,2}$. The homogeneous coordinates obey the equivalence relation 
$(z_1,z_2,\xi)\sim(\lambda z_1,\lambda z_2,\lambda^2 \xi)$, so that $\xi$ is a section
in $[-K_{\P^1}]$. Thus $h$ is a section of $[-2K_{\P^1}]$, i.e. a homogeneous polynomial
of degree four. Now consider an elliptic fibration over $\P^1$ described by an equation
of the form \eqref{weier}. From \eqref{def f} and \eqref{def g} it follows that 
$f$ is a section of $[-4K_{\P^1}]$ and $g$ is a section of $[-6K_{\P^1}]$, i.e. they
are homogeneous polynomials of degree $8$ and $12$, respectively. In this case
\eqref{weier} describes an elliptic $K3$ surface, see \eqref{k3}.

\subsection{The geometry of the D7-brane locus}
\label{d7locus}
\begin{figure}
\begin{center}
\includegraphics[height=2.5cm]{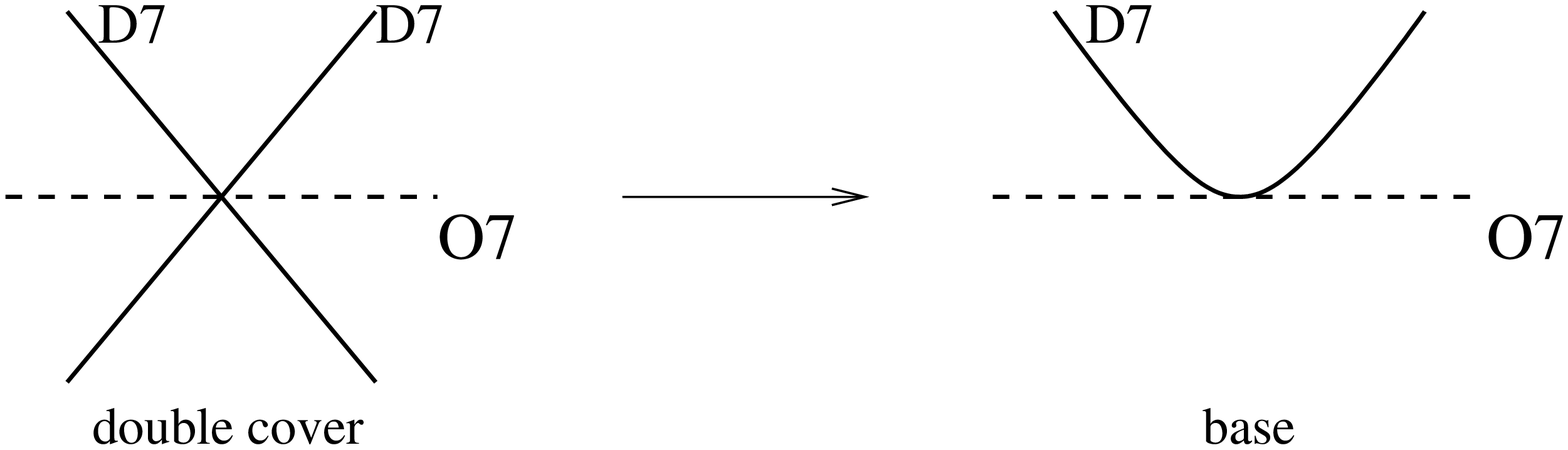}
\end{center}
\caption{\textsl{A situation in which two D7-branes intersect an O7-plane
in the same point produces a D7-brane touching the O-plane after modding out
the orientifold action and squaring the coordinate transverse to the O7-plane.
}}\label{2btof1}
\end{figure}

In the weak coupling limit, D7-branes are not generic hypersurfaces, but
they are described by an equation of the form
\begin{equation}\label{obstructions brane equation}
 {\cal D}: \quad \eta^2+12h\chi=0 \ ,
\end{equation}
where $h=0$ denotes the locus of the O7-plane. Throughout this work, we will refer to D7-branes 
that are described by an equation like \eqref{obstructions brane equation} as \emph{generic allowed} D7-branes. 
In the base space, \eqref{obstructions brane equation} forces the D7-brane to be a square when restricted
to the O-plane \cite{Braun:2008ua}. In the double cover, this means that any D-brane is mapped to a 
distinct\footnote{This holds only locally, as the D-brane, which is given by $\eta^2+12\xi^2\chi$ in the double cover, 
does not factorize in general} image brane in the vicinity of the O-plane, see Figure~\ref{2btof1}. It should
be stressed that this is not some artefact of the F-theory description, but can also be shown directly from the 
type IIB perspective \cite{Collinucci:2008pf}. We present a detailed discussion of the constraints a generic allowed 
D7-brane has to fulfill, as compared to a generic hypersurface of the same topology, in Section~\ref{sectobs3} in
the case of a complex two-dimensional base. This allows us to enumerate the obstructed deformations of a 
D7-brane. In this section we analyse the local geometry of a D7-brane in the vicinity of an O7-plane for the case of 
a complex two- and three-dimensional base.

\subsubsection{Two-dimensional base}

Let us investigate~\eqref{obstructions brane equation} in the vicinity of an intersection 
point. We parameterize the neighborhood of this point by complex coordinates $x$ and $z$. Without loss 
of generality, we take $h=z$ (i.e. the O7-plane is at $z=0$) and assume that the intersection 
is at $x=z=0$. This means that ${\cal D}=(\eta^2 + 12 h \chi)$ vanishes at $x=z=0$ 
and, since we already know that $h$ vanishes at this point, we conclude that 
$\eta(x=0,z=0) = 0$. Expanding $\eta$ and $\chi$ around the intersection 
point, 
$$\eta(x,w) = m_1 x + m_2 z + \dots\qquad\mbox{and}\qquad\chi(x,z) = n_0 + 
n_1 x + n_2 z + \dots \,,$$
we find at leading order\footnote{Note that $xz$ is subdominant w.r.t. to $z$, which is not true 
for $x^2$.}
\begin{equation}
 m_1^2 x^2 + 12n_0 z + \dots = 0 \ .\label{d72dbase}
\end{equation}
In the generic case $n_0 \ne 0$, this is the complex version of a parabola
`touching' the O-plane with its vertex. Thus, we are dealing with a double 
intersection point.\footnote{This is also clear from the fact that, if we were to introduce by hand 
a term $\sim x$ in (\ref{d72dbase}), our intersection point would split into two.} 
In the special case $n_0=0$, \eqref{d72dbase} is reducible and we are 
dealing with two D7-branes intersecting each other and the O7-plane at the 
same point. The former generic case hence results from the recombination of 
this D7-D7-brane intersection. In both cases, we have a double 
intersection point. In other words, the constraint corresponds to the 
requirement that all intersections between the D7-branes and O7-planes must 
be double intersection points. We will discuss the effect this has on D-brane
moduli in Section~\ref{sectobs3}.

Let us try to understand this configuration from the double cover perspective. 
Consider two D-branes at $x=\pm z$ and the involution $z \to - z$, which fixes the O-plane at $z=0$. 
After modding out the involution, our space looks locally like the upper half-plane. In order to make 
contact with the F-theory picture, we introduce a new coordinate $\tilde{z}=z^2$ and find that this 
situation is described by an O-plane at $\tilde{z}=0$ and a D-brane at $\tilde{z}=x^2$. 
Note that a single D7-O7 intersection in F-theory (which does not occur in the weak coupling limit),
corresponds to a single D7-brane that is mapped onto itself by the orientifold projection. This configuration, 
where the D-brane sits e.g.at $x=0$ is allowed in the presence of a second D-brane that coincides with the O-plane.
Note that the D7-brane locus in the double cover, given by
\begin{equation}
{\cal D}=\eta^2 + 12 \xi^2 \chi=0 \ ,
\end{equation}
has double-point singularities at $\eta=\xi=0$.

\subsubsection{Three-dimensional base}

In the case of a complex three-dimensional space, we can always choose
coordinates $x,y,z$ such that locally
\begin{align}
x\sim\eta,\hspace{.5cm}z\sim h,\hspace{.5cm}y\sim\chi\ \ .
\end{align}
Thus the O-plane is still located at $z=0$, whereas the D-brane locus is now given by
\begin{equation} \label{d73dbase}
 x^2+zy=0 \ .
\end{equation}
For any fixed $y\neq 0$, this is the same we found in the case of a complex two-dimensional base.
At $y=0$, however, something special happens. Note that we can redefine $z=a+ib$, $y=a-ib$ and
bring \eqref{d73dbase} in the canonical form of an $A_1$ singularity, see Table~\ref{ADEtable}. 
Note that the D7-brane in a two-dimensional base, \eqref{d72dbase}, is smooth. 
In the double cover Calabi-Yau, the D7-brane is described by
\begin{equation}
 x^2+z^2y=0 \ .
\end{equation}
As has been pointed out in \cite{Collinucci:2008pf}, this equation describes a singular 
variety: the $A_1$ singularity that occurs in the base has turned into a so-called Whitney 
umbrella in the double cover. Again, we find the same configuration as in the case of a complex two-dimensional 
base, two intersecting branes, as long as $y\neq 0$. The Whitney umbrella arises because these 
two lines merge as we approach $y=0$.

The fact that the D7-brane locus in the double cover is singular is problematic for two reasons.
The deformations of D7-branes are counted by cohomology classes of its worldvolume, 
$h^{2,0}_-$ for a complex three-dimensional base \cite{Jockers:2004yj} and $h^{1,0}_-$ for a complex 
two-dimensional base. These expressons, however, only were derived (and only make sense) for a smooth
D7-brane worldvolume. As we have seen, this is not the case. 
Second, the D3-brane tadpole gets a contribution that is proportional to the Euler characteristic of
the D7-brane worldvolume in the double cover, which is again not well-defined if the D7-brane is singular.

\section{Dualities}\label{gaugeenh}

In this section we give a brief review of the dualities between F-theory and 
M-theory, as well as the heterotic $E_8 \times E_8$ string. 

\subsection{M-theory and type IIA}

As M-theory compactified on $S^1$ is dual to type IIA in 10 dimensions, we 
can relate M-theory to type IIB by Compactifying on a further $S^1$ and 
applying T-duality. Thus, M-theory on $T^2$ corresponds to type IIB on 
$S^1$. The complexified type-IIB coupling constant is given by the complex 
structure of the torus. Furthermore, taking the torus volume to zero 
corresponds to sending the $S^1$ radius on the type IIB side to infinity. In 
other words, M-theory on $T^2$ with vanishing volume gives type IIB in 10 
dimensions, which we take as our working definition of F-theory. Considering the 
compactification of M-theory on an elliptically fibred manifold $X$ and using 
the above argument for every fibre, we arrive at type IIB with varying coupling 
on the base space. Of course we have to assume that the dualities we have used
are valid also non-perturbatively.

%%%%%%%%%%%%%%%%5
% more details ??????? give right references, see our papers
%%%%%%%%%%%%%%%%%%%5

The duality to M-theory serves to establish the fact that $X$ should be Calabi-Yau from
a further perspective. Compactifications of M-theory on a space $X$ of the form \eqref{weier}, i.e.
a compact complex K\"ahler manifold, only preserve some supersymmetry if $X$ is Calabi-Yau. This 
verifies the result of Section~\ref{double cover} from a different perspective.

The duality between M-theory and F-theory can be used to analyse F-theory vacua from the
perspective of M-theory, a tool that will be used (and sometimes tested) throughout this work. It
is important to keep in mind that F-theory only arises in the limit of vanishing fibre size.
It is clear from the type IIB side that the fibre volume is of no physical importance, so
M-theory has to get rid of its K\"ahler modulus before it can be dual to F-theory. The
limit in which the volume of the elliptic fibre of an M-theory compactification shrinks to zero 
size is called the \emph{F-theory limit}. Let us describe this in some
more detail: 
The sizes of effective curves $C_i$ of the elliptic Calabi-Yau manifold $X$ are given by 
\begin{equation}
j_i= \int_{C_i} J \ ,
\end{equation}
see also Appendix~\ref{homdivtoric}.
Denoting the elliptic fibre by $E$, we thus need to send $J$ in a limit in which 
\begin{equation}
  \int_{E} J \longrightarrow 0 \ .
\end{equation}
This means that $J$ is supposed to approach a boundary of the
K\"ahler cone of $X$. This fits the construction of F-theory we started with: while the size
of the fibre torus was irrelevant, size and shape of $B$ are of course physically relevant.

On the M-theory side, gauge enhancement cannot stem from massless open strings stretched 
between D-branes. The singularities that signal the appearance of extra massless states, however, 
arise by collapsing certain sets of two-spheres, as discussed in Section~\ref{singfibmono}. States 
that correspond to $M2$-branes wrapped on two-spheres have masses that are proportional to the sizes of 
those spheres. This means there are extra massless states in the spectrum when the two-spheres are 
collapsed. As the corresponding states are BPS, they will not receive corrections as the spheres shrink 
to zero size. In compactifications of M-theory to 6 dimensions or less, the
massless states that originate from $M2$-branes wrapped on collapsed two-spheres live on submanifolds
of the compactification manifold, which is of course identified with the locus of a brane in F-theory. 
Furthermore, M-theory contains a three-from potential, $C_3$, which gives rise to one 
massless $U(1)$ gauge bosons for every 2-cycle in $X$. Gauge symmetries in M-theory arise
in very much the same fashion as they do in IIA, where non-Abelian gauge symmetries stem from
wrapped $D2$-branes and $U(1)$s originate from the 3-form potential \cite{Katz:1996ht}, see also \cite{Aspinwall:1996mn}. 
From the perspective of type IIA string theory, the Cartan generators of exceptional groups originate from multi-pronged 
strings on the singular geometry, see \cite{Johansen:1996am,Gaberdiel:1997ud,DeWolfe:1998zf}.

As there is no conceptual difference between singularities of the $A$ and the $D$ type and 
exceptional singularities, it is clear that exceptional singularities in F-theory must give 
rise to the corresponding gauge groups. The appearance of singularities that correspond to non-simply-laced
gauge groups discussed in Section~\ref{sectionnonsimplaced} also leads to non-simply laced
gauge groups in a natural way: if two collapsed two-spheres are identified through monodromy then
so are the states of wrapped branes. This justifies the claims we made earlier: \emph{the non-Abelian part 
of the gauge group in F-theory compactifications is identical to the singularities of the elliptic fibration}
as indicated in Table~\ref{tatetable}.

\subsection[Heterotic $E_8\times E_8$]{Heterotic \boldmath$E_8\times E_8$}\label{dualhet}

The foundation of the duality between F-theory and the $E_8\times E_8$ 
heterotic string is the duality between M-theory compactified on $K3$ and the 
heterotic string compactified on $T^3$~\cite{Witten:1995ex}, see also
Section~\ref{wilson}. In the limit in 
which the fibre of $K3$ shrinks (the F-theory limit), one $S^1$ 
decompactifies so that we end up with heterotic $E_8\times E_8$ on $T^2$
\cite{Vafa:1996xn, Vafa:1997pm, LopesCardoso:1996hq, Lerche:1998nx, Lerche:1999de}.

In this duality, the K\"ahler modulus of the base and the complex
structure modulus of the fibre of $K3$ are mapped to the complex and K\"ahler 
structure moduli of the $T^2$ on the heterotic side. The precise relation between 
these parameters has been worked out in~\cite{LopesCardoso:1996hq}. The volume of the base
of $K3$ corresponds to the coupling of the heterotic string theory. The breaking 
of $E_8\times E_8$ that is achieved by Wilson lines on the heterotic side appears in the form of 
deformations of the Weierstrass equation away from the $E_8\times E_8$ singularity on the F-theory 
side. This is discussed in detail in Section~\ref{wilson}.

For compactifications of F-theory to less than 8 dimensions, we can use the beforementioned
duality fibrewise. This gives the statement that F-theory on a $K3$-fibred Calabi-Yau manifold (here the $K3$ fibre
must itself be elliptically fibred) is dual to the heterotic $E_8\times E_8$ string on an elliptically fibred Calabi-Yau 
threefold $H$ such that both have the same base $B$. This means that the elliptic Calabi-Yau fourfold $X$ has a base $B'$ which
is a $\P^1$ fibration over $B$. 

\begin{center}
\includegraphics[height=2cm]{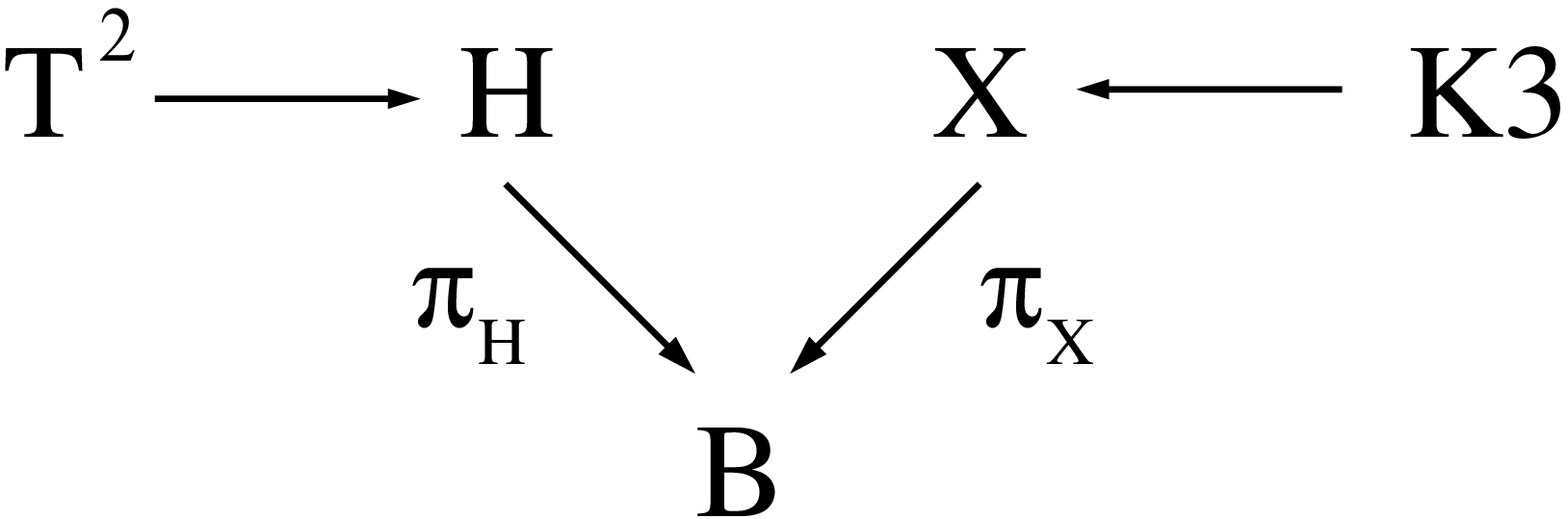}
\end{center}

On the F-theory side, the information about gauge enhancement is encoded in the degenerations of the
fibre $K3$. Over every point of $B$, this maps to a pair of Wilson lines on $T^2$ on the heterotic side. 
Fibering these data over $B$ gives the so-called spectral cover construction for vector bundles on elliptic
Calabi-Yau manifolds \cite{Friedman:1997yq, Donagi}. The point of view of spectral covers has recently
proven quite useful in the study of F-theory compactifications, see~e.g.~\cite{Donagi:2008ca, Donagi:2009ra, Marsano:2009gv, Tatar:2009jk}.

\section{Moduli}\label{sectmoduli}

From the perspective of the Weierstrass model, \eqref{weier}, moduli naturally arise as polynomial deformations: 
for a given base $B$, there can be many choices of the holomorphic sections $f$ and $g$. In simple examples, 
holomorphic sections of the anticanonical bundle of the base (or some power thereof) are nothing but homogeneous 
polynomials on the base. In this case, the number of sections can easily be counted by determining the number of 
monomials and subtracting the number of automorphisms of the base coordinates, see also Appendix~\ref{aplbundles}.

Let us work this out for the simple case of $K3$, introduced in \eqref{k3}. In this case $f$ and $g$ are homogeneous
polynomials of degree $8$ and $12$ on $\P^1$, respectively. As a homogeneous polynomial of degree $n$ on $\P^1$ has 
$n+1$ monomials, so that we need to specify $9+13=22$ complex numbers to specify the polynomials $f$ and $g$. In doing
so, we have used specific coordinates on $\P^1$. Any $SL(2,\C)$ transformation of the homogeneous coordinates, however, 
gives rise to an automorphism of $\P^1$. Thus we can use this $SL(2,\C)$ and set three coefficients to arbitrary values.
Furthermore, we can use the equivalence relation $(z_1,z_2)\sim(\lambda z_1,\lambda z_2)$ to eliminate one
further coefficient in $f$ and $g$. This can also be phrased differently: as all the physical information is
contained in $\tau$ as a function of the base, we can always rescale $f$ and $g$ such that is remains unchanged.
This rescaling is of course nothing but the one induced by sending $(z_1,z_2)\mapsto(\lambda z_1,\lambda z_2)$.
Thus we find that $22-3-1=18$ complex parameters completely determine \eqref{k3}.

As a second example, let us come back to the elliptic Calabi-Yau with base $B=\P^1\times\P^2$, introduced
in \eqref{p1p2}. In this case $f$ and $g$ are homogeneous 
polynomials of bidegree $(8,12)$ and $(12,18)$, respectively. A homogeneous polynomial of the coordinates
of $\P^1\times\P^2$ which has bidegree $(n,m)$ has 
\be
(n+1)\frac{(m+1)(m+2)}{2}
\ee
coefficients, see Appendix~\ref{modsections}. Thus, $819+2470=3289$ monomials are contained in $f$ and $g$. Any 
automorphism of $\P^1\times\P^2$ can clearly be
written as a composition of an automorphism of $\P^1$ and an automorphism of $\P^2$, so that the dimension of
this group is $3+8=11$ dimensional\footnote{We have collected some results concerning the automorphisms of
toric varieties in Appendix~\ref{toricAut}.}. We loose one further degree of freedom due to rescaling, so that
we finally find $3289-12=3277$ degrees of freedom.

The method we have just demonstrated can also be used to find the degrees of freedom of a single brane, i.e.
a factor of the discriminant locus. If we have a hypersurface $H$ in $B$ that is described by the vanishing of a
generic section of a bundle $[H]$, we can count the number of deformations by counting the number of 
holomorphic sections in $[H]$. 
The first adjunction formula \cite{Griffiths:1978}, see Appendix~\ref{adjunkapp}, 
gives the equality
\begin{equation}
 N_{H/B} = [H]|_H \ .
\end{equation}
Thus we actually do the usual thing by counting sections in $[H]$: we determine the deformations of a 
hypersurface by counting sections in the normal bundle. From this it follows directly that hypersurfaces
that are not described by completely generic sections in some bundle cannot have arbitrary deformations
in their normal bundle. In other words, their deformations naturally do not live in their normal bundle
in $B$, but some other bundle. Looking back at the D-brane locus in the weak coupling limit, \eqref{obstructions brane equation},
it is clear that the deformations of D-branes cannot come from their normal bundle in general. We will discuss
this in detail for the case of a two-dimensional base in Section~\ref{sectobs3}.

In the case of $K3$, the discriminant $\Delta$ has degree $24$, so that F-theory on $K3$ describes $24$ branes moving on 
$\P^1$. As we have shown above, this model has just $18$ complex degrees of freedom, so that the $24$ branes cannot move
independently. This happens because the discriminant locus is not given by a generic section in $[-12K_B]$ but must have 
the form $\Delta=4f^3+27g^2$. Naively, $24$ points on $\P^1$ can be displaced by section in the normal bundle, which is 
just $\C^{24}$ here. As $\Delta$ is a non-generic section in $[-12K_B]$, the normal bundle is, however, not the right object 
to look at. From the physics perspective, the fact that the $24$ branes can not all be mutually 
local prevents us from moving them freely, i.e. there are less open string states than naively expected. This
should become clear when we discuss the moduli space of $K3$ in Section~\ref{cyclesk3branes}.

Let us now switch to a different view of the geometric moduli of F-theory compactifications.
The deformations of the sections $f$ and $g$ are nothing but deformations of monomials appearing in the
Weierstrass model, \eqref{weier}. We have already argued that the elliptic fibrations that are relevant 
to F-theory should be such that \eqref{weier} is an elliptically fibred Calabi-Yau manifold. Thus we expect 
the moduli of F-theory compactifications to be connected to the moduli of elliptic Calabi-Yau manifolds, 
which can be split up in complex structure and K\"ahler moduli space\cite{Candelas:1990pi}.
As the elliptic Calabi-Yau spaces we compactify F-theory on are given as hypersurfaces defined by a Weierstrass model 
(or, more generally, complete intersection such that one of the defining polynomials is of the Weierstrass form), 
deformations of $f$ and $g$ are so-called polynomial deformations of the Calabi-Yau given by \eqref{weier}
and thus correspond to complex structure moduli. Although the number of polynomial deformations does not necessarily 
equal the number of complex structure moduli \cite{Green:1987rw}, they do agree for a huge number of examples. 
In particular, they agree for most of the examples discussed in this work.

Motivated by the last paragraph, we now describe the moduli of F-theory compactifications from the perspective of the
elliptically fibred Calabi-Yau n-fold $X$, as given by \eqref{weier}. This is at the heart of the description of
F-theory via M-theory, in which the elliptic Calabi-Yau $X$ is actually used to compactify 11D supergravity.
The moduli of $X$ split into $h^{n-1,1}$ complex structure and $h^{1,1}$ K\"ahler moduli\footnote{As we review in 
Section~\ref{k3mod}, this is slightly more subtle for $K3$ surfaces, where complex structure and K\"ahler moduli come from 
the same cohomology group.}.

Let us first discuss complex structure deformations. From the discussion of the previous section, we expect polynomial 
deformations of the Weierstrass model, and hence deformations of the branes contained in F-theory compactifications, to 
contribute to complex structure deformations. A second source of complex structure deformations of $X$ are complex 
structure deformations of the base space $B$.

Consider again the example of $B=\P^1\times \P^2$. As this space has no complex structure deformations, we 
expect all complex structure deformations of the elliptic fourfold \eqref{p1p2} to come from polynomial deformations of 
the Weierstrass model. By using the methods explained in Appendix~\ref{appendix} we find that $h^{3,1}(X)=3277$. This 
precisely matches the number of polynomial deformations computed above. Thus all complex structure deformations of $X$ 
correspond to deformations of branes in this case.

We now turn to the K\"ahler moduli of $X$. K\"ahler moduli can be thought of as coefficients of an
expansion of the K\"ahler form, $J$, into a basis of harmonic $(1,1)$-forms. Harmonic 1-forms
are dual to 2-cycles which nicely split up into 2-cycles of the base and the class of the elliptic
fibre. Here the minimality of the Weierstrass form, \eqref{weier}, is of great importance:
it has only one size related to the fibre which comes from the size of the ambient $\P_{1,2,3}$. All
components of the singular fibres that do not meet the section are blown down 
\cite{peters, Morrison:1996na, Morrison:1996pp}.

In conventional orientifold compactifications, there is a clear distinction between open and closed string moduli.
If we describe the moduli of F-theory compactifications through the geometric moduli of the elliptic Calabi-Yau
space $X$, this distinction is lost. The geometry of $X$ encodes both the shape of the base space $B$ and the
positions and types of various 7-branes that can be present. In particular, it is the complex structure moduli
space of $X$ which contains the moduli of the branes. The moduli space of complex structures of Calabi-Yau
spaces can in turn be described in terms of the periods, which are given by integrals of the holomorphic top 
form over the cycles of the middle dimensionality. Thus we expect a link between the deformations of 7-branes
and the middle homology of the corresponding elliptic Calabi-Yau manifold. Much of this work is devoted
to explore this idea.

\chapter{F-theory on \boldmath{$K3$}}\label{chapterk3}

In this section we explore compactifications of F-theory on $K3$. Because it
is the simplest F-theory compactification (apart from completely trivial ones),
many aspects of F-theory appear in a very pure form and can be easily analyzed.

\section[Geometry of the $K3$ surface]{Geometry of the \boldmath$K3$ surface}\label{k3mod}
In two complex dimensions there is, up to diffeomorphisms, just one compact 
Calabi-Yau manifold: $K3$. See \cite{peters} for the general theory and \cite{Aspinwall:1996mn} 
for the role of $K3$ in string dualities.

\subsection{Moduli space and second homology group}

The Hodge diamond of $K3$ is:
\be
\label{K3 hodge structure}
  {\arraycolsep=2pt
  \begin{array}{*{5}{c}}
    &&1&& \\ &0&&0& \\ 1&&20&&1. \\
    &0&&0& \\ &&1&&
  \end{array}}
\ee
%The Euler characteristic can be computed by integrating the top Chern class of $K3$, yielding the result $\chi(K3)=24$. 
%Let us demonstrate this: Calabi-Yau surface have no 1-cycles and a unique
%holomorphic top-form \cite{Candelas:1987is}. By Poincare-duality it follows that $K3$ furthermore has no 3-cycles.
%Hence we find that the following relation for the Euler characteristic: $\chi(K3)=2+2+h^{1,1}(K3)$. 
% show this

There is a natural inner product among the 2-forms of $K3$ given by
\be
\eta\cdot\gamma=\int_{K3}\eta\wedge\gamma \ .
\ee
On the dual homology cycles, this inner product counts the intersection numbers. The self-intersection number 
is given by the intersection of a cycle with another homologous cycle. For $K3$ surfaces, there is a direct 
relation between the self-intersection number of cycles and their topology. As the canonical bundle of $K3$ is 
trivial, the second adjunction formula applied to a curve $C$ embedded in $K3$ states that
\begin{equation}
 [K_C]^{-1}=N_{C/K3} \ .
\end{equation}
For a curve, the first Chern class of the anticanonical bundle is equal to the top Chern class of the
tangent bundle. Integrating it, we obtain the Euler characteristic of $C$. The self-intersection number
of $C$ in $K3$, on the other hand, in given by the number of points for which a section of the normal bundle
of $C$ in $K3$ must vanish. As the normal bundle of $C$ in $K3$ is a line bundle, this number is equal to the
integral of its first Chern class over $C$. The second adjunction formula (see Appendix~\ref{adjunkapp} then gives
\begin{equation}\label{selfintvsg}
\chi= 2-2g= C\cdot C \ .
\end{equation}
Note that this means that any curve which has a self-intersection number of more than 2 must be reducible.

The signature of the metric on the 2-forms of $K3$ can be determined by the Hirzebruch signature theorem, see
Appendix~\ref{sumchernclass}.
%%%%%%%% show this
As the result is $-16$, we know that there are $19$ negative definite forms and $3$ positive definite ones.
Hence the second homology class $H_{2}(K3,\mathbb{Z})$, equipped with the natural metric given by the intersection 
numbers between cycles, is an even (see \eqref{selfintvsg}) self-dual (Poincare duality) lattice with signature $(3,19)$, 
commonly denoted by $\Gamma_{3,19}$. By the classification of even self-dual lattices, the lattice $\Gamma_{3,19}$ is given 
by the direct sum of three copies of the hyperbolic lattice and two copies of the root lattice of $E_8$, albeit with negative 
definite norm on the $E_8$ summands. 
% cite somebody for this

The $E_{8}$ root lattice is the unique even unimodular lattice of rank 8.
Any element takes the form $\alpha=q_{I}E_{I}$, where $\{E_{I}\}_{I=1,...,8}$
is a basis of $\mathbb{R}^{8}$ satisfying $E_{I}\cdot E_{J}=-\delta_{IJ}$.
The coordinates $q_I$ have to be all integer or half-integer and must 
fulfill $\sum_{I=1,...,8}q_{I}=2\mathbb{Z}$ \cite{conwaysloane}. 

We choose the (non-unique) set of 8 simple roots
\begin{align}
\alpha_{1} & =\frac{1}{2}E_{1}+\frac{1}{2}E_{2}+...+\frac{1}{2}E_{8} & \alpha_{5} & =-E_{4}+E_{5}\nonumber \\
\alpha_{2} & =-E_{7}-E_{8} & \alpha_{6} & =-E_{3}+E_{4}\nonumber \\
\alpha_{3} & =-E_{6}+E_{7} & \alpha_{7} & =-E_{2}+E_{3}\nonumber \\
\alpha_{4} & =-E_{5}+E_{6} & \alpha_{8} & =-E_{7}+E_{8}.\label{eq:e8rootsystem}\end{align}

\begin{figure}[tt]
\begin{centering}
\includegraphics[width=7cm]{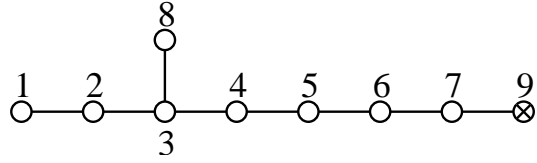}
\par\end{centering}

\caption{\textsl{The extended Dynkin diagram of $E_{8}$.}}\label{fig1}

\end{figure}

The structure of this basis is encoded in the \textit{Dynkin diagram}
of $E_{8}$. The \textit{extended Dynkin diagram}
is obtained by adding the (linearly dependent and thus non-simple) highest root \cite{helgason} (see Figure~\ref{fig1}).
\begin{align}
\alpha_{9} & =-2\alpha_{1}-4\alpha_{2}-6\alpha_{3}-5\alpha_{4}-4\alpha_{5}-3\alpha_{6}-2\alpha_{7}-3\alpha_{8}
  =-E_{1}+E_{2}.\label{eq:e8highestroot}
\end{align}
The coefficients in this expansion are known as the \textit{Coxeter labels}.

The reflections in the hyperplanes orthogonal to the 240 roots are
symmetries of the $E_{8}$ root lattice and generate the Weyl group
of type $E_{8}$. Its order is given by $4!\cdot6!\cdot8!=696729600$~\cite{conwaysloane}. 
The $E_8$ Weyl group contains a subgroup of order $8!\cdot 2^7$ consisting of all permutations of the coordinates and all even sign changes. This subgroup is the Weyl group of type $D_8$. The full $E_8$ Weyl group is generated by this subgroup and the block diagonal matrix ${\cal H}_4\oplus {\cal H}_4$ where ${\cal H}_4$ is the Hadamard matrix
\begin{equation}
    {\cal H}_4 = \tfrac{1}{2}\left(\begin{smallmatrix} 1 & 1 & 1 & 1\\ 1 & -1 & 1 & -1\\ 1 & 1 & -1 & -1\\ 1 & -1 & -1 & 1\\ \end{smallmatrix}\right)\:.
\end{equation}

There is a basis of $H_{2}(K3,\mathbb{Z})$ such that the matrix formed by the inner products of the basis vectors reads
\begin{equation}
U\oplus U\oplus U\oplus(-E_{8})\oplus(-E_{8}),\label{eq:intersectionmatrix}\end{equation}
where $E_{8}$ is the positive definite Cartan matrix of $E_{8}$ and
\[
U=\left(\begin{array}{cc}
0 & 1\\
1 & 0\end{array}\right).\]

We will denote the basis vectors spanning the three $U$ blocks by
$e_{i}$ and $e^{i}$, $i=1,2,3$. Accordingly, $e^{i}\cdot e_{j}=\delta_{j}^{i}$.
Using the notation introduced in the last section for the $E_8$ lattice, any integral 2-cycle can 
now be written as
\begin{equation}
p^{i}e^{i}+p_{i}e_{i}+q_{I}E_{I},\label{intk3}
\end{equation}
where $i=1,2,3$ and $I=1,...,16$. The $p_{i}$ as well as the $p^{i}$
are all integers, while the $q_{I}$ fulfill the relations $\sum_{I=1,...,8}q_{I}=2\mathbb{Z}$
and $\sum_{I=9,...,16}q_{I}=2\mathbb{Z}$ and furthermore have to be
\textit{all} integer or \textit{all} half-integer in each of the two
$E_{8}$ blocks.

\begin{figure}
\begin{center}
\includegraphics[height=6cm]{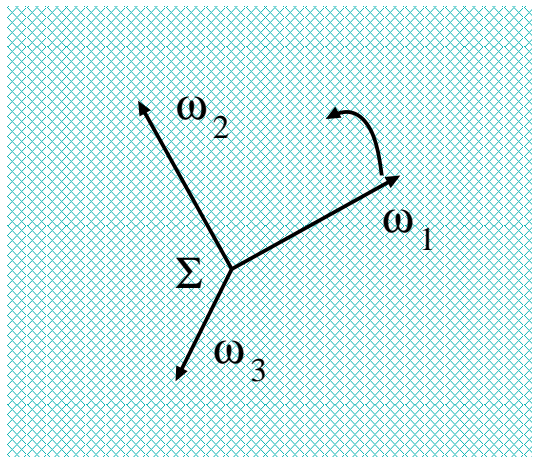}\label{figk3mod}
\caption{\textsl{The moduli of $K3$ surfaces correspond to rotations of a three-plane $\Sigma$, spanned by three 
positive-norm vectors $\omega_i$, in $H_2(K3)$.}}
\end{center}
\end{figure}

A point in the moduli space $M_{K3}$ of $K3$ is chosen, i.e. there is an implicitly defined Ricci flat metric,
by fixing the overallvolume of $K3$ and a positive signature 3-plane $\Sigma$ in
$H_{2}(K3,\mathbb{\mathbb{R}})\cong\mathbb{R}^{3,19}$, see Figure~\ref{figk3mod}.

We choose three real 2-cycles $\omega_{i}\in H_{2}(K3,\mathbb{\mathbb{R}})$,
$i=1,2,3$, which fulfill the constraints $\omega_{i}\cdot\omega_{j}=\delta_{ij}$ and span $\Sigma$. 
A real K\"ahler form $J$ and a holomorphic 2-form $\Omega^{2,0}$ for the $K3$ surface specified by $\Sigma$ 
are then given by $J=\sqrt{2\cdot \mbox{Vol}(K3)}\cdot\omega_{3}$ and $\Omega^{2,0}=\omega_{1}+i\omega_{2}$, 
respectively\footnote{Here and below, we use the same character for a 2-form, its associated
cohomology class and its Poincar\'{e}-dual 2-cycle.}. This description has an obvious $SO(3)$ symmetry
which acts by rotating the three 2-forms $\omega_{i}$. It leaves the $\Sigma$, and hence the metric,
invariant, but changes the complex structure. This gives $K3$ the structure of a Hyperk\"ahler manifold.

The Picard group, defined as
\be
\operatorname{Pic}(X)\equiv H^{1,1}(X)\cap H^{2}(X,\mathbb{Z}) \ ,
\ee
is given by the intersection of the lattice 
$H^{2}(X,\mathbb{Z})$ with the codimension-two surface 
orthogonal to the real and imaginary parts of $\Omega$. 
The dimension of $\operatorname{Pic}(X)$, also called Picard 
number, counts the number of algebraic curves and
vanishes for a generic $K3$ manifold.  

If we require $K3$ to admit an elliptic fibration with a section, i.e. an elliptic $K3$ described by 
a Weierstrass model, there are at least two algebraic curves embedded in $K3$ - the $T^2$ fiber and a section, the latter 
being equivalent to the base $\P^1$. Thus the space orthogonal to 
the plane defining the complex structure has a two-dimensional intersection 
with the lattice $H^{2}(K3,\mathbb{Z})$, which fixes two complex structure 
moduli\footnote{
This 
behavior is a specialty of $K3$ and is related to the fact that,
contrary to higher-dimensional Calabi-Yau spaces, $h^{1,1}\neq b^{2}$.
}.
One can show that the two vectors in the lattice corresponding to the base 
and the fiber form one of the $U$ factors in~\eqref{eq:intersectionmatrix}. 
%%%%% prove this
Thus, $\Omega$ has to be orthogonal to the subspace corresponding 
to this $U$ factor. The precise position of $J$, which lies completely in this 
$U$ factor, is fixed by the requirement that the fibre volume goes to zero
in the F-theory limit. The only remaining freedom is in the complex 
structure. As $\Omega^{2,0}$ must be orthogonal to $J$, it is confined to be a space-like 
two-plane in $\mathbb{R}^{2,18}$ with the inner product
\be
\label{K3 complex structure moduli space inner product}
U \oplus U \oplus -E_8 \oplus -E_8 \ .
\ee

There are $18$ complex structure deformations left in the elliptically 
fibred case: $\Omega$ may be expanded in twenty 2-forms, 
which leads to $20$ complex coefficients. However, there is 
still the possibility of an arbitrary rescaling of $\Omega$ by one complex 
number, as well as the complex constraint $\Omega\cdot\Omega=0$, so that we 
find an $18$-dimensional complex structure moduli space. 
Note that this is precisely the number of polynomial deformations of the
Weierstrass model description of $K3$, discussed in Section~\ref{sectmoduli}.

\subsection{Singularities}

As we have discussed in Section~\ref{ellsurf}, ADE singularities of complex surfaces arise
via vanishing two-spheres. By \eqref{selfintvsg}, cycles whose representatives are two-spheres
have self-intersection number $-2$. The converse also holds\footnote{This is non-trivial as there 
can be elements of $H_2(K3,\Z)$ that only have representatives that are the disjoint union of several curves.}, 
i.e. any element of the lattice $H_2(K3,\Z)$ that has self-intersection $-2$ has a representative which is a two-sphere \cite{peters}.
We will call these objects the \emph{roots} of the lattice $H_2(K3,\Z)$. 
As the volumes of cycles are measured by their projection onto the three-plane $\Sigma$, singularities 
develop at those points in moduli space at which there are roots that are orthogonal to $\Sigma$, 
i.e. elements $\gamma_I\in H_2(K3,\Z)$ that fulfill
 \begin{align}
\gamma_I\cdot \omega_i=0 \ , i=1..3  \hspace{1cm} \mbox{and} \hspace{1cm} \gamma_I\cdot\gamma_I=-2 \ .
 \end{align}
At a generic point in moduli space, no roots are orthogonal to $\Sigma$, so that our $K3$ surface is smooth.
If we happen to be at a point in moduli space that corresponds to a singular $K3$, we can resolve the singularity 
by rotating $\Sigma$ such that no more roots are orthogonal to it. 

As explained in the last section, we can decompose $\Sigma$ into the holomorphic 2-form $\Omega^{2,0}$ and the 
K\"ahler form $J$ in various ways. Thus rotations of $\Sigma$, and hence also resolutions of singularities, can be 
described as complex structure or K\"ahler deformations This choice gives rise to different complex structures 
but leaves the geometry otherwise untouched. This is the deeper reason why the two different methods of resolution 
presented in Section~\ref{ellsurf} gave the same result. The blow-up, which is equivalent to a K\"ahler deformation, 
has the same effect as the polynomial deformation, which is equivalent to a complex structure deformation, because
both correspond to the same deformation of the metric.

As the intersection pattern between the exceptional divisors characterizes the singularity, we can determine
the singularity that arises at a point $p$ in the moduli space of $K3$ by computing the intersection matrix
of the roots that are orthogonal to $\Sigma$ at $p$. This matrix is equal to minus the Cartan matrix of the corresponding 
Lie algebra. From the point of view of physics, this amazing correspondence between geometry and group theory
stems from the fact that gauge enhancement in compactifications of M-theory arises from singularities of
the compactification manifold. 

The beautiful isomorphism between singularities and Lie algebras can also 
be seen from a more modest perspective \cite{Aspinwall:1996mn}. Singularities of $K3$ that have
a crepant resolution, i.e. a resolution that respects the triviality of the canonical bundle, must correspond to
finite subgroups of $SU(2)$. This happens because $K3$ has holonomy group $SU(2)$, so that the holonomy that acts on a 
tangent vector upon encircling an isolated singularity must be a finite subgroup of $SU(2)$. The classification
of these groups happens to be the same as the classification of the ADE Lie algebras.

\subsection{Wilson line breaking and resolution of singularities}\label{wilson}

A different perspective on the moduli space of $K3$ can be gained by exploiting the duality
between M-theory on $K3$ and the heterotic $E_8\times E_8$ string on $T^3$. In this duality,
the Wilson lines, i.e. the gauge bundle, on $T^3$ is mapped to geometric deformations of $K3$.
This point of view is a very efficient tool to identify loci in the moduli space that feature
a specific singularities. In this section we give a brief introduction to Wilson line breaking
and show the similarity to the resolution of singularities of $K3$.

In gauge field theories based on a certain group, the symmetry can be broken by introducing 
Wilson lines associated with non-contractible loops of the underlying space-time geometry. This is
in one-to-one correspondence with Dynkin's method for finding maximal subgroups by deleting 
nodes in the extended Dynkin diagram. Altough we specifically discuss the case of $E_8$, the
structure of Wilson line breaking is of course similar for other groups.

The action of a Wilson line in $E_8$ (viewed as a vector in $\mathbb{R}^{8}$) on a root $\alpha$ is
\begin{equation}  \label{actionwilson}
\alpha\mapsto e^{2\pi i\alpha\cdot W}\alpha\:.\end{equation}
To find the sublattice of $E_{8}$ which corresponds to deleting a simple root $\alpha_{i}$,
we choose a Wilson line $W$ satisfying (see, e.g., \cite{key-51,key-51bis})
\begin{equation}
 \alpha_{i}\cdot W\not\in\mathbb{Z} \qquad\qquad \mbox{and}\qquad\qquad \alpha_{j}\cdot W\in\mathbb{Z} \qquad{\rm for }\: j\in\{1,...,9\}\setminus
\lbrace i \rbrace\:.
\end{equation}
Requiring this transformation to be a symmetry of the root lattice, we are left with the sublattice of roots satisfying 
$\alpha\cdot W\in\mathbb{Z}$ \cite{key-55,key-54}.

The $E_8\times E_8$ point in moduli space is realized when $\Sigma$ is located in the $U^{\oplus3}$ block spanned by $e_i,e^i$ ($i=1,2,3$)\footnote{Accordingly, $\omega_{i}^{E_{8}\times E_{8}}=a_{i}^{j}e_{j}+b_{k}^{i}e^{k}$, $i,j,k=1,2,3$, for real numbers $a_{i}^{j}$ and $b_{k}^{i}$ s.t. $\omega_{i}\cdot\omega_{j}=\delta_{ij}$.}.
%thus the simple roots of $\Lambda$, the sublattice of $\Gamma_{3,19}$ spanned by the roots in the orthogonal complement of $\Sigma$, give twice the Dynkin diagram of $E_{8}$.
Rotating the plane into the $E_8\times E_8$ block changes the singularity and eventually gives rise to a smooth $K3$. Singularities which may still be present after this rotation correspond to subgroups of $E_8\times E_8$. As we explain in detail in the following, one can relate the symmetry breaking by Wilson lines described above to the rotation of the $\Sigma$ plane in $H_2(K3,\mathbb{R})$.

Let us consider the rotation of $\Sigma$ from a point with $E_8\times E_8$ singularity to a point with $D_8\times E_8$ singularity as an
example. The Wilson line that realizes this breaking is $W_{I}=(1,0,0,0,0,0,0,0)$. This identifies a vector  $W=W_{I}E_{I}=(1,0,0,0,0,0,0,0)$ in the subspace of $H_{2}(K3,\mathbb{R})$ that corresponds to the first $E_{8}$ block in \eqref{eq:intersectionmatrix}. Let us now rotate $\Sigma$ in the direction of $W$. We can do this by rotating one of basis vectors of $U^{\oplus3}$ (where $\Sigma$ lives), e.g. $e^{1}$, in this direction: $e^{1}\rightarrow e^{1}+\beta\, W$, $\beta\in\mathbb{R}$. Once this rotation has been performed, $\Sigma$ is located in the subspace of $H_{2}(K3,\mathbb{R})$ spanned by
\begin{equation}
  e_{1}, \qquad e^{1}+\beta W, \qquad e_{2}, \qquad e^{2}, \qquad e_{3},  \qquad e^{3}\:.
\end{equation}
For a generic position of $\Sigma$ in this six dimensional space and for generic values of $\beta$, the lattice $\Lambda$ orthogonal to $\Sigma$ is of the type $D_{7}\times E_{8}$.\footnote{In M-theory on the $K3$ surface given by $\Sigma$ this leads to the gauge group $SO(14)\times U(1)\times E_{8}$.} Reinterpreting~\eqref{eq:e8rootsystem} as a set of simple roots of $\Gamma_{E_{8}\times E_{8}}$, this can be understood from the fact that the cycle corresponding to $\alpha_{1}$ as well as the cycle corresponding to the highest root (\ref{eq:e8highestroot}) acquire finite volume.
For $\beta=1$, we find that $\alpha_{9}+e_{1}$ is orthogonal to $e^{1}+\beta W$ (and hence to $\Sigma$) and we therefore have a further 
independent shrinking cycle. This results in a change of singularity type to $D_{8}\times E_{8}$.\footnote{This corresponds to gauge enhancement $SO(16)\times E_{8}$ in M-theory.} The fact that we found another shrinking cycle is due to the integrality of $\alpha_{9}\cdot W$. Thus the orthogonality of $\alpha_{9}+e_{1}$ to $e^1+\beta W$ for $\beta=1$ corresponds to the previously discussed condition for the highest root to survive after introducing the Wilson line $\beta W_I$.

This reasoning can be extended to a generic rotation of $\Sigma$ into the $E_{8}\times E_{8}$ block. The three vectors that span
$\Sigma$ correspond to the three Wilson lines that can be switched on on $T^3$ on the heterotic side. Let us be more explicit here.
A situation in which all Wilson lines vanish corresponds to a point with gauge symmetry $E_8 \times E_8$. Hence the three $\omega_i$
all lie in the three $U$ blocks spanned by $e^i, e_i, \ i=1,2,3$. Let us choose 
\begin{align}
\omega_1&=c_1 e_1+ d_1 e^1 \nn\\ 
\omega_2&=c_2 e_2+d_2 e^2 \nn\\
\omega_3&=c_3 e_3+d_3 e^3 \ .
\end{align}
The constants $c_i$ and $d_i$ should be chosen such that $\omega_i\cdot\omega_i=1$. Note that different $\omega_i$ are
automatically orthogonal.

Switching on three orthogonal\footnote{It is clear that we can always redefine three non-orthogonal Wilson
lines in such that they are orthogonal without altering their action on the roots, \eqref{actionwilson}.} Wilson lines 
$W_I^i$ in $E_{8}\times E_{8}$ on the heterotic side corresponds to the point in moduli space of $K3$ described by
\begin{align}\label{wilsontosing}
\omega_1&=c_1 e_1+ d_1 (e^1+W_I^1E_I) \nn\\ 
\omega_2&=c_2 e_2+d_2 (e^2+W_I^2E_I) \nn\\
\omega_3&=c_3 e_3+d_3 (e^3 +W_I^3E_I)\ ,
\end{align}
where summation over $I=1...16$ is understood. Note that we have to adjust the constants $c_i$ and $d_i$ to maintain
the condition $\omega_i\cdot\omega_i=1$. As the Wilson lines are mutually orthogonal, this also holds for the
$\omega_i$. 

If the three Wilson lines $W_I^i$ break $E_8\times E_8$ to a subgroup $G$, this group characterize
the singularities of $K3$ at the point in moduli space described by \eqref{wilsontosing}. This
allows us to use the techniques of Wilson line breaking of gauge groups to find a point in the
moduli space of $K3$ at which $K3$ develops specific singularities.

For elliptic $K3$ surfaces, the $SO(3)$ symmetry between the three $\omega_i$ is lost, i.e. we have to
choose which direction corresponds to the holomorphic 2-form $\Omega^{2,0}$ and which
to the K\"ahler form $J$. Furthermore, one of the $U$ blocks must be chosen to correspond to the fibre and the base of
the elliptic fibration. The two vectors spanning this $U$-block must be orthogonal to $\Omega^{2,0}$.
Let us choose e.g. $J=\omega_1$ and $\Omega^{2,0}=\omega_2+\iu \omega_3$.
As we discuss in detail in Section~\ref{sec9}, the Wilson line entering $J$ is set to zero in the F-theory 
limit. Hence the moduli space of F-theory on $K3$ is governed by only two Wilson lines. This is appropriate,
as F-theory on $K3$ is dual to the heterotic $E_8 \times E_8$ string on $T^2$, which allows only two
Wilson lines.

\section{K3 moduli space and F-theory}\label{cyclesk3branes}

In this section we give the F-theory interpretation of elliptic $K3$ surfaces. This
means that we map the complex structure moduli space of elliptic $K3$ surfaces to 
the movement of branes on $\P^1$. We have already discussed that configurations with coincident branes are
characterized by singularities which arise via collapsed two-spheres. The very same
2-cycles are used in the conventional parameterization of the complex structure moduli space of $K3$,
where one considers the period integrals,
\begin{equation}
\int_{C_I} \Omega^{2,0}=\pi_I \ .
\end{equation}
Geometrically, these periods are nothing but the volumes of $C_I$. Hence, the volumes of
2-cycles of $K3$ must determine the positions of the branes of F-theory. In this section
we show how this works in the weak coupling limit. In particular, we construct the relevant 
2-cycles from the monodromies of D-branes and O-planes.

\subsection{Cycles and branes}

In this section we want to gain a more intuitive understanding of the cycles
that are responsible for the brane movement. For this we picture the 
elliptically fibred K3 locally as the complex plane (in which branes are 
sitting) to which a torus has been attached at every point. We will construct 
the relevant 2-cycles geometrically. We have already seen that the cycles in 
question shrink to zero size when we move the branes on top of each other, 
so that these cycles should be correlated with the distance between the
branes. Remember that D7-branes have a non-trivial monodromy acting on the 
complex structure of the fibre as $\tau \to \tau +1$, which has 
\be
T = \left( \begin{array}{cc}
1 & 1 \\
0 & 1 \\
\end{array}
\right) 
\label{monodromy matrix T}
\ee
as its corresponding $SL(2,\mathbb{Z})$ matrix.
Similarly, O7-planes have a monodromy of $-T^4$, where 
the minus sign indicates an involution of the torus,
meaning that the complex coordinate $z$ of 
the torus goes to $-z$. Thus 1-cycles 
in the fibre change orientation when they are moved around 
an O-plane. 

\begin{figure}[!ht]
\begin{center}
\includegraphics[height=3.5cm,]{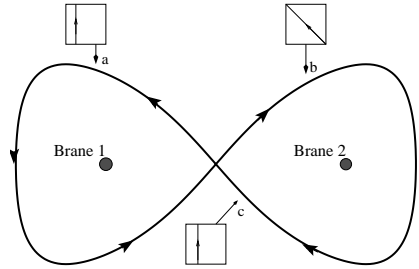}\caption{\textsl{The 
cycle that measures the distance between
two D-branes. Starting with a cycle in the $(0,1)$ direction 
of the fibre torus at point $a$, this cycle is tilted to 
$(-1,1)$ at $b$. Because we surround the second brane in the  
opposite way, the cycle in the fibre is untilted again so it 
can close with the one we started from.}}
\label{Cyclebetweentwobranes}
\end{center}
\end{figure}

If we want to describe a 2-cycle between two D-branes, 
it is clear that it must have one leg in the base and one
in the fibre to be distinct from the 2-cycles describing 
the fibre and the base. Now consider the 1-cycle being 
vertically stretched in the torus\footnote{Of course this 
notion depends on the $SL(2,\mathbb{Z})$-frame we consider, 
but anyway we construct the cycle in the frame where the 
branes are D-branes. In the end the constructed cycle will 
be independent of the choice of frame.}. If we transport 
this cycle once around a D-brane and come back to the same 
point, this cycle becomes diagonally stretched because of 
the T-monodromy. If we then encircle another brane in the 
opposite direction, the 1-cycle returns to its original form, so that it 
can be identified with the original 1-cycle. 
This way to construct a closed 2-cycle 
was already mentioned in~\cite{Sen:1996vd}. 
This 2-cycle cannot be contracted to a point since it 
cannot cross the brane positions because of the monodromy 
in the fibre. The form of the 2-cycle is illustrated in 
Figure~\ref{Cyclebetweentwobranes}. We emphasize that to get a non-trivial 
cycle, its part in the fibre torus has to have a vertical component.

\begin{figure}[!ht]
\begin{center}
\includegraphics[height=3.5cm,]{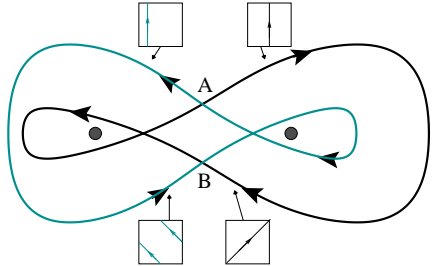}\caption{\textsl{The 
self-intersection number of a cycle between two D-branes. 
As shown in the picture, we may choose the fibre part of both cycles to be 
$(0,1)$ at $A$, so that they do not intersect at this point. At $B$ 
however, one of the two is tilted to $(1,1)$, whereas the other has undergone 
a monodromy transforming it to $(-1,1)$. Thus the two surfaces
meet twice in point $B$.}}\label{Cyclesbetweentwobranesintersecting}
\end{center}
\end{figure}

\begin{figure}[!ht]
\begin{center}
\includegraphics[height=3.5cm,]{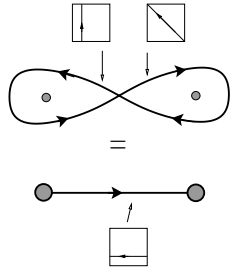}\caption{\textsl{The loop
between two D-branes can be collapsed to a line by 
pulling it onto the D-branes and annihilating the
vertical components in the fibre. All that remains
is a cycle which goes from one brane to the other
while staying horizontal
in the fibre all the time.}}
\label{looptoline}
\end{center}
\end{figure}

Next we want to compute the self-intersection number.
In order to do this, we consider a homologous cycle and 
compute the number of intersections with the original one.
By following the way the fibre part of the cycle evolves, one 
finds that the resulting number is minus two 
(see Figure~\ref{Cyclesbetweentwobranesintersecting}). The minus 
sign arises from the orientation. This is precisely what we expected 
from the previous analysis. To see the topology of the cycle more clearly,
it is useful to combine the two lines in the base stretching between the 
branes to a single line. This is shown in Figure~\ref{looptoline}. 
The component of the cycle in the fibre is then an $S^1$
that wraps the fibre in the horizontal direction, so that 
it shrinks to a point at the brane positions. 
Thus it is topologically a sphere, which fits 
with the self-intersection number of $-2$ and the
discussion of the previous section.

Now we want to determine the intersection number between 
different cycles and consider a situation with 
three D7-branes. There is one cycle between
the first two branes and one between the second and the third, 
each having self-intersection number $-2$. From 
Fig.~5 it should be clear
that they intersect exactly once.
If one now compares the way they intersect to the
figure that was used to determine the self-intersection number,
one sees that the two surfaces meet with one direction reversed, 
hence the orientation differs and we see that the mutual intersection 
number is $+1$. Thus we have shown that the intersection 
matrix of the $N-1$ independent cycles between $N$ D-branes is 
minus the Cartan Matrix of $SU(N)$.

\begin{figure}[!ht]
\begin{center}
\label{3branes}
\includegraphics[height=3.5cm,]{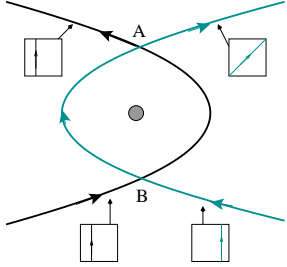}\caption{\textsl{Mutual 
intersection of two cycles.
Start by taking both cycles to have fibre part $(0,1)$ at $B$. The fact
that we closed a circle around the D-brane tells us that one of the two
has been tilted by one unit at $A$. Thus they meet precisely once.}}

\end{center}

\end{figure}

\begin{figure}[!ht]
\begin{center}
\includegraphics[height=3cm,]{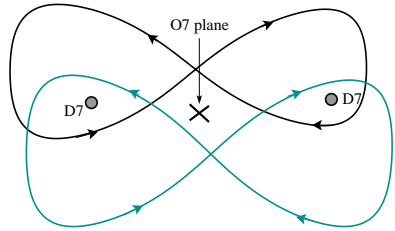}\caption{\textsl{Cycles that measure
how D-branes can be pulled onto O-planes.}}\label{DandO}
\end{center}
\end{figure}

We now want to analyze the cycles that arise in the
presence of an O7-plane. Two D7-branes in the vicinity of an
O7-plane can be linked by the type of cycle considered above. 
However, there are now two ways to connect the D-branes with each other: 
we can pass the O-plane on two different sides, as shown in Figure~\ref{DandO}.
By the same argument as before, each of these cycles has
self-intersection number $-2$. To get their
mutual intersection number, it is important to
remember the monodromy of the O-plane, which contains an involution of the torus 
fiber. Thus, the intersection on the right and the intersection on the left, 
which differ by a loop around the O-plane, have opposite sign. As a result, 
the overall intersection number vanishes.

\begin{figure}[ht!]
\begin{center}
\includegraphics[height=2cm,]{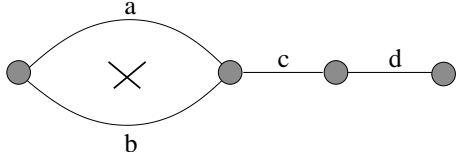}\caption{\textsl{Four D-branes and
an O-plane. The D-branes are displayed as circles and
the O-plane as a cross. To simplify the picture we have drawn
lines instead of loops.}}\label{four D and and an O}
\end{center}
\end{figure}

Now we have all the building blocks needed to discuss the 
gauge enhancement in the orientifold limit, which is $SO(8)^4$. 
We should find an $SO(8)$ for each O7-plane with four D7-branes on top of 
it. The cycles that are blown up when the four D7-branes move away from the 
O7-plane are shown in Figure~\ref{four D and and an O}. It is clear from the 
previous discussion that all cycles have self-intersection number $-2$ and 
cycle $c$ intersects every other cycle precisely once. Thus, collecting the 
four cycles in a vector $(a,b,c,d)$, we find the intersection form
\begin{equation}
D_4=\left(\begin{array}{cccc} -2 & 0 & 1 & 0\\
			   0 & -2 & 1 &0  \\
			   1 &  1 & -2 & 1 \\
			   0 & 0  & 1 & -2 \end{array}\right) 
\label{D4matrix} \ ,
\end{equation}
which is minus the Cartan matrix of $SO(8)$. It is also easy to
see that collapsing only some of the four cycles yields minus the Cartan
matrices of the appropriate smaller gauge enhancements.

Let us make a comment regarding F-theory on $K3$ away from the weak coupling
limit. Remember the representation of the 2-cycle that controls the distance between two D7-branes as
a line, Figure~\ref{looptoline}. The fibre component of this cycle is such that its shrinks at the position of the D-branes.
These are in turn characterized by their $(p,q)$ charge, which is $(1,0)$, so that the A cycle in Figure~\ref{torus} shrinks. 
Our construction of cycles hence is reminiscent of open strings in the presence of $(p,q)$ branes 
\cite{Johansen:1996am, Gaberdiel:1997ud, DeWolfe:1998zf}. 
Thus it is in principle clear how the cycles that make up the Dynkin diagrams of exceptional groups come about. In this
case one needs multi-pronged strings, so that there will be cycles that connect three instead of two branes in our picture. 
This gives us another point of view on the fact that elliptic $K3$ surfaces have only 18 complex degrees of freedom, but describe 
$24$ $(p,q)$ branes.

\subsection[$K3$ with four $D_4$ singularities]{\boldmath$K3$ with four \boldmath$D_4$ singularities}\label{so8}

In the last section we have discussed the cycles that govern the positions of
four D-branes grouped around an O-plane. As there is an $D_4$ singularity when
all four D-branes are on top of the O-plane, the intersection pattern of the
cycles is minus the Cartan matrix of $SO(8)$. This happens four times in the orientifold 
limit of $K3$, in which there are four $D_4$ singularities. We want to find the identify
which cycles in $H_2(K3,\Z)$ shrink to produce the orientifold limit.
For this we have to find a change of basis of $H_2(K3,\Z)$ such that the $D_4^{\oplus 4}$ 
becomes obvious. To find this change of basis, we first give the holomorphic 2-form $\Omega^{2,0}$
that describes an elliptic $K3$ with four $SO(8)$ singularities using the results of Section~\ref{wilson}.

We are interested in an elliptic $K3$ that is described by a Weierstrass equation, so that we have a section
which intersects each fibre once. As discussed above, this means that the symmetry between $J$ and $\Omega^{2,0}$
is broken. As fibre and base must be orthogonal to the holomorphic 2-form, $\Omega^{2,0}$ is confined to be an
expansion of cycles that live in the lattice
\be
U \oplus U \oplus -E_8 \oplus -E_8 \ ,
\ee
and $J$ is confined to fibre and base. Hence we can only use two\footnote{Note that we are not in F-theory limit until we send $J$ in
the limit in which the fibre shrinks. It is the great advantage of the Weierstrass model that it already anticipates
what will happen in the F-theory limit. We shall have more to say to this Section~\ref{sec9}.} Wilson lines, which fits with the fact 
that F-theory on $K3$ is dual to the heterotic $E_8\times E_8$ string on $T^2$.

Using the template of Section~\ref{wilson} we can then write the complex structure
of an elliptic $K3$ described by a Weierstrass model as
\begin{equation}\label{omegagen}
 \Omega=e^1+\tilde{U}e^2+\tilde{S}e_2-\left(\tilde{U}\tilde{S}+\frac{1}{2}W^2\right)e_1+W_IE_I \ .
\end{equation}
We have defined $W_I=W^1_I+\tilde{U}W^2_I$ and use $W^2$ as a shorthand for $(W^1_IE_I+\tilde{U}W^2_IE_I)^2$.

\begin{figure}[ht!]
\begin{center}
\includegraphics[height=6cm,]{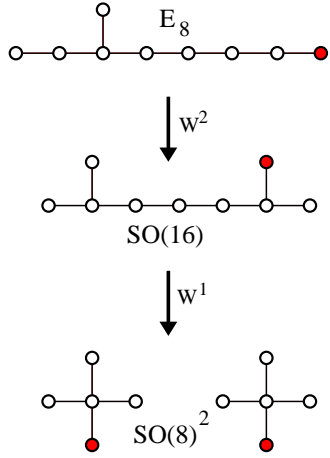}\caption{\textsl{The breaking pattern of $E_8$ to
$SO(8)\times SO(8)$ that is induced by $W_1$ and $W_2$. The highest roots are  marked in red.}}\label{e8toso8}
\end{center}
\end{figure}

Two Wilson lines that break $E_8 \times E_8$ down to $SO(8)^4$ are given by
\begin{align}
W^1=\left(0^4,\frac{1}{2}^4,0^4,\frac{1}{2}^4\right)\hspace{1cm}
W^2=\left(1,0^7,1,0^7\right) \ .
\label{Wilson lines SO8}
\end{align}
We have already shown in Section~\ref{wilson} that $W^2$ breaks $E_8 \times E_8$ down to 
$SO(16) \times SO(16)$. Furthermore, $W^1$ deletes the root $\alpha_5$, see \eqref{eq:e8rootsystem},
but leaves the highest root of $SO(16)$ untouched, see Figure~\ref{e8toso8}.

Plugging $W^1$ and $W^2$ into \eqref{omegagen} we find $\Omega$ at the $SO(8)^4$ point:
\begin{align}
\Omega_{SO(8)^4}=e^1+\tilde{U}e^2+\tilde{S}e_2-\left(\tilde{U}\tilde{S}
-1-\tilde{U}^2\right)e_1+W_I E_I \ .
\label{omegaso8}
\end{align}
Note that setting $\tilde{S}=2\iu$ and $\tilde{U}=\iu$ 
reproduces the complex structure given 
in~\cite{Gorlich:2004qm}.

We now show that the lattice vectors orthogonal to $\Omega_{SO(8)^4}$ span the lattice 
$D_4^{\oplus 4}$. Using the expansion~(\ref{intk3}), their coefficients have to 
satisfy
\begin{equation}
-\left(\tilde{U}\tilde{S}-1-\tilde{U}^2\right)p^1+p_1-W^1_{I}q_{I}+
\tilde{S}p^2+\tilde{U}(p_2-W^2_{I}q_{I})=0 \ .
\end{equation}
As we know that these lattice vectors must be orthogonal
to the complex structure for every value of $\tilde{U}$ and
$\tilde{S}$, we find the conditions
\begin{align}
p^1=0 &\hspace{1cm} p_1-W^1_{I}q_{I}=0 \nonumber \\
p^2=0 &\hspace{1cm} p_2-W^2_{I}q_{I}=0 \ .
\end{align}
These equations are solved by the following four groups of 
lattice vectors:

\vspace{.5cm}

\begin{center}
\begin{tabular}[h]{c|c|c|c|c}
& $A$&$B$&$C$&$D$\\
\hline
1&$E_{7}-E_{8}$&$-E_{15}+E_{16}$&
$-e_2-E_{1}+E_{2}$&$e_2+E_{9}-E_{10}$\\
2&$E_{6}-E_{7}$&$-E_{14}+E_{15}$ &
$-E_{2}+E_3$&$E_{10}-E_{11}$\\
3&$-e_1-E_{5}-E_{6}$&$e_1+E_{13}+E_{14}$&
$-E_{3}+E_{4}$&$E_{11}-E_{12}$\\
4&$E_{5}-E_{6}$&$-E_{13}+E_{14}$&
$-E_{3}-E_{4}$&$E_{11}+E_{12}$       

\end{tabular} 
\vspace{-1.7cm}
\begin{equation}
\label{AtoD}
\end{equation}
\end{center}

\vspace{1.5cm}

It is not hard to see that there are no mutual intersections between the 
four groups, and that the intersections within each group are given by the 
$D_4$ matrix~\eqref{D4matrix}. This serves as an explicit check 
that~(\ref{omegaso8}) is indeed the correct holomorphic 2-form of 
$K3$ at the $SO(8)^4$ point.

\begin{figure}[ht!]
\begin{center}
\includegraphics[height=2cm,]{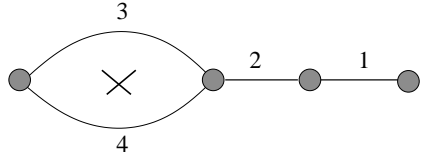}\caption{\textsl{The assignment between 
the geometrically constructed cycles between branes and the cycles of the 
table in the text. Note that the distribution of the cycles 1,3 and 4 is 
ambiguous.\label{1block}}}
\end{center}
\end{figure}

It should be clear that one can choose different linear combinations 
of the basis vectors in each block that still have the same inner 
product. This only means we can describe the positions of the D-branes by 
a different combination of cycles, which are of course linearly dependent on 
the cycles we have chosen before and span the same lattice.
We can make an assignments between the cycles in the table and the cycles 
constructed geometrically as shown in Figure~\ref{1block}.

\begin{figure}[ht!]
\begin{center}
\includegraphics[height=5cm,]{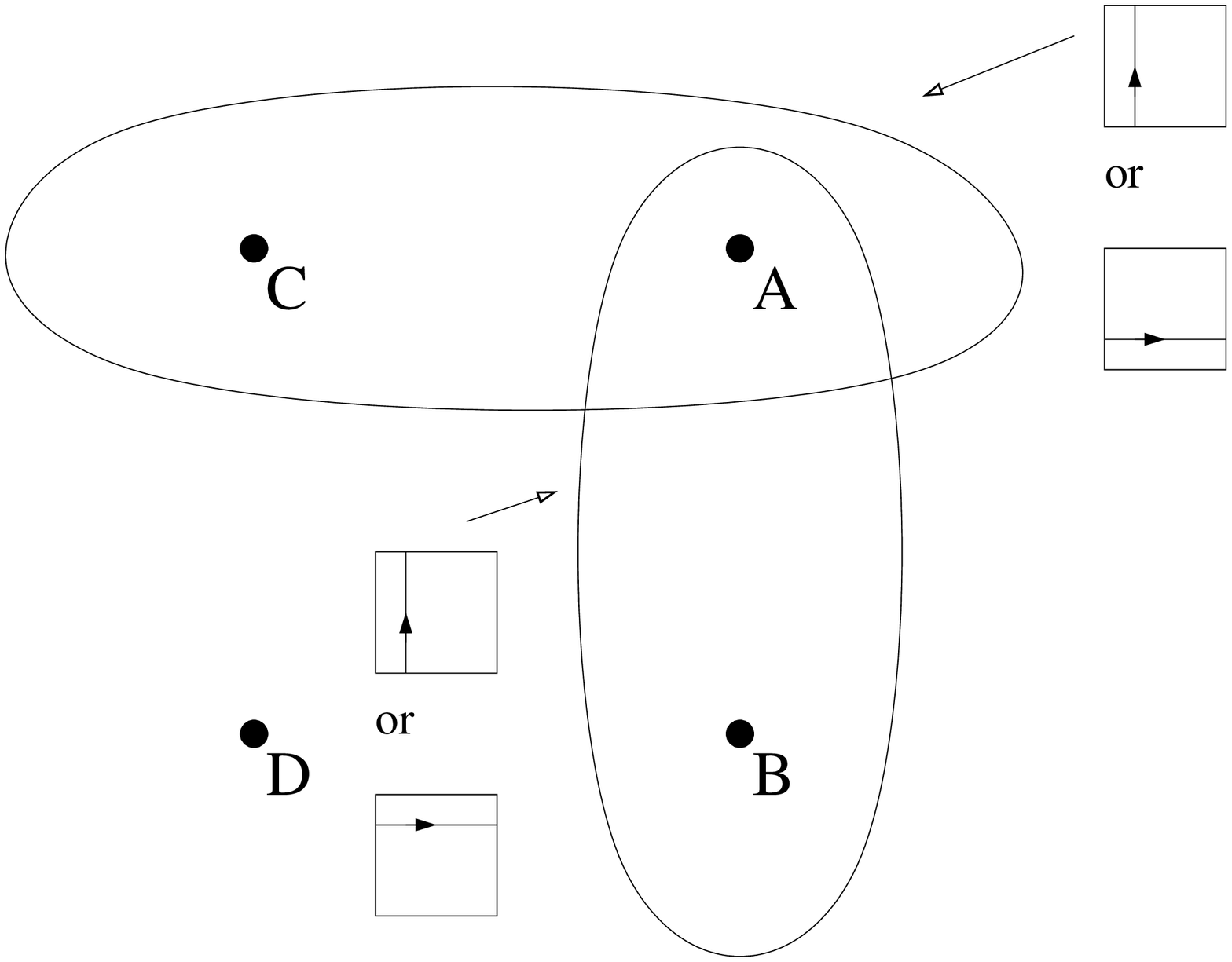}\caption{\textsl{The
torus cycles have to encircle two blocks to ensure trivial monodromy 
along the basis part of the cycles. Note that the cycles which are
orthogonal in the base and in the fibre intersect twice because of the
orientation change introduced in going around the O-plane.\label{fourt}}}
\end{center}
\end{figure}

When we are at the $SO(8)^4$ point, where $16$ of the $20$ cycles of $K3$ 
have shrunk, the only remaining degrees of freedom are the deformations of 
the remaining\footnote{As we explain in detail in Section~\ref{sec61}, this 
space is not quite the same as $T^4/Z_2$. For our purposes it is, however, okay 
to think of it as such.} $T^4/Z_2$. The four cycles describing these 
deformations have to be orthogonal to all of the $16$ brane cycles. There 
are four cycles satisfying this requirement,
\begin{align}
e^1+W_{I}^1E_{I} &\hspace{1cm} e^2+W^2_IE_I \nonumber \\
e_1 & \hspace{1cm} e_2 \ ,		\label{orthomega}
\end{align}
and the torus cycles must be linear combinations of them.
From the fibration perspective, the torus cycles are the cycles 
encircling two blocks (and thus two O-planes), so that the monodromy along 
the base part of those cycles is trivial. They can be either horizontal or 
vertical in the fibre, giving the four possibilities displayed in 
Figure~\ref{fourt}. Note that all of them have self-intersection zero and only 
those that wrap both fibre and base in different directions intersect 
twice. 

To find out which linear combination of the forms in (\ref{orthomega}) 
gives which torus cycle, we will consider a point in moduli space where 
the gauge symmetry is enhanced from $SO(8)^4$ to $SO(16)^2$. At this point, 
there are $16$ integral cycles orthogonal to $\Omega$ the intersection matrix 
of which is minus the Cartan matrix of $SO(16)^2$. Furthermore, we know 
that this situation corresponds to moving all D-branes onto two O7-planes. 
We will achieve this leaving two of the four blocks untouched, while 
moving the D-branes from the other blocks onto them. This means that we blow 
up one of the cycles in each of the blocks that are moved, while collapsing 
two new cycles that sit in between the blocks. Doing this we find three 
independent linear combinations of the cycles in~(\ref{orthomega}) that do 
not intersect any of the cycles that are shrunk. 

Before explicitly performing this computation, we choose a 
new basis that is equivalent to \eqref{orthomega}:
\begin{align}
\alpha\equiv2\left(e^1+e_1+W^1_IE_I\right) 
&\hspace{1cm}\beta\equiv2(e^2+e_2+W^2_IE_I)\nonumber \\
e_1 &\hspace{1cm} e_2 \ .	\label{basistcyc}
\end{align}
In this basis we can write $\Omega$ at the $SO(8)^4$ point as
\begin{align}
\Omega_{SO(8)^4}=\frac{1}{2}\left(\alpha+Ue_2+S\beta-USe_1\right) \ . 
\label{omegaUS}
\end{align}
We also have switched to a new parameterization in terms of 
$U$ and $S$. They will turn out to be the complex structures of the 
base and the fibre torus.

\begin{figure}[ht!]
\begin{center}
\includegraphics[height=2cm,]{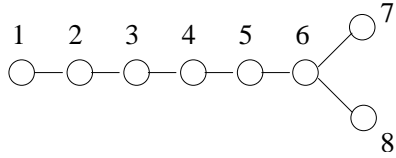}\caption{\textsl{The Dynkin diagram of 
$SO(16)$.}}
\label{dynkin16}
\end{center}
\end{figure}

Let us now go to the $SO(16)^2$ point by setting $W^1_I=0$ in 
(\ref{omegaso8}). After switching again from $\tilde{S}$ and $\tilde{U}$ to 
$S=\tilde{U}$ and $U/2=\tilde{S}-\tilde{U}$, we find
\begin{equation}
\Omega_{SO(16)^2}=e^1+\frac{U}{2}e_2+\frac{S}{2}\beta-\frac{US}{2}e_1 \ .  \label{omegaso16} 
\end{equation}
The $16$ integral cycles that are orthogonal to $\Omega$ are:
\begin{center}
\begin{tabular}[h]{c|c|c}
 & $E$&$F$\\  \hline
1&$-e_2-E_1+E_2$&$e_2+E_9-E_{10}$\\
2&$-E_2+E_3$&$E_{10}-E_{11}$\\
3&$-E_3+E_4$&$E_{11}-E_{12}$\\
4&$-E_4+E_5$&$E_{12}-E_{13}$\\
5&$-E_5+E_6$&$E_{13}-E_{14}$\\
6&$-E_6+E_7$&$E_{14}-E_{15}$\\
7&$-E_7+E_8$&$E_{15}-E_{16}$\\
8&$E_7+E_8$&$E_{15}+E_{16}$\\

\end{tabular}
\vspace{-2.8cm}
\begin{equation}
\label{EandF}
\end{equation}

\end{center}
\vspace{2cm}
They are labelled as shown in Figure~\ref{dynkin16}. Note that this means that 
we have moved block $A$ onto block $C$ and block $B$ onto block $D$.

One can check that, out of the basis displayed 
in (\ref{basistcyc}), only $\alpha$ has a 
non-vanishing intersection with some of the $16$ forms
above, whereas all of them are orthogonal to
$e_1,e_2,\beta$. Thus, $\alpha$ is contained in the cycle that
wraps the fibre in vertical direction and passes in between $A,B$ and $C,D$
(cf. Figure~\ref{fourt}). Furthermore, the non-zero intersection
between $e_1$ and $\alpha$ tells us that $e_1$ wraps the fibre 
horizontally (for this argument we used $e_1\cdot e_2=e_1\cdot\beta=0$).
Given these observations, it is natural to identify the four cycles 
(\ref{basistcyc}) with the four cycles displayed in Figure~\ref{fourt}. 
More specifically, we now know that $\alpha$ is vertical in the fibre and 
passes in between $A,B$ and $C,D$, while $e_1$ is horizontal in the fibre 
and passes in between $A,C$ and $B,D$. The four cycles characterize 
the shape of $T^4/Z_2$. Other possible assignments between 
the cycles of (\ref{basistcyc}) and those displayed in Figure~\ref{fourt} 
correspond to reparameterizations of the tori and are therefore 
equivalent to our choice. 

\begin{figure}[ht!]
\begin{center}
\includegraphics[height=3cm,]{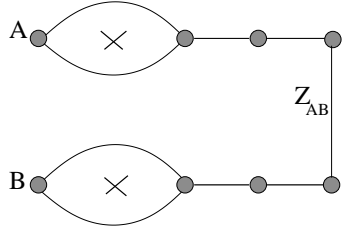}\caption{\textsl{The cycles
connecting the blocks. Note that all cycles in this
picture are built in the same way and thus all
lie horizontally in the fibre.\label{ZAB}}}
\end{center}
\end{figure}

We now have to assign the cycles $e_2$ and $\beta$ to the two remaining 
cycles of Figure~\ref{fourt}. For this purpose, we will explicitly construct 
the cycles $Z_{XY}$ between the four $SO(8)$ blocks. Since they can be 
drawn in the same way as the cycles between the D-branes, 
cf.~Figure~\ref{ZAB}, we see that all of them must have self-intersection 
number $-2$. We also know that their mutual intersections should be 
$Z_{XY}\cdot Z_{YZ}=1$. From what we have learned so far, all of them should 
be either orthogonal to $\beta$ and $e_1$ or orthogonal to $e_2$ and $e_1$. 
It is easy to check that the first case is realized by:
\begin{align}
Z_{AC}=E_{8}-E_1-e_2&\hspace{1cm}Z_{AB}=-(e_2+e^2)-E_1+E_8-E_9+E_{16}+e_1 
\nonumber \\
Z_{DB}=E_{16}-E_{9}-e_2&\hspace{1cm}Z_{CD}=e_2-e^2 \ .
\end{align}
One can show that the second case is not possible. This can be seen 
from the following argument:

If we can find $Z$-cycles that are orthogonal to $e_1$ and $e_2$,
we can decompose them as
\be
Z_{XY}=q_I E_I \ .
\ee
Note that the $e_i$ are now responsible for making the $Z$-cycles wind 
around the base torus, so that we do not loose any generality by omitting 
them in the decomposition above.
Because of the constraint $\sum q_{I}=2\mathbb{Z}$, we can only have
$Z_{AB}$ intersecting one of the cycles in block $A$ by putting
$q_I=\pm \frac{1}{2}$ appropriately for $I=5..8$. By the structure
of the lattice, \eqref{intk3}, this forces 
us to also set $q_I=\pm \frac{1}{2}$ for $I=1..4$. It is clear that
this will also make this cycle intersect with one of the cycles in block $C$, 
contradicting one of its defining properties. This means that we simply 
cannot construct the $Z$-cycles to be all orthogonal to $e_1$ and $e_2$. 

We can now use the intersections of the $Z$-cycles with
the complex structure at the $SO(8)^4$ point,
eq.(\ref{omegaUS}),
to measure their length and consistently distribute 
the four blocks on the pillow. We find that
\begin{align}
Z_{AB}\cdot\Omega_{SO(8)^4}=-U/2 & \hspace{1cm}
Z_{AC}\cdot\Omega_{SO(8)^4}=-\frac{1}{2} \nonumber \\
Z_{CD}\cdot\Omega_{SO(8)^4}=-U/2 & \hspace{1cm}
Z_{DB}\cdot\Omega_{SO(8)^4}=-\frac{1}{2} \ .
\end{align}

\begin{figure}[!ht]
\begin{center}
\includegraphics[height=7cm,]{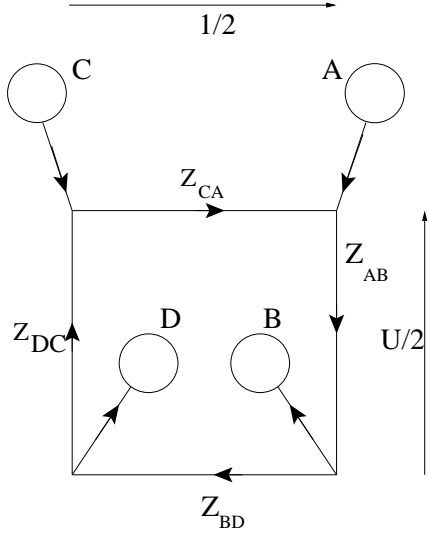}\caption{\textsl{A schematic picture 
of the cycles between the blocks. We have drawn arrows to indicate the 
different orientations. Note that the $Z$-cycles have intersection
$+1$ with the cycle they come from and $-1$ with the cycle they go to.
Note that they sum up to a torus cycle, $e_1$, telling us
which blocks are encircled.}\label{zcycles}}
\end{center}
\end{figure}

\begin{figure}[!ht]
\begin{center}
\includegraphics[height=8cm,]{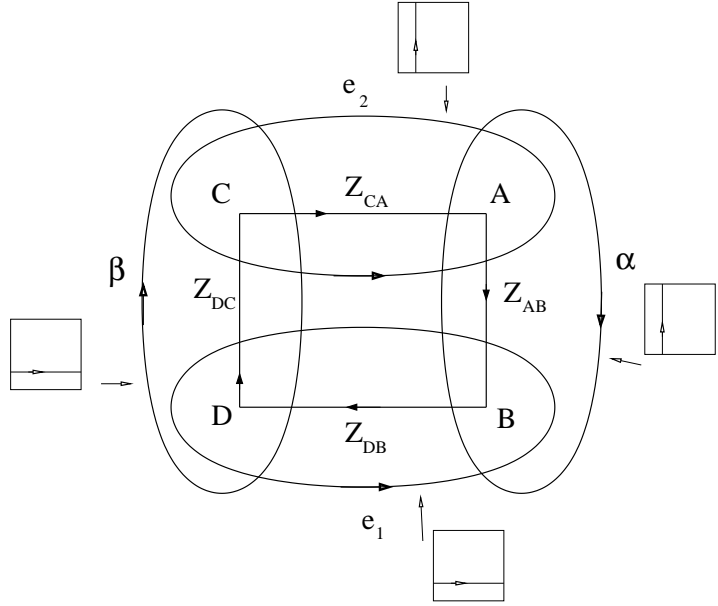}\caption{\textsl{The distribution
of the torus cycles.}\label{tcyc}}
\end{center}
\end{figure}

It is clear that we can add the two orthogonal
cycles $e_1$ and $\beta$ to the $Z$-cycles,
$Z\rightarrow Z+ne_1+m\beta$, without destroying their
mutual intersection pattern. However, this changes their length
by $n+Um$. This means that we can make the $Z$-cycles wind around the 
pillow $n$ times in the real and
$m$ times in the imaginary direction. Calling the real 
direction of the base (fibre) $x$ ($x'$) and the imaginary direction
of the base (fibre) $y$ ($y'$), we can now make the
identifications:
\begin{align}
e_1 \hspace{1ex}\mbox{winds around}\hspace{1ex}
x\hspace{1ex} \mbox{and}\hspace{1ex} x' \nonumber \\
e_2 \hspace{1ex}\mbox{winds around}\hspace{1ex}
x \hspace{1ex}\mbox{and}\hspace{1ex} y'\nonumber \\
\alpha \hspace{1ex}\mbox{winds around}\hspace{1ex}
y \hspace{1ex}\mbox{and}\hspace{1ex} y'\,\nonumber \\
\beta \hspace{1ex}\mbox{winds around}\hspace{1ex}
y \hspace{1ex}\mbox{and}\hspace{1ex} x'.
\end{align}
Alternatively one can find the positions of $e_2$ and
$\alpha$ by computing their intersections with
the $Z$-cycles:
\begin{align}
e_2\cdot Z_{CD}=e_2\cdot Z_{AB}&=-1,\hspace{1cm}
e_2\cdot Z_{AC}=e_2\cdot Z_{DB}=0\,,\nonumber \\
\alpha\cdot Z_{AC}=\alpha\cdot Z_{DB}&=-1,\hspace{1cm}
\alpha\cdot Z_{CD}=\alpha\cdot Z_{AB}=0 \ .
\end{align}
Note that these intersections change consistently when we let 
$Z\rightarrow Z+ne_1+m\beta$. We display the distribution
of the torus cycles in Figure~\ref{tcyc}.

At the orientifold point we can write the complex structure
of $T^4/Z_2$ as
\begin{align}
\Omega_{T^4/Z_2}=&\left(dx+Udy\right)\wedge\left(dx'+Sdy'\right) 
\nonumber \\
=& \hspace{1ex}dx\wedge dx'+Sdx\wedge dy'+Udy\wedge dx'+SUdy\wedge dy'\ .
\end{align}
In the above equation, $U$ denotes the complex structure of
the torus with unprimed coordinates, whereas $S$ denotes
the complex structure of the torus with primed
coordinates.
We have so far always switched freely between cycles and forms using the 
natural duality. We now make this identification explicit at the
orientifold point:\footnote{
We have normalized the orientifold such that $\int_{T^2/Z_2}dx\wedge 
dy=1/2$ and $\int_{T^2}\wedge dx'\wedge dy'=1$.
}
\begin{center}
\begin{tabular}[h]{cc}
$e_1=$& $-2dy\wedge dy'$ \\ 
$e_2=$& $2dy\wedge dx'$ \\ 
$\alpha=$& $2dx\wedge dx'$ \\ 
$\beta=$& $2dx\wedge dy'$ \ .
\end{tabular}
\end{center}
Thus we have shown that the parameters $U$ and $S$ in
\begin{align}
\Omega_{SO(8)^4}=\frac{1}{2}\left(\alpha+Ue_2+S\beta-USe_1\right) \ ,
\end{align}
do indeed describe the complex structure of the base
and the fibre torus. 

The findings of this section represent one 
consistent identification of the torus cycles and
the $Z$-cycles that connect the four
blocks. It is possible to add appropriate
linear combinations of $e_1$, $e_2$, $\alpha$
and $\beta$ without destroying the mutual 
intersections and the intersection pattern
with the $16$ cycles in the four $SO(8)$ blocks.
What singles out our choice is the form of $\Omega_{SO(8)^4}$ 
in~(\ref{omegaso8}) as well as the $SO(16)$ that was implicitly defined 
in~(\ref{omegaso16}).

%%%%%%%%%%%%%%%%%%%%%%%%%%%%%%%%%%%%%%%%%%%%%%%
\subsection{D-Brane positions from periods and the 
weak coupling limit revisited}\label{dbp}
%%%%%%%%%%%%%%%%%%%%%%%%%%%%%%%%%%%%%%%%%%%%%%%

In this section, we study deformations away from the $SO(8)^4$ point.
To achieve this, we have to rotate the complex structure
such that not all of the vectors spanning the 
$D_4^{\oplus 4}$ lattice are orthogonal
to it. In other words, we want to add terms
proportional to the forms in (\ref{AtoD}) 
to $\Omega_{SO(8)^4}$.
To do this, we switch to an orthogonal basis defined by
\begin{align}
\tilde{E}_1&=E_1+e_2 ,\hspace{1cm} \tilde{E}_{I}=E_I,& I=2..4, 10..12 
\nonumber\\
\tilde{E}_9&=E_9+e_2 ,\hspace{1cm} \tilde{E}_{J}=E_J+e_1/2,& J=5..8, 13..16 \ .
\end{align}
As we will see, each $\tilde{E}_I$ is responsible for moving
only one of the D-branes when we rotate the complex structure to 
\begin{equation}
\Omega=\frac{1}{2}\left(\alpha+Ue_2+S\beta-\left(US-z^2\right)
e_1+2\tilde{E}_{I}z_I\right) \ .
\label{Omegagen}
\end{equation}
Here $z^2$ denotes $z_Iz_I$.
Note that all of the $\tilde{E}_I$ are orthogonal to
$\Omega_{SO(8)^4}$, so that we only have to change the coefficient
of $e_1$ to maintain the constraint $\Omega\cdot\Omega=0$. 

We can use the information about the length
of the blown-up cycles to compute the new positions of the 
branes.
Let e.g. $z_1\neq 0$: This gives the first cycle of block 
$C$, $C_1=-e_2-E_1+E_2$, the length $\Omega\cdot C_1=z_1$, so that we 
move one brane away from the O-plane. As a result, the $SO(8)$ at block 
$C$ is broken down to $SO(6)$. At the same time, the sizes of $Z_{AC}$ and 
$Z_{CD}$ are changed to 
\begin{align}
Z_{AC}\cdot\Omega=-\frac{1}{2}+z_1,\hspace{1cm}
Z_{CD}\cdot\Omega=-\frac{U}{2}-z_1 \ .\label{genha}
\end{align}
Thus we can move the brane from block $C$ onto block $A$ by
choosing $z_1=\frac{1}{2}$, or onto block $D$ by choosing 
$z_1=-U/2$. This can also be seen from the overall gauge group which is
$SO(6)\times SO(10)\times SO(8)^2$ for these two values of $z_1$. 

As we have seen in the last paragraph, $z_1$ controls the 
position of one of the four D-branes located at block $C$,
as compared to the position of the O-plane at block $C$.
If we let all of the $z_I$ be non-zero, we find the 
following values for the lengths of the cycles in block~$C$:
\begin{center}
\begin{tabular}[h]{c|c}
$C_I$ & $C_I\cdot\Omega$\\ 
\hline
$C_1$& $z_1-z_2$ \\ 
$C_2$& $z_2-z_3$ \\ 
$C_3$& $z_3-z_4$ \\ 
$C_4$& $z_3+z_4$
\end{tabular}
\vspace{-1.6cm}
\begin{equation}
\label{lenghtscI}
\end{equation}
\end{center}
\vspace{1cm}

\begin{figure}[!ht]
\begin{center}
\includegraphics[height=7cm,]{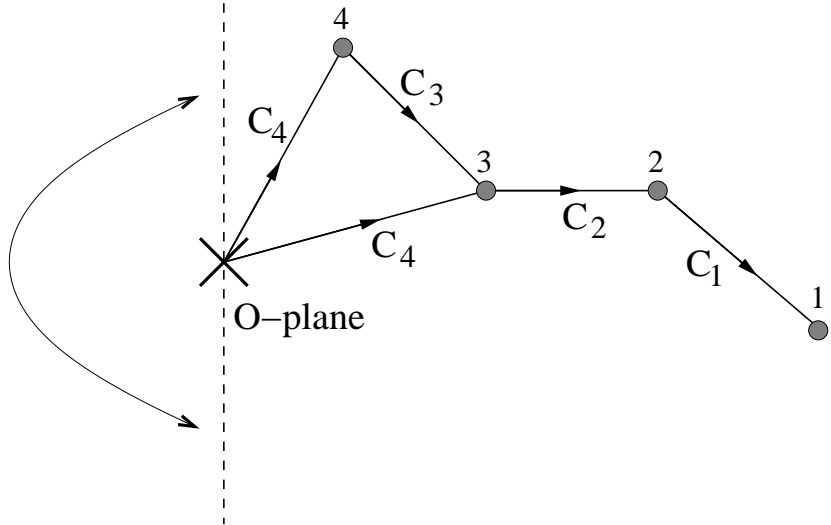}\caption{\textsl{The
positions of D-branes on $\mathbb{C}/Z_2$ are measured 
by complex line integrals along the cycles $C_I$. 
As indicated in the picture, $\mathbb{C}/Z_2$ is obtained 
from the complex plane by gluing 
the upper part of the dashed line to its lower part.
Due to the presence of the O-plane, the line integral along 
$C_4$ has to be evaluated as
indicated by the arrows. Using (\ref{lenghtscI}), one can 
see that the positions of the branes are given by the $z_I$.}}
\label{dpos}
\end{center}
\end{figure}

To determine the D-brane positions, it
is important to note that the D-branes
are moving on a pillow, $T^2/Z_2$.
We thus use a local coordinate system equivalent
to $\mathbb{C}/Z_2$. It is centered
at the position of the O-plane of block $C$ 
at one of the corners of the pillow.
The intersections of the cycles with $\Omega$
are line integrals along the base part of the 
cycles $C_I$ (see Figure~\ref{dpos}), multiplied 
by the line integral of their fibre part (which can
be set to unity locally). 
This is of importance for the length of $C_4$:
due to the orientation flip in the fibre when 
surrounding the O-plane, we have to evaluate both parts
of the line integral going from the O-plane to the
D-branes to account for the extra minus sign. 
This is indicated by the arrows that
are attached to the cycles in Figure~\ref{dpos}. 
It is then easy to see that associating the $z_I$
with the positions of the D-branes yields the correct 
results. Note that one achieves the same gauge enhancement
for $z_3=z_4$ and $z_3=-z_4$, because for both values
one of the $C_I$ is collapsed, cf.~(\ref{lenghtscI}). 
Thus the D-branes labelled $3$ and $4$ have to be at the
same position in both cases, which fits with the
fact that $z_I=-z_I$ holds due to the $Z_2$ action.

By the same reasoning, the remaining $z_I$ give the 
positions of the other D-branes measured 
relative to the respective O-plane. 
For example, the moduli $z_5$ to $z_8$ give the positions of the 
D-branes of block $A$ (see (\ref{AtoD})).
We have also shown that we can connect the four blocks
by following the gauge enhancement that arises when
we move a brane from one block to another, cf.~(\ref{genha}). 
This means 
that we can also easily connect the four coordinate systems
that are present at the position of each O-plane. 
We have now achieved our goal of explicitly mapping 
the holomorphic 2-form $\Omega$ to the positions of the 
D-branes. For this we have used forms 
dual to integral cycles. These are the cycles that 
support the M-theory flux which can be used to stabilize the 
D-branes. By using our results, it is possible to derive the
positions of the D-branes from a given complex structure 
(unless the solution is driven away from the weak coupling 
limit). We thus view this work as an important step towards 
the explicit positioning of D-branes by M-theory flux. 

The geometric constructions of this article
only make sense in the weak coupling limit, 
in which the monodromies of 
the branes of F-theory are restricted to those 
of D7-branes and O7-planes. 
It is crucial that the positions of the D-branes
and the shape of the base torus factorize in the weak
coupling limit, $S\rightarrow i\infty$. The shape 
of the base torus is measured by 
multiplying the cycles $\alpha$, $\beta$, $e_1$ and $e_2$
with $\Omega$. The result is independent of the positions
of the branes in the weak coupling limit, as the only potential
source of interference is the $z^2$ in $\alpha\cdot\Omega=US-z^2$,
which is negligible as compared to $US$.
Thus the branes can really be treated as moving
on $T_2/Z_2$ without backreaction in the weak
coupling limit.

Certain gauge groups, although present in F-theory, do not 
show up in perturbative type IIB orientifolds and thus cannot 
be seen in the weak coupling limit. The lattice of forms 
orthogonal to $\Omega$ only has the structure of gauge groups 
known from type IIB orientifold models when we let $S\rightarrow \iu \infty$ in (\ref{omegagen}). 
This comes about as follows: Starting from the $SO(8)^4$ point, we can cancel
all terms proportional to $W_I E_I$ in (\ref{omegagen})
when we are at finite coupling. In the limit
$S\rightarrow i\infty$, the fact that $\beta$ has
$S$ as its prefactor prevents the cancellation
of the term $W_I^2 E_I$ in $\Omega$. The 
presence of this term ensures that only
perturbatively known gauge enhancements 
arise.

\section{Enriques involutions and Weierstrass models}

In this section we consider the action of the Enriques involution\footnote{The Enriques 
involution is a fixed-point free holomorphic involution of $K3$ which is non-symplectic, i.e. it 
projects out the holomorphic 2-form \cite{peters,alexeev:2004}. It yields the Enriques surface as the 
quotient space.} on elliptic $K3$ surfaces. In particular, we are interested in its compatibility with the 
description of $K3$ by a Weierstrass model.

It is well-known that $T^{4}/\mathbb{Z}_{2}$, allows an Enriques involution \cite{key-65}. Our strategy is
to find the point in the moduli space of $K3$ at which it degenerates to $T^{4}/\mathbb{Z}_{2}$ and
then deform it to get to a $K3$ described by a Weierstrass model. It turns out that this deformation is not 
consistent with the {\it holomorphicity} of the Enriques involution, the obstacle being the single distinguished 
section of the Weierstrass model\footnote{A freely acting $\mathbb{Z}_2$-symmetry of the real metric manifold 
still exists, but it is not holomorphic in the complex structure of the Weierstrass model.}. 
This problem does not arise for elliptic $K3$ surfaces that are given in non-standard Weierstrass form, e.g. 
one with two distinguished sections \cite{Klemm:1996ts,Berglund:1998va}. In the F-theory limit all these
descriptions meet. In particular also the usual Weierstrass form becomes symmetric under the Enriques involution.

To describe this degeneration of $K3$ to $T^4/\Z_2$ in detail, we need to know which two-spheres shrink to
produce the 16 $A_1$ singularities. We first study the breaking of $E_8$ to $SU(2)^8$, so that we can use the map 
between Wilson lines and degenerations of $K$ discussed in Section~\ref{wilson}. 

\subsection[The lattice $E_{8}$ and its sublattice $A_{1}^{\oplus8}$]{The lattice \boldmath$E_{8}$ and its sublattice \boldmath$A_{1}^{\oplus8}$}\label{sec2}

As there is no distinction among the 16 $A_1$ singularities of $T^4/\Z_2$, we expect that the breaking
of $E_8$ to $SU(2)^8$ can be achieved by three equivalent Wilson lines.

In the following we will show that
\begin{equation}
W^{1}=(1,0^7), \hspace{.5cm}W^{2}=(0^4,-\frac{1}{2}^4)\hspace{.5cm}\mbox{and}\hspace{.5cm}
W^{3}=(0^2,-\frac{1}{2},\frac{1}{2},0^2,-\frac{1}{2},\frac{1}{2})\label{eq5}
\end{equation}
take us from $E_{8}$ to $A_{1}^{\oplus8}$.

It is easy to see that these three Wilson lines are equivalent, i.e. they are related by a
Weyl reflection\footnote{For example, if we apply ${\cal H}_4\oplus{\cal H}_4$ and the following element of the Weyl subgroup of type $D_8$, $(E_1,E_2,E_3,E_4,E_5,E_6,E_7,E_8)\mapsto(-E_3,E_8,E_4,-E_7,-E_1,E_6,E_2,-E_5)$, we get $W^1\mapsto W^3$, $W^2\mapsto W^1$ and $W^3\mapsto W^2$.}.

Let us start with $W^{1}$. This Wilson line removes $\alpha_{1}$,
giving us the Dynkin diagram~of~$D_{8}$.
\begin{figure}
\begin{centering}
\includegraphics[width=6cm]{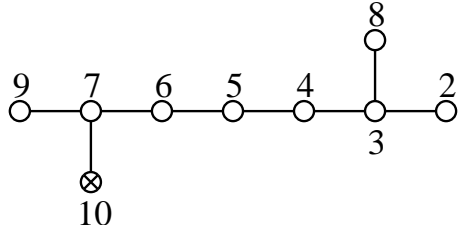}
\par\end{centering}
\caption{\textsl{The extended Dynkin diagram of $D_{8}$.}}\label{ExtDynD8}
\end{figure}
Adding the highest root of the $D_8$ lattice,
\begin{equation}
\alpha_{10}=-\alpha_{2}-2\alpha_{3}-2\alpha_{4}-2\alpha_{5}-2\alpha_{6}-2\alpha_{7}-\alpha_{8}-\alpha_{9}=E_{1}+E_{2}\:.\nonumber
\end{equation}
we obtain the extended Dynkin diagram of $D_8$ (see Figure~\ref{ExtDynD8}). 

\begin{figure}[tt]
\begin{centering}
\includegraphics[width=2cm]{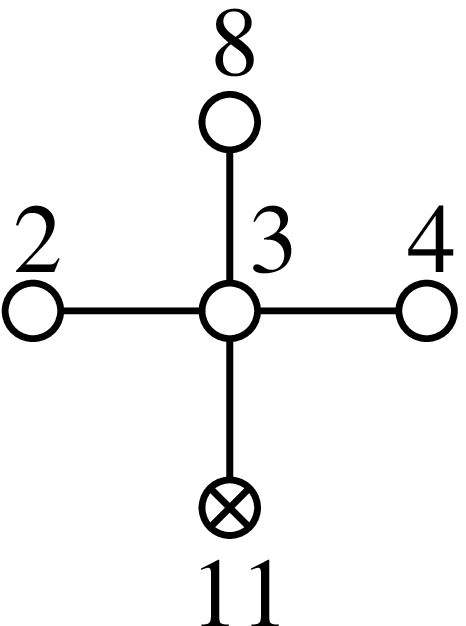}\hspace{2cm}\includegraphics[width=2cm]{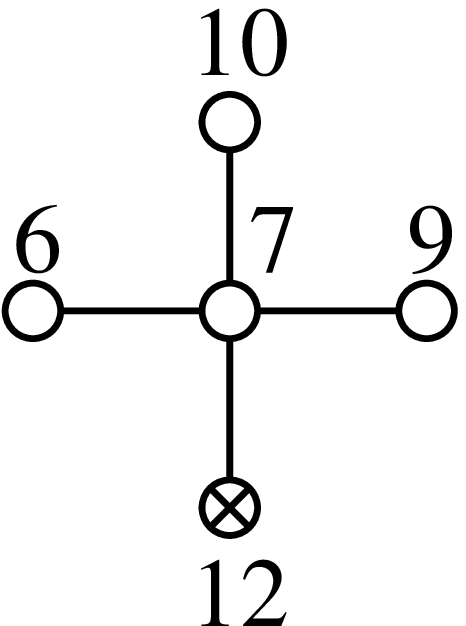}
\par\end{centering}

\caption{\textsl{Twice the extended Dynkin diagram of $D_{4}$.}}\label{fig2D4}

\end{figure}

Next, $W^{2}$ removes the node corresponding to $\alpha_{5}$. We are
left with two copies of the Dynkin diagram of $D_{4}$ (see Figure~\ref{fig2D4}), which we
extend by their respective highest roots
\begin{align}
\alpha_{11}&=-\alpha_{2}-2\alpha_{3}-\alpha_{4}-\alpha_{8}=E_{5}+E_{6}\nn \\
\alpha_{12}&=-\alpha_{6}-2\alpha_{7}-\alpha_{9}-\alpha_{10}=-E_{3}-E_{4}\ . \nonumber
\end{align}

Finally, $W^{3}$ removes $\alpha_{3}$
and $\alpha_{7}$, leaving us with 8 unconnected nodes corresponding
to the $A_{1}^{\oplus8}$ sublattice of $E_{8}$.\footnote{Here, the removed nodes are two instead of one; this is because the Wilson line acts on two simple groups.} 
The remaining simple roots are:
\begin{align}
\alpha_{2} &=-E_{7}-E_{8} &\quad &\alpha_{4} =-E_{5}+E_{6} &\quad & \alpha_{6} =-E_{3}+E_{4}&\quad & \alpha_{8}  =-E_{7}+E_{8}\nonumber \\
\alpha_{9} & =-E_{1}+E_{2}&\quad & \alpha_{10}  =E_{1}+E_{2}&\quad & \alpha_{11} =E_{5}+E_{6}&\quad & \alpha_{12} =-E_{3}-E_{4}\end{align}

\subsection[The $T^{4}/\mathbb{Z}_{2}$ orbifold limit of $K3$]{The \boldmath$T^{4}/\mathbb{Z}_{2}$ orbifold limit of \boldmath$K3$}\label{sec4}

We begin with some definitions regarding $T^{4}/\mathbb{Z}_{2}$. The non-trivial element
of $\mathbb{Z}_{2}$ acts as $-1$ on all the coordinates $x_i$ ($i=1,...,4$) 
of~$T^{4}$. After modding out, the 16 points of $T^{4}$ fixed
under the $\mathbb{Z}_{2}$-action lead to 16 $A_1$~singularities. Their locations are at
\begin{equation}\label{eq:SingLoc}
(x_1,x_2,x_3,x_4)=(\xi_1,\xi_2,\xi_4,\xi_4),\qquad \mbox{with} \qquad \xi_i=0,\frac12 \:.
\end{equation}

The 2-cycles of $T^4$ are all even with respect to $\mathbb{Z}_{2}$ and survive the orbifolding. An integral basis 
is given by the six 2-tori $\pi_{ij}$ corresponding to the $x_i$-$x_j$-plane. Their intersection numbers are 
\begin{equation}
\pi_{ij}\cdot\pi_{ml}=2\varepsilon_{ijml} \:.
\end{equation}
The corresponding Poincar\'e-dual 2-forms are
\begin{equation}
   \mbox{PD}[\pi_{ij}] = \epsilon_{ijpq}\,dx_p\wedge dx_q \:.
\end{equation}
As we will see in more details later, blowing up the 
16 $A_{1}$ singularities of $T^{4}/\mathbb{Z}_{2}$ gives rise to 16 $\mathbb{P}^1$s. They are orthogonal 
with respect to each other and to the torus-cycles $\pi_{ij}$. There is a natural choice of complex structure 
on $T^4/\mathbb{Z}_2$: %compatible with the Enriques involution (i.e. such that $\Omega\mapsto -\Omega$):
$z_1=x_1+\tau_1\,x_4$ and $z_2=x_2+\tau_2\,x_3$ \footnote{
The natural expressions for the K\"ahler form $J$ and the holomorphic 2-form $\Omega^{2,0}$ are then
$\Omega^{2,0} = dz_1 \wedge dz_2$ and $J = a_1 dz_1\wedge d\bar{z}_1 + a_2 dz_2\wedge d\bar{z}_2 + \mbox{Re}[b\,dz_1\wedge\bar{z}_2]$,
where $a_1,a_2\in \mathbb{R}$ and $b\in\mathbb{C}$. In terms of the Poincare-dual of the integral cycles $\pi_{ij}$, we have
\begin{align}
\Omega^{2,0} & = \pi_{34}+\tau_1 \pi_{13}+\tau_2\pi_{42}-\tau_1\tau_2\pi_{12} \nn\\
J & = \hat{a}_1  \pi_{23}  + \hat{a}_2 \pi_{14} + \mbox{Re}[b (\pi_{34}+\tau_1 \pi_{13}+\bar{\tau}_2\pi_{42}-\tau_1\bar{\tau}_2\pi_{12})],\nn
\end{align}
where we defined $\hat{a}_1=-2a_1 \mbox{Im}\tau_1$ and $\hat{a}_2=-2a_2 \mbox{Im}\tau_2$.
}.

It is well known that some $K3$ surfaces, including $T^{4}/\mathbb{Z}_{2}$, allow a fixed-point
free involution $\vartheta$ yielding an Enriques surface\footnote{Nikulin classified all involutions of $K3$
reversing the sign of $\Omega$ \cite{Nikulin:1986} and found that they can be labeled by three integers
$(r,a,\delta)$. Only one involution in this classification, $(10,10,0)\equiv\vartheta$,
has no fixed points.}. The action of $\vartheta$ on $T^{4}/\mathbb{Z}_{2}$ is given by \cite{key-65}
\begin{equation}
\vartheta: \qquad z_1\mapsto -z_1+\frac12,\qquad z_2\mapsto z_2+\frac12 \:.\label{ent4}
\end{equation}
Hence, $\pi_{14}$ and $\pi_{23}$ are even under $\vartheta$, while $\pi_{12}$, $\pi_{34}$, $\pi_{13}$ and $\pi_{42}$
are odd. From \eqref{eq:SingLoc} it is clear that the $A_1$ singularities are interchanged pairwise.
We will use the transformation properties of the cycles of $T^{4}/\mathbb{Z}_{2}$ under
this involution to identify them with specific cycles of the $K3$ lattice.

\begin{figure}[tt]
\begin{centering}
\includegraphics[width=7cm]{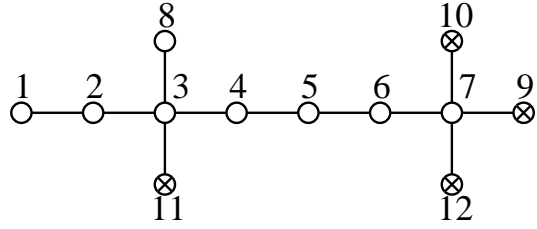}
\par\end{centering}

\caption{\textsl{The Dynkin diagram of the first $E_{8}$ extended by its highest root as well
as the highest roots of its sublattices of types $D_{4}$ and $D_{8}$.}}\label{FigDinkE8ext}

\end{figure}

\
We now discuss the cycles of $T^{4}/\mathbb{Z}_{2}$ from the $K3$ perspective.
The singular limit $T^{4}/\mathbb{Z}_{2}$ of $K3$ is obtained
by fixing the position of $\Sigma$ such that $16$ cycles with intersection
matrix $A_{1}^{\oplus16}$ shrink. We start with a $K3$ surface with
an $E_{8}\times E_{8}$ singularity and rotate $\Sigma$ to a $A_1^{\oplus 16}$ point. %Correspondingly, $\Sigma$ is located inthe $U^{\oplus3}$ block of $H^{2}(K3,\mathbb{R})$.
In Section~\ref{sec2} we have specified Wilson lines breaking $E_8$ to $A_1^{8}$. Using these Wilson lines and following the procedure detailed in 
Section~\ref{wilson}, we will arrive at the desired point in moduli space.

We introduce a set of simple roots $\gamma_{i}$, $i=1,...,8,13,...,20$, of $\Gamma_{E_{8}\times E_{8}}$
(cf. \eqref{eq:e8rootsystem}), the highest roots $\gamma_{9}$ and $\gamma_{21}$
of the $E_{8}$ root lattices as well as the highest roots $\gamma_{i}$,
$i=10,11,12,22,23,24$, of their respective sublattices of types $D_{4}$ and $D_{8}$ (see Figure~\ref{FigDinkE8ext}):
\begin{align}\label{eq:e8e8}
\gamma_{1} & =\frac{1}{2}E_{1}+...+\frac{1}{2}E_{8}  &\qquad& \gamma_{2} =-E_{7}-E_{8}  &\qquad& \gamma_{3} =-E_{6}+E_{7} \nonumber \\
\gamma_{4} & =-E_{5}+E_{6}  &\qquad& \gamma_{5} =-E_{4}+E_{5} &\qquad& \gamma_{6} =-E_{3}+E_{4} \nonumber \\
\gamma_{7} & =-E_{2}+E_{3}  &\qquad& \gamma_{8} =-E_{7}+E_{8} &\qquad& \gamma_{9} =-E_{1}+E_{2} \nonumber \\
\gamma_{10} & =E_{1}+E_{2}  &\qquad& \gamma_{11} =E_{5}+E_{6} &\qquad& \gamma_{12} =-E_{3}-E_{4} \nonumber \end{align}
\begin{align}
\gamma_{13} & =\frac{1}{2}E_{9}+...+\frac{1}{2}E_{16} &\qquad& \gamma_{14}  =-E_{15}-E_{16} &\qquad& \gamma_{15} =-E_{14}+E_{15}\nonumber \\
\gamma_{16} & =-E_{13}+E_{14} &\qquad& \gamma_{17} =-E_{12}+E_{13} &\qquad& \gamma_{18} =-E_{11}+E_{12}\nonumber \\
\gamma_{19} & =-E_{10}+E_{11} &\qquad& \gamma_{20} =-E_{15}+E_{16} &\qquad& \gamma_{21} =-E_{9}+E_{10}\nonumber \\
\gamma_{22} & = E_{9}+E_{10}  &\qquad& \gamma_{23} = E_{13}+E_{14} &\qquad& \gamma_{24} =-E_{11}-E_{12}\end{align}

On the basis of \eqref{eq5}, we choose the following Wilson-line vectors in 
$\Gamma_{E_{8}\times E_{8}}$ (The signs between the two $E_8$ factors will be justified in a moment 
by the properties of the $K3$ lattice under the Enriques involution):
\begin{align}\label{WLT4Z2}
W^1=(1,0^7,-1,0^7), \qquad W^2=(0^4,{-\frac12}^4,0^4,{\frac12}^4)\nonumber\\ W^3=(0^2,-\frac12,\frac12,0^2,-\frac12,\frac12,0^2,-\frac12,\frac12,0^2,-\frac12,\frac12)\:.
\end{align}

We start with $\Sigma$ living in the $U^{\oplus 3}$ space spanned by
$ \hat{e}_1$, $\hat{e}^1$, $\hat{e}_2$, $\hat{e}^2$, $\hat{e}_3$, $\hat{e}^3$.
The first step is to move $\Sigma$ in the direction of $W^1$ by the rotation $\hat{e}^{1}\rightarrow \hat{e}^{1}+ W^{1}$. 
The result is a $K3$ surface with $D_{8}\times D_{8}$ singularity. While $\gamma_{1}$, $\gamma_{9}$, $\gamma_{10}$, 
$\gamma_{13}$, $\gamma_{21}$ and $\gamma_{22}$ are blown up, the cycles
\begin{equation}
\gamma_{9}'  \equiv\gamma_{9}-\hat{e}_{1},\qquad
\gamma_{10}'  \equiv\gamma_{10}+\hat{e}_{1},\qquad
\gamma_{21}'  \equiv\gamma_{21}+\hat{e}_{1},\qquad
\gamma_{22}'  \equiv\gamma_{22}-\hat{e}_{1}
\end{equation}
collapse. $\gamma_{9}'$ ($\gamma_{21}'$) is the additional root appearing in the first (second) $E_8$ lattice. 
$\gamma_{10}'$ ($\gamma_{22}'$) are the corresponding highest roots of $D_{8}$.

Next, we rotate $\hat{e}^{2}\rightarrow \hat{e}^{2}+W^{2}$. Since the products of $\gamma_{2}$, $\gamma_{5}$,
$\gamma_{11}$, $\gamma_{14}$, $\gamma_{17}$ and $\gamma_{23}$ with the rotated basis vector $\hat{e}^{2}+W^{2}$
are all non-zero, these cycles acquire finite volume, while $\gamma_{i}$, $i=3,4,6,7,8,12,15,16,18,19,20,24$, 
and $\gamma_{i}'$, $i=9,10,21,22$, remain orthogonal to $\Sigma$. Out of the roots in \eqref{eq:e8e8}, however, 
we can take those that have an integer product with $W^{2}$ and construct the four further shrunk cycles
\begin{align}
\gamma_{2}'  \equiv\gamma_{2}+\hat{e}_{2},\qquad
\gamma_{11}'  \equiv\gamma_{11}-\hat{e}_{2},\qquad
\gamma_{14}'  \equiv\gamma_{14}-\hat{e}_{2},\qquad
\gamma_{23}'  \equiv\gamma_{23}+\hat{e}_{2}\:.
\end{align}
A set of simple roots for the orthogonal lattice $\Lambda$ is given by 
$\lbrace\gamma_{i}\rbrace_{i=3,4,6,7,8,15,16,18,19,20}$ and 
$\lbrace\gamma_{i}'\rbrace_{i=2,9,10,14,21,22}$. The intersection matrix of these 
simple roots is $D_{4}^{\oplus4}$. We therefore obtained a $K3$ surface with a $D_{4}^{4}$ singularity.

Finally, we rotate $\hat{e}^{3}\rightarrow \hat{e}^{3}+ W^{3}$, go to an $A_{1}^{\oplus16}$ point 
in $M_{K3}$. The roots that are removed from $\Lambda$ are $\gamma_{3}$, $\gamma_{6}$, $\gamma_{7}$, 
$\gamma_{8}$, $\gamma_{15}$, $\gamma_{18}$ and $\gamma_{19}$, $\gamma_{20}$, while the new shrinking cycles are
\begin{align}
\gamma_{6}'  \equiv\gamma_{6}+\hat{e}_{3},\qquad
\gamma_{8}'  \equiv\gamma_{8}+\hat{e}_{3},\qquad 
\gamma_{18}'  \equiv\gamma_{18}+\hat{e}_{3},\qquad
\gamma_{20}'  \equiv\gamma_{20}+\hat{e}_{3}.
\end{align}

To sum up, following the procedure outlined in the last section, we found that $K3$ can degenerate to $T^{4}/\mathbb{Z}_{2}$ if
$\Sigma$ is orthogonal to
%The roots in the in the orthogonal complement of $\Upsilon$ are
\begin{align}\label{16shrCycles}
\gamma_{2} & '=-E_{7}-E_{8}+\hat{e}_{2}, & \gamma_{14}' & =-E_{15}-E_{16}-\hat{e}_{2},\nonumber \\
\gamma_{4} & =-E_{5}+E_{6}, & \gamma_{16} & =-E_{13}+E_{14},\nonumber \\
\gamma_{6}' & =-E_{3}+E_{4}+\hat{e}_{3}, & \gamma_{18}' & =-E_{11}+E_{12}+\hat{e}_{3},\nonumber \\
\gamma_{8}' & =-E_{7}+E_{8}+\hat{e}_{3}, & \gamma_{20}' & =-E_{15}+E_{16}+\hat{e}_{3},\nonumber \\
\gamma_{9}' & =-E_{1}+E_{2}-\hat{e}_{1}, & \gamma_{21}' & =-E_{9}+E_{10}+\hat{e}_{1},\nonumber \\
\gamma_{10}' & =E_{1}+E_{2}+\hat{e}_{1}, & \gamma_{22}' & =E_{9}+E_{10}-\hat{e}_{1},\nonumber \\
\gamma_{11}' & =E_{5}+E_{6}-\hat{e}_{2}, & \gamma_{23}' & =E_{13}+E_{14}+\hat{e}_{2},\nonumber \\
\gamma_{12} & =-E_{3}-E_{4} & \gamma_{24} & =-E_{11}-E_{12}.
\end{align}
This set of cycles provides a primitive embedding of the $A_1^{\oplus 16}$ lattice into $\Gamma_{3,19}$.

The lattice orthogonal to the shrunk cycles $\Upsilon$ is given by integral combinations of the 
following six cycles:
\begin{equation}\label{1stBasisUps}
\hat{e}_{1}, \qquad 2(\hat{e}^{1}+ W^{1}), \qquad \hat{e}_{2}, \qquad 2(\hat{e}^{2}+ W^{2}), \qquad \hat{e}_{3}, \qquad 2(\hat{e}^{3}+ W^{3}) \:.
\end{equation}
The 3-plane $\Sigma$ lives in the subspace of $H_{2}(K3,\mathbb{R})$ spanned by these vectors so that the cycles in $\Upsilon$ 
in general have finite size. We want to identify this lattice with the $T^4/\mathbb{Z}_2$ lattice made up of the $\pi_{ij}$. We will use the transformation properties of the torus-cycles $\pi_{ij}$ under $\vartheta$ to identify them with elements of $\Upsilon$. 

Previously we have seen that the Enriques involution must map the singularities of $T^4/\mathbb{Z}_2$ to each other in pairs. We hence expect that the cycles on the left column in \eqref{16shrCycles} are mapped to the cycles on the right one. %Since the involution acts on the $K3$ lattice such that to exchange the two $E_8$ blocks (i.e. $E_I\leftrightarrow E_{I+8}$), the transformation properties of the $\hat{e}_i$ are fixed. 

Up to automorphism of $\Gamma_{3,19}$, the Enriques involution $\vartheta$ acts on the $K3$ lattice by interchanging the two $E_{8}$ 
as well as the two $U$-blocks, and as $-1$ on the remaining $U$-block \cite{peters} (see also \cite{Berglund:1998va}):
\begin{equation}\label{eq:enrinv}
\vartheta:\,e_{1} \mapsto-e_{1} \qquad e^{1} \mapsto-e^{1} \qquad
e_{2} \leftrightarrow e_{3} \qquad e^{2} \leftrightarrow e^{3} \qquad
E_{I} \leftrightarrow E_{I+8}.
\end{equation}

If we set $\hat{e}_i=e_i$ and apply the transformation \eqref{eq:enrinv} to the 16 cycles in \eqref{16shrCycles}, we do not obtain what we expect, i.e. that the 8 cycles in the left column in \eqref{16shrCycles} are mapped to the ones in the right column. To get this result, we need an Enriques involution such that the $\hat{e}_i$ have definite parity. A sensible identification is thus\footnote{One can check that this transformation provides an automorphism of the lattice $\Gamma_{3,19}$.} %Hence the two sets $\{\hat{e}_i,\hat{e}^i\}$ and $\{e_i,e^i\}$ are equivalent.
\begin{equation}
 \hat{e}_1 = e_1 \qquad \hat{e}_2= e_2-e_3 \qquad \hat{e}_3 = e^2+e^3 \qquad \hat{e}^1=e^1 \qquad \hat{e}^2=e^2 \qquad \hat{e}^3 = e_3 \:.
\end{equation}
Hence, the basis \eqref{1stBasisUps} of $\Upsilon$ becomes
\begin{equation}\label{2ndBasisUps}
e_{1}, \qquad 2(e^{1}+ W^{1}), \qquad e_{2}-e_3, \qquad 2(e^{2}+ W^{2}), \qquad e^2+e^{3}, \qquad 2(e_{3}+ W^{3}) \:.
\end{equation}

Note that the set of vectors \eqref{16shrCycles} could be guessed without any reference to a particular choice of Wilson lines by going directly to the Dynkin diagram language. Then, the involution property of the three orthogonal null vectors $\hat{e}_i$ of the $U^{\oplus 3}$-block that we add to rotate the shrinking cycles is determined by requiring the exchange of the two blocks.

We can now choose an integral basis of $\Upsilon$ with the properties of $\pi_{ij}$, i.e. whose elements 
have definite parity under $\vartheta$ and whose intersection matrix is\footnote{
The matrix $U(2)$ is equal to $\left(\begin{array}{cc} 0&2\\2&0\\ \end{array}\right)$.}
$U(2)^{\oplus 3}$:
\begin{equation}\label{eVSpi}\begin{array}{ccc}
e_{1},&\qquad& 2(e^{1}+e_{1}+ W^{1}),\\ e_{2}-e_{3},&\qquad& e^{2}-e^{3}+2(e_{2}-e_{3}+ W^{2}),\\ e^{2}+e^{3}, &\qquad& e_{2}+e_{3}+2(e^{2}+e^{3}+ W^{3})
\end{array}\end{equation}
One can check that this is an integral basis of $\Upsilon$. %(obtained by automorphism of $\Upsilon$ that is not an automorphism of $\Gamma_{3,19}$. ... ). 
The first two lines of \eqref{eVSpi} give two $U(2)$ blocks odd under $\vartheta$, while the last line gives a $U(2)$ block even under $\vartheta$. Therefore, we may identify $\pi_{23}$ and $\pi_{14}$ with the vectors $e^{2}+e^{3}$ and $e_{2}+e_{3}+2(e^{2}+e^{3}+ W^{3})$ of the last $U(2)$ block and $\pi_{12}$, $\pi_{34}$, $\pi_{13}$ and $\pi_{42}$ with the vectors $e_{1}$, $2(e^{1}+e_{1}+ W^{1})$, $e_{2}-e_{3}$ and $e^{2}-e^{3}+2(e_{2}-e_{3}+ W^{2})$ of the first two $U(2)$ blocks.

\subsection[$T^{4}/\mathbb{Z}_{2}$ as a double cover of $\mathbb{P}^1\times\mathbb{P}^1$]{\boldmath$T^{4}/\mathbb{Z}_{2}$ as a double cover of \boldmath$\mathbb{P}^1\times\mathbb{P}^1$}
\label{sec5}
In this section, we are going to study the connection of
$T^4/\mathbb{Z}_2$ with a smooth $K3$ from a geometric perspective. 
We show how to find the lattice $H_2(K3,\mathbb{Z})$ in a blow-up of~$T^4/\mathbb{Z}_2$. 

It is well-known that one can construct a smooth $K3$ as a double cover
\cite{peters,Sen:1997gv} over $\mathbb{P}^1\times \mathbb{P}^1$, branched 
along a curve of bidegree $(4,4)$:

\begin{equation}
\tilde{y}^{2}=h_{(4,4)}(\tilde{x}_{1},\tilde{x}_{2},\tilde{z}_{1},\tilde{z}_{2}).\label{eq:cyd}
\end{equation}

There are two algebraic cycles, each given by fixing a point on one of the $\mathbb{P}^1$s.
We call the associated Divisors $D_x$ and $D_z$.
The corresponding curves are tori and represent the generic fibres of two different elliptic
fibrations of the $K3$ surface given by \eqref{eq:cyd}. As (\ref{eq:cyd}) gives two values of $y$ for
a generic point on $\mathbb{P}^1\times \mathbb{P}^1$, we find $D_{x}\cdot D_{z}=2$.

\begin{figure}
\begin{centering}
\includegraphics[height=5cm]{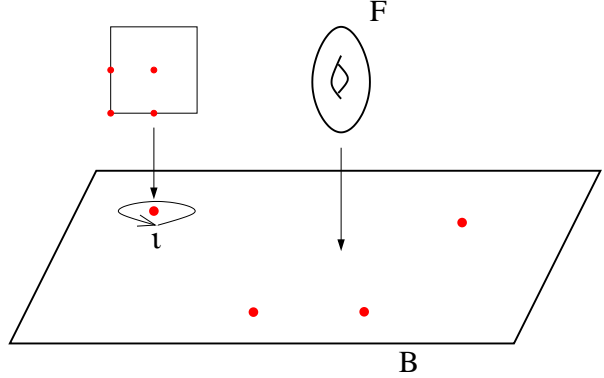}\label{fib}
\par\end{centering}
\caption{\textsl{The elliptic fibration $T^{4}/\mathbb{Z}_{2}\rightarrow B=T^2/\mathbb{Z}_2$,
has four singular fibres. Upon circling one of them, the fibre torus undergoes an involution 
$\iota$. Thus any section $B\hookrightarrow T^{4}/\mathbb{Z}_{2}$ has to pass through four singularities.}}

\end{figure}

Let us choose a particular form for $h_{(4,4)}$:
\begin{equation}
y^{2}=\prod_{k=1,..,4}(x-x_{k})\cdot(z-z_{k}).      \label{eq:alT4}
\end{equation}
For ease of exposition we have introduced the inhomogeneous coordinates
$x,y,z$. The surface defined by \eqref{eq:alT4} is easily recognized as 
$T^4/\mathbb{Z}_2$, as we explain in following: In the vicinity of the points $(y,x,z)=(0,x_{k},z_{h})$, 
it is given by $y^{2}=xz$, i.e. it has sixteen $A_{1}$ singularities.
Let us now describe this surface as an elliptic fibration. We project to the
coordinate $x$, so that each fibre torus is given by (\ref{eq:alT4}) with
$x$ fixed. The complex structure of the fibre torus is given by
the ratios of the branch points $z_k$. As these do not depend on $x$, the complex structure of
the fibre is constant. Over the four points in the base where $x=x_k$, we have $y=0$, so
that the fibre is $\mathbb{P}^1$ instead of $T^2$, see Figure~\ref{fib}. A
similar fibration is obviously obtained when projecting to the $z$-coordinate.

This is the very same structure one finds when projecting $T^4/\mathbb{Z}_2$ to any
$T^2/\mathbb{Z}_2$ suborbifold. Any of these projections can be promoted to an elliptic fibration
by choosing the complex structure of $T^4/\mathbb{Z}_2$ appropriately. Only two 
of them can, however, be seen algebraically in (\ref{eq:alT4}).
The divisors $D_{x}$ and $D_{z}$ correspond to multisections\footnote{A section is a
divisor that is not part of any fibre and intersects each fibre once. Correspondingly, a multisection or 
$m$-section intersects each fibre $m$ times.} (two-section) of these
two fibrations. They are tori and can be identified with $\pi_{23}$ and $\pi_{14}$. The
other $\pi_{ij}$ in $T^4/\mathbb{Z}_2$ cannot be seen algebraically.

Each of the two elliptic fibrations in (\ref{eq:alT4}) has four proper sections. 
Focussing again on the fibration given by projecting to the $x$-coordinate, they are given 
by mapping $x$ to $(y,x,z)=(0,x,z_k)$. Each of them passes through four $A_1$ singularities.
From the orbifold point of view, these sections are the usual 
divisors ($D_{i\alpha}=\{\zeta_i=\zeta_i^{\alpha,fixed}\}$), given by 
planes lying at the fixed loci of the orbifold action \cite{Lust:2006zh}.

We can understand how these sections arise in $T^4/\mathbb{Z}_2$. Fixing a projection, we have to
give a point in the fibre for every point of the base in a smooth manner. As the fibre undergoes
an involution when one surrounds one of the $x_k$ in the base, the sections have to pass through one
of the fixed points of this involution in the fibre. Again, not all of the sections that can be seen
this way in $T^4/\mathbb{Z}_2$ can be described algebraically in (\ref{eq:alT4}).

We label the sections $\sigma_{ij}^{k}$ by the two directions it spans in $T^4/\mathbb{Z}_2$ ($i,j$) 
and the fixed point in the fibre it passes through ($k$). Two $\sigma_{ij}^{k}$
that span different directions in $T^4/\mathbb{Z}_2$ are, of course, sections with respect
to different elliptic fibrations. The intersection numbers with the $\pi_{ij}$ are
\begin{equation}\label{eq:inT4}
\sigma_{ij}^{k}\cdot\pi_{lm}=\varepsilon_{ijlm}.
\end{equation}
As the intersections occur away from the singularities, \eqref{eq:inT4} will persist in a desingularized version of
$T^4/\mathbb{Z}_2$.

\begin{figure}
\begin{centering}
\includegraphics[height=9cm]{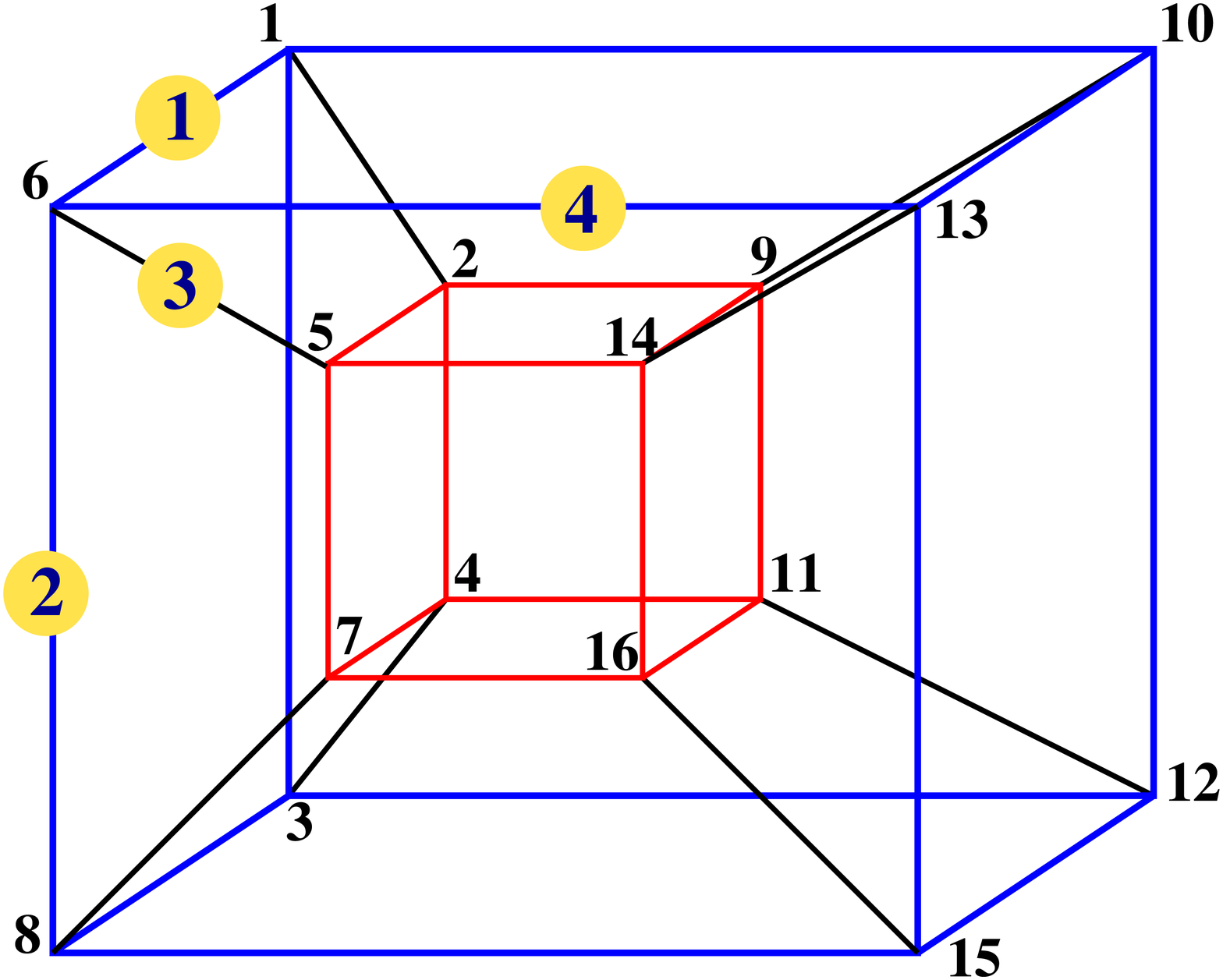}
\caption{\textsl{The set of sections $\sigma_{ij}^{k}$ and the $A_1$ singularities 
can be displayed as the two-dimensional faces and nodes of a four-dimensional 
hypercube. We picture this cube as two cubes of lower dimension whose nodes are connected 
as shown in the picture. We have numbered the four directions and the sixteen nodes, 
so that this figure can be used to determine which section meets which singularities.}}
\label{3cube} 
\end{centering}
\end{figure}

One way to visualize the geometry of $T^4/\mathbb{Z}_2$ is presented in Figure~\ref{3cube}. It will be
frequently used in the rest of this paper. At present, it serves to determine which singularities are 
met by which $\sigma_{ij}^{k}$ in the given labeling. 

\subsubsection{Divisors and cycles in the blow-up of $T^{4}/\mathbb{Z}_{2}$}
\label{sec6}

If we blow-up the sixteen $A_{1}$ singularities of $T^{4}/\mathbb{Z}_{2}$,
we introduce sixteen exceptional divisors $C_{\lambda}$ which satisfy $C_{\lambda}\cdot C_{\eta}=-2\delta_{\lambda\eta}$.
Naively, one would guess that the lattice of integral cycles of the blow-up
of $T^{4}/\mathbb{Z}_{2}$ is thus given by $A_{1}^{\oplus16}\oplus U(2)^{\oplus3}$.
But the blow-up of $T^{4}/\mathbb{Z}_{2}$ should be a smooth $K3$ surface, which
has $U^{\oplus3}\oplus(-E_{8})^{\oplus2}$ as its lattice of integral
cycles. The extra integral cycles are given by the preimages of the sections in the 
blow-up\footnote{Remember that a blow-up actually is a projection mapping the blown-up space to the space one starts with 
\cite{Griffiths:1978}.}.

To blow-up an $A_{1}$ singularity (locally given by $y^{2}=xz$
in $\mathbb{C}^{3}$) one introduces an extra $\mathbb{P}^{2}$ with
homogeneous coordinates $\xi_{i}$ and considers the set of
equations (see e.g. \cite{Griffiths:1978})
\begin{equation}
y^{2}  =xz,\qquad \xi_{1}y  =x\xi_{2},\qquad \xi_{1}z  =x\xi_{3}, \qquad \xi_{2}z  =y\xi_{3}
\end{equation}
in $\mathbb{C}^{3}\times\mathbb{P}^{2}$. The exceptional curve $C$ is a $\mathbb{P}^{1}$ given by
\begin{equation}
\xi_{2}^{2}  =\xi_{1}\xi_{3},\qquad\qquad x = y = z = 0.
\end{equation}
Its self-intersection is $C\cdot C=-2$.

The sections $\sigma$ are locally given by $y=x=0$. In the blown up
space $\mathbb{C}^{3}\times\mathbb{P}^{2}$ they are sitting at
\begin{equation}
x=y=0,\qquad\xi_{1}=\xi_{2}=0.%  ,\qquad\xi_{3}=1.
\end{equation}
This shows that the $\sigma_{ij}^{k}$ lead to smooth curves in the blown-up space. 
We furthermore deduce that the $\sigma$ intersect only those exceptional divisors that emerge 
at the four singularities they meet in $T^4/\mathbb{Z}_2$ before the blow-up. The even cycles 
of $T^{4}$, $\pi_{ij}$, are left completely unperturbed by the blow-up and cannot intersect
any of the exceptional divisors. We thus find the following intersections
in the smooth $K3$:
\begin{align}
C_{\lambda}\cdot C_{\eta} & =-2\delta_{\lambda\eta}, &\qquad& \pi_{ij}\cdot\pi_{ml} =2 \varepsilon_{ijml}, &\qquad& C_{\lambda}\cdot\pi_{ij} =0,\nn\\
\sigma_{ij}^{k}\cdot\pi_{ml} & =\varepsilon_{ijml}, &\qquad& \sigma_{jl}^{k}\cdot C_{\lambda} =1\hspace{0.2cm}\mbox{if}\hspace{1ex}i\in I_{jl}^{k}, &\qquad& \sigma_{jl}^{k}\cdot C_{\lambda} =0\hspace{0.2cm}\mbox{if}\hspace{1ex}i\not\in I_{jl}^{k}.\label{eq:inter}\end{align}

The index sets $I_{jl}^{k}$ can e.g. be determined from Figure~\ref{3cube} 
(remember that the $\sigma_{jl}^k$ correspond to the faces of the hypercube). As we know that the second homology of $K3$ is $22$-dimensional
and the cycles $C_{\lambda}$ and $\pi_{ml}$ are independent, it is clear
that we can use them as a basis for $H_{2}(K3,\mathbb{R})$. Thus
there exists an expansion of the cycles $\sigma_{jl}^{k}$ in terms
of this basis. Using the intersection numbers \eqref{eq:inter}, we 
conclude that\footnote{
This expression is consistent with those of the intersections between the $\sigma_{ij}^{k}$
that can be checked algebraically: If two faces do not meet
at all, their intersection number is clearly zero both algebraically and by \eqref{sigma}.
If they have one node in common, their intersection number is still zero
from (\ref{sigma}). This agrees with the algebraic model where one can check that
these two cycles miss each other in the blown-up $K3$. If two sections have two nodes in 
common, they can not be represented by algebraic subvarieties of \eqref{eq:alT4}. In this 
situation \eqref{sigma} determines their mutual intersection to be unity.}
\begin{equation}
\sigma_{ij}^{k}=\frac{1}{2}\cdot(\pi_{ij}-\sum_{\lambda\in I_{jl}^{k}}C_{\lambda}).\label{sigma}
\end{equation}

Before, we have shown that the $\sigma_{ij}^{k}$ are in fact elements
of the \emph{integral} homology of the smooth, blown-up $K3$. On the other hand we see
from \eqref{sigma} that they are not integral combinations of the $\pi_{ij}$ and $C_\lambda$. 
This tells us that the lattice of integral cycles consists of many more elements than
the ones in $A_{1}^{\oplus 16}\oplus U(2)^{\oplus3}$: it must also contain all
elements of the form \eqref{sigma}. It can moreover be shown that out of the $\sigma_{ij}^k$
and $C_\lambda$ one can construct a basis of integral cycles 
that has an intersection matrix with determinant minus one. 
As all self-intersections are even numbers, we have thus constructed an even 
unimodular lattice of signature $(3,19)$. This lattice must be $\Gamma_{3,19}=U^{\oplus3}\oplus(-E_{8})^{\oplus2}$,
the lattice of integral cycles of $K3$\cite{conwaysloane}. 

Note that the symmetries of $T^{4}/\mathbb{Z}_{2}$ are manifest in our construction. They simply
correspond to a relabeling of or a reflection along one of the four directions of the cube in~Figure~\ref{3cube}.

A similar construction of $\Gamma_{3,19}=H_2(K3,\mathbb{Z})$ has recently been discussed in 
\cite{Kumar:2009zc}. There it is exploited that $H_2(K3,\mathbb{Z})$ must be an unimodular lattice. 
This property of $H_2(K3,\mathbb{Z})$ is used to systematically 
enlarge $U(2)^{\oplus 3}\oplus A_1^{\oplus 16}$ to $\Gamma_{3,19}$. Our presentation differs in that 
we \emph{geometrically} identify the elements $\sigma_{ij}^k$ that enlarge the lattice
$U(2)^{\oplus 3}\oplus A_1^{\oplus 16}$ to $\Gamma_{3,19}$.

Related discussions of how to find integral cycles after blowing up singularities appear in \cite{Lust:2006zh} 
in the context of type IIB compactifications and in \cite{Nibbelink:2008tv} in the context of heterotic orbifolds (see also \cite{Heter1,Heter2,Heter3,Heter4}).

\subsubsection{Juxtaposition}
\label{sec7}
In the first part of this section, 
we have given a detailed description of the six finite size cycles of $T^4/\mathbb{Z}_2$ and of its
collapsed cycles. We have then identified them with holomorphic cycles in an algebraic model. Using the results of
Section~\ref{sec4}, we are able to match these cycles with the conventionally labelled $K3$ lattice of Section~\ref{k3mod} \footnote{
This shows that the embedding of $A_1^{\oplus 16}\subset H_2(K3,\mathbb{Z})$ obtained
in the first half of this paper is identical with the embedding of the $C_\lambda$ that is implicit from the last section.}:

The six torus cycles $\pi_{ij}$ are:
\begin{align}
\pi_{23}&=e^2+e^3&\qquad& \pi_{14}=e_2+e_3+2(e^2+e^3)+2W^3\nonumber \\
\pi_{12}&=e_1 &\qquad& \pi_{34}=2(e^1+e_1)+2W^1\nonumber\\ 
\pi_{13}&=e_2-e_3 &\qquad& \pi_{42}=e^2-e^3+2(e_2-e_3)+2W^2\:,
\end{align}
with $W^1$, $W^2$, $W^3$ given in \eqref{WLT4Z2}.

The exceptional cycles $C_\lambda$ are identified with the cycles in (\ref{16shrCycles}):
\begin{align}
C_{1} & =E_{1}+E_{2}+e_1 & C_{9} & =E_{9}+E_{10}-e_{1}\nonumber \\
C_{2} & =-E_{1}+E_{2}-e_1 & C_{10} & =-E_{9}+E_{10}+e_1\nonumber \\
C_{3} & =-E_{3}-E_{4} & C_{11} & =-E_{11}-E_{12} \nonumber \\
C_{4} & =-E_{3}+E_{4}+e^{2}+e^{3} & C_{12}& =-E_{11}+E_{12}+e^{2}+e^{3}\nonumber \\
C_{5}& =-E_{5}+E_{6} & C_{13} & =-E_{13}+E_{14}\nonumber \\
C_{6} & =E_{5}+E_{6}-e_{2}+e_{3} & C_{14}& =E_{13}+E_{14}+e_{2}-e_{3}\nonumber \\
C_{7} & =-E_{7}+E_{8}+e^{2}+e^{3} & C_{15} & =-E_{15}+E_{16}+e^2+e^3\nonumber \\
C_{8}& =-E_{7}-E_{8}+e_2-e_3 & C_{16} & =-E_{15}-E_{16}-e_2+e_3\:.\label{Ci}
\end{align}
% By writing down these assignments we have of course assumed that the
% two embeddings are in fact the same. What remains to be checked is the
% consistency of this assumption: both sides of the equations should 
% declare the same vectors to be lattice points in $U^3\oplus -E_8^{\oplus 2}$.
A non-trivial check of the identifications made above is to use \eqref{intk3} to show 
that all of the $\sigma_{ij}^k$ as given in (\ref{sigma}) are indeed elements
of the $K3$ lattice. The results are collected in Appendix~\ref{appa}.

We can now easily write down 
the roots of $E_8 \times E_8$ and the basis vectors $e_i, e^i$ of the three 
hyperbolic lattices in terms of the integral cycles we have found in the blow-up.
In terms of the standard labeling, they are given by
\begin{align}
&1:\frac{1}{2}\sum_{i=1}^8 E_i  =-\sigma_{12}^1-C_{3}-C_{8}+\pi_{13} &\qquad &2: -E_7-E_8  = C_{8}-\pi_{13} \nonumber \\
&3: -E_{6}+E_{7}  = \sigma_{23}^1 &\qquad &4:-E_{5}+E_{6}  = C_{5} \nonumber \\
&5:-E_{4}+E_{5}  = \sigma_{13}^1-\sigma_{23}^2+\pi_{23}+C_{6}-C_{4} &\qquad &6:-E_{3}+E_{4}= C_{4}-\pi_{23}\nonumber \\
&7:-E_{2}+E_{3}=\sigma_{23}^2 &\qquad &8:-E_{7}+E_{8}=C_{7}-\pi_{23}
\end{align}
for the first $E_8$ and by
\begin{align}
&1:\frac{1}{2}\sum_{i=9}^{16} E_i  =-\sigma_{12}^3-C_{16}-C_{11}+\pi_{12}-\pi_{13} &\qquad &\hspace{-2cm}2:-E_{15}-E_{16}=C_{16}+\pi_{13}  \nonumber \\  
&3:-E_{14}+E_{15}=\sigma_{23}^4 &\qquad &\nonumber\hspace{-2cm}4:-E_{13}+E_{14}=C_{13} \\
&5:-E_{12}+E_{13}= \sigma_{13}^2-\pi_{13}-\sigma_{23}^3+\pi_{23}+C_{14}-C_{12} \nonumber \\
&6:-E_{11}+E_{12}= C_{12}-\pi_{23}&\qquad &\hspace{-2cm}7:-E_{10}+E_{11}=\sigma_{23}^3 \nonumber\\
& & &\hspace{-2cm}8:-E_{15}+E_{16}=C_{15}-\pi_{23}
\end{align}
for the second $E_8$. We furthermore find that
\begin{align}
e_1&=\pi_{12} &\qquad& e^1=\sigma_{34}^2+C_{2}+C_{9}+\pi_{12} \nonumber \\
e_2&=\pi_{13}+e_3 &\qquad& e^2=\pi_{23}-e^3 \\
e_3&=\sigma_{14}^1-C_{7}-C_{4}-\sigma_{13}^4+\pi_{23}  &\qquad& e^3=\sigma_{42}^2-C_{16}+C_{8}+\sigma_{23}^1-\pi_{13}\:.\nonumber
\end{align}

\subsection{The Enriques involution}
\label{sec8}
In this section we will describe the Enriques involution in detail.

Let us first determine its action on the sixteen $A_1$ singularities and the corresponding exceptional 
divisors from its action on $H_2(K3,\mathbb{Z})$ 
\cite{peters,Berglund:1998va}:
\begin{equation}
\vartheta:\,e_{1} \mapsto-e_{1} \qquad e^{1} \mapsto-e^{1} \qquad
e_{2} \leftrightarrow e_{3} \qquad e^{2} \leftrightarrow e^{3} \qquad
E_{I} \leftrightarrow E_{I+8}\:. \label{actenr}
\end{equation}
From \eqref{Ci} we see that $C_\lambda\leftrightarrow C_{\lambda+8}$. Considering Figure~\ref{3cube}, this
means that the singularities are exchanged along the $3$-$4$-directions. This can be reproduced from the action 
of the Enriques involution on $T^4/\mathbb{Z}_2$, see \eqref{ent4}. We can also see the same behavior in the 
description of $T^4/\mathbb{Z}_2$ as a hypersurface, (\ref{eq:alT4}): 
By shifting and rescaling $x$ and $z$, we can always arrange that $x_1=-x_2$, $x_3=-x_4$ and $z_1=-z_2$, $z_3=-z_4$. 
The Enriques involution then acts as $\vartheta:(y,x,z) \mapsto (-y,-x,-z)$ \cite{peters}, so that the sixteen~$A_1$
singularities are exchanged as noted before.

\begin{figure}
\begin{center}
\includegraphics[height=4cm]{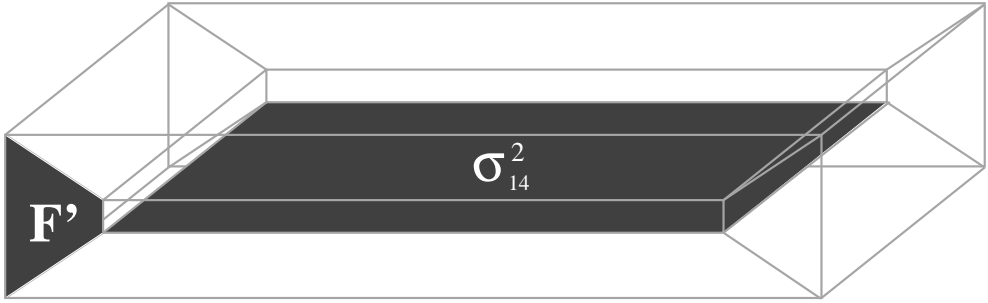}\caption{\textsl{Choosing the generic fibre to be in the homology class $\pi_{23}$, the
sections are in the $1$-$4$ direction (compare with Figure~\ref{3cube}). We have depicted the section $\sigma_{14}^2$
and the finite-size component of one of the singular fibres, $F'=\sigma_{23}^1$.}}\label{FandB}
\end{center}
\end{figure}

To fix an elliptic fibration of $T^4/\mathbb{Z}_2$, we first select $\pi_{23}=e^2+e^3$ as the homology class of the 
generic fibre. It is obviously invariant under the Enriques involution \eqref{actenr}. The sections are then given
by the $\sigma_{14}^k$, see Figures~\ref{3cube} and \ref{FandB}. 
They can be expressed in terms of the $K3$ lattice as
\begin{align}
\sigma_{14}^1 & =e_3+E_4+E_8\nonumber \\
\sigma_{14}^2 & =e_2+E_{12}+E_{16}\nonumber \\
\sigma_{14}^3 & =e_1+e_3+e^2+e^3 +\frac{1}{2}\left(E_1-E_2-E_3+E_4+E_5-E_6-E_7+E_8\right) \nonumber \\
& +\frac{1}{2}\left(-E_9-E_{10}-E_{11}+E_{12}-E_{13}-E_{14}-E_{15}+E_{16}\right) \nonumber \\
\sigma_{14}^4 & =-e_1+e_2+e^3+e^2 +\frac{1}{2}\left(-E_1-E_2-E_3+E_4-E_5-E_6-E_7+E_8\right) \nonumber \\
& +\frac{1}{2}\left(E_9-E_{10}-E_{11}+E_{12}+E_{13}-E_{14}-E_{15}+E_{16}\right).\label{sectenr}
\end{align}
The Enriques involution acts by exchanging them pairwise. This implies that the resulting Enriques surface is 
elliptically fibred with a two-section, i.e. $\tilde{B}\cdot \tilde{F}=2$. This result is expected from the 
general theory of Enriques surfaces \cite{peters}.
Note that the pairwise exchange of the sections under the Enriques involution can also be seen from (\ref{eq:alT4}).

\subsubsection{The standard Weierstrass model}\label{sec61}

We now want to make contact with a Weierstrass model with constant $\tau$. It takes the form \cite{Sen:1996vd}
\begin{equation}\label{weierorb}
y^2=x^3+\alpha_1 h^2xz^4+\alpha_2 h^3z^6=(x-\gamma_1 z^2h)(x-\gamma_2 z^2h)(x-\gamma_3 z^2h).
\end{equation} 
Here $\gamma_i$ and $\alpha_i$ are complex constants and $h$ is a homogeneous polynomial of the base coordinates of degree $4$.
Contrary to $T^4/\mathbb{Z}_2$, the surface described by this equation has four $D_4$ singularities.
There are three sections given by $y=0, x=\gamma_iz^2h$ that pass through the four $D_4$ singularities at $y=x=h=0$. The fourth 
section at $x^3=y^2, z=0$ does not hit any singularity. This section is a special feature of the Weierstrass model and we will 
denote it by $\hat{\sigma}$ in the following.

From what we have said, it is clear that $T^4/\mathbb{Z}_2$ cannot be described by a Weierstrass model. In fact, the section $\hat{\sigma}$ must be orthogonal to all shrinking cycles, and then, for $T^4/\mathbb{Z}_2$ it should belong to $\Upsilon$ (see \eqref{eVSpi}). But this is not possible, since this is a lattice with all intersection numbers being even, and the section $\hat{\sigma}$ should have intersection one with the fibre. %no elliptic fibre and holomorphic section exist whose mutual intersection is one. Thus there can not be a Weierstrass model description of this space.

\begin{figure}[tt]
\begin{center}
\includegraphics[height=55mm]{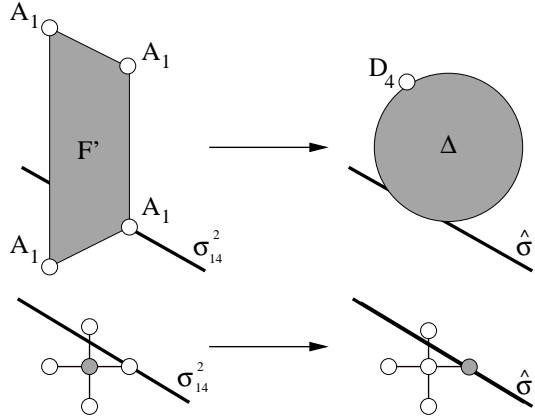}\caption{\textsl{When blowing up the $A_1$ singularity hit by the section $\sigma^2_{14}$ while collapsing the $F'$ component of the singular fibre, we produce a $D_4$ singularity. This $D_4$ singularity is not hit by the section $\sigma^2_{14}$, which is then identified with 
$\hat{\sigma}$. In this figure we display singularities and collapsed cycles in white and cycles of finite size in light grey. In the lower part of the figure we have drawn the intersection pattern between the cycles in a diagrammatic fashion. After the $F'$ component of the singular fibre is blown down and the $A_1$ singularity hit by the section is blown up, the collapsed cycles intersect according to the Dynkin diagram of $SO(8)$, so that this operation produces a $D_4$ singularity.}}\label{a1tod4}
\end{center}
\end{figure}

Intuitively, there is an obvious way how to get from $T^4/\mathbb{Z}_2$ to an elliptic $K3$ described by \eqref{weierorb}. First,
we choose one of the sections of $T^4/\mathbb{Z}_2$, say $\sigma_{14}^2$, that is to become $\hat{\sigma}$. We then blow up the 
singularities which are hit by this section while blowing down the finite-size components $F'$ of the four singular fibres. We 
have depicted this deformation in Figure~\ref{a1tod4}. 
The section $\sigma_{14}^2$, which is now identified with $\hat{\sigma}$, no longer intersects any singularities and the lattice 
of collapsed cycles is exactly $D_4^{\oplus 4}$. The other three sections are all forced to meet at the $D_4$ singularities.

After deforming $T^4/\mathbb{Z}_2$ to an elliptic $K3$ described by \eqref{weierorb}, the $\sigma_{14}^k$ remain sections
of the elliptic fibration. The symmetry among them that is present in $T^4/\mathbb{Z}_2$, however, is lost. This is what 
prevents the Enriques involution from acting on an elliptic $K3$ described by a Weierstrass model like \eqref{weierorb}.

We can make this more precise using our description of $T^4/\mathbb{Z}_2$ as a point in the moduli space of $K3$, that is, 
the position of the 3-plane $\Sigma$ with respect to the $K3$ lattice of integral cycles.
The prescription that comes from the previous consideration is that one has to move the plane $\Sigma$ such that the cycles intersecting the section $\sigma^2_{14}$ blow up, while the corresponding $F'_k=\sigma_{23}^k$ shrink to zero size.

The four sets of cycles that intersect as in Figure~\ref{a1tod4} are:
\begin{eqnarray}
 &C_1,C_2,C_3,C_4,F'_2\equiv\sigma_{23}^2=-E_2+E_3 \nn \\ &C_5,C_6,C_7,C_8,F'_1\equiv \sigma_{23}^1=-E_6+E_7 \nonumber\\
 &C_9,C_{10},C_{11},C_{12},F'_3\equiv \sigma_{23}^3=-E_{10}+E_{11} \nn\\&C_{13},C_{14},C_{15},C_{16},F'_4\equiv \sigma_{23}^4=-E_{14}+E_{15}\nonumber
\end{eqnarray}
Before rotating $\Sigma$, the cycles $C_\lambda$ are shrunk, while the $F'_k=\sigma_{23}^k$ have finite size. To go to the $D_4^{\oplus 4}$ point described by the Weierstrass model, the four cycles $C_4,C_7,C_{11},C_{16}$ must blow up, while the $\sigma_{23}^k$ must shrink. This requires the plane to be located in the subspace orthogonal to $\sigma_{23}^k$ ($k=1,...,4$) and $C_{\lambda}$ ($\lambda=1,2,3,5,6,8,9,10,12,13,14,15$). The latter is generated by:
\begin{equation}\label{Thlatt}\begin{array}{lcl}
\pi_{23}=e^2+e^3,&\qquad& \sigma^2_{14}=e_2+E_{12}+E_{16},\\ \pi_{12}=e_{1},&\qquad& \pi_{34}=2(e^1+e_1+W^1),\\ 
\pi_{13}=e_2-e_3, &\qquad& \pi_{42}=e^2-e^3+2(e_2-e_3+W^2)\:.
\end{array}\end{equation}
These integral cycles generate the lattice contained in this subspace and have intersection matrix\footnote{The corresponding embedding of the $D_4^{\oplus 4}$ lattice in the $K3$ lattice is equivalent (i.e. connected by an automorphism of the $K3$ lattice) to the one given 
in Section~\ref{so8}.}:
\begin{equation}
\left(\begin{array}{cccccc}
 0 & 1 & & & & \\ 1 & -2 & & & & \\ & & 0 & 2 & & \\ & & 2 & 0 & & \\ & & & & 0 & 2\\ & & & & 2 & 0\\
\end{array}\right)\:.
\end{equation}
To specify a complex structure compatible with the Enriques involution, we have to choose $\Omega$ as a linear combination of the odd cycles in \eqref{Thlatt} and $J$ as a linear combination of the even cycles in \eqref{Thlatt}. As $\pi_{23}$ is the only even cycle, $J$ must be proportional to it. This, however, violates the requirement $J\cdot J>0$.

\

We can also see the clash between the Weierstrass model description and the Enriques involution from a different perspective.
We start with the 3-plane $\Sigma$ in the lattice $\Upsilon$ (see \eqref{eVSpi}). %Since this is a lattice with all intersection numbers being even, no elliptic fibre and holomorphic section orthogonal to the shrinking $A_1$-cycles exist whose mutual intersection is one. Thus there can not be a Weierstrass model description of this space.
Since we want a complex structure compatible with a holomorphic Enriques involution, we take $\Omega$ in the odd subspace of $\Upsilon$.
We now try to make the rotation to an $SO(8)^4$ point, maintaining the symmetry under the Enriques involution.
Since the third Wilson line $W^3$ is symmetric under the 
Enriques involution, we can switch it off without destroying the symmetry. Note that this means that we have only changed $J$.
From the discussion of Section~\ref{sec2} to Section~\ref{sec4} it is clear that removing $W^3$ will result in a $K3$ with four $D_4$ singularities. 
Now the 3-plane $\Sigma$ lives in\footnote{Notice that, in contrast to \eqref{Thlatt}, these cycles do not generate the lattice orthogonal to the shrinking cycles.}
\begin{equation}\label{Thlatt2}\begin{array}{lcl}
\pi_{23}=e^2+e^3,&\qquad& e_2+e_3,\\ \pi_{12}=e_{1},&\qquad& \pi_{34}=2(e^1+e_1+W^1),\\ 
\pi_{13}=e_2-e_3, &\qquad& \pi_{42}=e^2-e^3+2(e_2-e_3+W^2)\:.
\end{array}\end{equation}
Comparing with \eqref{Thlatt}, we see that we have replaced $\sigma^2_{14}$ by $e_2+e_3$. This means that this time we have blown up the cycles $C_4,C_7,C_{12},C_{15}$ while shrinking the cycles $\sigma^k_{23}$. %Looking back at Fig.~\ref{3cube} one can check that there is no holomorphic section $\hat{\sigma}$ that hits none of the singularities.
When $\Sigma$ lives in \eqref{Thlatt2}, we can find a section that does not meet any singularities, e.g. $e_2-(e^2+e^3)$. However, there exists no choice for $\Omega$ such that $\Omega$ is odd and orthogonal to this section at the same time. In fact, these two conditions require $\Omega$ to live in a subspace with degenerate metric, as can be seen by looking back at \eqref{Thlatt2}. 
The complex structure that is demanded by the holomorphicity of the section $\hat{\sigma}$ and the complex structure demanded by the Enriques involution are not compatible.

In summary: Starting from $T^4/\mathbb{Z}_2$, there are two ways to rotate the 3-plane $\Sigma$ such as to get a Weierstrass model with 
$D_4^{\oplus 4}$ singularity. They have different behavior with respect to the Enriques involution: In the first case, we destroy the symmetry. 
In the second case, the symmetry is preserved, but there is no choice of complex structure that both admits a holomorphic section (which does not hit 
the singularities) and makes the Enriques involution holomorphic.

\subsubsection{A symmetric Weierstrass model}\label{sec62}

In the standard Weierstrass model description, in which the fibre is embedded as a hypersurface in  $\mathbb{P}_{1,2,3}$,
one always has one section $\hat{\sigma}$. By embedding the fibre in other spaces, it is
possible to obtain elliptic $K3$ surfaces with two or more sections \cite{Klemm:1996ts,Berglund:1998va}. 
In particular, it is known that embedding the fibre in $\mathbb{P}_{1,1,2}$ yields an elliptic $K3$ with
two sections which are permuted under the Enriques involution \cite{Berglund:1998va}. 
This elliptic $K3$ is given by an equation of the form
\begin{equation}
y^2=x^4+x^2z^2f_4+z^4f_8.\label{2weier}
\end{equation}
The $\mathbb{Z}_2$ transformation $(y,x,z)\mapsto (-y,x,-z)$ together with a holomorphic involution of the $\mathbb{P}^1$ base
has no fixed points and projects out the holomorphic 2-form, so that it provides an Enriques involution of $K3$.
The two holomorphic sections $\hat{\sigma}_1,\hat{\sigma}_2$ are given by $z=0, y=\pm x^2$ and are permuted 
under the Enriques involution. The j-function of this fibration is given by \cite{Berglund:1998va}:
\begin{equation}
\frac{1}{108}\frac{(f_4^2+12f_8)^3}{f_8(-f_4^2+4f_8)^2}.
\end{equation}
Let us discuss the limit in which the complex structure of the fibre is constant. To achieve this, we
take $f_8=f_4^2$. Setting $z=1$ and shifting $f_4$ by some multiple of $x^2$ to complete the square\footnote{Note that 
this is a bijective map between the coordinates $y,x,f_4$ and $y,x,f_4'$.}, we find the equation 
\begin{equation}
y^2=f_4'^2+x^4.
\end{equation}
Thus there are four $A_3$ singularities at the four points $f_4=x=y=0$. 

\begin{figure}
\begin{center}
\includegraphics[height=6cm]{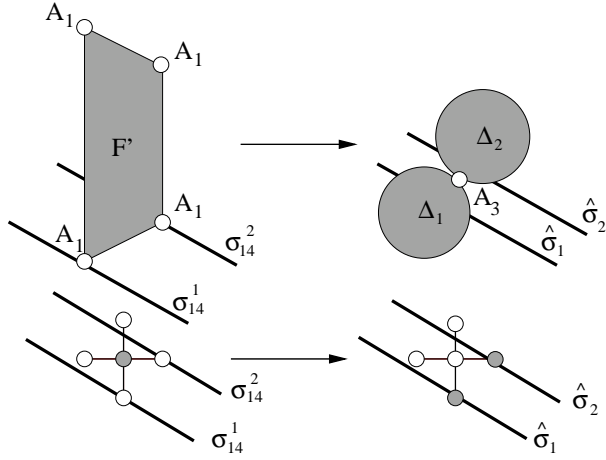}\caption{\textsl{When blowing up two $A_1$ singularities while shrinking the 
finite-size component of the singular fibre, we produce an $A_3$ singularity and two sections, $\hat{\sigma}_1$ and $\hat{\sigma}_2$, 
which do not hit any singularities. This works in a similar way as the deformation of $T^4/\mathbb{Z}_2$ to an $SO(8)^{\oplus 4}$
configuration, see Figure~\ref{a1tod4}. We again display singularities and collapsed cycles in white and cycles 
of finite size in light grey.}}\label{a1toa3}
\end{center}
\end{figure}

Let us find this configuration by deforming $T^4/\mathbb{Z}_2$. The strategy is similar to that employed
for the deformation of $T^4/\mathbb{Z}_2$ to a $D_4^{\oplus 4}$ configuration. In order to get two sections that
do not hit any singularities and that are interchanged by the Enriques involution we have to blow up
$C_\lambda$, $\lambda=3,4,7,8,11,12,15,16$. At the same time we shrink the cycles $\sigma_{23}^k$ to produce four $A_3$ singularities,
see Figure~\ref{a1toa3}.

This can be realized by\footnote{We have chosen $J$ and $\Omega$ in a six-dimensional subspace of the 10-dimensional space
orthogonal to the~12~$A_3^{\oplus 4}$ cycles. The lattice of cycles orthogonal to a generic $\Sigma$, i.e. orthogonal to the six
basis cycles of \eqref{a34}, then has a dimension 
which is bigger than~12. By examining this lattice, one can check that in spite 
of this the singularity is still $A_3^{\oplus 4}$. Another way to see this is through the associated 
Wilson-line breaking.}%: the Wilson-line appearing in $J$ ($W^{3'}=(0^3,\frac12,0^3,\frac12,0^3,\frac12,0^3,\frac12)$) breaks $SO(8)$ without preserving its rank.}Maximal subgroup??? 
%Rob give Wilson-line more on this kind of breaking ?
\begin{eqnarray}
J&=&b\,\pi_{23}+f\pi_{14}-\frac{f}{2}\sum_\lambda C_\lambda=b\,\pi_{23}+f\left(\sigma_{14}^1+\sigma_{14}^2\right), \nonumber \\
\Omega&=&\pi_{34}+U\,\pi_{13}+S\,\pi_{42}-U\,S\,\pi_{12}.   \label{a34}
%\omega&=&s_1\pi_{12}+s_2\pi_{34}+s_3\pi_{13}+s_4\pi_{42}.
\end{eqnarray}
Here $f$ gives the volume of the elliptic fibre. 
The two sections $\hat{\sigma}_{1}=\sigma_{14}^1$ and $\hat{\sigma}_{2}=\sigma_{14}^2$ are orthogonal to $\Omega$ and do not intersect any
of the collapsed cycles. $J$ and $\Omega$ have the right transformation properties under the Enriques involution.

\section{F-theory Limits}\label{sec9}

There is more than one way to construct an elliptic Calabi-Yau (n+1)-fold that describes a type IIB orientifold compactification 
on a Calabi-Yau n-fold\footnote{Running Sen's weak coupling limit \cite{Sen:1996vd,Sen:1997gv,Sen:1997kw} backwards, a general
procedure to construct an F-theory Calabi-Yau 4-fold, given a generic type IIB setup with D7-branes and O7/O3-planes, was obtained in \cite{Collinucci:2008zs,Blumenhagen:2009up}.}. In this section we will first discuss the F-theory limits
of the examples given above. Even though they all describe type IIB vacua with constant dilaton, the corresponding M-theory 
backgrounds are quite different. They only unify in the F-theory limit. We will illustrate this fact for the simple examples 
described in this paper and consider an elliptically fibred Calabi-Yau two-fold, i.e. $K3$, whose fibre 
has a constant complex structure. We then argue that a similar result holds for configurations in which the
dilaton varies. We reconsider our results in Section~\ref{sec:ftheory}.

We now consider two different types of Weierstrass models with constant $\tau$ and the $T^4/\mathbb{Z}_2$ limit of $K3$:
\begin{itemize}
\item \textit{An elliptically fibred $K3$ with one distinguished section.} There are four points on the base $\mathbb{P}^1$ where 
the 2-fold develops a $D_4$ singularity. The cycles corresponding to fibre and section are $F=\pi_{23}$ and 
$\hat{\sigma}=\sigma_{14}^2$. The K\"ahler form and the complex structure live in the space \eqref{Thlatt}. They can be chosen as\footnote{The most general expression for $J$ also includes two deformations in $\langle \pi_{12},\pi_{34},\pi_{13},\pi_{42}\rangle$; we do not include these here, as they are not relevant for the 7-dimensional gauge group and go to zero in the F-theory limit \cite{Braun:2008pz, Valandro:2008zg}.}:
\begin{eqnarray}
J&=&b\,\pi_{23}+f\sigma_{14}^2, \nonumber \\
\Omega&=&\pi_{34}+U\,\pi_{13}+S\,\pi_{42}-U\,S\,\pi_{12}.
\end{eqnarray}
This point in moduli space is the one reached from $T^4/\mathbb{Z}_2$ by the rotation of $J$ described in Section~\ref{sec61}.
\item \textit{An elliptically fibred $K3$ with two distinguished sections.} Again there are four points on the base where
the two-fold develops a singularity. This time, however, this is an $A_3$ singularity, as described in Section~\ref{sec62}.
The K\"ahler form and the complex structure can be given by
\begin{eqnarray}
J&=&b\,\pi_{23}+f\left(\sigma_{14}^1+\sigma_{14}^2\right), \nonumber \\
\Omega&=&\pi_{34}+U\,\pi_{13}+S\,\pi_{42}-U\,S\,\pi_{12}.
\end{eqnarray}
\item \textit{The space $(T^2\times T^2)/\mathbb{Z}_2$, i.e. the $T^4/\mathbb{Z}_2$ limit of $K3$.} This manifold has 
16 $A_1$ singularities. One choice for the K\"ahler form and the complex structure is
\begin{eqnarray}
J&=&b\,\pi_{23}+f\pi_{14}, \nonumber \\
\Omega&=&\pi_{34}+U\,\pi_{13}+S\,\pi_{42}-U\,S\,\pi_{12}.
\end{eqnarray}
\end{itemize}
We notice that the only difference between the three cases is the expression for the K\"ahler form $J$.

Compactifying M-theory on these manifolds gives different 7-dimensional spectra, as all three have different singularities. In particular, we obtain the gauge groups $SO(8)^4$ in the first case, $SO(6)^4\times U(1)^4$ in the second case and $SU(2)^{16}$ in the third case. In the dual type IIB model on $S^1_B\times T^2/\mathbb{Z}_2$, we have four D7-branes on top of each O7-plane wrapping $S^1_B$. However, in the second and third case the gauge group is broken by Wilson lines along the~$S^1_B$~\footnote{The deformations of $J$ inside the $SO(8)$ cycles are mapped to the 8th component of the type IIB vectors (see \cite{Valandro:2008zg} for details).}. On the type IIB side, the F-theory limit is given by $R_B\rightarrow \infty$. 
In this limit the $S^1_B$ decompactifies and the Wilson lines become trivial, leaving $SO(8)^4$ as the 8-dimensional gauge group.

\

Let us have a more detailed look at the F-theory limit from the M-theory perspective, i.e., we send the fibre size to zero and see how the 
K\"ahler form and the complex structure behave.

\begin{itemize}
 \item In the first case, the F-theory limit is very simple: since the fibre $F$ is orthogonal to $\Omega$, 
its size is given by
\begin{equation}\label{limitweier}
\rho(F) = \int_F J = F\cdot J = f.
\end{equation}
This vanishes in the F-theory limit $f\rightarrow 0$, and the K\"ahler form becomes $J\rightarrow b\,\pi_{23}$.

Note that we find some further shrinking cycles in this limit: the cycles $C_4,C_7,C_{11},C_{16}$ only have a finite size due to their intersection with
$\sigma_{14}^2$ in $J$. Letting $f\rightarrow 0$ they collapse so that the intersection pattern of shrunk cycles is now
four times the \emph{extended} Dynkin diagram of $SO(8)$. This is expected from a general perspective: The component of the fibre that
has finite size and the four associated collapsed cycles have the extended Dynkin diagram of $SO(8)$ as their intersection pattern (see Figure~\ref{a1tod4}). Their sum, i.e the singular fibre, is homologous to the generic fibre, see e.g. \eqref{sigma}. Sending the volume of the 
generic fibre to zero, all five cycles have to collapse. 

\item The second case differs only through the term proportional to $f$ in the K\"ahler form~$J$. In the
limit~$f\rightarrow 0$ we thus reach the same point in the moduli space of $K3$.

\item The same happens for $T^4/\mathbb{Z}_2$. Our choice of $J$ and $\Omega$ has of course been completely arbitrary.
Using Figure~\ref{3cube}, we can easily discuss the most general case: $J$ is then given as
\begin{equation}\label{KaelerFormT4Z2}
 J=f\pi_{ij}+b\,\pi_{ml},
\end{equation}
with four different indices $i,j,m,l$. The holomorphic 2-form $\Omega$ lives in the space spanned by the
$\pi_{pq}$ that have zero intersection with the K\"ahler form \eqref{KaelerFormT4Z2}. Besides the sixteen cycles $C_\lambda$ we find that
all of the four $\sigma_{ml}^k$ (with $k=1,...,4$) are collapsed when $f\rightarrow 0$. Theses 20 cycles
intersect precisely according to the extended Dynkin diagram of $SO(8)^{\oplus 4}$, as expected. 
\end{itemize}

Let us discuss the F-theory limit from an even more general perspective and consider situations
in which $\tau$ is not constant over. For this we use the description of Section~\ref{wilson}.
Let us choose $e_1$ to denote the fibre and start at a configuration with an $E_8\times E_8$ singularity.
Let us take $J$ to be given by
\begin{align}
  J&= c e_1 + fe^1 \ ,
\end{align}
so that the volume of the fibre is $f$. Let us furthermore choose $\Omega$ to be in the
subspace spanned by $e_2, e^2, e_3$ and $e^3$, so that it is orthogonal to $J$ and the fibre.
We can rotate the three-plane $\Sigma$ to any position by rotating the three basis vectors
$e^i$ to $e^i+W^i_IE_I$. The singularity we will find is equal to the breaking induced
by the three Wilson lines $W^i$.
The K\"ahler form is 
\begin{align}
  J&= c e_1 + f(e^1+W^1_IE_I) \ ,
\end{align}
The F-theory limit is now given by $f\rightarrow 0$.
This limit has to be such that the norm of $J$, given by
\be
J\cdot J= 2cf -f^2W^1_IW^1_I \ .
\ee
stays positive. This means that $W^1$ is completely irrelevant in the F-theory limit and only the gauge symmetry 
breaking induced by $W^2$ and $W^3$ plays a role. Note that we cannot let $W^1\sim f^{-1}$ because of the positivity
bound on the norm of $J$. Remember that the Weierstrass model description of $K3$ corresponds to switching on only Wilson 
lines only in the expression for $\Omega$, but none in the expression for $J$. As the F-theory limit sets to zero the Wilson 
line in the expansion of $J$, the gauge symmetry breaking, which is equivalent to the structure of the singularity, stays the 
same for the Weierstrass model whereas it always reduces to the gauge symmetry breaking induced by the Wilson lines in $\Omega$ 
in other cases. Hence the Weierstrass model is the minimal choice.

We also found another result: Before the F-theory limit, the Enriques involution is consistent only with $T^4/\mathbb{Z}_2$ and 
the symmetric Weierstrass model. {\it In the F-theory limit the Enriques involution is also consistent with the standard Weierstrass model.} The
point here is that one can have different M-theory versions of F-theory vacua, so that some symmetries of the F-theory model may
only show up in the F-theory limit. The Weierstrass model seems like the natural choice for F-theory compactifications:
the F-theory limit has its simplest form from its perspective, see \eqref{limitweier}, and, as we have shown above, the singularity 
structure is unperturbed. The example of the Enriques involution is special because a symmetry that is present after 
performing the F-theory limit cannot be anticipated from the Weierstrass model. Thus its minimality can come with a price.

\chapter{F-theory on elliptic Calabi-Yau threefolds} \label{chapter3}

In this chapter, we discuss F-theory compactifications on elliptic three-folds and 
present a hands-on approach to the parameterization of the brane moduli space.
In particular, we find a parameterization in terms of the periods of the elliptically fibred
Calabi-Yaus in the dual F-theory picture.
These compactifications correspond to type IIB orientifold models on complex surfaces.

Type IIB orientifolds with two compact complex dimensions arise from involutions acting on $K3$. Involutions of $K3$ have been classified by Nikulin~\cite{Nikulin:1979,Nikulin:1983,Nikulin:1986}. The resulting base spaces $B$ are Fano surfaces or (blow-ups of) Hirzebruch surfaces.
The fixed point locus of the involution defines the O-plane.
The corresponding F-theory model is defined on the Calabi-Yau threefold that is constructed as an elliptic fibration over the base space $B$. 
The monodromy points of the fibration give the location of the branes in $B$ and thus the motion of branes corresponds to complex structure deformations of the fibration, i.e.\ complex structure deformations of the threefold. Note that the branes are located on holomorphic hypersurfaces and therefore their positions are described by holomorphic polynomials, i.e.\ the branes are given by divisors.
F-theory models on Calabi-Yau threefolds have been first discussed in~\cite{Morrison:1996na,Morrison:1996pp}.

By performing the weak coupling limit we can make contact with the corresponding orientifold model~\cite{Sen:1996vd,Sen:1997gv}.
The monodromy of D-branes and O-planes acts on the 1-cycles in the fibre torus. If one combines such a 1-cycle with a real surface 
encircling the brane in an appropriate way, one is able to construct a non-trivial 3-cycle. The deformation of such 3-cycles 
characterizes the deformation of the corresponding branes.

We are able to geometrically construct all 3-cycles in the way described above. We start at the orientifold point in moduli space, where the O-plane coincides with four D-branes, and construct the 3-cycles of the threefold describing the motion of the O-plane in the base. Then we move one D-brane after the other off the O-plane and construct the emerging cycles corresponding to their motion. By counting the degrees of freedom we see that we find all 3-cycles. 

We will now give an overview of this work including the main results of each
section.

In Section~\ref{localConstruction} we start by investigating the recombination
of branes in a small neighborhood around an intersection point.
Geometrically, the recombination process blows up the nodal
point\footnote{At a nodal point an embedded Riemannian
surface is locally described by the equation $xy=0$ with $x,y \in \C$. This
gives rise to two intersecting hypersurfaces situated at $x=0$ and $y=0$.} at the
intersection, which generates a 1-cycle of the recombined brane with
non-vanishing size. This 1-cycles is the boundary of a disc: a relative 2-cycle. 
By fibering this disc with the 1-cycle of $T^2$ that degenerates on the boundary, 
these 1-cycles are shown to be in one-to-one correspondence to F-theory 3-cycles 
in the case of D-branes. In the case of O-planes, each such 1-cycle corresponds to 
two 3-cycles of the underlying F-theory threefold. The periods associated to these 
F-theory cycles determine the recombination of D-branes and O-planes.

We start addressing global issues in Section~\ref{DonO}. For simplicity, we
first restrict ourselves to the subset of elliptically fibred Calabi-Yau
spaces $Z$ which correspond to type IIB models at the orientifold point.
This means that each O-plane coincides with a stack of four D-branes. From
the geometric point of view, the brane locus is described by a Riemannian
surface $\cal D$ embedded in the base space $B$. According to the analysis of
Section~\ref{localConstruction}, the 1-cycles of $\cal D$ correspond to
3-cycles of $Z$ which parameterize the motion of the O-plane. In particular, a
self-intersection point of this O-plane develops if any of these 3-cycles
shrinks. We will refer to these cycles as 'recombination cycles'. We find that,
in addition to recombination cycles, there exist F-theory 3-cycles which are
associated to the 2-cycles of the base space $B$ of the type IIB model.
Locally these cycles can be visualized as products of the 2-cycles in
the base and a 1-cycle of the $T^2$ fibre. Combining them with the
recombination cycles, we have enough periods to parameterize the complex
structure of the type IIB orientifold model. Thus we have constructed all
3-cycles of $Z$ which have non-zero volume at the orientifold point.

A first step towards the generic situation is taken in Section~\ref{D7-branes without obstructions}.
First, we separate only one of the D-branes from the D-brane stack, leaving three
D-branes on top of the O-plane. Geometrically, the situation is appropriately
described by two hypersurfaces of the same degree embedded in $B$, which
generically intersect each other in isolated points. Fixing the O-plane in $B$,
we then identify loci in the base which correspond to F-theory
3-cycles. In contrast to the cycles governing the deformations of the O-plane, we
end up with cycles of two different kinds. The first kind of cycle is a relative
2-cycle stretched between a 1-cycle of the D-brane and a 1-cycle of the O-plane.
It locally measures the distance between D-brane and O-plane.
The second kind of cycle is again a relative 2-cycle. By contrast to the first kind, 
its boundary is not formed by 1-cycles in D-brane and O-plane but by two lines 
connecting a pair of O-plane-D-brane intersection points. One may think of
this 2-cycle as measuring both the distance between D-brane and O-plane 
and the distance between two of their intersection points.

Next, we consider more general D-brane configurations, demanding only that at least 
one D-brane remains on top of the O-plane. In this case, D-brane deformations are still 
associated to deformations of generic hypersurfaces.
To be more explicit, the D-brane locus $\eta^2+h\chi=0$, in the
notation of \cite{Sen:1997gv}, can be restricted to be of the form
$\eta=hp$, which yields $h(hp^2+\chi)=0$. This
corresponds to a situation, in which one D-brane coincides with the
O-plane, but all other D-branes are recombined into the generic surface $\chi'=hp^2+\chi=0$.
In this case we can construct a complete base of $H_3(Z)$ by iteratively moving single 
D-branes independently off the O-plane and letting them recombine at their intersection 
points according to the results in Section~\ref{localConstruction}.

In Section~\ref{moreDoffO} we finally discuss the most general case of a `naked' O-plane and a fully
recombined single D-brane. The latter can only have double intersections 
with the O-plane and is hence no longer given by a generic hypersurface \cite{Braun:2008ua, Collinucci:2008pf}. 
We compute the number of moduli in three ways. First we count the number of deformations that are contained in a 
polynomial of the form $\eta^2+h\chi=0$ for any given base space. We show that the difference between this
number and the number of moduli of a generic hypersurface is given by the number of double-intersection points

We then reproduce this counting by describing deformations of the brane through
sections of the canonical bundle, as suggested in \cite{Beasley:2008dc}. Finally, we
show that our construction yields precisely the right number of cycles to explain the
degrees of freedom from the perspective of the elliptic threefold.

\section{Local construction} \label{localConstruction}

We are interested in the recombination and displacement of O7-planes
and D7-branes in complex two-dimensional type IIB orientifolds. In the picture
given by F-theory, these moduli are encoded in complex structure deformations of
the corresponding elliptically fibred Calabi-Yau threefold. As these complex
structure deformations originate from 3-cycles we are interested in finding
these. To begin our analysis, we start constructing the threefold cycles in the
weak coupling limit locally from the topology of the D-branes and O-planes and
our knowledge of the fibration. In a similar fashion as in Section~\ref{cyclesk3branes}
we will describe these cycles as a fibration of a 1-cycle in the fibre over some
(real) surface in the base. Instead of considering the whole threefold, we will
first consider O7-plane/D7-brane configurations in flat space. As we know the
elliptic fibration over these configurations, this will give a local picture of
the elliptically Calabi-Yau threefold in which we can identify some of the
3-cycles.

\subsection{Recombination of two intersecting D7-branes} \label{recomD}

\begin{figure}
\begin{center}
\includegraphics[height=4cm, angle=0]{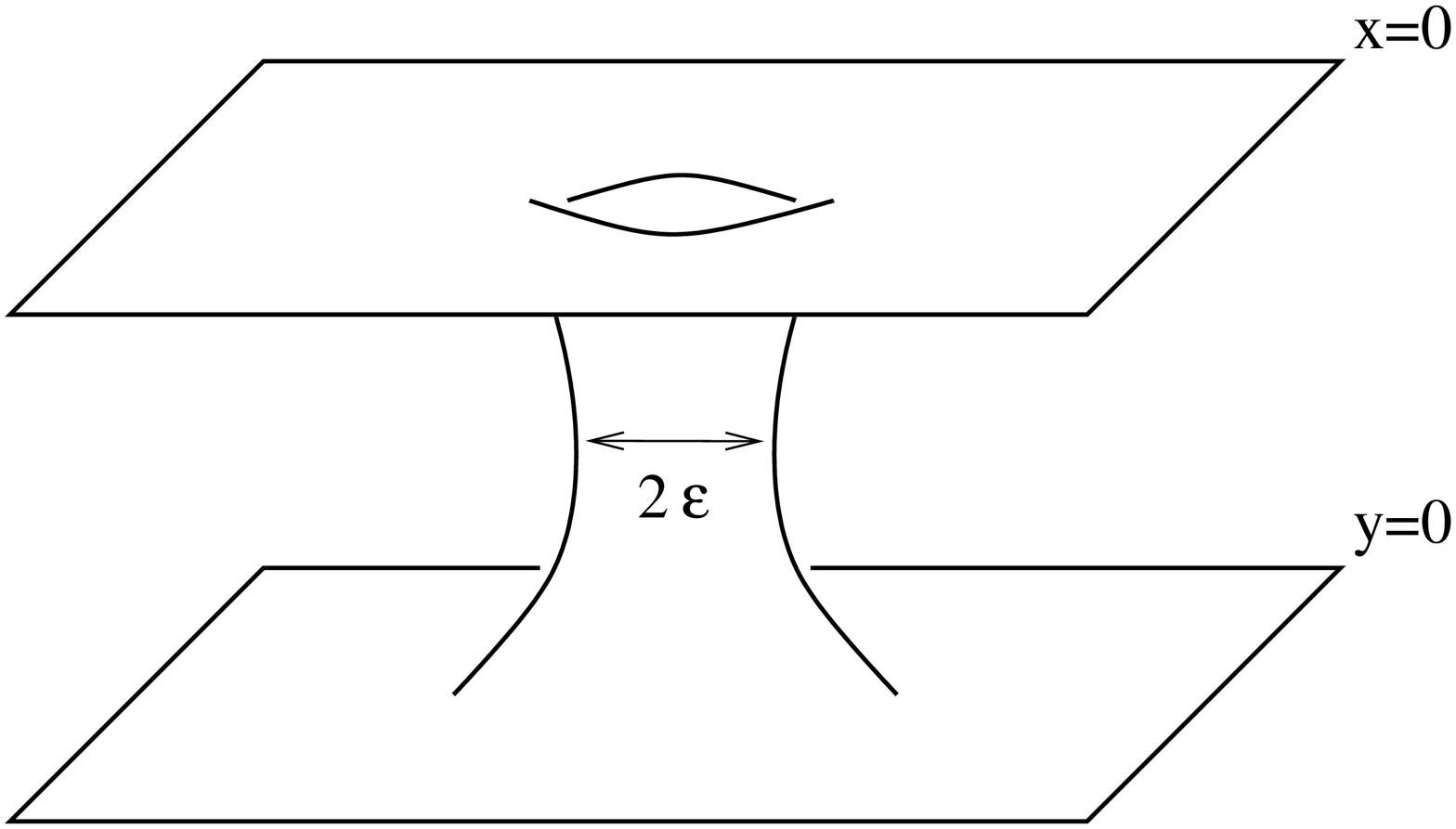}
\end{center}
\caption{\textsl{The surface formed by the recombined D-brane, as described by
\eqref{2D}. The parameter $\epsilon$ determines the radius of the circle
that sits at the narrowest point.}}
\label{hyperbolo}
\end{figure}

Consider two D-branes in a complex $2$-dimensional base space the intersection point of which 
is well-separated from other branes and O-planes. This situation is described by the equation
\begin{equation}
xy=0 \ ,
\end{equation}
which factorizes into $x=0$ and $y=0$.
The recombination is characterized by the deformation
\begin{equation}
xy=\tfrac{1}{2} \epsilon^2  \ ,  \label{2D}
\end{equation}
after which the equation no longer factorizes.
Far away from the intersection, for $|x|\gg\epsilon$ or
$|y|\gg\epsilon$, the recombined brane is still approximated by two branes at $y=0$ and
$x=0$. We take $\epsilon$ to be real. To understand the topology of the recombined D-brane, we introduce new coordinates
\begin{equation}
x=r\exp{\mbox{i}\phi} \ , \hspace{1cm} y=\rho\exp{\mbox{i}\psi} \ .
\end{equation}
In these coordinates, eq.\eqref{2D} reads
\begin{equation}
r\rho=\tfrac{1}{2} \epsilon^2 \ , \hspace{1cm} \phi+\psi=0 \ . \label{2D_polar}
\end{equation}
The equation
\begin{equation}
r=r_0
\end{equation}
characterizes an $S^1$ in the recombined brane, parameterized by $\phi \in [0,2\pi)$. When $r_0$ varies between zero and infinity, this loop sweeps out the whole recombined D-brane. The length of this loop, which is given by $2\pi\sqrt{r_0^2+\left(\frac{\epsilon^2}{2r_0}\right)^2}$, diverges when $r_0$ tends to zero or infinity, corresponding approximately to a circle in a brane at $x=0$ or $y=0$. It takes its minimum value, $2\pi \epsilon$, for $r_0= \tfrac{1}{\sqrt{2}} \epsilon$. The
topology of the recombined brane is thus given by a `throat' that connects two
asymptotically flat regions, as shown in Figure~\ref{hyperbolo}.

\begin{figure}
\begin{center}
\includegraphics[height=4cm, angle=0]{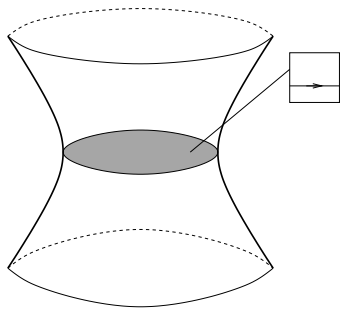}
\end{center}
\caption{\textsl{By taking a disc which has its boundary on a D-brane and
adding the horizontal cycle in the fibre torus at every point, we obtain a non-trivial 3-cycle.}}
\label{3S}
\end{figure}

When $\epsilon\rightarrow 0$, the `minimal loop' $r_0=\tfrac{1}{\sqrt{2}} \epsilon$ collapses and
eq.~\eqref{2D} factorizes, corresponding to two intersecting D-branes. We can
now construct the $3$-cycle that controls this process from the F-theory point of
view:
We recall that, over every point of the four-dimensional base space in which the brane is embedded, we have a torus fibre and that the $(1,0)$-cycle of this torus shrinks at the D7-brane locus. Consider a disc in the base space the boundary of which is the $1$-cycle of the D7-brane world volume discussed above (e.g.\ with $r_0=\tfrac{1}{\sqrt{2}} \epsilon$). The relevant $3$-cycle is obtained by taking the $(1,0)$-cycle of the fibre torus at every point of this disc.
This is illustrated in Figure~\ref{3S}. One easily convinces oneself that this cycle is a $3$-sphere.
It is obvious from the above that the volume of this $3$-cycle, divided by the square root of the fibre 
volume to keep it finite in the F-theory limit, characterizes the 
recombination process.\footnote{The F-theory limit of M-theory is characterized by the limit of zero size of the elliptic fibre. Since the complex structure moduli of the threefold should be independent of the size of this 2-cycle, we have to rescale the volume of 3-cycles by appropriate powers of its size.}

\subsection{Recombination of two intersecting O7-planes}\label{recO}

In the following we will locally construct 3-cycles in F-theory that
correspond to the movement of O-planes in its Type IIB dual. In order to simplify
the monodromy structure, we will consider the (singular) case of four D-branes coinciding
with the O-plane. In this way, the only monodromy appearing is an involution of
the fibre torus.

The recombination of two O7-planes is described by the same equation~\eqref{2D} as in
the D7-brane case. Thus two recombined O7-planes will
also form a surface which contains a throat supporting a circle of minimal
circumference. 

\begin{figure}
\begin{center}
\includegraphics[width=8cm]{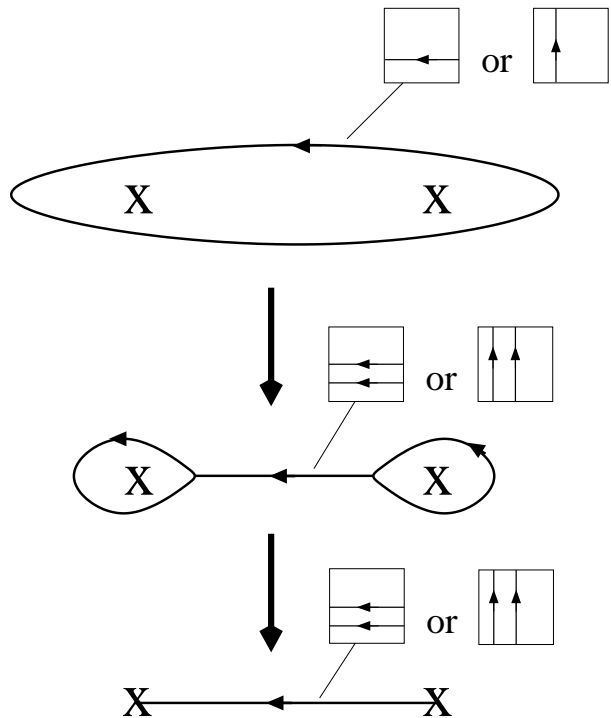}
\end{center}
\caption{\textsl{O-planes in a complex two-dimensional base give rise to cycles that have an arbitrary component in
the fibre and encircle the positions of the two O-planes. As shown in the figure, one can subsequently deform these 
cycles so that their base component becomes a line connecting the two O-planes. As the fibre component changes its
orientation upon circling one of the O-planes, the fibre component of the resulting line is twice that of the
original loop.}}
\label{o7cyc}
\end{figure}

To describe the cycle that controls the recombination of the O7-plane, let
us first recall its construction in the case of a complex one-dimensional base
space. In this case, the O-planes are merely points in the base. We can construct
a non-trivial cycle by taking a loop that circles two O-planes, together with an
arbitrary component in the fibre. As is shown in Figure \ref{o7cyc}, we can collapse
this cycle to a line that starts at one of the O-planes and ends at the other one.

Keeping the construction in the case of a complex one-dimensional base in mind, we can repeat the construction done 
for the D-brane: we take a disc ending on the O-plane in the base and one of the two fibres in Figure \ref{o7cyc}
to construct a 3-cycle. Just as in the D-brane case, the size of this cycle will describe the recombination
process of two intersecting O-planes.

\section{F-Theory models at the orientifold point} \label{DonO}

\begin{figure}
\begin{center}
\input{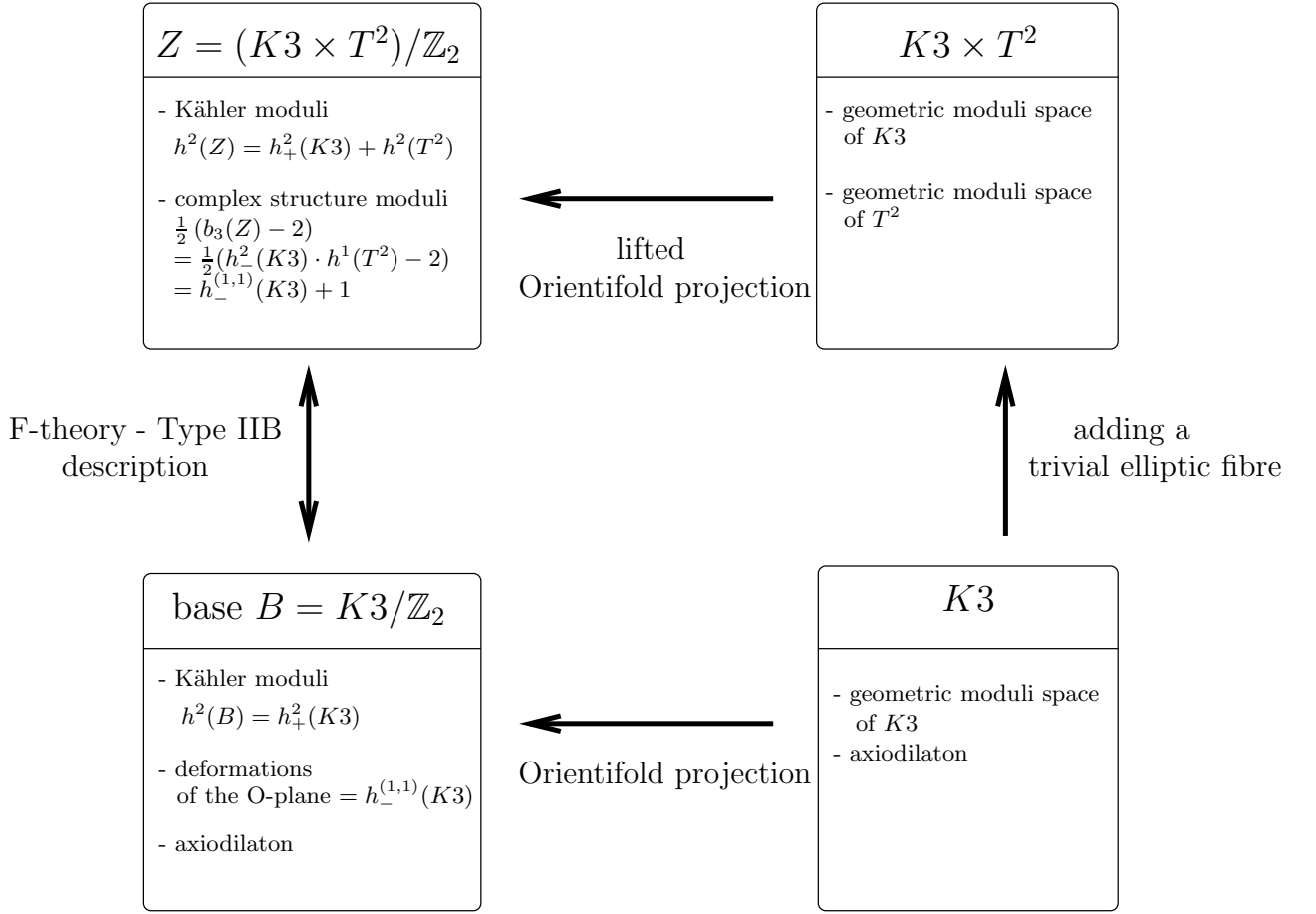}
\end{center}
\caption{\textsl{Type IIB orientifolds of $K3$ may be described by F-theory on
$Z=(K3\times T^2)/\Z_2$. The complex structure deformations of $Z$ correspond to
deformations of the O-plane and the axiodilaton.}} \label{scheme}
\end{figure}

In the following, we discuss F-theory compactifications on elliptic threefolds that can be
constructed as orientifolds. In particular, we demand that all D-branes coincide with the O-plane in this
section. We are going to describe these models from several perspectives, summarized
 in Figure \ref{scheme}.

\subsection{The type IIB perspective}\label{orientiIIB}

Let us start with the well-known type IIB perspective. Besides the various form-fields, the moduli
space of type IIB on $K3$ contains the geometrical moduli space of $K3$ and the complexified
string coupling, also known as the axiodilaton (see e.g. \cite{Aspinwall:1996mn}). The geometric
moduli can be elegantly described as the rotations of a three-plane of positive norm inside
$H^2(K3)$. The three positive-norm vectors that span this three-plane can then be used to construct
the K\"ahler form $J$ and the holomorphic two-form $\Omega^{(2,0)}$. As is common for Calabi-Yau threefolds,
the geometric moduli of $K3$ can then be mapped to K\"ahler and complex structure deformations.

Apart from the inner parity of the various degrees of freedom coming from the involution of the
world-sheet, orientifolding includes an involution $\iota$ of space-time. Furthermore, this involution
has to map the holomorphic two-form $\Omega^{(2,0)}$ to minus itself, so that the quotient space
$B$ is not Calabi-Yau. Involutions of this kind are known as \emph{non-symplectic} in the mathematics
literature and have been classified by Nikulin \cite{Nikulin:1986}. We have summarized the main
results in Appendix \ref{nikulinClassification}. The classification of non-symplectic involutions of
$K3$ implies that $B$ is rational, so that it is a del Pezzo surface $dP_i$, a Hirzebruch surface $\Hirz[n]$,
or a blow-up of a Hirzebruch surface. For a short review of rational surfaces see Appendix~\ref{complexSurfaces}.

Under the action of $\iota$, the cohomology groups of $K3$ decompose into eigenspaces:
\begin{equation}
H^{(p,q)}(K3)=H^{(p,q)}_+(K3)\oplus H^{(p,q)}_-(K3)\ .
\end{equation}
The geometric moduli of this orientifold model were discussed in detail in
\cite{Brunner:2003zm}. The complex structure deformations that are compatible
with $\iota$ are in one-to-one correspondence with elements of $H^{(1,1)}_-(X)$.
As the K\"ahler form of $K3$ is even under $\iota$, compatible K\"ahler deformations
can be parameterized by $H^{(1,1)}_+(K3)$. Since $\iota$ is a non-symplectic involution, we have
$H^{(1,1)}_+(K3)=H^2_+(K3) \simeq H_2^+(K3)=H^2(B)$.

The fixed point locus of $\iota$ is the orientifold plane $O$.
It is given by the vanishing locus of a section of $[-2K_B]$, where $[K_B]$ denotes the
canonical bundle\footnote{For a divisor $D$ we denote the corresponding line bundle by $[D]$.}
of the base space \cite{Sen:1996vd}. In other words, the O-plane is equivalent to $-2K_B$ as a
divisor. This is necessary to ensure that the double cover is a Calabi-Yau space. To cancel the D7-brane
charge, the homology class of all the D-branes has to equal four times the
homology class of the O-plane. In this section we choose to align four D-branes with the O-plane. The only
geometric deformations of this configuration are hence given by the K\"ahler deformations of the base and
the deformations of the O-plane. These must be equivalent to the deformations
of $K3$ compatible with the orientifolding.

Let us illustrate this in the simple example of $B=\C P^2$. The sections of $[-2K_{\C P^2}]$ are
given by homogeneous polynomials of degree 6. We can count the degrees of freedom that correspond to
deformations of this polynomial: there are 28 independent monomials and hence 28 complex coefficients.
One of these can be set to unity by an overall rescaling. In addition, the embedding of $O$ in
$\C P^2$ is only defined up to automorphisms of $\C P^2$. This automorphism group is complex 8-dimensional
(see Appendix \ref{toricAut}), eliminating $8$ degrees of freedom. We thus end up with $19$ complex degrees of
freedom. As $b_2(\C P^2)=1$, we find that $h^{(1,1)}_-=19$, giving the right number of degrees of freedom.
We have collected some more examples in Table \ref{summExam} at the end of the present section.

\subsection{Deformations of the O-plane}

Deformations of a Riemannian surface, such as the O-plane $O$, are
given by holomorphic sections of its normal bundle $[\nbo]$. The dimension of the
space of holomorphic sections of $[\nbo]$ is commonly denoted by $h^0(\nbo)$.

A Riemannian surface, such as the O-plane, has $3g(O)-3$ complex structure deformations. Let us explain why this is also the number of deformations in the embedding.
From the adjunction formula we have $K_O=\nbo+K_B$.\footnote{Here and in the following the restriction of $K_B$ to $O$ is implicit.}
As $O$ is linearly equivalent to $-2K_B$, we have $\nbo = -2K_B$ and thus find $\nbo = 2K_O$.
Serre duality then tells us that
\begin{equation}
H^0(\nbo)= H^0(2K_O)= H^1(T_O)^* \ ,
\end{equation}
in other words we find
\begin{equation}
h^0(\nbo) = 3g(O)-3\ . \label{moduliRiemann}
\end{equation}

For later convenience, we show how to derive \eqref{moduliRiemann} using the Riemann-Roch-Theorem \cite{Griffiths:1978}
\begin{equation}
h^0(\nbo) = h^0(K_O-\nbo)+\text{deg } \nbo - g(O) + 1
\end{equation}
where $K_O$ is the canonical divisor of $O$ and $g(O)$ the genus of $O$.
The degree of a line bundle $L$ is the number of zeros of a generic section of
$L$. If $\text{deg } \nbo > \text{deg } K_O = 2g(O)-2$ it follows that $h^0(K_O-\nbo)
= 0$. In this case
\begin{equation}
h^0(\nbo) = \text{deg } \nbo - g(O) + 1. \label{dimHolSections}
\end{equation}
Since a section of the normal bundle $[\nbo]$ is nothing but a
deformation of the Riemannian surface, $\text{deg } \nbo$ is just the
self-intersection number of $O$.
As $O$ is linearly equivalent to $-2K_B$, we find that its self-intersections number is
$O\cdot O=4 K_B\cdot K_B$. Furthermore, we find from $K_O = -K_B$ the Euler characteristic of $O$
to be $\chi_O = \nbo \cdot K_O= -2K_B^2$.
Hence we obtain
\begin{equation}
\text{deg } \nbo = 4 K_B\cdot K_B = -2 \chi_O = 4g(O)-4 > 2g(O)-2 \text{   for }
g(O) \geq 2 \ .
\label{selfintersection}
\end{equation}
Thus the requirement for Eq.\ \eqref{dimHolSections} is satisfied and we find \eqref{moduliRiemann}.

So far, we have neglected the fact that some deformations of the O-plane are equivalent
to applying an automorphism $A \in \Aut(B)$ of the base $B$ and as such do not
represent valid complex degrees of freedom. Thus the number of deformations of the O-plane, $\text{Def }O$,
is given by
\begin{equation}
\dim_\C\text{Def}O = 3g-3 - \dim_\C \aut_B\ ,
\end{equation}
where $\aut_B$ denotes the Lie-algebra of $\Aut(B)$. In the
case of a toric variety, this quantity can be found by the procedure
explained in Appendix \ref{toricAut}.

\subsection{F-theory perspective}\label{fthpers}

Having discussed the moduli from the type IIB perspective, we now turn to the
F-theory description of the same situation. Since we have taken all D-branes
to be aligned with the O-plane, the axiodilaton is constant along $B$.
The corresponding F-theory description thus must be such that the complex structure of the
elliptic fibre is constant. Before orientifolding, there are no $SL(2,\Z)$ monodromies and
the F-theory threefold is simply the product $K3\times T^2$. The orientifolding introduces
a monodromy that acts as an involution on the fibre $T^2$. It occurs upon encircling the O-plane
locus. We can describe this situation by lifting the involution $\iota$ to an involution $\tilde{\iota}$
on $K3 \times T^2$ by defining
\begin{equation}
\tilde{\iota}(x,z)=(\iota x, -z)\ , \label{Finvolution}
\end{equation}
where $x \in X$ and and $z \in T^2$. Modding out $\tilde{\iota}$ yields the
F-theory compactification on $Z=(K3\times T^2)/\Z_2$.

The homology groups of $K3 \times T^2$ are
\begin{eqnarray*}
H_1(K3 \times T^2) &=& H_1(T^2) \ , \\
H_2(K3 \times T^2) &=& H_2(K3) \oplus H_2(T^2)\ , \\
H_3(K3 \times T^2) &=& H_2(K3) \otimes H_1(T^2)\ ,
\end{eqnarray*}
since $H_1(K3) = 0$.
Keeping the cycles even under $\tilde{\iota}$ yields the homology of $Z$:
\begin{eqnarray}
H_2(Z) &=& H_2^+(K3) \oplus H_2(T^2)\ , \nonumber \\
H_3(Z) &=& H_2^-(K3) \otimes H_1(T^2) \ . \label{homologyZ}
\end{eqnarray}

F-theory on $Z$ emerges from M-theory on the same manifold in the limit of
vanishing fibre size. The geometric moduli of M-theory on $Z$ are deformations of $X\times T^2$
that respect the $\Z_2$ action. Hence the K\"ahler and complex structure deformations of $Z$ are
linked to even cycles of $K3 \times T^2$. The K\"ahler moduli of $B$ and the volume of the elliptic
fibre become the K\"ahler moduli of $Z$. As the fibre size tends to zero in the F-theory limit, it does
not give rise to a physical modulus in F-theory, so that we find the same number of K\"ahler moduli as for the $K3$-orientifold $B$.

The 3-cycles of $Z$ originate from the odd 2-cycles of $K3$.
The number of complex structure moduli, $\dim_\C {\cal M}^{CS}$, of a Calabi-Yau space is given by the
number $h^{(2,1)}=\frac{1}{2}\left(b^3-2\right)$ \cite{Candelas:1990pi}. If we choose a
symplectic basis $(A^a, B_b)$, the complex structure moduli space is locally parameterized by the $h^{(2,1)}$
independent periods:
\begin{equation}
z^a=\int_{A^a}\Omega,\hspace{1cm} \Pi_b(z)=\int_{B_b} \Omega \ ,
\end{equation}
where $\Omega$ is the holomorphic 3-form. In the present case, we only
consider complex structure deformations that do not destroy the structure $Z=(K3\times T^2)/\Z_2$ (by resolving the orbifold singularities, for instance).
We can think of this restriction as fixing
a number of periods. The relation between complex structure deformations and $b^3$,
$\dim_\C {\cal M}^{CS}=\frac{1}{2}\left(b^3-2\right)$ thus holds for an orbifold like $Z$ as well.

In the present case we find that
\begin{equation}
\dim_\C {\cal M}_Z^{CS}=\frac{1}{2}\left(b^3(Z)-2\right)=\frac{1}{2}\left(2b^{2}_{-}(K3)-2\right)=b^{2}_-(K3) -1\ .
\end{equation}
As the holomorphic two-form of $K3$ and its complex conjugate are always
odd for the involutions considered we have $b^2_-(K3)=h^{(1,1)}_-(K3)+2$, so that we find
\begin{equation}
\dim_\C {\cal M}_Z^{CS}=h^{(1,1)}_-(K3) +1\ . \label{csZ}
\end{equation}

In F-theory, the present situation is described by an elliptic threefold in which
there is a $D_4$ singularity along a curve that is equivalent to $-2K_B$. This
curve is the location of the O-plane. The complex structure deformations of the
threefold that preserve the singularity structure correspond to deformations
of the O-plane and the value of the axiodilaton. This is expressed in~\eqref{csZ}:
the number of complex structure deformations equals the number of deformations of the
O-plane, $h^{(1,1)}_-(K3)$, plus one. This fits nicely with the aforementioned result that
deformations of the double cover $K3$ compatible with the orientifold involution originate
from cycles of $K3$ that are odd under the orientifold involution.

There is a subtle point worth mentioning here. In the present case, $Z$ has four $A_1$ singularities along the
location of the O-plane. The manifold constructed from the Weierstrass model that corresponds to the orientifold $B$ has however a $D_4$ singularity over the location of the O-plane. Even though for both the fibre torus
undergoes the same monodromies \cite{Friedman-Morgan(1994)}, they give rise to different physics
for compactifications of M-theory. However, in the F-theory limit both manifolds coincide, as
they are connected by blow-ups and blow-downs of the singular fibres, leading to the same type IIB model. 
See Section~\ref{sec9} for a discussion of this and related issues.

\subsection{3-cycles at the orientifold point}
In the following, we will discuss how the 3-cycles of $Z$ emerge from the topology
of the O-plane. Let us first return to the example of $B=\C P^2$. In this case
the O-plane locus has $19$ deformations, so that we expect the corresponding threefold
to have $20$ complex structure moduli yielding $42$ independent 3-cycles.

The local construction of F-theory 3-cycles from the O-plane topology, presented in Section \ref{recO},
suggests that we obtain two F-theory cycles
for each 1-cycle of the O-plane $O$. Since $\dim H_1(O)=2g(O)$, where $g(O)$ is
the genus of $O$, we expect $4g(O)$ F-theory 3-cycles. As mentioned before, $O$
is given by a defining polynomial $h$ of degree 6 in the $\C P^2$ case,
yielding a curve of genus $g(O)=10$. We thus obtain 40 cycles in $H_3(Z)$ instead of 42.
However, from the global point of view there is exactly one other cycle that could be lifted
to a F-theory 3-cycle, namely the cycle corresponding to the
hyperplane divisor $H$ of $\C P^2$. Naively, we can add a leg with two possible orientations
in the fibre to $H$ so that the lift yields two extra cycles. Including these, we get the
right number of 42 cycles. We have collected some more examples in Table \ref{summExam}.

There is one potential difficulty to this construction. The
hyperplane divisor will generically intersect the O-plane $O$. Since the
fibre degenerates on $O$ it is not a priori clear how to lift $H$ properly.
As it will become clear later on, $H$ can indeed be lifted to an F-theory cycle
but for now this remains a conjecture motivated by the counting.

It is now natural to conjecture that for any base space $B$, the number of non-degenerate
cycles in $H_3(Z)$ is given by
\begin{equation}
b_3(Z)=4g(O)+2 b_2(B)\ . \label{conjecture}
\end{equation}

The Lefschetz fixed point theorem  \cite{Griffiths:1978} allows us to relate the topology
of the O-plane to the topology of $Z$ in the general case. For a $K3$ surface it reads
\begin{equation}
2+b_2^+(X)-b_2^-(X)=\chi(O) \ . \label{Lefshetz2}
\end{equation}
From this is follows directly that
\begin{equation}
b_3(Z)=2b_2^-(K3)=4-2\chi(O)+2b_2^+(K3) \ .
\end{equation}
On the other hand, we know that the fixed point locus is given by the disjoint union of
a curve of genus $g$ and $k$ spheres \cite{Nikulin:1986}, see also Appendix \ref{nikulinClassification}.
Thus we have
\begin{equation}
\chi(O)=2-2g+2k \ ,
\end{equation}
which yields
\begin{equation}
b_3(Z)=4g+2b_2(B)-4k \ .     \label{relation}
\end{equation}
This equation can also be derived directly using \eqref{Nikulin}. For the cases in which the O-plane is given
by a single smooth complex surface we have $k=0$, so that (\ref{conjecture}) indeed holds. Note that this is
the case in the example of $B=\C P^2$ discussed before.

\begin{table}
\begin{center}
\begin{tabular}{|c||c|c|c|c|c|c|}
\hline
surface $B$ & $dP_0$ & $dP_1$ & $dP_2$ & $dP_3$ &
$\Hirz[0]$ & $\Hirz[2]$ \\
\hline
$g(O)$ & 10 & 9 & 8 & 7 & 9 & 9 \\
$b_2(B)$ & 1 & 2 & 3 & 4 & 2 & 2 \\
$h^{(1,1)_-}$ & 19 & 18 & 17 & 16 & 18 & 17 \\
$\dim_\C$ Def($O$) & 19 & 18 & 17 & 16 & 18 & 17 \\
\hline
$b_3(Z)=2h^{(1,1)}_-+4$ & 42 & 40 & 38 & 36 & 40 & 38 \\
\hline
$4g(O)+2b_2(B)-4k$ & 42 & 40 & 38 & 36 & 40 & 40 \\
\hline
\end{tabular}
\end{center}
\caption{\textsl{As discussed in the main text, the moduli of type IIB orientifold models on $K3$
can be described in different ways, see also Figure \ref{scheme}. This table contains some numerical
examples for simple base spaces $B$. The calculation of the appearing quantities is explained in the
text. Note that $dP_0= \C P^2$, $\Hirz[0] = \C P^1 \times \C P^1$ and $dP_1 =\Hirz[1]$.}}
\label{summExam}
\end{table}

The results for different surfaces are given in Table \ref{summExam}. It is interesting to note that in the case of
$\Hirz[2]$ the degrees of freedom in the type IIB picture do not fit the number
of complex structure moduli in the F-theory picture. Indeed, in this case
one can show that there is one complex structure deformation of the
Calabi-Yau threefold which is not realized as a polynomial deformation of the
Weierstrass model (see e.g. \cite{Green:1987rw} for a discussion of this
phenomenon in the physics literature).

It is nice to see that (\ref{relation}) is invariant under blow-ups of the base:
As
\begin{equation}
K_O=K_B+O=-K_B\ ,
\end{equation}
we deduce the Euler characteristic of $O$ to be
\begin{equation}
\chi(O) = -K_O\cdot O = -2 (-K_B)\cdot (-K_B)\ .
\end{equation}
Now consider the blow-up $\pi:\tilde{B} \rightarrow B$ at a generic point
$u\in B$. This means we add an exceptional divisor $E$ so that
$b^2(\tilde{B})=b^2(B)+1$. The behavior of the anticanonical divisor
under blow-ups is~\cite{Griffiths:1978}
\begin{equation}
-K_{\tilde{B}}=\pi^*(-K_B)- (\dim_\C B-1)E = -K_B-E \ .
\end{equation}
The exceptional divisor can always be chosen to satisfy \cite{Safarevic}
\begin{eqnarray}
E^2=-1 && E \cdot T_i = 0 \text{ for all toric divisors } T_i \in \text{Div}(B) \ .
\end{eqnarray}
In particular, this implies $-K_B \cdot E = 0$. It is now straight forward to
determine the Euler characteristic of $\tilde O$:
\begin{equation}
\chi(\tilde O)=-2\left(-K_B-E\right)\cdot\left(-K_B-E\right)=\chi(O)+2 \ .
\label{EulerBlowUp}
\end{equation}
Since the Euler characteristic is given by the relation $\chi=2-2g+2k$, \eqref{EulerBlowUp}
implies that $g-k\rightarrow g-k-1$ under blow-ups of $B$ at generic
points. We thus need to show that $b_3(Z)\rightarrow b_3(Z)-2$ under blow-ups of the base.
As $b_2^-+b_2^+$ is fixed and the blow-up increases $b_2^+$ by one, $b_2^-$ must
decrease accordingly. Hence we find $b_3(Z)\rightarrow b_3(Z)-2$, so that \eqref{relation}
remains valid.

Formula \eqref{relation} can be given a further interpretation in terms of the double cover $K3$. Let us start with the case $k=0$ and consider a 1-cycle of the O-plane. This 1-cycle is trivial inside the base. Thus, there exists a real disk in the base whose boundary coincides with this 1-cycle. If we now go to the double cover, we end up with two disks that are glued together at their boundary, which is located at the 1-cycle of the O-plane. This gives rise to a two-sphere on $K3$ which is a non-trivial 2-cycle since the corresponding O-plane 1-cycle was non-trivial. Clearly, this 2-cycle is odd under the involution on $K3$ since the involution changes the orientation of the 2-cycle. When we add the fibre, each combination of this 2-cycle with a 1-cycle in the fibre gives rise to a 3-cycle in the threefold. Hence, we get twice as many 3-cycles on the threefold as there are 1-cycles on the O-plane, i.e.\ $2g$ many. This construction is a further motivation
for the cycles that were constructed locally in Section \ref{recO}. If one builds the double cover of $\C^2$ branched along the vanishing
locus of an equation of the form \eqref{2D} one finds the space $\C^2/Z_2$ blown up at the origin. The exceptional cycle of this blow-up is
an odd 2-cycle under the orientifold projection. Its image under the orientifold projection yields precisely the base part of the 3-cycle that controls the O-plane motion in F-theory.

Let us now turn to the contribution coming from $H_2(B,\mathbb{Z})$ and consider a 2-cycle of $B$. Recall that $H_2(B,\mathbb{Z})=H_{2\,+}(K3,\mathbb{Z})$.
Since we assumed $k=0$, we have $H_{2\,+}(K3,\mathbb{Z})^*/H_{2\,+}(K3,\mathbb{Z}) = \mathbb{Z}_2^a$, cf.\ Appendix \ref{nikulinClassification}, and every basis 2-cycle of $H_{2\,+}(K3,\mathbb{Z})$ can be understood as the sum of two (maybe intersecting) basis 2-cycles of $H_{2}(K3,\mathbb{Z})$ that are exchanged by the involution. Then, the difference of these basis 2-cycles gives an element in $H_{2\,-}(K3,\mathbb{Z})$, which can be combined on $K3\times T^2$ with one of the fibre 1-cycles to build an even 3-cycle that descends to the threefold. By this we obtain two 3-cycles of the threefold for each 2-cycle in B. This explains the second contribution in \eqref{conjecture}. For $k=0$, we have $a= b_2(B)$.

Now let us discuss the case of nonzero $k$. This means that we now additionally have $k$ non-trivial rigid two-spheres in $B$ that are part of the fix point locus, i.e.\ which are filled out by the O-plane. Let us consider one of them. Clearly, this cycle has only one pre-image in $K3$, which is left fixed by and thus even under the involution map. Furthermore, we can write \eqref{Nikulin} as
\begin{equation}
 b_2^+(K3)=r=a+2k
\end{equation}
so that it follows that there must be a second even cycle for any fixed $S^2$.
All of these cycles do not lead to any 3-cycle on the threefold and do not contribute to \eqref{conjecture}.
As the quantities $r,a,g$ and $k$ are actually not independent, but related by \eqref{Nikulin}, we find  \eqref{relation}.

Note that from this discussion we see that we can decompose the second cohomology class $H_2(K3,\mathbb{Z})$ of $K3$ into three parts, corresponding to
\begin{itemize}
\item $k$ spheres consisting of fix points plus $k$ further cycles, all of them being even under the involution,
\item $2a$ pairs of cycles which are interchanged
\item $2g$ spheres which are invariant up to an orientation reversal.
\end{itemize}
The fact that these cycles give the correct number of 2-cycles of $K3$, i.e.\
\begin{equation}
 2k + 2a + 2g = 22 \
\end{equation}
follows directly from \eqref{Nikulin}.
\section{D7-branes without obstructions}\label{D7-branes without obstructions}

\subsection{Pulling a single D-brane off the orientifold plane}\label{1doffO}

In this section we now want to leave the orientifold point by moving one D-brane off the O-plane.
The most general form of the hypersurface $\cal D$ which is the position of the D-branes is given by~\cite{Sen:1997gv}
\begin{equation}\label{genald7}
{\cal D}: \quad \eta^2+12 h \chi = 0 \ ,
\end{equation}
where $h,\eta$ and $\chi$ are sections in $[-2K_B], [-4K_B]$ and $[-6K_B]$, respectively.
We will call a D7-brane described by an equation of the form above a \emph{generic allowed} D7-brane.
Note that the equation $h=0$ describes the position of the O-plane $O$. At the orientifold point, the O-plane
coincides with four D-branes, so that $\eta=h^2$ and $\chi=h^3$ and Eq.~\eqref{genald7} reads
\begin{equation}
{\cal D}: \quad h^4 = 0 \ .
\end{equation}
We can now vary the sections $\eta$ and $\chi$ in order to deform the D-branes. Choosing $\eta = h^2$
and $\chi=\tfrac{1}{12} h^2p$, where $p$ is a section in $[-2K_B]$, then yields
\begin{eqnarray}
{\cal D}: \quad  h^3(h+p)= 0 \ .
\end{eqnarray}
The surface $\cal{D}$ consists of four components: three D-branes still coincide
with $O$ while one is deformed and thus separated from the O-plane.
The deformation is given by the generic section $p$ which is of the same degree as $h$.
This fits nicely with the fact that infinitesimal deformations of a surface
correspond to sections in the normal bundle of $O \subset B$.

Let us first return to the example of $\C P^2$. We can count the number of deformations
of the single D-brane that is moved off the O-plane by counting the monomials
of $p$ and subtracting the one complex degree of freedom of overall rescaling.
Note that fixing the O-plane in $B$ generically breaks the
automorphism group of $B$ completely and thus its dimension does not reduce the
number of degrees of freedom. In the present case, $p$ will be a homogeneous polynomial of degree
6 yielding $\frac{1}{2}(6+1)(7+1)-1=27$ complex degrees of freedom.

The number of deformations can also be obtained by analysing sections in the normal bundle of the
D-brane. As the D-brane we are considering is linearly equivalent to $[-2K_B]$, the analysis
of the previous section leading to Eq.~\eqref{moduliRiemann} applies. As the genus of the
D7-brane is given by $10$ in the case of $B=\C P^2$, we can immediately confirm that the
number of deformations is given by $27$. Following an argument similar
to the one presented in Section \ref{fthpers} we thus expect to find $54$ 2-cycles that govern the
displacement of a single D-brane that is equivalent to $[-2K_B]$ from the O-plane in $\C P^2$.
It is clear that a similar computation can be performed for other base spaces.

\subsubsection{3-cycles between O-plane and D-brane}
\label{3cycfrom1cycDbrane}

\begin{figure}
\begin{center}
\includegraphics[height=3cm]{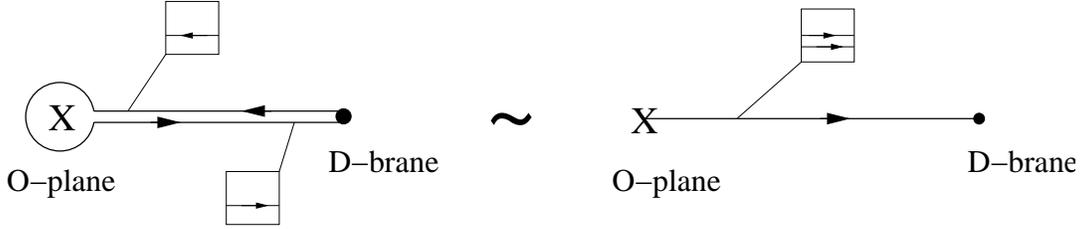}
\end{center}
\caption{\textsl{A possible representative of cycles determining the distance
between O-plane and D-brane in the case of F-theory compactified on $K3$ is given
by a loop that starts and ends at a D-brane and encircles the O-plane. When deforming it,
it looks line a line connecting the D-brane to the O-plane that has twice the horizontal cycle
of the fibre torus as its component in the $T^2$-fibre.}} \label{K3cycle}
\end{figure}

We now construct the 3-cycles that describe the process of moving
a single D-brane off the O-plane. Let us first discuss the analogue of these cycles for F-theory
compactified on $K3$, where O-plane and D-brane are points rather than complex lines.
We can link the two by a path that begins at the D-brane, encircles the O-plane and then ends at
the same D-brane. To construct a 2-cycle, we add the horizontal fibre to every point of this curve,
see Figure~\ref{K3cycle}.

The existence of this cycle can also be demonstrated by the following argument:
two D-branes in the vicinity of an O-plane can be connected by a cycle in two ways: the cycle can pass the O-plane on one
side or the other \cite{Braun:2008ua}. We can find a cycle that connects just one of the two D-branes to the
O-plane by forming the sum (or difference) of these two cycles. The resulting cycle has self-intersection
number $-4$ and can be deformed to any of the two representatives discussed in Figure~\ref{K3cycle}.

Coming back to F-theory compactified on an elliptically fibred Calabi-Yau threefold, we can
generalize this construction as follows. We choose a 1-cycle $A \in H_1({\cal D})$ of the
D-brane and a representative ${\cal A }\subset {\cal D}$. Since we are considering the
case in which the D-brane has the same topology as the O-plane, we can find a
corresponding cycle and representative on the O-plane that coincides with $\cal A$
when D-brane and O-plane are on top of each other. Now we apply the above construction to every point
$p \in {\cal A}$. In other words, we fiber $\cal A$ with the 2-cycles of Figure \ref{K3cycle}.
In this way we obtain $2g({\cal D})$ 3-cycles that measures the distance between the
D-brane and the O-plane.

\subsubsection{3-cycles from intersections between D-brane and O-plane}
\label{cycintd7o7}

\begin{figure}
\begin{center}
\includegraphics[height=3cm]{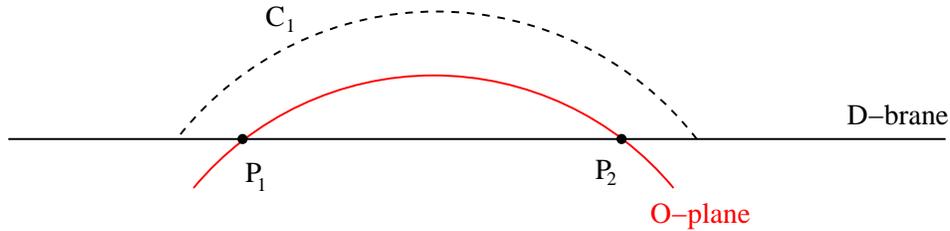}
\end{center}
\caption{\textsl{A 2-dimensional cut through the relative 2-cycle
$C_1$ of type II. Note that $C$ is in fact a half sphere surrounding the
intersections $P_1$ and $P_2$.}} \label{halfspherecycle}
\end{figure}

Another type of cycle can be constructed as follows. Consider two intersection points $P_1$ and $P_2$
of the D-brane with the O-plane (see Figure~\ref{halfspherecycle}). Since $\cal D$ is connected, we can find a loop
$l \subset {\cal D}$ that surrounds both intersection points. This immediately
implies that $l$ can only be contracted if $P_1$ coincides with $P_2$.
We know from Section~\ref{localConstruction} that the disc in B the boundary of which is $l$ can be lifted to a 3-cycle in the threefold $Z$. Indeed, we again fiber the 1-cycle of the torus that degenerates at the D-brane over the disk. This 3-cycle cannot be contracted due to the presence of the O-plane. The involution on the fibre which is part of the monodromy of the O-plane prevents the disk from passing through the O-plane position -- the fibre will simply be ill-defined if the disk intersects the O-plane. Clearly, this 3-cycle has again the topology of a three-sphere and its volume is proportional to the distance between the intersection points with the O-plane.
We can understand this 3-cycle also in another way. We can fiber the $K3$ 2-cycle of Figure~\ref{K3cycle} over the line on the D-brane that connects the two intersection points with the O-plane.

\begin{figure}
\begin{center}
\includegraphics[height=3cm]{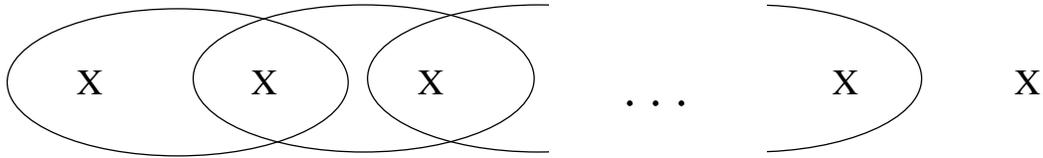}
\end{center}
\caption{\textsl{Boundaries $\partial C_i$ and intersection points with th O-plane $X$ in ${\cal D}$.}}
\label{intersections}
\end{figure}

Let us now count the number of independent cycles that can be constructed in this way.
Suppose there are $I$ intersection points $P_i$ on ${\cal D}$. Let $C_i$ be a disc such that $\partial C_i=l_i$ is a loop
on ${\cal D}$ which surrounds the intersection points $P_i$ and $P_{i+1}$. The boundaries $\partial C_i$
are elements of the first homology group of the D-brane with the O-plane cut out, $H_1({\cal D}\setminus {\cal D}\cap O)$.
Note that $C_i$ and $C_{i+1}$ will generically intersect only in (two) points that are located on the D-brane world volume since
the D-brane and the cycles have codimension two (cf.\ Figure \ref{intersections}). This yields $I$ such
loops\footnote{We identify $i=I+1$ with $i=1$.} and each loop gives a relative 2-cycle $C_i$. The $I$ cycles we construct in this
way are not linearly independent. We can construct the union
\begin{equation}
C = \bigcup_{i=0}^{\frac{I-4}{2}}C_{2i+1} \ .
\end{equation}

The boundary $\partial C$ of the relative 2-cycle $C$ surrounds all intersection points on $\cal D$ except for $P_{I-1}$ and $P_{I}$.
Since D is compact, this is equivalent to saying that $\partial C$ surrounds just $P_{I-1}$ and $P_I$.
Thus, $C$ is relatively homologous to $C_{I-1}$ and hence $C_{I-1}$ is not independent of the others.
In the previous argument we just used half of the $C_i$, namely those where $i$ is odd. We showed that one cycle can be expressed as
a linear combination of the others. The same argument goes through for the
complementary subset of $C_i$ where $i$ is even. Having constructed $I$ 2-cycles $C_i$, we are now left with $I-2$ independent F-theory 3-cycles. This is illustrated in Figure~\ref{intersections}.

Putting everything together, the number of 3-cycles we obtain by the constructions presented is $2g({\cal D})+I-2$.
As discussed before, the intersection number $I$ of D-brane and O-plane is
nothing but the self-intersection number of $O$. From Eq.~\eqref{selfintersection} we
thus know that $I=4g(O)-4$. Adding the $2g(O)$ cycles coming from the 1-cycles of the D-brane and the
$I-2=4g(O)-6$ cycles coming from the intersections with the O-plane, we arrive at the total number of
$2g(O)+I-2=2(3g(O)-3)$. These cycles determine the infinitesimal separation of the D-brane from the O-plane.
This fits exactly with the number of general deformations of Riemann surfaces obtained before. We conclude that
the union of both kinds of 3-cycles parameterize the location of the D-brane.

Coming back to the example of $B=\C P^2$, we find that $g({\cal D})=10$ and $I=36$,
so that we can construct $54$ cycles. This precisely fits the expectation expressed
at the beginning of this section.

\subsection{More general configurations}

We now want to generalized the discussion to the case of multiple D-branes separated
from the O-plane. When we move multiple branes off the O-plane and let them recombine,
we can no longer describe the resulting D-brane locus in terms of sections in the normal bundle
of the O-plane. Furthermore the worldvolume of the D-brane will in general be no longer be a
generic hypersurface as it is forced to have double intersections with the
O-plane \cite{Braun:2008ua,Collinucci:2008pf}. Namely, $\cal D$ is given by Eq.~\eqref{genald7}.
However, the D-brane curve will still be a generic hypersurface if we consider
one of the following \mbox{configurations}:
\begin{itemize}
\item[i)]{We leave one D-brane on the O-plane. This corresponds to choosing $\eta
= hp$, where $p$ is a section in $[-2K_B]$. This yields
\begin{equation}
{\cal D}: \quad h\underbrace{(hp^2+12\chi)}_{=\chi'} = 0.
\end{equation}
Since $\chi$ is generic, $\chi'$ is. Note that $\chi'$ is a section in
$[-6K_B]$ and therefore describes three recombined D-branes
generically not coinciding with the O-plane. The resulting D-brane locus is
appropriately described by a generic hypersurface in $B$, the zero locus of
$\chi'$, as required. The polynomial $h$ shows that one D-branes still sits on
the O-plane.}

\item[ii)]{We move all D-branes in stacks of two. This case corresponds to the
choice $\chi=0$ so that the D-brane is described by $\eta^2 = 0$ and hence a
generic section in $[-4K_B]$.}
\end{itemize}

In this section we analyse configurations in which the D-brane degrees of
freedom are associated to the deformation moduli of a generic hypersurface of
lower degree. For the whole analysis, we completely fix the O-plane and focus
on the D-brane degrees of freedom.

We start again with the example where $\C P^2$ is the
base space and analyse the general case afterwards. Assume that in addition to the
O-plane we have a D-brane that is linearly equivalent to $[-2mK_B]$, i.e. whose locus in
$B$ is described by the vanishing of a homogeneous polynomial of degree $n=6m$. Due to the D7
tadpole cancellation condition, this means that $4-m$ D-branes still coincide with the O-plane. As the
presence of the O-plane generically breaks the whole automorphism group of $\C P^2$, this
D7-brane has
\begin{equation}\label{degd7}
\tfrac{1}{2}(n+1)(n+2)-1
\end{equation}
complex degrees of freedom.

In the last section we found that a D7-brane that is equivalent to $[-2K_B]$ has $g({\cal D})+\frac{I}{2}-1$ complex degrees of freedom and gives rise to $2g({\cal D})+I-2$ 3-cycles in the threefold. In the remainder of this section, we show that this statement holds for any D7-brane that is equivalent to $[-2mK_B]$ for any $m$. The reader primarily interested in results may wish to skip the rest of this section and continue with Section~\ref{moreDoffO}.

Let us now try to find an analog of Eq.~\eqref{degd7} for an arbitrary base
space $B$. We will formulate all relevant quantities in terms of the
self-intersection of the anti-canonical divisor of the base, $(-K_B)\cdot (-K_B)=S_B$. For any
F-theory model, the worldvolume of the O-plane is equivalent to the divisor $-2K_B$. We consider
the situation with $m$ recombined D-branes $\cal D$ and $4-m$ D-branes coinciding with
the O-plane. In order to apply the Riemann-Roch Theorem, cf.\ Eq.~\eqref{dimHolSections},
we need to know the degree of the canonical divisor of $\cal D$, denoted by $K_{\cal D}$.
It is given by
\begin{equation}
\deg K_{\cal D} = 2g({\cal D})-2.
\end{equation}
The Euler characteristic of $\cal D$ is
\begin{equation}
\chi({\cal D})=\left[-K_B+2mK_B\right]\cdot\left[-2mK_B\right]=-2m(2m-1)S_B \ ,
\end{equation}
so that the genus is given by
\begin{equation}
g({\cal D})=m(2m-1)S_B+1 \ .
\end{equation}
Thus we find the degree of $K_{\cal D}$ to be
\begin{equation}
\deg K_{\cal D} =2m(2m-1)S_B \ .
\end{equation}
The self-intersection number of $\cal D$ is $\deg
N_{\cal D} = (2m)^2S_B$, so that the condition $\deg K_{\cal D} < \deg
N_{\cal D}$ is satisfied. Hence we can use Eq.~\eqref{dimHolSections} to find
the number of valid deformations:
\begin{equation}\label{genmatch}
\begin{aligned}
h^0(N_{\cal D}) &= (2m)^2S_B -
m(2m-1)S_B =
m(2m+1)S_B \\
&=1+m(2m-1)S_B+2m S_B-1=g({\cal D}) + \frac{I}{2}-1 \ .
\end{aligned}
\end{equation}
In the last line we have used that the number of intersections between the D-brane and
the O-plane is
\begin{equation}
I=\#\left(O \cap {\cal D}\right) =
(-2K_B)\cdot(-2mK_B)=4m S_B \ .
\end{equation}
We thus expect to find $2g+I-2$ 3-cycles that govern the deformation of the D-brane locus.

From the relation derived in the last paragraph it is clear how the 3-cycles that
control the motion of a D-brane arise:
On the one hand, we can build a 3-cycle from every 1-cycle
of the D-brane, using the construction given in Section~\ref{recomD}.
On the other hand, we can build cycles that measure the distance between intersections
of the D-brane with the O-plane, as discussed in Section~\ref{3cycfrom1cycDbrane}.

This can also be understood from the perspective of the $K3$ double cover as we now
explain qualitatively. It is known that for a smooth D-brane in the double cover, i.e. one that
does not have double intersections with the O-plane, the deformations are given by
1-cycles of the D-brane that are odd under the involution~\cite{Jockers:2004yj}.
As this is the situation discussed in this section, we should be able to link
the 3-cycles we have constructed to odd 1-cycles of the D-brane in the
double cover.

A 1-cycle of a D-brane ${\cal D}$ in $B$ that has been moved off the O-plane
generically does not intersect the O-plane. Furthermore, we can always deform the 1-cycle such that its winding number is zero with respect to the O-plane. 
Therefore, this
1-cycle has two pre-images in the double cover $K3$ which are interchanged by the involution.
The sum of both is even under the involution and therefore descends to the 1-cycle of ${\cal D}$
we started with. The difference of both, however, is odd under the projection and should refer to a
deformation of the D-brane. This suggests a fact already discussed: 1-cycles of ${\cal D}$ are related to 3-cycles
of the Calabi-Yau threefold. Furthermore, we can consider a line on ${\cal D}$ connecting two intersection
points of ${\cal D}$ with the O-plane and go to the double cover $K3$.  This line then becomes two lines joined at their
end points, i.e.\ a non-trivial closed 1-cycle on the double cover D-brane that is odd under the involution. Thus,
there should be a second kind of 3-cycles which are closely related to the intersections of the D-brane with the O-plane,
supporting our claim that we can construct 3-cycles from intersections between the D-brane and the O-plane.

As the intersections between O-plane and D-brane are points on a complex curve, one naively expects each intersection
point to correspond to a complex degree of freedom. From the relation between moduli and 3-cycles it follows that there
should roughly be $2I$ 3-cycles that stem from intersection points between D-brane and O-plane. This is, however, not the
case, as we only found half of that. Thus there seems to be some mismatch here. The solution to this puzzle is that fixing
the periods of the one-cycles of the D-brane leaves only few degrees of freedom for the intersection points with the O-plane
to vary. Hence the displacement of the intersection points is governed not only by the 3-cycles that are attached to intersection
points but also through the 3-cycles that stem from the 1-cycles of the D-brane. 
%We explain the geometric origin of this in Appendix~\ref{abelSection}.

Compared to Section~\ref{1doffO}, our cycle analysis is complicated by the fact that the O-plane can in principle pierce a disc that ends
on a 1-cycle.\footnote{As we will discuss in more detail later, this is ultimately related to the structure of
branch cuts on the D-brane when building the double cover.} The monodromy of the O-plane then prevents the construction
of a 3-cycle as its fibre part is transformed to a different cycle upon encircling the O-plane locus. To tell if we can find a disc that ends on a given loop
on the D-brane and does not intersect the O-plane, we need to check that the winding number of this loop
around the O-plane vanishes.
We can define the winding number on the first homology of the D7-brane with the intersection points with the
O-plane cut out, $H_1({\cal D},O \cap {\cal D})$, and then project to the subspace of zero winding number. 
Since $H_1({\cal D},O \cap {\cal D})$ has the dimension $2g+I-1$ and there are elements of $H_1({\cal D},O \cap {\cal D})$ that have a non-zero winding number, this projection leads to a subspace of dimension $2g+I-2$, i.e.\ we find $2g+I-2$ independent cycles we can use to construct non-trivial 3-cycles of the elliptic fibration. This reproduces the
result that is expected from an analysis of the degrees of freedom of a D-brane.

Let us again come back to the example of $\C P^2$. In this case one easily finds the
numbers $g({\cal D})=(n-1)(n-2)/2$ and $I=6n$, so that
\begin{equation}
\underbrace{\frac{1}{2}(n-1)(n-2)}_{=g({\cal
D})}+\underbrace{3n-1}_{=\frac{I}{2} -1}=\underbrace{\frac{1}{2}(n+1)(n+2)-1}_{
= \dim_\C \text{Def}({\cal D})},\label{matchcp2}
\end{equation}
which exactly matches the number of degrees of freedom, as given by \eqref{degd7}.

Note that we can use the reasoning presented in this section also for the
situation discussed at the beginning of this section, in which a
single D-brane is moved off the O-plane. Although we used different
3-cycles in both cases, the results agree as expected. The two sets of
cycles just give a different basis of the third homology group of the threefold.

We can also give an inductive construction of 3-cycles in terms of (relative) 1-cycles
as long as all D-branes are described by completely generic hypersurfaces. We start with
the case in which the D-brane locus and the O-plane locus coincide. These cycles were
discussed at length in Section~\ref{DonO}. New cycles appear when the first D-brane is moved
away from the O-plane, namely the cycles given in Section~\ref{3cycfrom1cycDbrane}. We used
in this analysis that the D-brane is given by a section in the normal bundle of the O-plane.

We can now independently move two D-branes
off the O-plane, both given by sections in the normal bundle of $O$. Additionally to the
cycles for described in Section~\ref{recomD}, there are intersections
between the two D-branes. Thus, the D-brane locus is a nodal Riemann surface $\overline{\cal D}$
with $I_{DD}$ nodes, where $I_{DD}$ denotes the number of intersection points of the two D-branes.
By generic deformations of these singular intersection points, the D-branes recombine
at these nodes as described in Section~\ref{recomD}, yielding a smooth Riemann surface $\cal D$. Note
that the genus $g({\cal D})$ of $\cal D$ is identical to the arithmetic
genus $p(\overline{\cal D})$ of $\overline{\cal D}$. For the arithmetic genus
the following identity holds~\cite{Hori:2003ic}
\begin{equation}
p(\Sigma) = \delta+1+\sum_{i=1}^{k}(g_i-1), \label{arithmeticgenus}
\end{equation}
where $\Sigma$ is the nodal Riemann surface with $\delta$ nodes and $k$
irreducible components which have the geometric genus $g_i$, respectively. In
our case of interest, Eq.~\eqref{arithmeticgenus} reduces to
\begin{equation}
g({\cal D})=p(\overline{\cal D})=2g({\cal D}_0)+I_{DD}-1,
\end{equation}
where ${\cal D}_0$ denotes the single D-brane. Using Eq.~\eqref{genmatch} we can
immediately give an expression for the number of independent deformations of
$\cal D$:
\begin{eqnarray}
\dim_\C \text{Def}({\cal D})&=&h^0(N_{\cal D})=g({\cal D})+\frac{I_{OD}}{2}-1
\nonumber \\
&=& 2g({\cal D}_0)+I_{DD}-2+I_{OD_0} \nonumber \\
&=& 2\dim_\C \text{Def}({\cal D}_0)+I_{DD}. \label{recomDbrane}
\end{eqnarray}
Here $I_{OD_0}$ denotes the intersection number between $O$ and ${\cal D}_0$
and we used that $I_{OD}=2I_{OD_0}$. The first term in the last line in
Eq.~\eqref{recomDbrane} are the degrees of freedom obtained
by moving the D-branes ${\cal D}_0$ independently. The second part, namely
$I_{DD}$, gives the number of recombination parameters. Each
intersection point gives exactly two cycles. In fact, these are
locally the recombination cycles obtained in Section~\ref{recomD}.
In the same way we can discuss the case of three D-branes moved off the O-plane.

\section{D7-branes with obstructions} \label{moreDoffO}

\subsection{D-brane obstructions} \label{Obstructions}\label{sectobs3}
\begin{figure}
\begin{center}
\includegraphics[height=2.5cm]{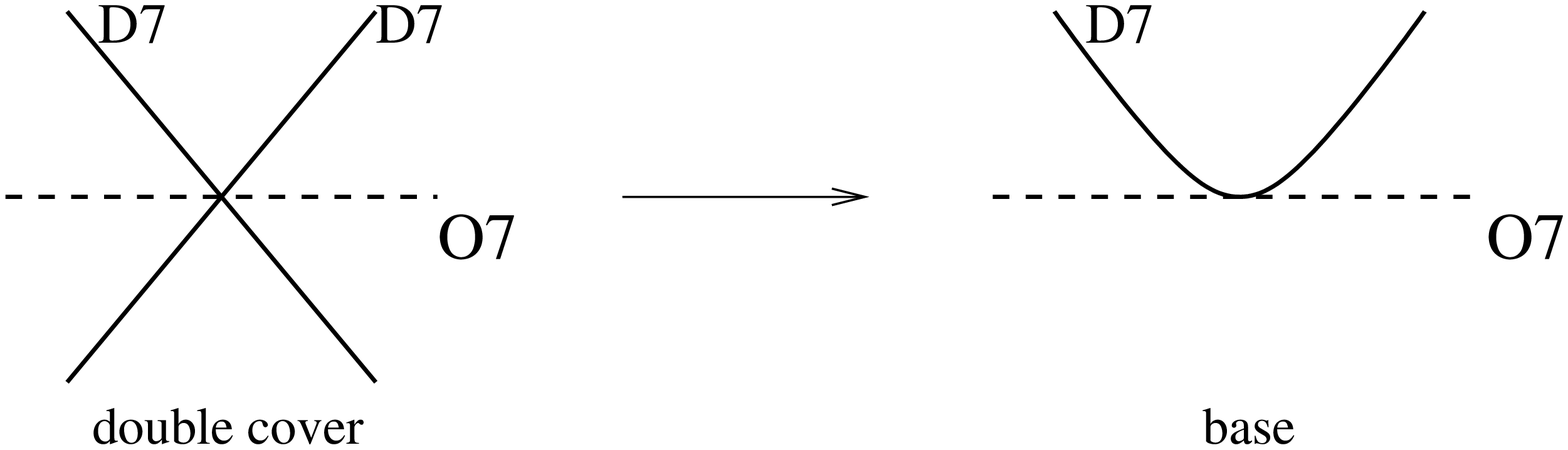}
\end{center}
\caption{\textsl{A situation in which two D7-branes intersect an O7-plane
in the same point produces a D7-brane touching the O-plane after modding out
the orientifold action and squaring the coordinate transverse to the O7-plane.
}}\label{2btof}
\end{figure}

In the weak coupling limit the D7-brane locus is not given by the zeros
of a generic polynomial, but by the zeros of a polynomial of the form
\begin{equation}\label{obstructed_Weierstrass}
 {\cal D}: \quad \eta^2+12 h \chi = 0 \ .
\end{equation}
We refer to D7-branes that are described by an equation of this form as \emph{generic allowed}
D7-branes. As has recently been discussed \cite{Braun:2008ua,Collinucci:2008pf},
this form forces the D7-brane to have double intersections
with the O7-plane. From the perspective of F-theory this means that the D7-brane
forms a parabola touching the O-plane in the origin, see Figure~\ref{2btof}.

Let us try to understand this configuration from the double cover perspective.
Consider two D-branes at $x=\pm z$ and the involution $z \to - z$, which fixes the O-plane at $z=0$.
After modding out the involution, our space looks locally
like the upper half plane. In order to make contact with the F-theory picture,
we introduce a new coordinate $\tilde{z}=z^2$ and find that this situation is
described by an O-plane at $\tilde{z}=0$ and a D-brane at $\tilde{z}=x^2$.
Note that a single D7-O7 intersection in F-theory (which does not occur in the weak coupling limit),
corresponds to a single D7-brane that is mapped onto itself by the
orientifold projection. This configuration, where the D-brane sits e.g.\ at $x=0$ is allowed
in the presence of a second D-brane that coincides with the O-plane.

In \cite{Braun:2008ua} it was observed that the difference between the degrees
of freedom of a generic allowed D7-brane and the degrees of freedom of a
generic hypersurface of the same degree is given by half the number of intersections
between the D7-brane and the O7-plane\footnote{Here we of course count the topological
intersections between two generic surfaces that are homologous to the D7-brane and the O7-plane.}.
We checked this explicitly only for base spaces $\C P^2$ and $\C P^1 \times \C P^1$. Here we
extend this analysis to all possible base spaces. We use the fact that a generic hypersurface
${\cal D}_\textrm{gen}$ that is linearly equivalent to $2mK_B$ has
\begin{equation}
\dim_\C \text{Def}({\cal D}_\textrm{gen})=m(2m+1)K_B\cdot K_B=m(2m+1)S_B \label{defhyp}
\end{equation}
deformations. To simplify equations we again use $S_B$ as a shorthand for $(-K_B)\cdot (-K_B)=S_B$.
As in the last section we keep the O-plane fixed so that the automorphism group of the base is completely
broken. As the number of double intersections is given by $2mS_B$, the double intersections lead
to $2mS_B$ constraints, so that we expect the number of deformations encoded in Eq.~\eqref{genald7}
to be
\begin{equation}\label{def_obstructedbrane}
\dim_\C \text{Def}({\cal D}) = m(2m+1)S_B-2mS_B=m(2m-1)S_B\ .
\end{equation}

The degrees of freedom in the expression (\ref{genald7}) are given by the
number of monomials in $\eta$ and $\chi$, minus an overall rescaling
and the redundancy that corresponds to shifting $\eta$ by $h\alpha$, $\alpha$ being
a polynomial of appropriate degree. To compute the number of monomials, we note that
we can take the polynomials $\eta$, $\chi$ and $\alpha$ to define hypersurfaces on
their own and compute the number of their deformations using Eq.~\eqref{defhyp}.
The number of monomials is then given by the number of deformation plus one. For a D7-brane that is
equivalent to $-2mK_B$, $\eta$, $\chi$ and $\alpha$ are sections of $[-mK_B]$,
$[-(2m-2)K_B]$ and $[-(m-2)K_B]$, respectively. Thus the degrees of freedom in Eq.~\eqref{genald7} are given in this case by
\begin{equation}\label{def_obstructedbrane2}
\begin{aligned}
\dim_\C \text{Def}({\cal D}) & = \underbrace{\tfrac{1}{2}m(m+1)S_B+1}_{M(\eta)}+\underbrace{(m-1)(2m-1)S_B+1}
_{M(\chi)} \\
&-\underbrace{\left(\tfrac{1}{2}(m-2)(m-1)S_B+1\right)}_{M(\alpha)}-1
\\ &=m(2m-1)S_B \ ,
\end{aligned}
\end{equation}
which coincides with \eqref{def_obstructedbrane}. 

For a generic hypersurface ${\cal D}_\textrm{gen}$ one can actually show that in the double cover its deformation space is isomorphic to $H^{(1,0)}_-({\cal D}'_\textrm{gen})$, where ${\cal D}'_\textrm{gen}$ is the corresponding preimage of ${\cal D}_\textrm{gen}$ in the double cover \cite{Jockers:2004yj}:
\begin{equation}\label{H(1_0)_generic}
\begin{aligned}
h^{(1,0)}_-({\cal D}'_\textrm{gen})&=g({\cal D}_\textrm{gen})+\tfrac{1}{2}I_{{\cal D}_\textrm{gen}-O7}-1  \\
&=m(2m-1)S_B+1+2m S_B-1  \\
&=m(2m+1)S_B \ ,
\end{aligned}
\end{equation}
where the first of the above equalities follows from the Riemann-Hurwitz theorem \cite{Griffiths:1978}. 
We want to stress that the above statement is not true any more for a generic allowed brane ${\cal D}$ and its double cover ${\cal D}'$. 
More precisely, we see from \eqref{def_obstructedbrane2} that the number of degrees of freedom is exactly $g({\cal D})-1$. A comparison with the computation in \eqref{H(1_0)_generic} suggests that the cycles in $H^{(1,0)}_-({\cal D}')$ which are related to the double intersections do not give rise to deformations of ${\cal D}$. 

Let us now perform an analogous computation for the generic allowed brane
${\cal D}$ and its double cover ${\cal D}'$ to confirm this observation.
As we already discussed in Figure~\ref{2btof} and below
\eqref{obstructed_Weierstrass}, ${\cal D}'$ is a smooth brane apart
from its self-intersections (which occur at every intersection
point with the O-plane). Removing these singular points from ${\cal D}'$, 
we obtain a smooth Riemann surface with punctures on which the
$Z_2$ orbifold projection acts freely. Subsequently, we compactify this
punctured Riemann surface in the obvious way, by adding one point per
puncture. The result is a smooth compact Riemann surface with free
$Z_2$ action, which we continue to call ${\cal D}'$ by abuse of notation. 
While this smooth Riemann surface is not realized as a submanifold of the
double-cover Calabi-Yau, its $Z_2$ projection ${\cal D}$ is still our
familiar generically allowed D-brane given as a submanifold of the base
$B$. By standard arguments \cite{Jockers:2004yj}, its allowed deformations
correspond to $Z_2$-odd sections of the canonical bundle $K_{{\cal D}'}$
of ${\cal D}'$, which is understood as a smooth Riemann surface as
explained above. Thus, repeating the calculation of \eqref{H(1_0)_generic},
the number of deformations is given by
\begin{equation}
h^{(1,0)}_-({\cal D}')=g({\cal D})-1 =m(2m-1)S_B \ ,
\end{equation}
where we have again used the Riemann-Hurwitz theorem, but now for a freely
acting involution. This agrees with our previous results. The advantage
of this new derivation is that we are now able to specify which bundle
over ${\cal D}$ encodes these deformations. Indeed, while the even
sections of $K_{{\cal D}'}$ correspond to sections of $K_{\cal D}$, the
odd sections of $K_{{\cal D}'}$ can be understood as sections of
a `twisted' canonical bundle $\tilde{K}_{\cal D}$ over ${\cal D}$. The
latter is is defined as the $Z_2$ projection of $K_{{\cal D}'}$, where
the $Z_2$ action on ${\cal D}'$ is supplemented by a `$-1$' action on the
fibre. Locally, in a small neighbourhood of an intersection point
with the O-plane, this is still the canonical bundle of ${\cal D}$,
in agreement with the discussion of \cite{Beasley:2008dc,Beasley:2008kw,Cordova:2009fg}.

\subsection{Recombination for double intersection points} \label{RecObstr}
Now we want to understand the number of 3-cycles from the threefold perspective and explain why the number of 3-cycles is reduced by $I$ when compared with our results in Section~\ref{D7-branes without obstructions}. As mentioned above, we need to understand the winding numbers of the D-brane 1-cycles relative to the O-plane. First we discuss this locally for a single 1-cycle and then in Section \ref{ObstructionsCycles} analyze the global situation.

\begin{figure}
\begin{center}
\includegraphics[height=4cm, angle=0]{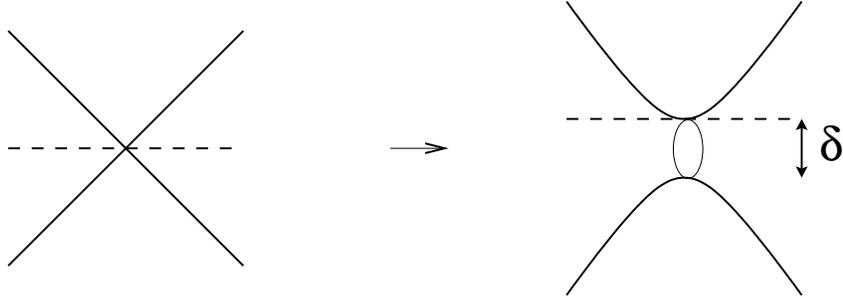}
\end{center}
\caption{\textsl{Recombination of the two D-branes at the common intersection point with the O-plane leads to a double intersection point of the recombined D-brane with the O-plane.}}
\label{recombprocessobstr}
\end{figure}

\begin{figure}
\begin{center}
\includegraphics[height=4cm, angle=0]{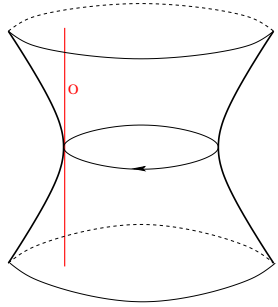}
\end{center}
\caption{\textsl{After recombination, the O-plane has a double intersection point with the recombined D-brane.}}
\label{recombinationobstructed}
\end{figure}

\begin{figure}
\begin{center}
\includegraphics[height=4cm, angle=0]{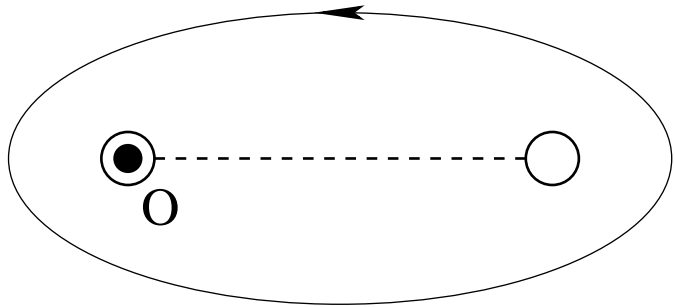}
\end{center}
\caption{\textsl{Recombination at a double intersection point blows up the branch cut between the two branch points. One of the branch points is the point where the O-plane meets the recombined D-brane (illustrated by the black dot). The loop around the branch cut illustrates the non-trivial 1-cycle encircling the throat.}}
\label{branchcutoplane}
\end{figure}

Let us again consider the recombination of two intersecting D-branes, cf.\ Section~\ref{recomD}, but now in the presence of the O-plane at the intersection point. Furthermore, we assume that we have already moved the fourth D-brane off the O-plane such that we describe the D-branes by Eq.~\eqref{genald7}. If we set in the local model
\begin{equation}
\begin{aligned}
 h & =z \ , \\
 \eta & =x \ , \\
 \chi & = \tfrac{1}{12} (z - \delta) \ ,
\end{aligned}
\end{equation}
we have the situation of an O-plane at $z=0$ and a D-brane given at
\begin{equation}\label{brane_recomb_obstr}
 x^2 = z (z-\delta) \ .
\end{equation}
For $\delta=0$, it parameterizes the situation of two D-branes at $x \pm z =0$ and an O-plane at $z=0$, all of them intersecting at the origin, as shown in the left picture in Figure~\ref{recombprocessobstr}. If we now give $\delta$ a non-zero value, we get to the recombined situation to the right in Figure~\ref{recombprocessobstr}.
Here the diameter of the throat connecting the two D-branes is given by $\delta$.
After recombination, the O-plane touches the D-brane tangentially at $x=z=0$, see Fig.~\ref{recombinationobstructed}.
With help of Eq.~\eqref{brane_recomb_obstr}, we can picture the D-brane as the $z$-plane with a branch cut between the two branch points at $z=0$ and $z=\delta$, as shown in Figure~\ref{branchcutoplane}.
The intersection with the O-plane is also given by $z=0$, and therefore, the D-brane 1-cycle encircles the O-plane intersection exactly once. However, since the O-plane is just a plane parameterized by $z=0$, any loop on the D-brane world-volume encircling $z=0$ encircles, when understood as a curve in $B$, the O-plane exactly once, too. Thus, the D-brane 1-cycle has a winding number of one relative to the O-plane.

Since the D-brane 1-cycle wraps the O-plane exactly once, we cannot build a 3-cycle out of it as done in Section~\ref{recomD}.
Because of the involution of the O-plane the fibre would simply be ill-defined. From the double cover perspective, it is clear that
the 3-cycle construction fails for a D-brane 1-cycle with an odd wrapping number:
In the double cover, the lift of this 1-cycle is not closed and does not define a D-brane 1-cycle in the double cover.

\subsection{Threefold cycles, obstructions and the intersection matrix} \label{ObstructionsCycles}

\begin{figure}
\begin{center}
\includegraphics[height=8cm, angle=0]{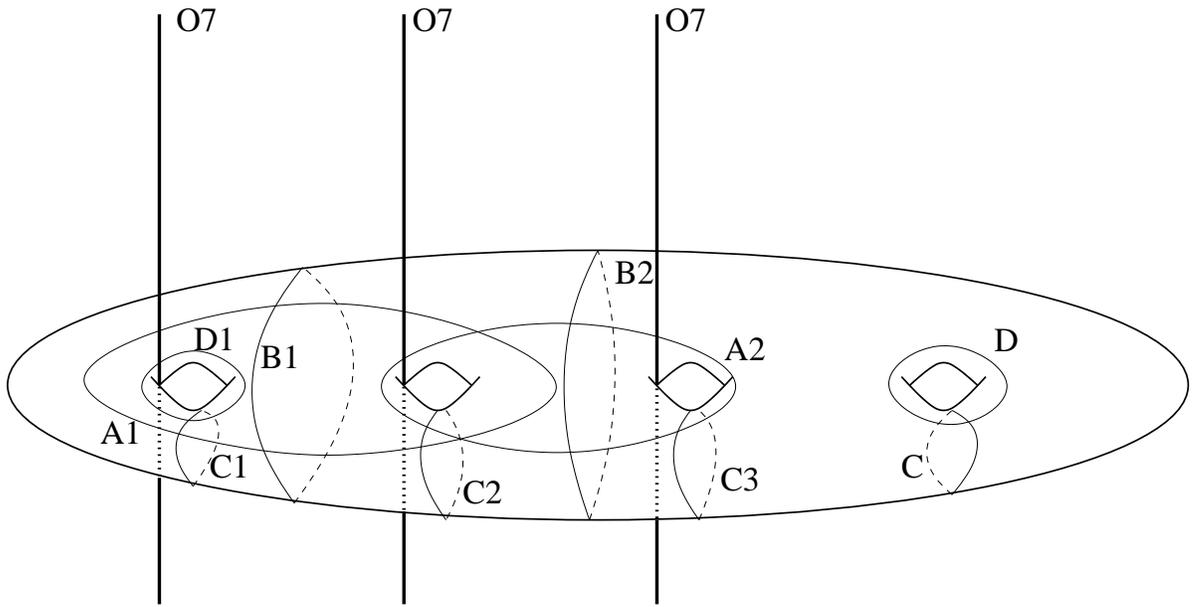}
\end{center}
\caption{\textsl{Here we see the various curves on the D-brane which lead to 3-cycles of the threefold. The curves C and D give rise to 3-cycles in the usual way, as explained in Section~\ref{recomD}. The curves A1 and A2 have an even wrapping number and therefore also lead to non-trivial 3-cycles. The dual cycles are constructed from the curves C1, C2 and C3 and are linearly equivalent to the 3-cycles over B1 and B2, which are non-trivial due to the O-plane monodromy.}}
\label{CyclesObstructions}
\end{figure}

In this section we outline how threefold three-cycles arise from the topology of a generic allowed D-brane, analogously 
to Section~\ref{1doffO}. We do not provide a rigorous proof but rather sketch the construction of threefold 
3-cycles from the 1-cycles of a generic allowed D-brane in the presence of a `naked' O-plane.
Using \eqref{genmatch} we can express the number of degrees
of freedom of a generic allowed brane in terms of its genus:
\begin{equation}
\dim_\C \text{Def}({\cal D})=m(2m-1)S_B -1=g({\cal D})-1.
\end{equation}
We thus expect that we can construct $2g-2$ 3-cycles.
All intersections of the D-brane with the O-plane are double intersections, which makes it impossible to construct cycles in the spirit of Section~\ref{cycintd7o7}. However, we still can build 3-cycles related to the D-brane 1-cycles, as we explain now. For this, we first choose a symplectic basis for the 1-cycles of the D-brane and then discuss how the basis 1-cycles lead to 3-cycles in the threefold, depending on the wrapping number of these 1-cycles with respect to the O-plane.
We illustrate the curves on the D-brane that lead to 3-cycles in Fig.~\ref{CyclesObstructions}.
Note that Fig.~\ref{CyclesObstructions} represents a local picture of the D-brane, ignoring the topology of $B$ and of the O-plane. 

Consider a symplectic pair of D-brane 1-cycles.
First assume that both 1-cycles have zero wrapping number with respect to the O-plane. If this occurs, we can construct a 3-cycle over each of them as explained in Section~\ref{recomD}, giving a symplectic pair of 3-cycles which has no intersection with any of the other 3-cycles. An example of such a symplectic pair is given by C and D in Fig.~\ref{CyclesObstructions}.

Next consider a symplectic pair of 1-cycles where one of the 1-cycles has wrapping number one while the paired cycle has still wrapping number zero. An example of such a pair is given by C1 and D1 in Fig.~\ref{CyclesObstructions}, where D1 has wrapping number one and C1 has wrapping number zero.
If we try to use the same technique as in Section~\ref{recomD} and fibre the usual 1-cycle over a disc ending on D1, we cannot close this 3-cycle since the monodromy of the O-plane inverts the orientation of the fibre 1-cycle. One might have the idea to use a loop corresponding to 2D1, which surrounds the O-plane twice. Since the orientation of the fibre 1-cycle is inverted twice, the corresponding 3-cycle is closed. However, this 3-cycle must be trivial since it is by construction symmetric under orientation reversal, i.e.\ the 3-cycle is equivalent to minus itself. 

This result has several consequences. First of all, we see that if we have a symplectic pair of 1-cycles where both have an even wrapping number, we can always add 2D1 such that the wrapping numbers are zero and apply the construction method of Section~\ref{recomD} to find two 3-cycles. This should be seen in contrast to Section~\ref{D7-branes without obstructions}, where one had to restrict to the D-brane 1-cycles that have zero wrapping number and thereby reduced the number of appropriate 1-cycles by one.
Secondly, the construction of 3-cycles from cycles with odd wrapping number must be modified. 
Note that if both 1-cycles of a symplectic pair have an odd wrapping number, we can replace one of them by the sum of both, leading to a symplectic pair of 1-cycles where only one wrapping number is odd.
Thus, the remaining case is that exactly one of the two D-brane 1-cycles has an odd wrapping number.\footnote{The number of such symplectic pairs is actually $8 S_B$, as we explain now. If we consider Eq.~\eqref{genald7} with $\chi=\tfrac{1}{12} h \alpha^2$, this corresponds to two D-branes at $\eta \pm h \alpha$ which then can be recombined at each of the $8 S_B$ intersection points. With the result of Section~\ref{RecObstr} we conclude that there are $8 S_B$ D-brane 1-cycles with odd wrapping number. These 1-cycles correspond to the $I/2=8 S_B$ double intersections of the D-brane with the O-plane.}

As discussed above, over a D-brane 1-cycle with odd wrapping number no 3-cycle can be defined due to the involution that is part of the O-plane monodromy. However, we can use the sum of two such 1-cycles which is represented by a curve encircling two of the double intersection points with the O-plane.
Examples of such curves are denoted in Fig.~\ref{CyclesObstructions} by ${\rm A}1={\rm D}1-{\rm D}2$ and ${\rm A}2={\rm D}2-{\rm D}3$. 
Note that since 3-cycles constructed out 2D1, 2D2 oder 2D3 are trivial, the third combination ${\rm D}1+{\rm D}3$ does not lead to any independent 3-cycle but to the one constructed out of ${\rm A}1+{\rm A}2$.
Thus, this construction gives us one 3-cycle less than there are D-brane 1-cycles with odd wrapping number.

Now we turn to the 1-cycles that are dual to those with odd wrapping number, represented by C1, C2 and C3 in Fig.~\ref{CyclesObstructions}. Since these 1-cycles have even wrapping number we can construct 3-cycles $C_i$ out of them. However, there is a linear dependence between them, as we show now. Consider the curves B1 and B2 in Fig.~\ref{CyclesObstructions}. They both lead to non-trivial 3-cycles $B_i$ due to the intersection points of the D-brane with the O-plane. Since the involution of the O-plane inverts the orientation of the fibre 1-cycles, the 3-cycle $C_i$ comes back to minus itself when once encircling the O-plane.
More precisely, if we denote the corresponding 3-cycles by $C_i$ and $B_i$, due to the monodromy of the O-plane there is
\begin{equation}
  2 C_1 = B_1 \ , \quad 2 C_2 = B_2-B_1 \ , \quad  2 C_3 = - B_2 \ ,
\end{equation}
leading to
\begin{equation}
  2 C_1 + 2 C_2 + 2 C_3 = 0 \ .
\end{equation}
On a general D-brane, the linear dependence analogously reads
\begin{equation}
 2 \sum_{i=1}^{I/2} C_i = 0 \ ,
\end{equation}
where the sum runs over all 3-cycles $C_i$ which are coming from D-brane 1-cycles dual to those with odd wrapping number. Thus, we find that the total number of 3-cycles coming from the D-brane sector is two less than the number of 1-cycles of the D-brane, leading to $2g({\cal D})-2 = \chi({\cal D}) = 56 S_B$ 3-cycles. This coincides with the result we achieved in Section~\ref{Obstructions}.

Let us now discuss the intersection matrix of the 3-cycles discussed here. As already stated above, the intersection matrix of the 3-cycles that come from 1-cycles with even wrapping number is the standart symplectic one. For the 3-cycles which are constructed out of curves of type A and B, the corresponding 3-cycles $A_i$ and $\frac{1}{2} B_j = \sum_{i=1}^j C_i$ also form a symplectic basis.\footnote{Note that the $A_i$ do not intersect each other due to the monodromy of the O-plane.} Thus, we find the symplectic basis for the 3-cycles of the threefold that come from the D-brane sector.

\chapter{F-theory on fluxed \boldmath$K3\times K3$}
\label{fluxonk3xk3}

In this chapter we consider flux stabilization of M/F-theory on $K3\times K3$.
We are able to demonstrate how fluxes stabilize D7 branes or stacks of D7 branes in a completely 
explicit fashion. This allows us to select a specific desired gauge group by the choice 
of flux numbers. As explained above, this is the same procedure required for 
the flux stabilization of non-Abelian gauge symmetries in the 
(non-perturbative) F-theory context, which has recently attracted significant 
attention in the context of GUT model building 
\cite{Beasley:2008dc,Beasley:2008kw,Donagi:2008ca,Donagi:2008kj,Donagi:2009ra}. We therefore
expect that straightforward generalizations of our methods will be useful both for more
complicated D7 brane models as well as for their non-perturbative F-theory cousins. 

Moduli stabilization by fluxes in M-theory on $K3\times K3$ has been studied extensively in the
past, especially in relation with the type IIB dual 
(see, e.g.~\cite{Gorlich:2004qm,lmr05,drs99}). In our work we derive the flux potential for the geometric
moduli from dimensional reduction. We express it in a form manifestly invariant under the $SO(3)$
symmetry of the $K3$ moduli space. In this form, it is immediate to see how the minimization condition
is translated into a condition on fluxes and on geometric data of the two $K3$'s. We find all
Minkowski minima, both supersymmetric and non-supersymmetric. 
An analogous explicit search for (supersymmetric) flux vacua has been reported in~\cite{ak05}. Our
results are more  general since we do not restrict  ourselves to
attractive $K3$ surfaces, where a maximal number of integral 2-cycles are holomorphic.\footnote{At  
a technical level, this means that only a discrete set of values are allowed
for the various complex structure moduli. There is then also only a 
very restricted set of fluxes which are suitable for stabilizing such points.}

Our analysis of moduli stabilization is also more explicit than the previous works on $K3\times K3$,
since we use a parameterization of D7-brane motion by the size of integral 2-cycles, as derived 
in Chapter~\ref{chapterk3}. Thus, at least in the weak coupling limit, we have a simple geometric interpretation for 
every integral basis cycle. Using the choice of flux numbers, this gives us 
full control over the positions of 4 O7 planes and 16 D7 branes moving on a 
$\mathbb{CP}^1$ base (corresponding to type IIB on $T^2/\mathbb{Z}_2$). % as well as the complex 
%structure of the second `non-F-theory' $K3$. 

Our techniques can be used to study the stabilization of all the gauge groups that can be realized
by F-theory on $K3$. It turns out that tadpole cancellation is very restrictive and allows only very
special flux choices. 

We begin our analysis in Section~\ref{sec:fluxpotential} with a derivation of the $K3\times K3$ flux
potential, which closely follows the generic Calabi-Yau derivation 
of~\cite{gvw99, bb96, hl01}. We emphasize the fact that, 
due to the hyper-K\"ahler structure of $K3$, its geometric moduli space can 
be visualized by the motion of a three-plane in the 22-dimensional space of 
homology classes of 2-cycles. This three-plane is spanned by the real and 
imaginary parts of the holomorphic 2-form and by the K\"ahler form. The
resulting $SO(3)$ symmetry of the geometric moduli is manifest in the
expression for the scalar potential we arrive at. The three-dimensional theory also has a number of
gauge fields, and the flux induces mass terms for some of them, which we derive explicitly. This
breaking of gauge symmetries can be understood in the dual type~IIB picture as the gauging of some
shift symmetry in the flux background.

In Section~\ref{sec:modstab} we analyse the minima of the flux potential. 
To preserve four-dimensional Poincar\'e invariance, we consider 4-form 
fluxes that belong to $H^2(K3)\otimes H^2(K3)$.
A flux of this form gives rise to a linear map between the spaces of 2-cycles of
the two $K3$'s: integrating the flux on a 2-cycle of one $K3$ we get a 2-form on the other 
$K3$ (which is Poincar\'e dual to a 2-cycle). Minkowski vacua arise if the flux 
maps the three-planes determining the metric of the $K3$'s onto each other. 
We derive the conditions the flux matrix has to satisfy in order 
for two such planes to exist and to be completely fixed by the choice of 
fluxes. Furthermore, we clarify the more restrictive conditions under which 
the plane determined by the flux is consistent with the F-theory limit. In 
this case, the plane cannot be fixed completely. The unfixed moduli 
correspond to Wilson lines around the $S^1$ of the type IIB model which 
decompactifies in the F-theory limit. These degrees of freedom are not part 
of the moduli space of type~IIB compactified to four dimensions, as they 
characterize the (unphysical) constant background value of one component of 
the four-dimensional vector fields. In fact, the corresponding propagating 
degrees of freedom become part of the four-dimensional vector fields 
(see~\cite{Valandro:2008zg} for a comprehensive analysis of the duality map between the 
4d fields of M-theory on $K3\times K3$ and type IIB string theory on 
$K3\times T^2/\mathbb{Z}_2$). 

The main point of this chapter, the explicit stabilization of D-brane positions, 
is the subject of Section~\ref{sec:movebrain}. After recalling the parameterization of D7-brane 
motion in terms of M-theory cycles derived in Chapter~\ref{chapterk3}, we provide 
explicit examples of flux matrices which fix situations with gauge symmetries 
$SO(8)^4$, $SO(8)^3\times SO(6)$ and $SO(8)^3\times SO(4)\times SU(2)$. In all cases we also fix 
the complex structure moduli of the lower $K3$. The first case corresponds to the orientifold, where
4 D7 branes lie on top of each O7 plane. In the second case, one  D7~brane is moved away from an
O~plane. Finally, in the third case, a stack of two D7~branes is separated from one of the O~planes.
In these examples almost all the K\"ahler moduli (which correspond to deformations of the lower $K3$
and do not  affect the positions of the D7 branes) are not stabilized. When one of them is
stabilized, a K\"ahler modulus of the upper $K3$ is stabilized, too. As mentioned before, this
corresponds to some gauge field becoming massive. To clarify this point, we 
present two examples where one of the D7 branes is fixed at a certain 
distance from its O~plane: In the first example, one further K\"ahler modulus 
is fixed, breaking the $U(1)$ gauge group. This phenomenon of gauging by 
fluxes is common in flux compactifications\cite{aaf03,aaft03,hkl06,Jockers:2004yj}. In the
second example, we stabilize the single D7 brane without fixing further 
K\"ahler moduli and hence without gauge symmetry breaking. We also provide 
an example where almost all moduli are fixed. In this case, only the fibre  
volume of F-theory, the volume moduli of the two $K3$s, and three metric 
moduli of the lower $K3$ remain undetermined.

Section~\ref{sec:susyvac} contains a brief discussion of supersymmetry. Generically, we obtain 
$\Nn=0$ vacua of no-scale type. For specific, non-generic choices of the flux 
matrix, we find three-dimensional ${\cal N}=2$ or ${\cal N}=4$ supersymmetry.
To determine the amount of surviving supersymmetry, it suffices to know the 
eigenvalues of the flux matrix restricted to the two three-planes.

\section[$K3$ Flux Potential]{\boldmath$K3$ Flux Potential}
\label{sec:fluxpotential}
In this chapter we compactify M-theory to three dimensions on $K3\times K3$ 
and analyse the effects of 4-form flux. The main new points of our 
presentation are the following: We maintain a manifest $SO(3)\times SO(3)$ 
symmetry of the moduli space of $K3\times K3$ in the calculation of the 
potential in Section~\ref{sec:potential}. Furthermore, we explicitly derive 
the flux-induced masses for the vector fields arising from the 3-form 
$C_3$ in Section~\ref{sec:vectormass}.

\subsection[M-Theory on $K3\times K3$]{M-Theory on \boldmath$K3\times K3$}
The compactification of M-theory on a generic four-fold is described in 
detail in\cite{bb96}. Here we specialize to the case of $K3\times K3$. To 
distinguish the two $K3$'s, we write the compactification manifold as 
$K3\times \Kt$. Correspondingly, all quantities related to the second $K3$ 
will have a tilde.

The relevant M-theory bosonic action is\cite{cjs78}
\begin{align}\label{MthAction}
  \begin{split}
    S_\text{M} &= \frac{2\pi}{\ell_M^9} \left\{ \int \d^{11}x \sqrt{-g} 
    \left( R -\frac{1}{2} |F_4|^2
    \right)- \frac{1}{6} \int  C_3\wedge F_4\wedge F_4 \right\}\\
    &\quad +\left(\frac{2\pi}{\ell_M^3}\right) \left(\int 
    C_3\wedge I_8(R) + \int
      \d^{11} x \sqrt{-g}\, J_8\!\left(R\right)\right)\ ,
  \end{split}
\end{align}
where $\ell_M$ is the eleven-dimensional Planck length, $F_4=dC_3$, and 
$I_8(R)$ and $J_8\!\left(R\right)$ are polynomials of degree 4 in the 
curvature tensor \cite{dlm95,t00}. When we compactify on $K3\times \Kt$, we obtain a 
three-dimensional theory with eight supercharges, i.e.\ $\Nn=4$ in three 
dimensions. This can be inferred from the fact that each $K3$ has holonomy
group $SU(2)$ and correspondingly two invariant spinors. 

Let us analyse the geometric moduli. $K3$ is a hyper-K\"ahler manifold: its 
metric is defined by three 2-forms $\omega_i$ in $H^2(K3)$ plus the overall 
scale. $H^2\!\left(K3\right)$ is a 22-dimensional vector space equipped with 
a natural scalar product,\footnote{
Throughout
this work, we freely identify forms, their cohomology classes, the 
Poincar{\'e}-dual cycles and their homology classes.
}
\begin{align}\label{K3modmetric}
  v \cdot w \equiv \int_{K3} v \wedge w \qquad \forall\; 
  v,w\in H^2\!\left(K3\right)\,,
\end{align}
which has signature $(3,19)$, i.e.\ there are three positive-norm directions. 
The three vectors $\omega_i$ defining the metric must have positive norm and 
be orthogonal to each other. Hence they can be normalized according to 
$\omega_i \cdot \omega_j = \delta_{ij}$. The K\"ahler form and 
holomorphic 2-form and can then be given as
\begin{align}
  J&= \sqrt{2 \nu}\,\omega_3\,, & \omega&=\omega_1 +\I \omega_2\,.
\end{align}
This definition is not unique: we have an $S^2$ of possible 
complex structures and associated K\"ahler forms. Each of them defines the 
same metric, which is then invariant under the $SO(3)$ that rotates the 
$\omega_i$'s.

The motion in moduli space can now be visualized as the motion of the 
three-plane $\Sigma$ spanned by the $\omega_i$'s, which is characterized 
by the deformations of the $\omega_i$ preserving orthonormality. The 
corresponding $\delta \omega_i$ are  in the subspace orthogonal to 
$\Sigma$, which is 19-dimensional. Together with the volume, this
gives $3\cdot 19+1=58$ scalars in the moduli space of one $K3$. The same 
parameterization can be used for the second $K3$, where the corresponding 
scalars are $\nut$ and the components of $\delta\omegat_j$. Altogether one 
finds $58+58=116$ scalars from the metric on $K3\times \Kt$. Furthermore, 
since $K3$ has no harmonic 1-forms, there are no 3d vectors coming from 
the metric.

\subsection{The Scalar Potential\label{sec:potential}}

We now allow for an expectation value for the field strength $F_4$
of the form 
\begin{align} \label{4formflux}
  \left<F_4\right> \equiv G_4= G^{I\Lambda} \eta_I\wedge\etat_\Lambda\,,
\end{align}
where $\{\eta_I,\etat_\Lambda\}$ (with $I,\Lambda=1,...,22$) is an integral 
basis of $H^2(K3)\times H^2(\Kt)$. The flux satisfies a (Dirac) 
quantization condition\footnote{
The 
precise quantization condition for a generic fourfold $Y$ is $\ell_M^{-3}\,
[G_4]-\frac{p_1}{4} \in H^4(Y,\mathbb{Z})$, where $p_1$ is the first 
Pontryagin class\cite{w96}. Since $\frac{p_1}{2}$ is even for $Y=K3\times K3$, the 
quantization condition becomes simply $\ell_M^{-3}\,[G_4] \in 
H^4(Y,\mathbb{Z})$.
}:
$\ell_M^{-3} G^{I\Lambda}\in\mathbb{Z}$. In the following, we will always 
denote this type of flux by $G_4$ while the generic 4-form field strength 
will be $F_4$. 

The flux potential for the moduli is found by reducing the M-theory action.
In the presence of fluxes, the solution to the equations of motion is a 
warped product of a Calabi--Yau fourfold and a three-dimensional non-compact 
space\cite{bb96,drs99,gkp01}. In the following, we neglect backreaction and
work with the undeformed Calabi--Yau space $K3\times \Kt$ as the internal 
manifold. The underlying assumption is that, in analogy to\cite{gkp01}, 
for any zero-energy minimum of the unwarped potential a corresponding 
zero-energy warped solution will always exist. After Weyl rescaling, the 
potential is given by\cite{hl01}
\begin{align}\label{ModPotential}
  V= \frac{4\pi}{\ell_M^9}\frac{1}{4\Vv^3}\left(\,\int_{\mathrlap{
  \mspace{20mu}K3\times\Kt}}\d^8\xi
  \sqrt{g^{(8)}} |G_4|^2 - \frac{\ell_M^6}{12}\, \chi\right) \,,
\end{align}
where $\chi$ is the Euler number of the compact manifold. For
$K3\times\Kt$, it is $\chi=24^2=576$. Given our previous discussion of $K3$ 
moduli space, we expect that \eqref{ModPotential} will be invariant under 
$SO(3)\times SO(3)$ rotations once we express the metric in terms of 
$\omega_i$ and $\omegat_j$. 

In the absence of spacetime-filling $M2$ branes, the cancellation of 
$M2$-brane-charge on the compact manifold $K3\times \Kt$ requires\cite{bb96}
\begin{align}\label{TadpoleCanc}
 \frac{1}{2\,\ell_M^6}\int G_4\wedge G_4 =  \,\frac{\chi}{24} \:.
\end{align}
This allows us to express the second term in~\eqref{ModPotential} through 
the flux. It is convenient to set $\ell_M=1$ and to introduce a 
volume-independent potential $V_0$ by writing $V=\frac{2\pi}{\Vv^3} \, V_0$. 
Here $\Vv=\nu\nut$ is the volume of $K3\times\Kt$. Our result now reads 
\begin{equation}\label{V0expr}
 V_0 = \frac12 \int_{\mathclap{K3\times\Kt}} (G_4\wedge \ast G_4 - 
G_4\wedge G_4)\,,
\end{equation}
with $G_4$ given by \eqref{4formflux}. 

On $K3$, each $\eta_I$ can be split into a sum of two vectors, parallel and 
perpendicular to the 3-plane $\Sigma$:
\begin{align}\label{SplitEtaI}
  \eta_I = \sum_i(\eta_I\cdot \omega_i) \,\omega_i + \Pp[\eta_I]=
  \eta_I^\parallel+\eta_I^\perp \:.
\end{align}
Here $\Pp$ is the projector on the subspace orthogonal to $\Sigma$. The first
term, which corresponds to the projection on $\Sigma$, has been given in a 
more explicit form using the orthonormal basis $\omega_i$ of the $\Sigma$ 
plane for later convenience. The two terms of~\eqref{SplitEtaI} represent a 
selfdual and an anti-selfdual 2-form\cite{gvw99}, allowing us to write the 
Hodge dual of a basis vector as
\begin{align}\label{HodgeonSplit}
  \ast_{\scriptscriptstyle K3}\,\eta_I =  \eta_I^\parallel-\eta_I^\perp \,. 
\end{align}
The same applies to $\widetilde{K3}$. 

If we insert \eqref{4formflux} and \eqref{SplitEtaI} into the 
expression~\eqref{V0expr} for $V_0$ and we use the relation  
\eqref{HodgeonSplit} for the action of the Hodge~$\ast$, we find
\begin{align}
  \begin{split}
    V_0 &= - \left\{\eta_I^\parallel\cdot \eta_J^\parallel
      \left(\left(G^{I\Lambda}\etat_\Lambda\right)^\perp \!\cdot
      \left(G^{J\Sigma}\etat_\Sigma\right)^\perp\right)\right.\\
  &\quad \mspace{35mu} \left.+ \left(\left(\eta_IG^{I\Lambda}\right)^\perp 
      \!\cdot\left(\eta_J G^{J\Sigma}\right)^\perp\right) 
  \etat_\Lambda^\parallel\cdot
        \etat_\Sigma^\parallel \right\} \,.
  \end{split}
\end{align}
Since 
\begin{align}
  \eta_I^\parallel\cdot \eta_J^\parallel&= \sum_i \left(\eta_I\cdot 
  \omega_i\right) \left(\eta_J\cdot \omega_i\right) \,,
\end{align}
we can write $V_0$ as
\begin{align}
  \begin{split}
    V_0 &=-\left\{ \sum_{i} \Ppt[G^{I\Lambda}\,(\eta_I\cdot\omega_i)\,
    \etat_\Lambda] \cdot \Ppt[
      G^{J\Sigma}\,(\eta_J\cdot\omega_i)\,\etat_\Sigma] \right.\\ 
    & \left.\mspace{153mu}+ \sum_{j}\Pp[G^{I\Lambda}\,(\etat_\Lambda\cdot
      \omegat_j)\,\eta_I] \cdot \Pp[
      G^{J\Sigma}\,(\eta_\Sigma\cdot\omegat_j) \,\eta_J] \right\} \,.
  \end{split}
\end{align}

To write it in a more compact form, we define two natural homomorphisms 
$G:H^2(\Kt) \rightarrow H^2(K3)$ and $G^a:H^2(K3)\rightarrow H^2(\Kt)$ by
\begin{align}
  G\, \tilde{v} &=\int_\Kt G_4\wedge \tilde{v} = (G^{I\Lambda} 
  \Mt_{\Lambda\Sigma} \tilde{v}^\Sigma)\, \eta_I\,, & G^a v &=\int_{K3} 
  G_4\wedge v= (v^J M_{JI} G^{I\Lambda}) \etat_\Lambda \,.
\end{align}
where $v=v^J\eta_J\,\in H^2(K3)$, $\tilde{v}=\tilde{v}^\Sigma\etat_\Sigma\, 
\in H^2(\Kt)$ and $M_{IJ}$, $\Mt_{\Lambda\Sigma}$ represent the metrics in 
the bases $\eta_I$, $\etat_\Lambda$. The operator $G^a$ is the adjoint 
of $G$, i.e. $(v\cdot G \tilde{v})=(G^a v,\tilde{v})$. The matrix 
components of these operators are $G^I_{\phantom{I}\Sigma}\equiv G^{I\Lambda} 
\Mt_{\Lambda\Sigma}$  and $\left(G^a\right)^\Sigma_{\phantom{\Sigma}I}
\equiv \left(G^T\right)^{\Sigma J}M_{JI}$. 

The moduli potential is then given by 
\begin{align}\label{potentialV}
  V = -\frac{2\pi}{\Vv^3} \left( \sum_i \left\| \Ppt[G^a\omega_i] \right\|^2 
  + \sum_j \left\| \Pp[G\,\omegat_j] \vphantom{\Ppt}\right\|^2\right) \,.
\end{align}
As expected, it is symmetric under $SO(3)$ rotation of the $\omega_i$'s and 
of the $\omegat_i$'s\footnote{
The 
projectors $\Pp$ and $\Ppt$ are obviously symmetric as they project onto the 
space orthogonal to all the $\omega_i$'s.}.

This potential is positive definite since the metrics for $H^2(\Kt)$ and 
$H^2(K3)$ defined in \eqref{K3modmetric} are negative definite on the subspace 
orthogonal to the $\omega_i$'s and  the $\omegat_i$'s. We note also that the 
volumes of the two $K3$'s are flat directions parameterizing the degeneracy 
of the absolute minimum of the potential, in which $V=0$.

We can also rewrite this potential expressing the projectors through the $\omega$'s:
\begin{align}\label{potentialV2}
  V = \frac{2\pi}{(\nu\nut)^3} \left( -\sum_i \left\| G^a\omega_i \right\|^2 
  - \sum_j \left\| G\,\omegat_j \right\|^2  + 2 \sum_{i,j}(\omegat_j\cdot 
    G^a \omega_i)(\omega_i \cdot G\, \omegat_j)\right) \,.
\end{align}
This is again manifestly symmetric under $SO(3)$ rotations.

\subsubsection{The Potential in Terms of \boldmath$W$ and $\check{W}$\label{VWWhat}}

The scalar potential can also be expressed in terms of two superpotentials. For a $CY_4$, it reads \cite{hl01}
\begin{align}
 V=\frac{e^K}{\Vv^3} \mathcal{G}^{\alpha\bar{\beta}} D_\alpha W D_{\bar{\beta}} \overline{W} +
 \frac{1}{\Vv^4}\left(\frac12 \check{\mathcal{G}}^{mn}\partial_m \check{W}\partial_n \check{W} -
   \check{W}^2\right) \,.
\end{align}
Here $K= -\ln \int_{CY_4} \Omega\wedge \overline{\Omega}$ and  $W$ and $\check{W}$ are given by
\begin{align}
 W&=\int_{CY_4} \Omega \wedge G_4\,,  & \check{W}&=\frac{1}{4} \int_{CY_4} J\wedge J \wedge G_4\,.
\end{align}
The complex structure moduli are labelled by $\alpha=1,\dotsc,h^{3,1}$, while $m=1,\dotsc,h^{1,1}$
counts the K\"ahler moduli.

For $K3\times K3$, we get a similar but not identical form. Note fist that the above
potential depends on $h^{1,1}+2h^{3,1}$ real moduli. This is the
dimension of the metric moduli space of a $CY_4$. But it is not the case for $K3\times \Kt$, whose
moduli space has dimension 
\begin{align}
 2 \times 58 = 2 \left( 3(h^{1,1}(K3)-1)+1\right)\,.
\end{align}
The moduli are the volume and the deformations of the $\omega_i$'s that are orthogonal to
all the $\omega_i$'s and whose number is then $h^2(K3)-3=h^{1,1}-1$. On the other hand ,
\begin{align}
  \begin{split}
    h^{1,1}\!\left(K3\times \Kt\right) +2h^{3,1}\!\left(K3\times \Kt\right)&\\ 
    &\mspace{-80mu}= 2 \left(\vphantom{\frac{a}{a}}h^{1,1}\!\left(K3\right) 
      +  2 h^{2,0}\!\left(K3\right)h^{1,1}\!\left(K3\right)\right) = 2\times  60\,. 
  \end{split}
\end{align}
This is again a reflection of the fact that for $K3$, only the three-plane itself is geometrically
meaningful: The two ``missing'' moduli correspond to the rotation of $J$ into real and imaginary
parts of $\omega$.

By an explicit computation one can get the new form of the potential:
\begin{align}\label{WWhpotential}
  \begin{split}
    V &= V_{G_{3,1}} + V_{G_{2,2}} \\
    &=\frac{e^{K}}{\Vv^3}  \mathcal{G}_{(0)}^{\alpha\bar{\beta}} D_\alpha W D_{\bar{\beta}} \overline{W}
      + \frac{1}{\Vv^4}\left(\frac12 \check{\mathcal{G}}^{mn}\partial_m \check{W}\partial_n \check{W}^2 -
        \check{W}^2\right)\,.
  \end{split}
\end{align}
The second term, $V_{G_{2,2}}$ is the same as for the $CY_4$ (note that
$m=1,...,h^{1,1}(K3)+h^{1,1}(\Kt)$). The only difference is in $V_{G_{3,1}}$: In the $CY_4$ case it
is given by the integral of  $G_{3,1}\wedge G_{1,3}$, where the subscript denotes the Hodge
decomposition. In that case it is also equal to the primitive part $G_{3,1}^{(0)}\wedge
G_{1,3}^{(0)}$, since  $G_{3,1}$ is automatically primitive. On $K3\times \Kt$, it is not primitive
and one must remove from $G_{3,1}$ the piece proportional to $J$. This is what the
metric $\mathcal{G}_{(0)}$ does. It is 
given by  
\begin{align}
\mathcal{G}_{(0)} = \left(\begin{array}{cc}
 -\frac{\int_{K3}\chi_\alpha\wedge \bar{\chi}_{\bar{\beta}}}{\int_{K3} \omega\wedge \bar{\omega}} & \\ &
	 -\frac{\int_{\Kt}\tilde{\chi}_\rho\wedge \bar{\tilde{\chi}}_{\bar{\sigma}}}{\int_{\Kt}
     \omegat\wedge \bar{\omegat}} 
\end{array}\right)\,,
\end{align}
where $\{\chi_\alpha\}$ is a basis for (1,1)-forms orthogonal to $\omega_3$.

The supersymmetry condition for the vacua can be written in terms of these two superpotentials. In
this case they assume the standard form (see for example \cite{gvw99,hl01,lmr05})
\begin{align}
 D_\alpha W &= 0\,, & W &= 0\,,& \partial_m \check{W} &= 0 \,.
\end{align}
The first two conditions say that the $G_4$ is a (2,2)-form, while the last one implies $G_4$ is
primitive.

\subsubsection{More general fluxes}

In \eqref{4formflux} we have only considered fluxes $G_4$ with two legs on 
each  $K3$. More generally the flux could be of this form:
\begin{align}\label{4formfluxbis}
   \left<F_4\right>=G_4 + \mathcal{G} \rho + \widetilde{\mathcal{G}} \rhot\,,
\end{align}
where $\rho$ and $\rhot$ are the volume forms on $K3$ and $\Kt$.\footnote{
The 
normalization is $\int_{K3} \rho   =\int_{\Kt} \rhot  =1$.
}
To obtain the general potential, we need to compute $\ast \left<F_4\right>$. 
Using our previous result for $\ast G_4$ and the Hodge duals 
\begin{align}
  \ast\rho = \frac{\nut}{\nu} \,\rhot \qquad\mbox{and}\qquad \ast \rhot = 
  \frac{\nu}{\nut} \, \rho \,
\end{align}
of $\rho$ and $\rhot$, we find 
\begin{align}
  \begin{split}
    V_\text{new} &= \frac{\pi}{(\nu\nut)^3}\int_{\mathclap{K3\times\Kt}} 
   \left(F_4\wedge \ast F_4 - F_4\wedge F_4 \right)  \\
    &=  \frac{2\pi}{(\nu\nut)^3}\left\{V_0 + \frac{1}{2} \mathcal{G}^2 
    \left(\frac{\nut}{\nu}\right)  +\frac{1}{2} \widetilde{\mathcal{G}}^2 
    \left(\frac{\nu}{\nut}\right) - \mathcal{G} \widetilde{\mathcal{G}} 
    \right\}\,.
  \end{split}
\end{align}
With the substitutions $\mathcal{V}=\nu\nut$ and $\xi=\sqrt{\frac{\nut}
{\nu}}$, the potential can be concisely written as
\begin{align}
  V_\text{new} = \frac{2\pi}{\mathcal{V}^3} \left\{V_0+ \frac{1}{2} 
  \left(\mathcal{G} \,\xi -\widetilde{\mathcal{G}}\, \frac{1}{\xi}\right)^2 
  \right\} \,.
\end{align}
This potential is still positive definite and has minima at points where it 
vanishes, but it now has only one unavoidable flat direction, the overall 
volume of $K3\times \Kt$. The ratio of the volumes is fixed at 
$\xi^2=\widetilde{\mathcal{G}}/\mathcal{G}$.

\subsection{Gauge Symmetry Breaking by Flux\label{sec:vectormass}}
In our context, F-theory emerges from the duality between 
M-theory on $K3\times\Kt$, with $\Kt$ being elliptically fibred, and 
type IIB on \mbox{$K3\times T^2/\mathbb{Z}_2\times S^1$}. The F-theory limit 
consists in taking the fibre volume to zero on the M-theory side, and in 
taking the radius of the $S^1$ to infinity on the type IIB side (see 
Section~\ref{sec:ftheory} for the details of this limit). Before analysing 
the effect of gauge symmetry breaking by fluxes, we recall the different 
origins of four-dimensional gauge fields in type IIB and in F-theory. 

Type~IIB theory on $K3\times T^2/\mathbb{Z}_2$ contains 16 vectors from 
gauge theories living on D7 branes and 4 vectors from the reduction of 
$B_2$ and $C_2$ along 1-cycles of $T^2/\mathbb{Z}_2$. In three 
dimensions, one then has 20 three-dimensional gauge fields and 20 scalars 
corresponding to Wilson lines along the $S^1$. In the F-theory limit, these 
scalars combine with the vectors to give the required 20 four-dimensional 
vector fields. 

In M-theory on $K3\times\Kt$, vectors arise from the reduction of the 
3-form $C_3$ along 2-cycles in $K3$ or $\Kt$. Since we are in three 
dimensions we have the freedom to dualize some of these vectors, treating 
them as three-dimensional scalars. To match the type~IIB description, the 
correct choice is to treat only the fields coming from the reduction of 
$C_3$ on 2-cycles of $\Kt$ as vectors\footnote{A 
simple intuitive argument for this choice can be given by comparing the 
seven-dimensional theories coming from M-theory on $\Kt$ and type IIB on 
$T^2/\mathbb{Z}_2\times S^1$. In M-theory, we have seven-dimensional vectors 
associated with 2-cycles stretched between the pairs of degeneration loci 
of the fibre (which characterize D7 branes). In type IIB, the corresponding 
vectors come directly from the D7-brane worldvolume theories. The fact that 
they are associated with branes rather than with pairs of branes is simply a 
matter of basis choice in the space of $U(1)$s.}.
This reduction gives 22 vectors in three dimensions. However, since $\Kt$ is 
elliptically fibred, there are two distinguished 2-cycles: the base and the 
fibre. They require a special treatment in the F-theory limit and, as a 
result, three-dimensional vectors arising from these two cycles do not become 
four-dimensional gauge fields in the F-theory limit. Instead, one of them 
corresponds to the type~IIB metric with one leg on the $S^1$, while the 
other is related to $C_4$ with three legs on $T^2/\mathbb{Z}_2\times 
S^1$\cite{Valandro:2008zg}. We will not consider these two vectors in the following and 
focus on the remaining 20 three-dimensional vectors associated with the 
reduction of $C_3$ on generic 2-cycles of $\Kt$. 

Each of these vectors absorbs a three-dimensional scalar (corresponding to a 
Wilson line degree of freedom on the type IIB side) to become a 
four-dimensional vector. These 20 scalars come from the metric moduli space 
of $\Kt$. More precisely, 18 arise from the variations $\delta\omegat_3$ of 
the K\"ahler form in directions orthogonal to the three-plane and to the 
base-fibre subspace\footnote{
For 
an elliptically fibred $\Kt$, two directions of the three-plane are orthogonal 
to base and fibre subspace, while $\omegat_3$ has a component along the 
base-fibre subspace. This explains the above number of independent 
variations as $18=22-(3+2-1)$. The variation of $\omegat_3$ within the 
base-fibre subspace corresponds to part of the metric in the F-theory limit. 
}. 
The two remaining scalars come from variations $\delta\omegat_1$ and 
$\delta\omegat_2$ of the holomorphic 2-form which lie in the base-fibre 
subspace and are orthogonal to $\omegat_3$. For a detailed analysis of the 
matching of fields on both sides of the duality, see~\cite{Valandro:2008zg}.

Given these preliminaries, it is now intuitively clear why F-theory flux 
generically breaks gauge symmetries: The flux induces a potential for the 
metric moduli, making them massive. This applies, in particular, to those 
moduli which become vector degrees of freedom in four dimensions. Hence, 
the full four-dimensional vector becomes massive by Lorentz 
invariance\footnote{
Correspondingly 
in type IIB,  putting 2-form flux on certain cycles of wrapped D7 branes 
breaks the gauge symmetry of the brane\cite{hkl06,aaf03,aaft03,Jockers:2004yj}.}.

To derive the vector mass term explicitly, we begin by writing $C_3$ in 
the form
\begin{align}\label{eq:C3KK}
  C_3&=C_1^I\wedge \eta_I + \Ct_1^\Sigma\wedge\etat_\Sigma+C_3^\text{flux}\,. 
\end{align}
Here $\eta_I$ and $\etat_\Sigma$ are basis 2-forms on the $K3$ factors, 
$C_1^I$ and $\Ct_1^\Sigma$ are 1-form fields in three dimensions, and 
$C_3^\text{flux}$ is the contribution responsible for the 4-form flux 
(which is only locally defined). As before, the flux is given by $G_4=
G^{I\Sigma}\eta_I\wedge\etat_\Sigma=\d C_3^\text{flux}$. 

In the reduction of the action, the $\int \left|F_4\right|^2$ term leads to 
the flux term $\int \left|G_4\right|^2$ (which is irrelevant for our present 
discussion) and to kinetic terms for $C_1^I$ and $\Ct_1^\Sigma$. The metric 
for the kinetic terms is given by
\begin{align}
  &\int _{\mathclap{K3\times\Kt}} \eta_I\wedge*_8\eta_J=\nut \int_{K3}
  \left(\eta_I^\parallel \wedge \eta_J^\parallel -\eta_I^\perp\wedge 
  \eta_J^\perp \right)=\nut \, g_{IJ} \,,\\ 
  &\int_{\mathclap{K3\times\Kt}} \etat_\Lambda\wedge*_8\etat_\Sigma =\nu
  \int_{\Kt}\left(\etat_\Lambda^\parallel \wedge \etat_\Sigma^\parallel -
  \etat_\Lambda^\perp\wedge \etat_\Sigma^\perp \right)=\nu\, 
  \gt_{\Lambda\Sigma}\,. 
\end{align}
We have split off the volume dependence, so that $g_{IJ}$ and 
$\gt_{\Lambda\Sigma}$ are dimensionless. Note that these metrics are positive 
definite since the subspace orthogonal to the three-plane has 
negative-definite metric. Note also that there is no kinetic mixing between 
the $C_1^I$ and the $\Ct_1^\Sigma$ since $\int \eta_I\wedge*_8\etat_\Sigma=0$.

We now turn to the Chern--Simons term $\int C_3\wedge F_4\wedge 
F_4$. Evaluating this term with $C_3$ of the form~(\ref{eq:C3KK}), we see that 
the contribution $\int C_3^\text{flux} \wedge F_4\wedge F_4$ vanishes: 
$C_3^\text{flux}$ has three legs on $K3\times\Kt$, so $F_4\wedge F_4$ would
need to have three legs on $\mathbb{R}^{1,2}$ and five legs on $K3\times\Kt$. 
This is, however, inconsistent with~(\ref{eq:C3KK}). The other contributions 
give
\begin{align}
  \int  C_3\wedge F_4 \wedge F_4 &= \int_{\mathbb{R}^{1,2}}2 G_{I\Sigma} 
  \left(C_1^I\d \Ct_1^\Sigma +\Ct_1^\Sigma \d C_1^I  \right) \,.
\end{align}
Thus, the flux matrix $G_{I\Sigma}= M_{IJ}G^{J\Lambda} \Mt_{\Lambda\Sigma}$ 
couples $C_1^I$ and $\Ct_1^\Sigma$. (Note that flux proportional to the 
volume forms of $K3$ and $\Kt$ would, in addition, lead to couplings 
$\sim C_1^I\d C_1^J$ and $\sim \Ct_1^\Sigma\d \Ct_1^\Lambda$.) 

We have now arrived at the three-dimensional effective action
\begin{align}\label{eq:3dactionflux}
  \begin{split}
    S_C^{(3)}&= \int_{\mathbb{R}^{1,2}} \left\{\nut g_{IJ}\, \d C_1^I \wedge 
    *\d C_1^J +\nu \gt_{\Lambda\Sigma}\, \d
    \Ct_1^\Lambda \wedge *\d \Ct_1^\Sigma\vphantom{\frac{2}{3}}\right.\\
    &\quad\mspace{40mu}\left.+\frac{2}{3}G_{I\Lambda} \left(C_1^I\wedge 
    \d \Ct_1^\Lambda+\Ct_1^\Lambda \wedge \d  C_1^I\right)\right\} \,. 
  \end{split}
\end{align}
As explained before, only the vectors $\Ct_1^\Sigma$ become four-dimensional 
vectors in the F-theory limit\cite{Valandro:2008zg}. It is convenient to dualize the 
remaining vectors $C_1^I$, replacing them by scalars $C_0^I$. To this 
end, we turn the equation of motion, 
\begin{align}
  \d \left(* \d C_1^I + \frac{2}{3} \frac{1}{\nut} g^{IJ} G_{J\Sigma} \d 
  \Ct_1^\Sigma\right)\,=\,0\,,
\end{align}
into a Bianchi identity by defining $C_0^I$ through
\begin{align}
  * \d C_1^I+\frac{2}{3}\frac{1}{\nut}  g^{IJ} G_{J\Sigma} \Ct_1^\Sigma\,=\, 
  \d C_0^I \,.
\end{align}
It follows that the $C_0^I$ have to transform non-trivially under the gauge 
transformations of the $\Ct_1^\Sigma$:
\begin{align}\label{eq:dualgauge}
  \Ct_1^\Sigma&\longrightarrow  \Ct_1^\Sigma +\d \Lambda_0^\Sigma\,, & 
  C_0^I &\longrightarrow C_0^I +\frac{2}{3}\frac{1}{\nut} g^{IJ}G_{J\Sigma} 
  \Lambda^\Sigma_0\,.
\end{align}
In other words, the vectors $\Ct_1^\Sigma$ gauge shift symmetries of the 
scalars $C_0^I$, with the charges determined by the flux. 

The equation of motion of $C_0^I$ follows formally from $\d \d C_1^I=0$, the 
Bianchi identity of $C_1^I$. Since $**=-1$ on $\mathbb{R}^{1,2}$, we find 
\begin{align}
0=\d\d C_1^I=-\d ** \d C_1^I&=\d * \left(\d C_0^I -\frac{2}{3}\frac{1}{\nut} 
g^{IJ}G_{J\Sigma} \Ct_1^\Sigma\right)\,.
\end{align}
We now want to find a gauge invariant action from which this equation of 
motion can be derived. Such an action is given by 
\begin{align}
  S_C^\text{dual}&= \int_{\mathbb{R}^{1,2}} \d^3 x \sqrt{-g_3}\left\{ \nu \left|\d \Ct_1^\Sigma\right|^2 
  + \nut \left|\d C_0^I - \frac{2}{3} \frac{1}{\nut} g^{IJ} G_{J\Sigma} 
  \Ct_1^\Sigma \right|^2\right\}  \,. 
\end{align}
The corresponding Einstein-Hilbert term has the usual volume prefactor and can 
be brought to canonical form by a Weyl rescaling of the three-dimensional metric. This gives 
the kinetic term of the vectors a prefactor $(\nu\nut)\nu$, which we can 
absorb in a redefinition of $\Ct_1^\Sigma$. The resulting mass matrix has 
the form
\begin{align}
  m^2_{\Sigma\Lambda}&\sim \frac{1}{\left(\nu\nut\right)^3} G_{I\Sigma} 
G_{J\Lambda} g^{IJ} \,,
\end{align}
which is positive semidefinite since $g^{IJ}$ is a positive definite metric. 
The number of gauge fields which become massive is determined by the rank of 
the flux matrix. Comparing with Eq.~(\ref{potentialV}), we see that the 
masses are of the same order as the masses of the flux stabilized geometric 
moduli. This confirms the intuitive idea put forward at the beginning of 
this section: The vectors $\Ct_1^\Sigma$ and some geometric moduli are 
combined in the F-theory limit to produce four-dimensional vectors. For this to work in 
the presence of fluxes, both the three-dimensional vectors and scalars need to have the 
same flux-induced masses.

\section{Moduli Stabilisation\label{sec:modstab}}

In this section we turn to the flux stabilization of moduli. First, we will analyse under which
conditions the potential~(\ref{potentialV}) has minima at $V=0$, and whether there are flat
directions. Then we will see which restrictions we have to impose in order to map the M-theory
situation to F-theory, and discuss possible implications for gauge symmetry breaking.

Let us first comment on the flux components which are proportional to the volume forms. In what
follows, we do not consider these components, in other words, we set
$\mathcal{G}=\widetilde{\mathcal{G}}=0$. The reason is that we want to end up with a
Lorentz-invariant four-dimensional theory. By going through the M-theory/F-theory duality
explicitly, one can see that this requires that the flux needs to have exactly one
leg in the fibre torus and hence two legs along each $K3$. Thus, we can without loss of generality
use a flux in the form of Eq.~(\ref{4formflux}), and the associated potential~(\ref{potentialV}).

\subsection{Minkowski Minima}
Clearly, the potential~(\ref{potentialV}) cannot stabilize the volumes $\nu$ and $\nut$. They are
runaway directions in general, and flat directions exactly if the term in brackets vanishes. This
term is a sum of positive definite terms, so each of these must vanish if we want to
realize a minimum with vanishing energy. Since each term contains a projection onto the subspace
orthogonal to the three-planes spanned by the $\omega_i$'s and $\omegat_i$'s, the bracket clearly
vanishes if and only if the flux homomorphisms map the three-planes into each other, though not
necessarily bijectively:
\begin{align}\label{eq:planetoplane}
  G \,\Sigmat &\subset \Sigma\,, & G^a \Sigma \subset \Sigmat\,.
\end{align}
Note that what is required is not merely the existence of three-dimensional subspaces which are
mapped to each other, but that both subspaces are positive-norm. If the metrics were positive definite,
this condition would be trivial since any real matrix can be diagonalized by choosing appropriate
bases in $H^2\!\left(K3\right)$ and $H^2\!\left(\Kt\right)$.

We will now show that the conditions~\eqref{eq:planetoplane} are equivalent to the conditions that the
map $G^a G$ is diagonalizable and all its eigenvalues are real and non-negative%
\footnote{Note that $G^a G$ maps
$H^2\!\left(\Kt\right)$ onto itself, so it makes sense to speak of eigenvalues and eigenvectors.
Note also, however, that although $G^a G$ is a selfadjoint operator, this does not imply that its
eigenvalues are real since the metric is indefinite. We have collected some facts about linear
algebra on spaces with indefinite metric in Appendix~\ref{app:linalg}.}.

Let us assume that there exist two three-planes $\Sigma$ and $\Sigmat$ such that the relations \eqref{eq:planetoplane} hold.
The cohomology groups can be decomposed into orthogonal subspaces, $H^2\!\left(K3\right)=\Sigma\oplus R$ and
$H^2\!\left(\Kt\right)=\Sigmat\oplus \Rt$, such that the metric \eqref{K3modmetric} defined by the wedge product is positive
(negative) definite on $\Sigma$ and $\Sigmat$ ($R$ and $\Rt$). The
conditions~(\ref{eq:planetoplane}) imply that $G$ and $G^a$ are block-diagonal, i.e.~we also
have $G\Rt\subset R$ and $G^a R\subset \Rt$.
It is then obvious that the selfadjoint operator $G^aG$ obeys\footnote{Similarly $G\,G^a$ obeys $G\,G^a \Sigma \, \subset \Sigma$.}
\begin{equation}\label{eq:planetoplane2}
  G^a G \Sigmat \, \subset \Sigmat\,. %\qquad \mbox{ and similarly } \qquad G\,G^a \Sigma \, \subset \Sigma \:.
\end{equation}
As each block is selfadjoint relative to definite metrics, $G^a G$ is diagonalizable with real and non-negative eigenvalues.
% Furthermore, for any $\et\in\Sigmat$
% and $\vt\in\Rt$ we have 
% \begin{align}
%   (G^a G \et \cdot \et)&= (G\et \cdot G\et) \geq 0\,,\\
%   (G^a G \vt \cdot \vt)&= (G\vt \cdot G\vt) \leq 0\,,
% \end{align}
% so $G^a G$ is indeed positive semidefinite. The same reasoning applies to $G G^a$. 

We now show that the converse also holds. Assume that $G^a G$ is diagonalizable
with non-negative eigenvalues\footnote{In this case, because of the non-degeneracy of the inner product, there alway exists a basis of non-null eigenvectors.}. This defines a decomposition of $H^2(\Kt)$ in  $\Sigmat\oplus \Rt$,
where $\Sigmat$ is the three-dimensional subspace given by the eigenvectors with positive norm.
The fact that $G^aG$ maps $\Sigmat$ into itself implies that $G$ maps positive norm vectors into
positive norm vectors: Indeed, give $\et\in \Sigma$, 
\begin{align}
   (G\et \cdot G\et)=(G^a G\et \cdot \et) \geq 0\,,
\end{align}
If $G^aG|_{\Sigmat}$ is invertible (non-zero eigenvalues in $\Sigmat$), then we can define
$\Sigma$ as the image of $G|_{\Sigmat}$. The fact that $\Sigmat$ is invariant under $G^aG$
implies that the image of $G^a|_\Sigma$ is $\Sigmat$ and \eqref{eq:planetoplane} is proved.
The case in which $G^aG$ has non-trivial kernel does not present any complication. Since the kernel of $G^aG$
coincides with the kernel of $G$ \footnote{%One might worry that for an $\et$ in the kernel of $G^a G$, it
%might happen that $G\et\neq 0$ is a lightlike vector in $H^2\!\left(K3\right)$. However, s
Since $G$ and $G^a$ are adjoint to each other, there is an orthogonal decomposition $H^2\!\left(K3\right)=\im G
\oplus {\rm Ker}\, G^a$. Take $\et \in {\rm Ker}\, G^aG$; since $G \et$ is both in $\im G$ and in ${\rm Ker} \, G^a$, it is the zero vector, proving that $\et \in {\rm Ker}\, G$.},
the image of $G|_{\Sigmat}$ is no more three dimensional. One then defines $\Sigma$ as the image of $G|_{\Sigmat}$
plus the positive norm vectors in the kernel of $G^a$.

To summarize, the conditions~(\ref{eq:planetoplane}) are equivalent to the condition that $G^a G$ is
diagonalizable and all its eigenvalues are real and non-negative. In this case (see Appendix~\ref{app:linalg})
the matrices for $G^aG$, $G$ and $G^a$ take the form
\begin{align}\label{GaGCanonicalForm}
 (G^aG)_d = \diag \!\left(a_1^2,a_2^2,a_3^2,b_1^2,\dotsc,b_{19}^2\right) \nn\\ G_d=G^a_d= 
	\diag \!\left(a_1,a_2,a_3,b_1,\dotsc,b_{19}\right) 
\end{align}
in appropriate bases, where $a_i^2$ are the eigenvalues of $G^aG$ relative to positive norm eigenvectors, while $b_c^2$ are relative to negative norm eigenvectors.

\

Finally, we want to see whether there are flat directions. The potential
has a flat direction, if there are infinitesimally different positions of the three-planes 
$\Sigmat,\Sigma$ which give Minkowski minima. Given a flux (such that $G^aG$ is diagonalizable with positive eigenvalues), the minima correspond to $\Sigmat$ ($\Sigma$) generated by the positive norm eigenvectors of $G^aG$ ($G\,G^a$). If all the eigenvalues are different from each other, there can be only three positive norm eigenvectors, and the minimum is isolated. If a positive norm and a negative norm eigenvector have the same eigenvalue, e.g. $a_1=b_1$, then a flat direction arises: Any three-dimensional space spanned by $\vt_{a_2}$, $\vt_{a_3}$ and $\vt_{a_1}'=\vt_{a_1}+ \epsilon \,\ut_{b_1}$ ($\epsilon \ll 1$) will give a different $\Sigmat$ that still satisfies the conditions \eqref{eq:planetoplane}. It is easy to see that an analogous flat direction develops for $\Sigma$. Note that if some $a_i$ are degenerate then the rotation of the vectors does not move the three-planes.

This shows that flat directions of the potential are absent if and only if the sets of eigenvalues
$\left\{a_i^2\right\}$ and $\left\{b_a^2\right\}$ are pairwise distinct.

\subsection{F-Theory Limit\label{sec:ftheory}}

We are interested in stabilizing points in the moduli space of $K3\times \Kt$ which can be mapped to
F-theory. This means we require that $\Kt$ is an
elliptic fibration over a base $\mathbb{C}\Pp^1$, and that the fibre volume vanishes.

The first requirement means that $\Kt$ needs to have two elements $\widetilde{B}$ (the
base) and $\widetilde{F}$ (the fibre) in the Picard group, i.e.~two integral (1,1)-cycles, whose
intersection matrix is
\begin{equation}\label{eq:blockBF}
\left(\begin{array}{cc}
	-2 & 1 \\ 1 & 0\\
\end{array}\right)\,.
\end{equation}
Note that by a change of basis from $\left(\widetilde{B},\widetilde{F}\right)$ to
$\left(\widetilde{B}+\widetilde{F},\widetilde{F}\right)$, this intersection
matrix is equivalent to one $U$ block in the general form~(\ref{eq:intersectionmatrix}) of the metric in an
integral basis. 

As  (1,1)-cycles, $\widetilde{B}$ and $\widetilde{F}$ must be orthogonal to the holomorphic 2-form. In
our case this means that the $\widetilde{\Sigma}$ plane has a two-dimensional subspace orthogonal orthogonal to
$\left<\widetilde{B},\widetilde{F}\right>$. This subspace is spanned by the real and imaginary part of 
the holomorphic 2-form $\omegat=\omegat_1+\I\omegat_2$.
On the other hand, $\left<\widetilde{B},\widetilde{F}\right>$ contains the third positive-norm
direction, so $\omegat_3$ cannot be also orthogonal to 
$\left<\widetilde{B},\widetilde{F}\right>$. For the following discussion it is convenient to
consider directly the K\"ahler form $\jt$ instead of 
$\nut$ and $\omegat_3$ separately. The K\"ahler form can be parametrized as
\begin{align}\label{eq:jtelliptic}
  \jt&= b \widetilde{B} + f\widetilde{F} + c^a \ut_a \,,
\end{align}
where $\ut_a$ is an orthonormal basis (i.e.~$\ut_a\cdot \ut_b=-\delta_{ab}$) of the space orthogonal to
$\widetilde{F}$, $\widetilde{B}$ and $\omegat$. This is the most general form of $\jt$ for an
elliptically fibred $\Kt$.

\

Now we turn to the second requirement: the fibre must have vanishing volume. This is what is called
the {\it F-theory limit}. For the K\"{a}hler form~(\ref{eq:jtelliptic}), we find the volumes of the
fibre and the  base to be\footnote{
More generally the volume of a 2-cycle $C_2$ is given by the projection on the three-plane $\Sigma$, multiplied by the $K3$ volume:
\begin{equation}
  \rho\left( C_2 \right) = \nu^{1/2} \sqrt{\sum_{i=1}^3 (\omega_i\cdot C_2)^2} = \nu^{1/2} \Vert C_2|_{\Sigma} \Vert \:.\nonumber
\end{equation}
}
\begin{align}
  \rho\!\left(\widetilde{F}\right)&=\jt\cdot\widetilde{F} = b\,, &
  \rho\!\left(\widetilde{B}\right)&=\jt\cdot\widetilde{B} = f -2b \,.  
\end{align}
Hence, the F-theory limit involves $b\to 0$, and in this limit, the base volume will be given by
$f$. On the other hand, the volume of the entire $\Kt$ is
\begin{align}
  \frac{1}{2}\, \jt\cdot\jt &= b\left(f-b\right) - \frac{1}{2} c^a c^a\,.
\end{align}
This volume is required to be positive, so we get a bound on the $c^a$,
\begin{align}\label{eq:caconstraint}
  \frac{1}{2}c^a c^a &< b\left(f-b\right)\,.
\end{align}
Thus, in the F-theory limit we have to take the $c^a$ to zero at least as fast as $\sqrt{b}$.
Once the limit is taken, the volume of $\Kt$ vanishes and the K\"ahler form is given by
\begin{align}\label{jFth}
 \jt = f \widetilde{F}\,,
\end{align}
regardless of the initial value of the $c^a$.
 Note that the constraint~(\ref{eq:caconstraint}) is consistent with the
intuitive picture of the fibre torus shrinking simultaneously in both directions: The $c^a$ measure
the volume of cycles which have one leg in the fibre and one in the base, %(and that are orthogonal
%to the complex structure $\omegat$)
so they shrink like the square root of the fibre volume $b$.

The K\"ahler moduli space is reduced in the F-theory limit: We lose not only the direction along
which we take the limit, but also all transverse directions except for the base volume $f$, which becomes
the single K\"ahler modulus of the torus orbifold. In the duality to type IIB on $K3\times
T^2/\mathbb{Z}_2\times S^1$, the $c^a$ parametrize Wilson lines of the gauge fields along the $S^1$
as long as the fibre volume is finite. In the F-theory limit, which corresponds to the radius of the
$S^1$ going to infinity, the Wilson lines disappear from the moduli space. The propapagating degrees of freedom related to them 
combine with the three-dimensional vectors from the 3-form $C_3$ reduced along 2-cycles of
$\Kt$ to form four-dimensional vectors (cf.~Section~\ref{sec:vectormass}, see also~\cite{Valandro:2008zg}). 

From this perspective, we see that it is important not to fix the modulus controlling the size of
the fibre. In fact, if we leave it unfixed, we have a line in the M-theory moduli space
corresponding to this flat direction of the potential. Of this line, only the point at infinity
($b=0$) corresponds to F-theory. This limit is singular in the sense that the F-theory
point is not strictly speaking in the moduli space of $\Kt$, but on its boundary. 
As we show below, this point is at infinite distance from every other point in the moduli space of $\jt$, and it actually
corresponds to the decompactification limit in type IIB.

\subsubsection{F-theory limit in the moduli space of M-theory}

In this section we show that the point in the moduli space of M-theory that corresponds to the
F-theory limit is at infinite distance from any other point, as expected for a decompactification
limit.

Let us fix two directions of the three-plane $\widetilde{\Sigma}$ to form the holomorphic 2-form,
let us say $\omegat=\omegat_1+\I\omegat_2$, so $\jt=\sqrt{2\nu}\,\omegat_3$. We are left with 20
moduli: the 19 $\delta\omegat_2^m$ deformations of $\omegat_3$ and the volume $\nut$. 
These remaining 20 moduli can be parametrized with the 20 deformations of 
$\jt$ in $H^{1,1}(\widetilde{K3})$: 
\begin{align}\label{eq:jtagain}
  \jt = b \widetilde{B} + f \widetilde{F} + c^a \ut_a\,,  \qquad\qquad \mbox{with }
  \ut_a \mbox{ a basis }\bot \left<\widetilde{F},\widetilde{B},\omegat_1,\omegat_2\right> \,.
\end{align}
So we are essentially left with the K\"ahler moduli space.

The metric on this moduli space is ($i,j$ run over $\left\{b,f,c^a\right\}$)
\begin{align}\label{eq:modmetric}
 g_{ij} = -\partial_i\partial_j \log \left(\int \jt\wedge \jt\right) =-\partial_i\partial_j \log
 \left(2\,b(f-b)-c^a c^a\right)\,. %\qquad \mbox{where }\hat{t}^2=\sum_{a=1}^{18}
% (\hat{t}^a)^2 
\end{align}

We want to use this metric to compute the distance between one general point of the moduli space and
a point corresponding to the F-theory limit. As discussed in Section~\ref{sec:ftheory}, $b$ and $f$
give the volumes of fibre and base, and the F-theory limit involves $b\to 0$ while respecting the
bound~(\ref{eq:caconstraint}). We will consider a curve parameterized by $\epsilon$, 
\begin{align}\label{eq:modpath}
 b &= b_0\epsilon^2\,, &f &= \text{const.}\,, & c^a&=c^a_0 \epsilon\,,
\end{align}
where $c^a_0 c^a_0 = 2 \alpha b\left(f-b\right)$ and $\alpha\in\left[0,1\right)$ parameterizes the
degree to which the bound is saturated. Note that the parameterization~(\ref{eq:jtagain}) is simple,
but not exceedingly convenient. In particular, one might worry that the volume of $\Kt$ vanishes in
the limit of $\alpha\to 1$, even though base and fibre volume stay finite. However, before that
limit is reached, one can reparameterize the basis cycles such that the new $c^a$ are again
zero, while $f$ is now smaller than before. The limit $\alpha\to1$ is then the same as $\epsilon\to
0$.

The metric distance of the F-theory point from any other point ($\epsilon_0$) is given by $\int_{\epsilon_0}^0ds$, where
\begin{equation}
 ds = \sqrt{g_{ij} \dot{X}^i\dot{X}^j}\,d\epsilon \:.
\end{equation}
$X^i$ are $b,f,c^a$ and $\dot{X}^i$ are the derivatives of $X^i$ with respect to $\epsilon$.
% The metric~(\ref{eq:modmetric}) is not block-diagonal, so any explicit calculation becomes
% cumbersome. However, it is still straightforward to show that the F-theory point is infinitely far
% away: From the parameterization~(\ref{eq:modpath}) we see that the volume is essentially
% proportional to $\epsilon^2$, 
% \begin{align}
%   \frac{1}{2} \jt\cdot\jt &= \epsilon^2 \left( b_0 \left(f-\epsilon^2 b_0\right)-\frac{1}{2} c^a_0
%     c^a_0\right)\,.
% \end{align}
By explicit calculation, one can show that all terms in the sum under the square root are of order $\epsilon^{-2}$ in the limit
$\epsilon\to 0$, times some finite coefficient. Hence, the metric distance from any finite point
$\epsilon_0$ to $\epsilon=0$ is 
\begin{align}
 \int^0_{\epsilon_0} \d s = \int^0_{\epsilon_0} \frac{\d\epsilon}{\epsilon} \cdot\left(\text{term
     finite for $\epsilon\to 0$}\right)\,,
\end{align}
i.e.~it diverges logarithmically.

\subsubsection{Fluxes in the F-theory limit}

To see which fluxes are compatible with the F-theory limit, we first note that there must be no flux along
either $\widetilde{B}$ or $\widetilde{F}$ because Lorentz invariance of the four-dimensional theory
requires that the flux must have exactly one leg along the fibre. This means that in a basis of
$H^2\!\left(\Kt\right)$ consisting of $\widetilde{B}$, $\widetilde{F}$ and orthogonal forms, the
flux matrix must be of the form
\newfont{\krass}{cmr17 scaled 3500}
\begin{align}
  G^{I\Sigma} &=
  \begin{pmatrix}
    0 &0 &  \\
    \vdots & \vdots & \smash{\raisebox{-5.5ex}{\text{\krass *}}} \\
    0 &0 &
  \end{pmatrix}\,.
\end{align}
This leads to a $G^a G$  with two rows and columns of zeroes, 
\begin{align}\label{eq:GaGF}
  G^aG &= \begin{pmatrix} 0 &0 & \dots &0 \\0 &0 & \dots &0 \\ \vdots & \vdots
    &\mspace{-20mu}\mathrlap{\smash{\raisebox{-6.5ex}{\krass *}}}&\\ 0& 0&  &\\
    \end{pmatrix}\,,
\end{align}
hence the direction along which we take the F-theory limit is automatically flat.

To discuss the matrix form of $G$, it is convenient to choose an equivalent basis for $H^2(K3)$, i.e. a basis containing
$B$ and $F$ and 20 orthogonal vectors.
%For simplicity, we will in the following choose a basis in $H^2\!\left(K3\right)$ in which the first
%two vectors are essentially equivalent to a $U$ block of~(\ref{eq:UUUE8E8}), i.e.~they are
%orthogonal to the rest and the  metric on this subspace is of signature $(1,1)$.
We then restrict to fluxes of the type 
\begin{align}
  G^{I\Sigma}=\begin{pmatrix} \label{eq:Fthflux}
    \begin{matrix}\textstyle 0&0\\0&0\end{matrix} & \text{\Large 0}\\
        \text{\Large 0}\rule[14pt]{.5pt}{0pt}  & G^{I\Lambda}_\text{F-th}
    \end{pmatrix}\,,
\end{align}
although this is not the most general form. Here, $G^{I\Sigma}_\text{F-th}$ is a $20\times20$
matrix which we will also call $G^{I\Sigma}$ for simplicity. %Note that
%these fluxes are consistent with $\Kt$ to be elliptically fibred.

\section{Brane Localization\label{sec:movebrain}}

One of our aims is to find a flux that fixes a given configuration of
branes. The results obtained so far allow us to do that: As we have discussed in Section~\ref{cyclesk3branes}, the
positions of the D7 branes are encoded in the complex structure $\omegat=\omegat_1+\I \omegat_2$ of
$\widetilde{K3}$. This can be understood as follows: We can find certain cycles which
measure the distance between branes. A given brane configuration can thus be characterized by the
volumes of such cycles. Most relevant for the low-energy theory is the question whether there are
brane stacks (corresponding to gauge enhancement) which is signalled by the vanishing of
interbrane cycles. So choosing a given brane configuration determines a set
of integral cycles which are to shrink, i.e.~which should be orthogonal to the complex structure%
\footnote{These are cycles with one leg in the base and one in the fibre and which are
orthogonal to $\omegat_3$ once we take the F-theory limit \eqref{jFth}.}. We want to find what is the flux that
fixes such a complex structure.

The flux needs to satisfy some constraints: It must be integral and it must satisfy the
tadpole cancellation condition~(\ref{TadpoleCanc}). The first condition means that the entries of
the flux matrix $G^{I\Sigma}$ in a basis of integral cycles must be integers.
The tadpole cancellation condition translates into a condition on the trace of $G^a G$,
\begin{align}\label{TadpoleCancCond}
  \tr G^a G &= \tr G^T\!\! M G \Mt = 48\,.
\end{align}
Of course, we also require that the flux gives Minkowski minima, i.e.~$G^aG$ needs to have only
non-negative eigenvalues. These conditions turn out to be rather restrictive, and a scan of all
$20\times20$-matrices is computationally beyond our reach. Fortunately, the block-diagonal structure
alluded to above allows us to restrict to smaller submatrices of size $2\times 2$ or $3\times 3$,
where an exhaustive scan is feasible.

\subsection{D-Brane Positions and Complex Structure \label{sec:BraneCS}}

In the weak coupling limit, in which the F-theory background can be described by
perturbative type IIB theory, the complex structure deformations of the upper $K3$ have an 
interpretation in terms of the movement of D-branes and O-planes on $\mathbb{C}\Pp^1$ \cite{Sen:1997gv}.
From the perspective of the elliptically fibred $\widetilde{K3}$, D-branes and O-planes are 
points on the $\mathbb{C}\Pp^1$ base where the $T^2$ fibre degenerates. The positions of these
points are encoded in the complex structure of $\widetilde{K3}$: The 18 complex structure
deformations\footnote{These are the deformations of $\omegat_1$ and $\omegat_2$ in the space orthogonal
to $\widetilde{F}$ and $\widetilde{B}$.} 
specify the 16 D-brane positions, the complex structure of $T^2/\mathbb{Z}_2 \sim \mathbb{C}\Pp^1$, and the value of the 
axiodilaton. The map between the two descriptions is worked out in detail in Section~\ref{dbp}. 
When several D-branes coincide, the $\widetilde{K3}$ surface develops singularities which reflect
the corresponding gauge enhancement \cite{Witten:1995ex,Sen:1996vd,Aspinwall:1996mn,Gaberdiel:1997ud,Lerche:1999de,a00ag04}.
These singularities can also be seen to arise when the volume of certain cycles shrinks to zero:
\begin{align}
\int_{\gamma_i}\omegat=\int_{\Kt}\gamma_i\wedge\omegat=\gamma_i\cdot\omegat \longrightarrow 0\,.
\end{align}
Note that these cycles have one leg on the base and and one leg on the fibre torus, so
$\gamma_i\cdot \jt=0$ \footnote{Since we are interested in the F-theory limit, we will only consider
  vacua corresponding to $\jt$ being in the block 
  $\left< \widetilde{F},\widetilde{B}\right>$.}. Hence their volume is given by $\sqrt{\nut} \left|
  \gamma_i\cdot\omegat\right|$. 
If the $\gamma_i$ are integral cycles (for the structure of integral cycles on $K3$ see
Section~\ref{k3mod}) with self-intersection $-2$, their shrinking produces a singularity that
corresponds to a gauge enhancement.
Since these cycles can be thought of as measuring distances between branes, this is equivalent to
D-branes that are coinciding. The Cartan matrix that displays the gauge enhancement is
given by the intersection matrix of the shrinking $\gamma_i$.

Let us consider an $SO(8)^4$ point. From the
D-brane perspective this corresponds to putting four D-branes on each of the four O-planes. In terms of the basis given 
in Section~\ref{k3mod}, the complex structure of $\widetilde{K3}$ is given by
\begin{align}
\omegat_{SO(8)^4}=\frac{1}{2}\left(\alpha+Ue_2+S\beta-USe_1\right)\,.
\end{align}
For the sake of brevity we have introduced\footnote{Note that although $e_i$, $e^i$, $E_I$, $\alpha$
and $\beta$ are forms on $\Kt$, we omit the tildes to avoid unnecessary notational clutter.}
\begin{align}\label{eq:alphabeta}
  \alpha&\equiv2\left(e^1+e_1+W^1_IE_I\right)\,,&&& \beta &\equiv2(e^2+e_2+W^2_IE_I)\,,
\end{align}
where
\begin{align}
  W^1&=\left(0^4,\frac{1}{2}^4,0^4,\frac{1}{2}^4\right)\qquad \text{and} \qquad
  W^2=\left(1,0^7,1,0^7\right)
\end{align}
describe the mixing of cycles from the $U$ and $E_8$ blocks. Note that they can be interpreted as
Wilson lines, breaking $E_8\times E_8$ to $SO(8)^4$ in the duality to heterotic string theory. 
The parameter $U$ describes the positions of the four O-planes, which is equivalent to the
complex structure of the $T^2$ in type IIB before orientifolding. The dilaton, which is constant
in this configuration, is given by the complex structure of the fibre torus, $S$.

We can now move away from the $SO(8)^4$ configuration by rotating $\omegat$. A convenient parameterization is
given by 
\begin{align}
\omegat=\frac{1}{2}\left(\alpha+Ue_2+S\beta-\left(US-z^2\right)
e_1+2\widehat{E}_{I}z_I\right) \ ,\label{Omegagensec6}
\end{align}
with shifted $E_8\times E_8$ block vectors  $\widehat{E}_I= E_I +W_I^1 e_1 +W_I^2 e_2$. 
Explicitly, they are
\begin{align}\label{eq:Ehat}
  \begin{aligned}
    \widehat{E}_1&=E_1+e_2 ,\hspace{1cm} \widehat{E}_{I}=E_I,&& I=2..4,
    10..12 \,,\\
    \widehat{E}_9&=E_9+e_2 ,\hspace{1cm} \widehat{E}_{J}=E_J+e_1/2,&&
    J=5..8, 13..16 \, . 
  \end{aligned}
\end{align}
The $\widehat{E}_I$ are orthogonal to $\alpha$ and $\beta$ and still satisfy $\widehat{E}_I\cdot
\widehat{E}_J=-\delta_{IJ}$. As we have discussed in Section~\ref{dbp}, the $z_I$ are the positions of the branes 
relative to their respective O-planes in the double cover of  $\mathbb{C}\Pp^1\simeq T^2/Z_2$. 

Now we can deduce the brane positions and the gauge enhancement from
a given expansion of the holomorphic 2-form $\omegat$ (which is equivalent to knowing
the complex structure of $\widetilde{K3}$).
We can either match any expansion of $\omegat$ in the basis given in Section~\ref{k3mod} to
(\ref{Omegagensec6}), or we can compute the intersection numbers between $\omegat$ and the cycles given
in Table~$\ref{AtoD}$ to find the periods of the cycles of $\widetilde{K3}$. In this way we obtain the
value of the dilaton and the D-brane and O-plane positions. Note that contrary to the
basis given by $\alpha, \beta, e_1, e_2$ and the cycles in Table~(\ref{AtoD}), the basis we used in
the expansion~(\ref{Omegagensec6}) is not an integral basis (as the $\widehat{E}_I$ are half-integral).

\subsection[Fixing D7-brane Configurations by Fluxes]{Fixing D7-brane Configurations by Fluxes}

We are now ready to outline a systematic procedure for choosing a flux which fixes a given D7-brane gauge group.
In particular, we will be interested in non-Abelian gauge enhancement. The Cartan matrix of the underlying Lie-Algebra
is given by the intersection matrix of the lattice of shrinking 2-cycles. Thus, we need to understand which fluxes
make a particular subspace of 2-cycles shrink.
We will take these cycles as part of the basis orthogonal to $\left<\widetilde{B},\widetilde{F}\right>$ discussed at the end of Section \ref{sec:modstab}. Then we consider the orthogonal lattice, i.e. the lattice made up of (integral) cycles orthogonal to the shrinking ones (and to $\left<\widetilde{B},\widetilde{F}\right>$). Choosing an integral basis for this lattice completes the basis of cycles of $H_2(\Kt)$ orthogonal to $\left<\widetilde{B},\widetilde{F}\right>$. Note that in this basis the metric on $H_2(\Kt)$ is block-diagonal, with a negative definite block for the subspace of shrinking cycles. We also choose a basis of integral cycles of $H_2(K3)$ such that the metric has two blocks with the same dimensions as on the $\Kt$ side.

In this basis it is easy to write down a flux that fixes $\omegat$ orthogonal to the shrinking cycles: It can be taken to have the block-diagonal form\footnote{Actually, it is enough that $G^aG$ is of this form.}
\begin{equation}\label{Gprocedure}
\left(\begin{array}{cc}
 G_\perp &  \\ & G_{\rm shk}  \\ 
\end{array}\right)\:.
\end{equation}
Thus, when diagonalizing $G^aG$, the positive norm eigenvectors are in the first block and hence orthogonal to the shrinking cycles.

One has finally to check whether there are more shrinking cycles than those we imposed.

\subsection[Fixing an $SO(8)^4$ Point]{Fixing an \boldmath$SO(8)^4$ Point\label{sec:so8stab}} 
In this section, we will follow the procedure described in the previous section to construct
a flux that fixes the F-theory moduli corresponding to four
D7 branes on top of each O7 plane. This $SO(8)^4$ configuration is realized when there are
sixteen shrinking cycles whose intersection matrix is $D_4^4$. These shrinking cycles are given by the four blocks
$A,B,C,D$ as defined in~(\ref{AtoD}). The basis of the orthogonal lattice is given by
$\alpha,e_1,\beta,e_2$ (see Eq.~(\ref{eq:alphabeta})). Since the only
nonvanishing intersections in this set are $\alpha\cdot e_1=\beta\cdot e_2=2$, the
intersection matrix is
\begin{align}
  \Mt = \begin{pmatrix}
    \begin{matrix}
      0&2\\2&0\\
    \end{matrix} &&\\
    &\begin{matrix}
      0&2\\2&0\\
    \end{matrix}&\\
    && D_4^4
  \end{pmatrix}\,.
\end{align}
For $K3$ we choose the same basis. Note that we are ignoring the $U$
block spanned by base and fibre.

Then we take the $20\times 20$ flux matrix %(again ignoring the F-theory zero $2\times 2$ block)
with respect to these bases to be
\begin{align}\label{eq:so8flux}
G^{I\Lambda}=\left(\begin{array}{ccc}
G_1 &&\\ &G_2 &\\ && \mathbf{0}_{16}\\
\end{array}\right)\,,
\end{align}
where $G_1$ and $G_2$ are $2\times 2$ blocks (which form the $G_\perp$ of \eqref{Gprocedure})
and the zero block is $16\times 16$ ($G_{\rm shk}$ of \eqref{Gprocedure}). If $G_1$ and $G_2$
satisfy the condition to have minima, then one $\omegat_j$ is fixed along the space
$\left<\alpha,e_1\right>$,  while the other is fixed in the space $\left<\beta,e_2\right>$. This
immediately gives a complex structure $\omegat$ that is orthogonal to the $D_4^4$ blocks
$A,B,C,D$ and hence realizes an $SO(8)^4$ point.

An explicit example of an integral flux that satisfies the tadpole cancellation condition
(tr$\,G^aG=48$) and fixes an $SO(8)^4$ point is given by: 
\begin{align}\label{G1G2so8}
  G_1 = \left(\begin{array}{cc}
      1&1\\1&1\\
    \end{array}\right)\,, \qquad\qquad\qquad
  G_2 = \left(\begin{array}{cc}
      1&1\\1&3\\
    \end{array}\right)\,.
\end{align}
The corresponding blocks for $G^aG$ are
\begin{align}
  (G^aG)_1 = \left(\begin{array}{cc}
      8&8\\8&8\\
    \end{array}\right)\,, \qquad\qquad\qquad
  (G^aG)_2 = \left(\begin{array}{cc}
      16&24\\8&16\\
    \end{array}\right)\,,
\end{align}
and the corresponding eigenvalues are
\begin{align}\label{eigenvalSO(8)4}
  \lambda_{\omegat_1} = 16\,, \qquad \lambda_{\tilde{u}_1} = 0\,,  \qquad\
  \lambda_{\omegat_2} = 8(2+\sqrt{3})\,,  \qquad \lambda_{\tilde{u}_4} =
  8(2-\sqrt{3})\,.
\end{align}
We see that their sum is precisely $48$, as required by tadpole cancellation, and that they are all
non-negative, as required by the minimum condition. Moreover, the ones corresponding to positive norm
eigenvectors are different from those relative to negative norm eigenvectors, as required by the
stabilization condition. 

The positive norm eigenvectors of the two matrices give $\omegat_1$,
$\omegat_2$:
\begin{align}
 \omegat_1 &= \frac{\alpha}{2} +\frac{e_1}{2}\,,
 &\omegat_2 &=  3^{1/4}\,\frac{\beta}{2} +\frac{1}{3^{1/4}}\,\frac{e_2}{2}\,.
\end{align}
From the comparison of $\omegat=\omegat_1 + \I \omegat_2$ with the general form~(\ref{Omegagensec6}), we
see that indeed the complex structure is fixed at a (non-integral) point where $z_I=0$, and that the complex
structures of base and fibre are given by
\begin{align}
  U&= \frac{1}{\sqrt[\leftroot{2}4]{3}}\, \I \,, & S&=\sqrt[\leftroot{2}4]{3} \,\I \,.
\end{align}
Since $S$ is the type IIB axiodilaton, we have stabilized the string coupling at a moderately small
value of $3^{-1/4}\cong 0.76.$ However, we can probably realize smaller coupling by considering
generic $4\times 4$ matrices rather than the $2\times 2$ block structure of Eq.~(\ref{eq:so8flux}).  

This flux fixes also the deformations of $\omega_1$ and $\omega_2$. On the other hand,
$\omega_3$ and $\omegat_3$ are eigenvectors of $G\,G^a$ and $G^aG$
relative to zero eigenvalues. Then their deformation along all negative
eigenvectors relative to zero eigenvalues are left unfixed.
In type IIB, this corresponds to leaving unfixed K\"ahler moduli of
$K3\times
T^2/\mathbb{Z}_2$, while fixing the complex structure and the D7-brane
positions. The unfixed deformations of $\omegat_3$ correspond to gauge
fields in type IIB that remain massless \cite{Valandro:2008zg}.
In the studied case, two of the $19\times 2$ deformations of $\omega_3$ and $\omegat_3$, the ones
along $\tilde{u}_4$, are fixed (as $\lambda_{\tilde{u}_4} = 8(2-\sqrt{3})$ is different from zero). Fixing a
deformation of $\omegat_3$ corresponds to giving a mass to the corresponding
gauge field in type IIB dual. In fact, this flux corresponds to the type IIB flux
that makes one four-dimensional vector massive \cite{aaf03,aaft03,hkl06,Jockers:2004yj}. One can see this also from the M-theory point of view: One three-dimensional vector gets a mass from fluxes. This vector combines with the deformation of $\omegat_3$ to give a four-dimensional massive vector.

Finally we note that the lower $K3$ is generically non-singular, as $\omega_3$
will generically not be orthogonal to the $E_8$ block cycles.

\

As a second example we will reproduce one of the solutions given in \cite{ak05}
by using our methods. As it is discussed there, attractive $K3$ surfaces are classified in terms of
a matrix
\begin{equation}
 Q = \left(\begin{array}{cc}
      p\cdot p&p \cdot q\\p \cdot q&q \cdot q\\
    \end{array}\right)\, ,
\end{equation}
in which $p$ and $q$ are integral 2-forms. 
The holomorphic 2-form of $\Kt$ is then given by
\begin{equation}
\omegat=\pt+\tau \qt \,.\label{Ompq}
\end{equation}
Of the 13 pairs of attractive $K3$'s given in \cite{ak05}, we will
discuss the one defined by
\begin{equation}
 Q = \left(\begin{array}{cc}
      8 & 4 \\ 4 & 8 \\
     \end{array}\right),\qquad \qquad
\widetilde{Q}=\left(\begin{array}{cc}
      4 & 2 \\ 2 & 4 \\
    \end{array}\right)\, .\label{QtildeQ}
\end{equation}
This pair has the advantage that both $K3$'s have an orientifold interpretation
which means that we can expand $\pt$ and $\qt$ in terms of $e_1$ , $e_2$, $\alpha$ and $\beta$
(and similarly, for the lower $K3$, $p$ and $q$ in terms of $e_1'$ , $e_2'$, $\alpha'$ 
and $\beta'$). Clearly, there are many ways to do this which correspond to different embeddings
of the lattice spanned by $p$ and $q$ into the lattice spanned by $e_1$ , $e_2$, $\alpha$ 
and $\beta$. We make the following choice:
\begin{align}
  \begin{aligned}
    p=&e'_1+2\alpha'+2\beta'\,,   &\tilde{p}=&e_1+\alpha+\beta\,,\\
    q=&e'_2+2\beta'	 \,,  &\tilde{q}=&e_2+\beta\, .\label{pqexp}
  \end{aligned}
\end{align}
According to \cite{ak05}, stabilization at this point occurs through the flux
\begin{equation}
G=\frac{1}{2}\left(\gamma\omega\wedge\overline{\omegat}+\overline{\gamma}\,\overline{\omega}
\wedge\omegat\right)
\end{equation}
with $\gamma=1+\frac{i}{\sqrt{3}}$. In the basis given by $\alpha,e_1,\beta,e_2$ and $\alpha',e'_1,\beta',e'_2$,
%$\tilde{\alpha},\tilde{e}_1,\tilde{\beta},\tilde{e}_2$, 
the flux matrix reads
\begin{align}
 G^{I\Lambda}= \begin{pmatrix}
    2 &  2 & 2 & 0   \\
    1 & 1 & 1 & 0  \\
    0 &  0 & 2 & 2 \\
    -1 & -1  & 0 & 1
  \end{pmatrix}\, .
\end{align}

The positive norm eigenvectors of $G^aG$ are given by $\omegat_1=(1,1,\frac{1}{2},-\frac{1}{2})$ and 
$\omegat_2=(0,0,1,1)$. Rescaling the second one so that they both have the same
norm, we arrive at $\tilde{\omega}=\omegat_1+i\frac{\sqrt{3}}{2} \omegat_2$. This is precisely the same result as 
what one obtains from inserting (\ref{pqexp}) into (\ref{Ompq}). 

The eigenvalues of $G^a G$ are $\lambda_{\omegat_1} = \lambda_{\omegat_2}=24$, $\lambda_{\tilde{u}_1}=\lambda_{\tilde{u}_2}=0$. In the last section we will see that this corresponds to an $\Nn=1$ (4d) vacuum. Moreover, in this case all the K\"ahler moduli of both $K3$'s are left unfixed by fluxes, as all the eigenvalues $b_a$ are equal to zero.

\subsection{Moving  Branes by Fluxes\label{sec:so6stab}}

Now we want to see how to change the flux~\eqref{eq:so8flux}, with $G_1$ and $G_2$ given by
\eqref{G1G2so8}, to fix a different D7-brane configuration in which some D7 branes have been moved
away from the orientifold planes. In particular, we will find fluxes that fix configurations
where we move one or two branes off one of the stacks, breaking one $SO(8)$ to
$SO(6)$ or $SO(4)\times SU(2)$. In the following we will consider only the $C$-block.
The cycles belonging to blocks $A,B,D$ will remain shrunk.

\subsubsection*{\boldmath$SO(8)^3\times SO(6)$}
\label{1stExampleBrMv}
Moving one D7~brane from one stack in type IIB corresponds to blowing up one of the 4 cycles of this block. % this is achieved by
%blowing up one of the 16 cycles %$\widetilde{A}_h,\widetilde{B}_h,\widetilde{C}_h,\widetilde{D}_h$ 
%$A_h,B_h,C_h,D_h$.%, as explained at page \pageref{1stExampleBrMv}.
For the first example, % \label{1stExampleBrMv} 
consider the complex structure determined by~(\ref{Omegagensec6}) with $z_1=d$ and al other $z_I=0$.
One can check that all cycles given in Table~\ref{AtoD} except $C_1$ remain orthogonal to $\omegat$.
Looking at Figure~\ref{1block}, it is clear that this means we have moved one D-brane away from the
O-plane, as claimed. Thus $SO(8)$ is broken to $SO(6)$.
At the same time the cycles that remain shrunk in block $C$ have an intersection matrix that
is equivalent to minus the Cartan matrix of $SO(6)$. This means that we have effectively
crossed out the first line and the first column of the Cartan matrix of $SO(8)$ by removing
$C_1$ from the set of shrunk cycles:
\begin{align}
  \begin{pmatrix}
    -2 & 1 & 0 & 0   \\
    1 & -2 & 1 & 1  \\
    0 &  1 & -2 & 0 \\
    0 & 1  & 0 & -2
  \end{pmatrix}
  \longrightarrow 
  \begin{pmatrix}
    -2 & 1 &1  \\
    1 & -2 & 0 \\
    1  & 0 & -2 
  \end{pmatrix}\,.
\end{align}
We want an integral basis in which shrunk and blown-up cycles do not intersect each 
other. %However we still want the basis cycles to be orthogonal to $\omegat$ at an $SO(8)$ point. 
To achieve this we keep the shrunk cycles $C_2$, $C_3$, $C_4$ and instead of $C_1$
we take the integral cycle $2\widehat{E}_1=2\left(e_2+E_1\right)$ (see (\ref{eq:Ehat})) to describe the
brane motion in block $C$.
We find the intersection matrix    
\begin{align}\label{1stExampleIntMatrix}
  \begin{pmatrix} 
    -4 & 0 & 0 & 0  \\
    0 & -2 & 1 & 1  \\
    0 &  1 & -2 & 0 \\
    0 & 1  & 0 & -2
  \end{pmatrix}\,. 
\end{align} 
We choose an analogous basis for the lower $K3$.

\

The basis $\alpha,e_1,\beta,e_2,2\widehat{E}_1,C_2,C_3,C_4,A,B,D$, is the one that gives the flux matrix the block-diagonal form \eqref{Gprocedure},
with the shrinking cycles given by $C_2,C_3,C_4,A,B,D$ and the orthogonal ones by $\alpha,e_1,\beta,e_2,2\widehat{E}_1$.
%integral, and so any
%integer $G^{I\Lambda}$ gives an integral flux. Clearly, the flux that stabilizes this situation
%cannot be block-diagonal as in the $SO(8)^4$ case, (\ref{eq:so8flux}), but has to 
Such a block-diagonal flux matrix generally gives $\omegat$ a component along
$\widehat{E}_1$. An example is given by: % suitable flux that fixes such an $\omegat$ is
\begin{align}
  G^{I\Lambda} = \begin{pmatrix}
    1 & 1 & & & & \\
    1 & 1 & & & & \\
    & & 1 & 1 & 0 & \\
    & & 1 & 3 & 1 & \\
    & & 0 & 1 & 0 &\\
    & & & & & \mathbf{0}_{15}\\
\end{pmatrix}\,.
\end{align}
where the $3\times 3$ block is with respect to the cycles $\beta,e_2,2\widehat{E}_1$ for
both $K3$'s. From the type IIB perspective, we are also turning on fluxes on the D7~branes.

$G^aG$ satisfies the tadpole cancellation condition. The eigenvalues corresponding to the first
block are the same as in Eq.~(\ref{eigenvalSO(8)4}), the ones in the second block are   
\begin{align}
  \lambda_{\omegat_2} = 24.6\,, \qquad\ \lambda_{\tilde{u}_2} = 5.5\,,  \qquad \lambda_{\tilde{u}_4}
  = 1.9 \,. 
\end{align}
They are all positive and different from each other. The positive norm eigenvectors give
$\omegat_1$ and $\omegat_2$: 
\begin{align}
 \omegat_1 = \frac{\alpha}{2} + \frac{e_1}{2} \qquad,\qquad \omegat_2 = 0.9 \,\frac{\beta}{2} + 1.3
 \frac{e_2}{2} + 0.3\, \widehat{E}_1\:.
\end{align}
The corresponding $\omegat$ is orthogonal the $S^2$ cycles with intersection matrix
$SO(6)\times SO(8)^3$, but it is not orthogonal to the cycle $2\widehat{E}_1$ which is now blown up, at a
volume $\rho\!\left(2\widehat{E}_1\right)=0.6\,\sqrt{\nut}$. This
corresponds to the motion of one D7~brane away from the orientifold plane of block $C$. 
Note that again the coupling is moderately weak, $g=1/1.61 = 0.6$. 

We also note that, with respect to our $SO(8)^4$-example, we have fixed one more deformation
of $\omega_3$ and one of $\omegat_3$.
%The first one corresponds to the fact that D7-brane fluxes can fix K\"ahler moduli.
The stabilization of the extra
$\omegat_3$ deformation is the signal of a mass for the gauge field on the D7~brane that has been
moved. This mass is explained in type IIB by the fact that D7 fluxes gauge some shift symmetries by
vectors on the branes. Since the $U(1)$ on the brane is broken, the resulting gauge group is
$SO(8)^3\times SO(6)$ \cite{aaf03,aaft03,hkl06,Jockers:2004yj}.

\subsubsection*{\boldmath$SO(8)^3\times SO(6)\times U(1)$}
\label{1stExampleBrMvbis}
In the example studied above, we have given a flux that fixes the desired brane configuration.
Moreover it fixes one further deformation of $\omega_3$ and one of $\omegat_3$, with respect to 
the $SO(8)^4$ example presented before. This is related to the fact
that the rank of the $3\times 3$ block has been increased to $3$; so we get two negative norm eigenvectors with
non-zero eigenvalues. But we can choose a different flux, such that the number of negative
norm eigenvectors relative to non-zero eigenvalues does not change with respect to the $SO(8)^4$
case:
\begin{align}
  G^{I\Lambda} = \begin{pmatrix}
    1 & 1 & & & & \\
    1 & 1 & & & & \\
    & & 1 & 1 & 0 & \\
    & & 1 & 3 & 1 & \\
    & & 0 & 0 & 0 &\\
    & & & & & \mathbf{0}_{15}\\
\end{pmatrix}\:,
\end{align}
where the $3\times 3$ block is still with respect to the cycles $\beta,e_2,2\widehat{E}_1$.
% for $\widetilde{K3}$ and for three integral cycles of $K3$ that have the same intersection matrix.

Again, $G^aG$ satisfies the tadpole cancellation condition. The eigenvalues relative to the first block are the same as in
Eq.~(\ref{eigenvalSO(8)4}). The eigenvalues of the second block are
\begin{align}
  \lambda_{\omegat_2} = 27.3\:, \qquad\ \lambda_{\tilde{u}_2} = 4.7 \:, \qquad \lambda_{\tilde{u}_4} = 0\:.
\end{align}
They are all non-negative and different from each other. The positive norm eigenvectors give
$\omegat_1$ and $\omegat_2$: 
\begin{align}
 \omegat_1 = \frac{\alpha}{2} + \frac{e_1}{2} \qquad,\qquad \omegat_2 = 0.8 \,\frac{\beta}{2} + 1.4
 \frac{e_2}{2} + 0.3\, \widehat{E}_1 \:.
\end{align}
As before, the corresponding $\omegat$ is orthogonal the $S^2$ cycles with intersection matrix
$SO(6)\times SO(8)^3$, but it is not orthogonal to the cycle $2\widehat{E}_1$ which is now blown up, at a
volume $\rho\!\left(2\widehat{E}_1\right)=0.6\,\sqrt{\nut}$. Again, one D7~brane is moved from the orientifold plane of block $C$. 
%Note that again the coupling is moderately weak, $g=1/1.61 = 0.6$. 

In this case, we do not break any further $U(1)$. In fact, the flux we
turned on contributes to the gauging of an isometry that has been
gauged also in the $SO(8)^4$ case. This can be easily understood in the M-theory context, where the relevant gauge field is one of the $\Ct^\Lambda_{1\mu}$.

% We finally note that in this case the negative norm eigenvector relative to the zero eigenvalue is
% parallel to an integral vector.

\subsubsection*{\boldmath$SO(8)^3\times SO(4)\times SU(2)$}
As a further example\label{2ndExampleBrMv}, let us choose $z_1=z_2=d$ and all other $z_I=0$. We
now find that $\omegat\cdot C_2=d$. For all other cycles in
Table~(\ref{AtoD}) the 
intersection with $\omegat$ still vanishes, so we have blown up a different cycle than in the
previous examples. From the assignment between cycles and forms it is clear that we have moved two
branes away from the O-plane. As $C_1$ remains shrunk,
these branes are on top of each other. From the type IIB perspective, we thus expect the
gauge symmetry $SO(4)\times SU(2)$. Examining the intersection matrix of the shrunk cycles $C_1$,
$C_3$ and $C_4$ we indeed find a diagonal matrix with entries $-2$. This happens because we have
blown up the cycle $C_2$ and thus deleted the second row and second column from the Cartan matrix
of
$SO(8)$: 
\begin{align}
  \begin{pmatrix}
    -2 & 1 & 0 & 0   \\
    1 & -2 & 1 & 1  \\
    0 &  1 & -2 & 0 \\
    0 & 1  & 0 & -2 
  \end{pmatrix}
  \longrightarrow 
  \begin{pmatrix}
    -2 & 0 & 0\\
    0 & -2 & 0 \\
    0 & 0  & -2 
  \end{pmatrix}\,.
\end{align}
The result is minus the Cartan matrix of $SO(4)\times SU(2)$, as expected.
As before, we need a basis of integral cycles in which shrunk and blown-up
cycles do not intersect. % and which is still orthogonal to $\omegat_{SO(8)^4}$. 
To construct it, we replace the cycle $C_2$ with the cycle $\widehat{E}_1+\widehat{E}_2= e_2+E_1+E_2$. It has
self-intersection $-2$, so that the intersection matrix in the new basis of cycles which we use for D-brane motion 
in the $C$ block is 
\begin{align}\label{2ndExampleIntMatrix}
  \begin{pmatrix}
    -2 & 0 & 0 & 0   \\
    0 & -2 & 0 & 0  \\
    0 &  0 & -2 & 0 \\
    0 & 0  & 0 & -2 
  \end{pmatrix}\,.
\end{align}
In this basis, a flux that stabilizes the desired gauge group is given by:
\begin{align} 
  G^{I\Lambda} &= 
  \begin{pmatrix}
    1&1 &&&& \\
    1&1 &&&&\\ 
    && 1& 1& 1 &\\
    &&1&3&1\\
    &&1&1&2&\\
    &&&&& \mathbf{0}\\
  \end{pmatrix}\,,
\end{align}
where now the $3\times 3$ block is with respect to the cycles $\beta,e_2,\widehat{E}_1+\widehat{E}_2$. The
eigenvalues corresponding to this block are: 
\begin{align}
  \lambda_{\omegat_2} = 19.6\,, \qquad\ \lambda_{\tilde{u}_2} = 11.2\,,  \qquad \lambda_{\tilde{u}_4}
  = 1.2\,.
\end{align}
They are all positive and different from each other. $\omegat_1$ and $\omegat_2$ are given by:
\begin{align}
  \omegat_1 &= \frac{\alpha}{2} + \frac{e_1}{2} \,,& \omegat_2 & =  1.5\,\frac{\beta}{2} +
  0.8\,\frac{e_2}{2} - 0.3\, \left(\widehat{E}_1+\widehat{E}_2\right)\:.
\end{align}
The corresponding $\omegat$ is orthogonal the $S^2$ cycles with intersection matrix
$SO(4)\times SU(2) \times SO(8)^3$, but it is not orthogonal to the cycle
$\widehat{E}_1+\widehat{E}_2$ which is now blown up. 

Also in this example, we have fixed one further deformation of $\omega_3$ and one of $\omegat_3$.
This in particular breaks the gauge group on the two D7~branes from $U(2)$ to $SU(2)$.

\subsection{Fixing almost all Moduli}
In the previous examples we have considered fluxes that stabilize the D7-brane positions and part of the
metric moduli of $K3$, while leaving some geometric moduli unfixed. This was due to the large amount
of zero eigenvalues of $G^aG$. In what follows, we will present an example of an integral flux that
satisfies the tadpole cancellation condition and fixes almost all geometric moduli. The
remaining unstabilized moduli are the size of the fiber in $\Kt$, as prescribed by the F-theory
limit, three deformations of $\Sigma$, and the two volumes of $K3$ and $\Kt$.  

To write down the flux we will choose two different bases of integral cycles in $H^2(K3)$ and in $H^2(\Kt)$. The
second one is the same as in the example $SO(8)^4$, while for $H^2(K3)$ we choose an integral basis
with intersection matrix %$\clubsuit$write explicitly the BASIS?$\clubsuit$
\begin{equation}
\left(\begin{array}{cccc}
\begin{array}{cc} 0&1\\1&0\\ \end{array} &&& \\
&\begin{array}{cc} 0&1\\1&0\\ \end{array} && \\
&&\begin{array}{cc} 0&1\\1&0\\ \end{array} & \\
&&& D_4^4\\
\end{array}\right)\,.
\end{equation}
In these bases, we choose the flux matrix to be
\begin{equation}
G^{I\Lambda}=\left(\begin{array}{ccc}
%\begin{array}{cc} 0&0\\0&0\\ \end{array} &&& \\
\begin{array}{rr} 1&-1\\-1&1\\ \end{array} && \\
&\begin{array}{rr} 1&-1\\-1&1\\ \end{array} & \\
&& G_{(4)}^4\\
\end{array}\right)\,.
\end{equation}
where 
\begin{equation}
G_{(4)}=\left(\begin{array}{rrrr}
-1 & -1 &0 &0 \\ 0 & 0 & 1 & 1 \\ 0 & 0 & 1 & 0 \\ 0 & -1 & 0 & 0 \\ 
\end{array}\right) \:.
\end{equation}

This flux satisfies tr$\,G^aG=8+8+4\times 8 =48$. Moreover, the $G_{(4)}$ blocks have eigenvalues equal
to 2, while the $2\times 2$ blocks have eigenvalues equal to $0$ for the positive norm eigenvectors
and $8$ for the negative norm eigenvectors. In the next section, we will see that the resulting minimum is supersymmetric ($\Nn=2$ in 4d). The eigenvalues relative to positive norm eigenvectors are such that all moduli are fixed apart from the
deformations of the $\omega_i$'s and the $\omegat_j$'s in the first U-block\footnote{This is a singular example, as now the lower $K3$ is singular.}.

\section{SUSY Vacua\label{sec:susyvac}}

Finally, we want to study the question of supersymmetric vacua. This question has been analyzed
for M-theory on a generic eight-dimensional manifold in~\cite{bb96,drs99}.
In the presence of fluxes a supersymmetric solution
is a warped product of $\mathbb{R}^{1,2}$ and some internal manifold which is conformally
Calabi--Yau \cite{bb96}. %In the following, we will consider the K\"ahler form and 
%complex structure of the underlying Calabi--Yau manifold.
The flux $G_4$ must be primitive ($J\wedge G_4=0$) and of Hodge type $(2,2)$ with respect to the K\"ahler form and the
complex structure of the underlying Calabi--Yau\footnote{
In the following, all the quantities of the internal manifold are relative to the unwarped Calabi--Yau metric.
}. Given a metric with $SU(4)$ holonomy, there is only one associated K\"ahler form $J$
and one  holomorphic 4-form $\Omega$. Moreover there are only two invariant Majorana--Weyl spinors, which implies $\Nn=2$ supersymmetry in the three-dimensional theory.

In our case, $K3\times K3$ has holonomy $SU(2)\times SU(2)$. As we have seen previously, for each $K3$ factor, 
the metric is invariant under the $SO(3)$ that rotates the $\omega_i$'s. This means that, given the metric
of $K3\times K3$, there is an $S^2\times S^2$ of
possible complex structures and associated K\"ahler forms.
Moreover, the holonomy $SU(2)\times SU(2)$ implies that the number of globally defined Majorana--Weyl spinors is four,
corresponding to $\Nn=4$ supersymmetry in three dimensions. The $R$-symmetry is the $SO(4)\simeq SO(3)\times SO(3)$
that rotates the four real spinors and the corresponding $S^2\times S^2$ of complex structures.
When this symmetry is broken to the $SO(2)$ which rotates the real and imaginary part of $\Omega$,
then we have $\Nn=2$ supersymmetry. On the other hand, if it is completely broken we have $\Nn=0$.

A minimum is supersymmetric if we can associate with the metric a K\"ahler form $J$ and a complex structure $\Omega$, such that $G_4$ is primitive and of Hodge-type (2,2). This means that there must be a choice of $\omega_i$ and $\omegat_j$, let us say $J=\sqrt{2\nu}\,\omega_3+\sqrt{2\nut}\,\omegat_3$ and  $\Omega=\omega\wedge\omegat$ (with $\omega = \omega_1+\I \omega_2$ and $\omegat = \omegat_1+ \I \omegat_2$), such that $G_4 \wedge J =0$ and $G_4\wedge \Omega =G_4\wedge \bar{\Omega} =0$. In our formalism, this is equivalent to:
\begin{itemize}
  \item Primitivity, $G_4 \wedge J = 0$ :
    \begin{align}\label{primitCond}
      G\,\omegat_3&=0\,, & G^a \omega_3&=0\,.
    \end{align}
    In terms of the eigenvalues of $G^aG$ this means $a_3=0$. We see that the primitivity condition translates to the existence of a non-trivial kernel of $G^aG|_{\Sigmat}$ and $G\,G^a|_{\Sigma}$. The vectors in the kernels make the K\"ahler form.
  \item $G_4=G_4^{(2,2)}$:
    \begin{equation}\label{22Cond}
	0 = (\omega \cdot G\omegat) = a_1 - a_2 \,\,\,.
    \end{equation}
    This means $a_1=a_2\equiv a$.
\end{itemize}

To summarize, the necessary and sufficient condition for the flux to preserve SUSY in the minimum is that 
$G$ (when restricted to the block $\Sigmat,\Sigma$) takes the form
\begin{align}
  G\big|_{\Sigmat}&=\begin{pmatrix} a&&\\&a&\\&&0\end{pmatrix} \,.
\end{align}
For $a=0$, the $SO(4)$ $R$-symmetry is unbroken and the minimum preserves all the $\mathcal{N}=4$ supersymmetries.
For $a\neq0$, only an $SO(2)$ subgroup of the $R$-symmetry is preserved
and we have $\mathcal{N}=2$ supersymmetries in three dimensions.

We note that in the case of fluxes which are compatible with the F-theory limit, the condition
$a_3=0$ is always satisfied and so one has simply to check that the other two eigenvalues are equal
to each other or possibly zero.

\newpage
\chapter*{Summary and Outlook}

%The present work adresses the question of moduli stabilization in compactifications of type IIB string theory with 
%D7-branes.
%We use the framework of F-theory, which allows to discuss the geometric moduli of the compactification manifold and the moduli 
%of D7-branes one the same footing. Although moduli stabilization can studied with ease in this description, the D7-brane
%moduli are encoded in a complicated way, making it difficult to directly address phenomenological questions. This motivates a 
%detailed study of how different configurations of D7-branes are expressed in F-theory compactifications. Building on simple
%and intuitive examples, our strategy is to increase complexity as we proceed.

This work addresses the moduli of D7-branes in type IIB orientifold compactifications from the perspective of
F-theory. In F-theory, the moduli of the geometry and the moduli of the D-branes are on a equal footing:
both are encoded in the moduli of a higher-dimensional elliptically fibred Calabi-Yau manifold.
This opens up an elegant way to study the flux stabilization of D-branes. To address phenomenological questions, 
one has to translate the complex structure deformations of the elliptic Calabi-Yau manifold of F-theory back to 
the positions and the shape of the branes. In this thesis, we have approached this problem in an intuitive and 
constructive way. 

As it is the simplest non-trivial example, we have first discussed F-theory compactifications on $K3$, where
the D7-branes are points moving on a two-sphere. After showing how 2-cycles of $K3$ can be constructed from the
knowledge of the fibration in the vicinity of the D7-branes and O7-planes, we have been able to give the map between 
the periods of $K3$ and the corresponding D7-brane configurations in Chapter~\ref{chapterk3}. A powerful tool in
this context is the relation between singularities of $K3$ and the gauge symmetry of coincident D-branes.
We have reviewed how this beautiful correspondence can be inferred from the duality to the heterotic $E_8\times E_8$ string.
This enabled us to systematically construct $K3$ surfaces with given singularities. In particular, we have studied
the relation between an elliptic $K3$ described by a Weierstrass model and $T^4/\Z_2$ in detail. While both
manifolds describe F-theory at the orientifold point, they have a different geometric structure. We have used our
findings to explain why $T^4/\Z_2$ allows an Enriques involution, while this is not possible for an elliptic $K3$
described by a Weierstrass equation. Furthermore, we have studied the behavior of $K3$ in the F-theory limit,
in which the volume of the fibre torus shrinks to zero. We have shown that there can be many different $K3$ surfaces which 
go to the same point in moduli space in this limit. This behavior has a very clear description in terms of the
heterotic dual: M-theory on $K3$ is dual to the heterotic string on $T^3$, so that three Wilson lines can be present.
F-theory, on the other hand, is dual to the heterotic string on $T^2$, which allows only two Wilson lines. We have
verified that the F-theory limit changes the geometry of $K3$ such that one of the three Wilson lines becomes irrelevant.

We have shown that our strategy of constructing cycles from the topology of D-branes and O-planes is also
viable in the case of elliptic threefolds in Chapter~\ref{chapter3}. Even though the analysis becomes much
more complicated, we can still formulate all results in great generality as all structure descends from 
simple topological properties of the base. Furthermore, we have used the fact that all base spaces are 
quotients of $K3$, which have been classified in the mathematics literature. As D7-branes and O7-planes can 
intersect in a complex two-dimensional base we also encounter obstructions in the deformations of the D7-branes 
as compared to a generic hypersurface. We have shown that the number of obstructed deformations equals the number 
of double intersections between D7-branes and O7-planes. Different configurations of D7-branes and O7-planes give 
rise to various singularities of the elliptic threefold. It would be very interesting to use our results to 
describe singularities of elliptic threefolds in terms of shrinking cycles.

The next level of complication is given by elliptic fourfolds. Although some of the techniques used in
Chapter~\ref{chapter3} can easily be generalized to this case, many details still remain a subject of future investigation. 
The main obstacle is that D7-branes and O7-planes are no longer complex curves but complex surfaces, which are much
more complicated objects. An approach that might prove valuable in this context is the duality to heterotic 
compactifications. Using the duality between the moduli of $K3$ on the F-theory side and Wilson lines on $T^2$ 
on the heterotic side fibrewise, one obtains a description of D7-branes in terms of the so-called spectral cover. 
As we have worked out the map between the cycles of $K3$ and the moduli of D7-branes, one can use the spectral cover
to retrieve information on how this map is fibred. Even though the duality between the heterotic string
and F-theory only works for F-theory on $K3$-fibred fourfolds, there is evidence that spectral
covers can also be useful in a more general context \cite{Donagi:2009ra, Blumenhagen:2009yv}.

We have finally shown how to use the map between 2-cycles of $K3$ and D7-brane moduli to study 
flux stabilization of F-theory on $K3\times K3$ in Chapter~\ref{fluxonk3xk3}. We have derived the potential 
induced by the fluxes and, using the fact that the 4-form flux can be considered as a linear map between the cohomology
groups of the two $K3$ surfaces, found a simple geometric condition for a flux to minimize the potential. 
This condition enabled us to fix any configuration of D7-branes by choosing an appropriate flux. We have also 
shown how to determine the amount of surviving supersymmetry directly from the flux matrix.
The M-theory fluxes dual to Poincar\'e-symmetry-preserving type IIB fluxes do not stabilize the size of the fibre, so 
that we always have a flat direction in the M-theory moduli space. Of this line, only one point corresponds to a four 
dimensional vacuum, the one associated with the F-theory limit.

\section*{Acknowledgements}

I am greatly indebted to Arthur Hebecker for supervising this work. His generous 
support and his dedication to and knowledge about physics were an outstanding inspiration. 
His ability to get to the core of problems and the challenges posed through his questions both made
major contributions to this work. Most importantly, he and his blackboard were always available 
for discussing what was on my mind.

I thank Michael G. Schmidt for valuable advice, many interesting conversations and for 
agreeing to be the co-referee for this thesis. 

I would like to thank my collaborators, Rainer Ebert, Sebastian Gerigk, Christoph L\"udeling,
Michele Trapletti, Hagen Triendl and Roberto Valandro for the enjoyable work we did together and for
the many things that I was able to learn from them. 

I am very grateful to all the people at the Institute for Theoretical Physics in
Heidelberg for the friendly atmosphere and the numerous discussions. In particular, I would like to thank 
Julian Behrend, Felix Br\"ummer, Thomas Dent, Ivan Donkin, Svend Domdey, Mischa Gerstenlauer, Christian Gross, Daniel Gr\"unewald, 
Tae-Won Ha, Sebastian Halter, Benedict v. Harling, Simon K\"ors, Sven Krause, Andreas v. Manteuffel, Stefan Groot-Nibbelink, 
Tatsuya Noguchi, Jan-Martin Pawlowski, Patrick Pl\"otz, Richard Schmidt, Lily Schrempp, Gianmassimo Tassinato, Martin Trappe, 
Sergei Winitzki, Nico Wintergerst and Robert Ziegler. I also thank Sonja Bartsch, Cornelia Merkel and Melanie Steiert for 
the smiles that were on their faces whenever I entered their office.

I am indebted to the Center for the Fundamental Laws of Nature at Harvard University, in particular
the group of Frederik Denef, for hospitality during my visit. 

I benefittet from discussions, advice and explanations from Vincent Bouchard, Volker Braun, Ching-Ming Chen, 
Andres Collinucci, Frederik Denef, Mboyo Esole, Florian Gmeiner, James Gray, Thomas Grimm, Denis Klevers, Maximilian Kreuzer, Johanna Knapp, 
Paul Koerber, Luca Martucci, Christoph Mayrhofer, Harun Omer, Filipe Paccetti Correia, Sakura Sch\"afer-Nameki, Harald Skarke, 
Washington Taylor, Nils-Ole Walliser, Timo Weigand, Alexander Westphal and Akin Wingerter.

I would like to thank Markus Banagl, Maximilian B\"ormann, Aron Fischer, Florian Gaisendrees, Fillip Levikov, Richard L\"oser, Falk L\"owner, 
Konstantin Sch\"adler and Oliver Strasser for various enlightening conversations and hospitality whenever I made a visit to 
the Institut of Mathematics in Heidelberg. 

I also thank Johannes Albrecht, Christian Schultz-Coulon, Marc Deissenroth, Michael Henke and Ulrich Uwer for interesting discussions
about experimental implications of physics beyond the Standard Model.

I am very grateful to $\ast$, my family and my friends for love, support and tolerating the moodswings
that go along with the emotional rollercoaster ride of trying to solve problems in theoretical physics. 
Sadly, my father did not live to see the outcome.

Last but not least, I would like to thank the Heidelberg Graduate School of Fundamental physics of funding and 
Jeanette Braun, Sven Krause, Christoph L\"udeling and Roberto Valandro for help with the manuscript. 

\begin{appendix}

\chapter[Appendix]{}\label{appendix}
In this appendix we review some of the mathematical concepts and techniques and examples that are
relevant for this thesis. We will not explain most things in depth and with mathematical rigour but
rather give the tools, guided by examples, necessary to carry out computations. We take some background on K\"ahler manifolds and
(co)homology for granted, see \cite{Candelas:1987is, Nakahara:2003nw, Hori:2003ic, Bouchard:2007ik} for an introduction.
For the mathematically inclined reader, we recommend \cite{BottandTu} and \cite{Huybrechts}.

\section{Characteristic Classes}
\label{sumchernclass}

Characteristic classes are a very elegant way of quantifying topological properties of vector bundles \cite{milnorstash}. 
They make their appearance in many areas of theoretical physics, see e.g.~\cite{bertlmann, GS, Nakahara:2003nw} 
for an introduction to characteristic classes and their relation to anomalies in gauge field theories. In particular, their 
appearance in the Atiyah-Singer index theorem allows to compute indices with ease. See \cite{Roe} for an introduction to the 
index theorem. A very beautiful approach to index theorems is supersymmetric quantum mechanics \cite{Gaume}.

Let us start by introducing the Chern class.
The total {\bf Chern class} of a vector bundle $V$ is given as a formal sum of even forms:
\begin{align}
c(V)&=\det\left(1+\frac{1}{2\pi}F\right)=1+\frac{1}{2\pi}\Tr F+... \\
 &=1+c_1+c_2+....\ ,
\end{align}
Here $F$ is the curvature (=field strength) of the bundle and $c_n$ is a $2n$-form. An important property of the Chern classes is
{\bf naturality}, i.e.  if $f$ is a map from $Y$ to $X$ and $E$ is a vector bundle over $X$, we have $c(f^{-1}E)=f^*c(E)$,
where  $f^*$ denotes the pull-back. An important consequence of this is the {\bf splitting principle}. Assume there
exists a space $X_s$, called the split manifold, and a map $F:X_s\mapsto X$ such that the pull-back of the vector bundle $E$ splits into a
sum of line bundles $F^{-1}E=L_1\oplus...\oplus L_n$ and $F^*$ embeds the cohomology groups of $X$ in $X_s$. 
Naturality of the Chern classes then ensures that any polynomial
identity in the Chern classes can be shown by pretending that the vector bundles in question can be
written as direct sums of line bundles. It can be shown that one can find a split manifold
for every vector bundle \cite{BottandTu}, so that all computations can be carried out on the
level of the split bundle.

In terms of the eigenvalues of $\frac{1}{2\pi}F$, which we denote by $r_i$, the Chern class reads
\be
c(V)=\prod_i(1+r_i) \ .
\ee
The $r_i$ are known as the {\bf Chern roots}. If we pull back the vector bundle $V$ to a split bundle,
the $r_i$ are the first Chern classes of the line bundles $L_i$. 
From this, the Whitney product formula follows:
\be
c(V_1\oplus V_2)=c(V_1)c(V_2) \ .
\ee

The Euler characteristic of a holomorphic vector bundle which has an n-dimensional manifold $M$ as its base 
can be computed by integrating the top Chern class:
\be
\chi(V)=\int_M c_{n/2}(V) \ .
\ee
If we take $V$ to be the holomorphic tangent bundle of a manifold $M$, $\chi$ gives the ordinary Euler 
characteristic of $M$.

The {\bf Chern character} of a vector bundle, $\ch(F)$, is likewise defined as
\begin{equation}
\ch(F)=\Tr\e^F=\sum_i\e^{r_i} \ .
\end{equation}
It can be expressed in terms of the Chern classes as
\be
\ch(F)=R+c_1+\frac{1}{2}\left(c_1^2-2c_2\right)+... \ .
\ee

The {\bf Hirzebruch L-genus} is
\be
\Lc(F)=\prod_i \frac{r_i}{\tanh (r_i)}=1+\frac{1}{3}\left(c_1^2-2c_2\right)+... \ .
\ee
For an even dimensional manifold, the intersection product defines a metric on the space
of differential forms of the middle dimensionality. Integrating the Hirzebruch L-genus gives the signature
of this metric.

The {\bf A-roof genus} is
\be
\hat{A}=\prod_i \frac{r_i/2}{\sinh (r_i/2)}=1-\frac{1}{24}\left(c_1^2-2c_2\right)+... \ .
\ee

The {\bf Todd class} is
\be
\Td(F)=\prod_i \frac{r_i}{1-\e^{-r_i}}=1+\frac{1}{2}c_1+\frac{1}{12}\left(c_1^2+c_2\right)+\frac{1}{24}c_1c_2+...
\ee
Integrating the Todd class yields the {\bf arithmetic genus} 
\be
\chi_0=1-h^{1,0}+h^{2,0}-... \ . 
\ee

The higher arithmetic genera can be obtained through the integrals
\be
\chi_n=\int \left(\frac{1}{n!}\partial_x^n \prod_i (1+x\e^{-r_i})\frac{r_i}{1-\e^{-r_i}}\right)_{x=0} \ .
\ee

\section{Toric varieties}\label{toricvarsect}

Toric geometry is a collection of manifolds, called toric varieties, and techniques in the realm of algebraic geometry.
Toric varieties have a very simple geometric structure, all of which is given by combinatorics. In this
section we describe the construction of toric varieties by using homogeneous coordinates. After we have introduced the notion of 
line bundles and their associated divisors, divisor rings of toric varieties are reviewed. The standard references on toric geometry 
are \cite{Oda:1988} and \cite{Fulton}. See also \cite{Hori:2003ic ,Bouchard:2007ik, Skarke:1998yk} and \cite{Denef:2008wq} for an 
introduction from a physicists perspective.

\subsubsection{Fans and cones}

\begin{figure}
\begin{tabular}{c@{\hspace{3cm}}c}
\includegraphics[height=4cm]{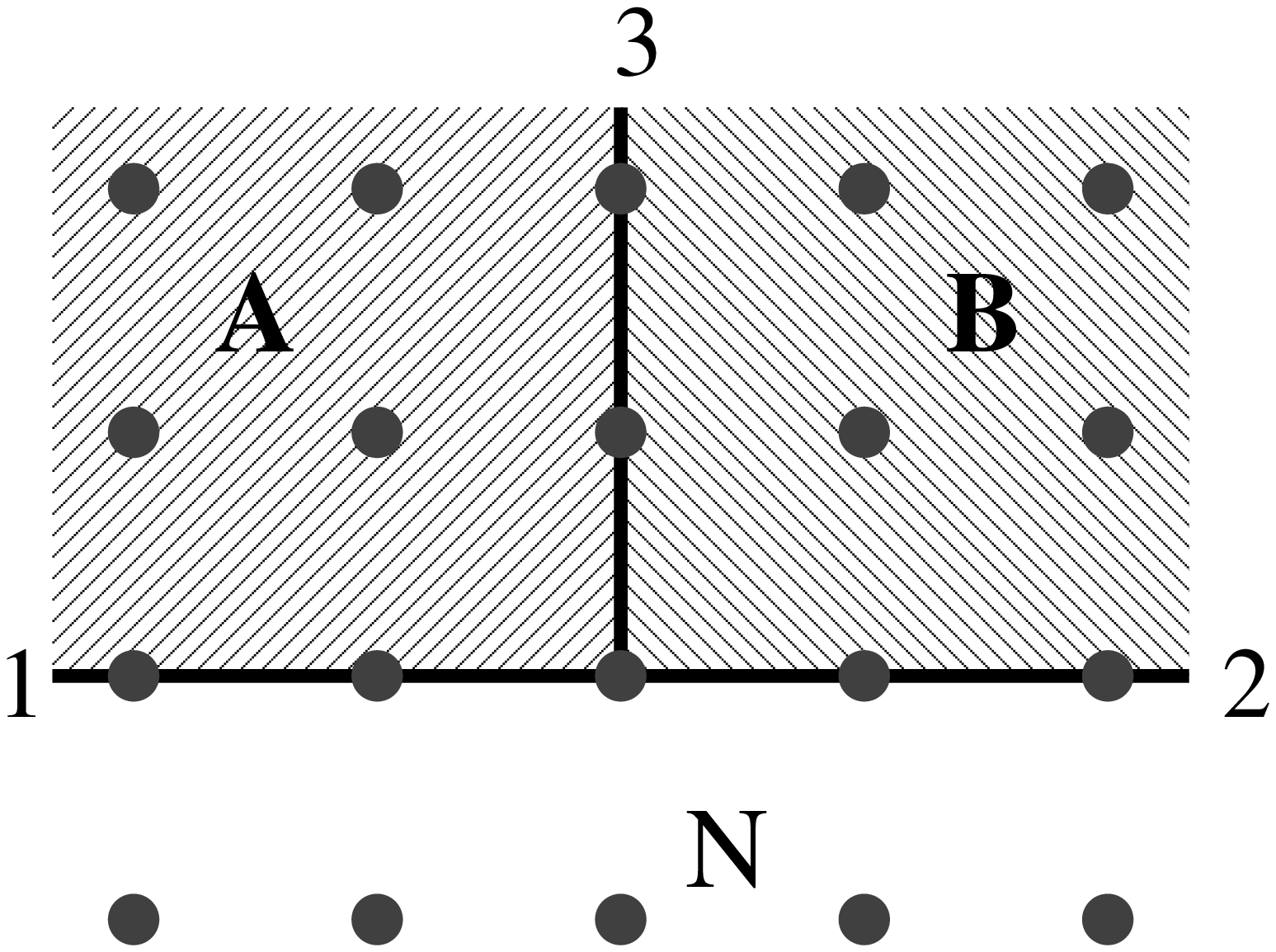}  &
\includegraphics[height=3cm]{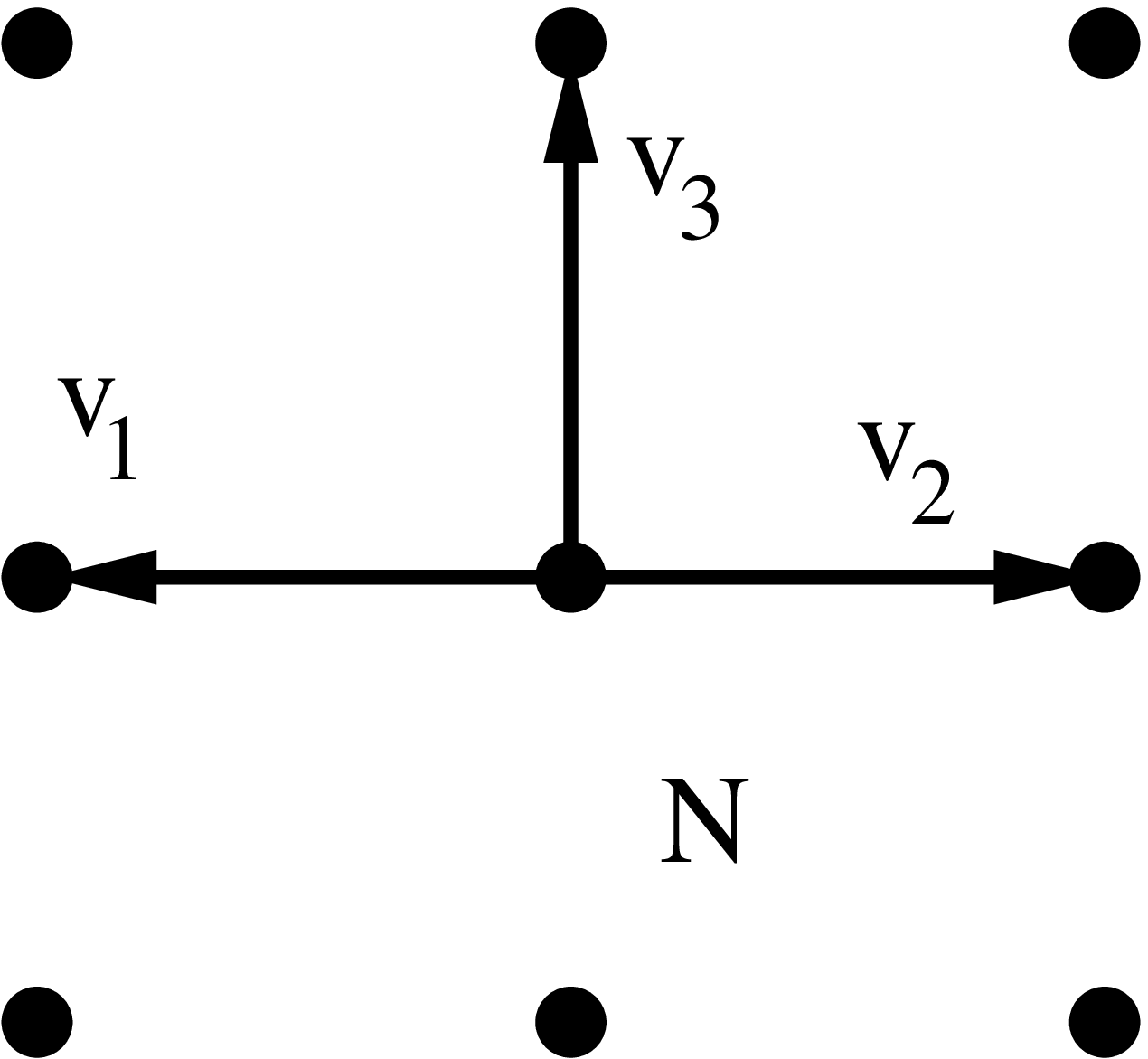}  \\
\end{tabular}
\caption{\textsl{A simple fan that is built out of strongly convex rational polyhedral cones is shown on the left.
Here we have that $\Sigma(2)=\{A,B\}$ and $\Sigma(1)=\{1,2,3\}$. On the right,
we have depicted the associated lattice vectors $v_i$.}} \label{fan1} 
\end{figure}

Let us start with the construction of toric varieties through fans using homogeneous coordinates. From this
perspective, toric varieties look very much like a generalization of complex projective spaces. Furthermore, the fans can be used
to read off many important properties of the toric variety they describe in an intuitive fashion. The basic building blocks 
of fans are {\bf strongly convex rational polyhedral cones}, which we will refer to as cones in the following. Such a cone $c$ can be 
written as the set 
\be
c=\{ \sum_{i=1..n} r_1v_i | r_i \geq 0 \} \ ,
\ee
generated by a finite set of vectors $v_i$. To be \emph{strongly} convex, $c$ must furthermore satisfy $-c\cap c = \{0\}$. 
The vectors $v_i$ are required to be elements of a lattice $N$, called the toric lattice.
Naturally, any face of a cone will again be a cone. A {\bf fan} $\Sigma$ is a collection of cones such that each face of a cone 
is contained in it, and the intersection of two cones is a face of both. The set of k-dimensional cones in a fan $\Sigma$ is denoted by 
$\Sigma(k)$. A simple example of a fan is shown on the left in Figure~\ref{fan1}.
We can uniquely associate a shortest lattice vector to any one-dimensional cone $v_i$. We denote the components
of $v_i$ by $v_i^{\mu}$. The lattice vectors of our example are shown on the right in Figure~\ref{fan1}.

The toric variety corresponding to a fan $\Sigma$ is obtained as follows. We first associate a complex coordinate $z_i$ 
to every one-dimensional cone $v_i$. If there are $n$ lattice vectors, the toric variety $X(\Sigma)$ associated to the fan 
$\Sigma$ is given by
\be\label{toricv}
X(\Sigma)=\left(\C^n-Z(\Sigma)\right)/G \ ,
\ee
where $Z$ is called {\bf exceptional set} or {\bf Stanley-Reisner ideal} and $G$ is an abelian group. 
Let us first describe how to obtain $Z(\Sigma)$. If a set of one-dimensional cones $\rho_i, i\in I$ does not span
a cone in $\Sigma$, the solution to $z_i=0$ for all $i \in I$, taken as a subspace of $\C^n$, is contained in 
$Z(\Sigma)$. In the example presented in Figure~\ref{fan1}, we hence have to subtract $\{z_1=z_2=0\}$.

The group $G$ is a subgroup of the group of maps
\be
\gamma(t_1,...,t_n): (z_1,..,z_n)\mapsto (z_1t_1,...,z_nt_n) \ ,
\ee
parameterized by $n$ numbers $t_i \in \C^*$. Here $\C^*=\C-{0}$.
The information encoded in the fan appears in the further map
\be \label{phimap}
\phi: (t_1,...,t_n)\mapsto(\prod_{j=1}^{n}t_j^{v_j^1},...,\prod_{j=1}^{n}t_j^{v_j^m}) \ .
\ee
The kernel of this map, i.e. the $t_i$ that are mapped to $(1,...,1)$ by $\phi$
define those maps $\gamma$ that make up the group $G$. 

These definitions are rather ad hoc and we refer the reader to the literature for an explanation 
of the details of this construction.

Let us make the recipe given above less abstract by coming back to our example. There we have
\be
\phi: (t_1,t_2,t_3)\mapsto(t_1t_2^{-1},t_3) \ .
\ee
The kernel is given by $t_1=t_2$ and $t_3=1$. Hence the equivalence relation we have to divide $\C^3-\{z_1=z_2=0\}$ by 
is $(z_1,z_2,z_3)\sim(\lambda z_1,\lambda z_2,z_3)$. Hence we recognize this toric variety as $\P^1\times \C$.

Note that whenever we find that some of the lattice vectors $v_i$ are linearly equivalent, $\sum_i Q_{K i} v_i=0$, $G$ 
induces the equivalence relation
\be\label{linrelvecttor}
(z_i)\sim(\lambda_{K}^{Q_{K i}} z_i) \ .
\ee
The $Q_{K i}$ are called {\bf charges}. Toric varieties appear naturally in gauged linear sigma models (GLSM), in which the
$Q_{K i}$ correspond to the charges of fields. See \cite{Hori:2003ic, Denef:2008wq} for an introduction to the
GLSM approach to toric varieties.

The abelian group $G$ always splits into a part isomorphic to $(\C^*)^g$ 
and a finite piece. In many cases it is convenient to describe toric varieties by the charges. 
From this perspective, toric varieties are generalizations of (weighted) projective spaces. 

If we have $n$ one-dimensional cones embedded in a $m$ dimensional vector space, we can find $n-m$ linear relations
among them. \eqref{linrelvecttor} then tells us that $G$ has the dimension $n-m$. Hence the dimension of the toric 
variety $T(\Sigma)$ is $n-(n-m)=m$. A fan in a m-dimensional space, which we will refer to as a m-dimensional fan, 
always yields a toric variety of complex dimension $m$.

Toric varieties have K\"ahler metrics which descend from the K\"ahler metric of $\C^n$. See \cite{Huybrechts} for
the properties of K\"ahler manifolds.

\subsubsection{Affine coordinates}

In (weighted) projective spaces one may find local coordinate charts by choosing
{\bf affine coordinates}. Let us first consider the example of $\P^1$, which is given by
\be 
\frac{\left(\C^2-\{z_1=z_2=0\}\right)}{(z_1, z_2)\sim(\lambda z_1, \lambda z_2)} \ .
\ee
Its fan is given by dropping the two-dimensional cones $A$ and $B$ as well as $v_3$ from the
fan in Figure~\ref{fan1}. We can cover $\P^1$ by two charts: For all $z_1\neq 0$ we may define 
$\hat{z}_2=z_2/z_1$, which serves as a coordinate on the patch $U_1=\{z_1\neq 0\}$. The patch 
$U_2=\{z_2\neq 0\}$ is defined by demanding $z_2\neq 0$, so that we can choose $\hat{z}_1=z_1/z_2$
as a coordinate. The coordinate transformation which is defined on $U_1 \cap U_2$ is given 
by $\hat{z}_1=\hat{z}_2^{-1}$. A similar strategy can be used to find coordinate charts for all toric varieties. One
uses the homogeneous coordinates to form expression which are invariant under the scaling,
i.e. have charge zero. These expressions will generically not be defined on the whole
variety in question, i.e. they will only give rise to local coordinate charts.

\subsubsection{Singularities}

\begin{figure}
\begin{center}
\includegraphics[height=7.5cm]{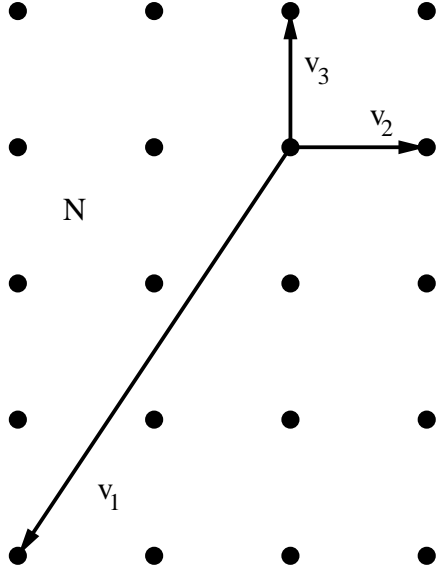}  
\caption{\textsl{The fan of $\P_{1,2,3}$. This time, we have just drawn the lattice vectors generating its one-dimensional
cones.}}\label{fan3} 
\end{center}
\end{figure}

Let us discuss another example of a toric variety, see Figure~\ref{fan3}.
The lattice vectors $v_i$ have the coordinates $v_1=(-2,-3)$, $v_2=(1,0)$ and $v_3=(0,1)$.
There is the relation $v_1+2v_2+3v_3=0$, from which we deduce that $G$ induces the equivalence relation
$(z_1,z_2,z_3)\sim (\lambda z_1,\lambda^2 z_2,\lambda^3 z_3)$. By explicitly writting down $\phi$ one
can check that $G$ does not contain any further elements.
As there is no cone which is spanned by all three $v_i$, we find that the exceptional set is $Z=\{z_1=z_2=z_3=0\}$.
Hence this example gives the toric description of the weighted projective space $\P_{1,2,3}$.

We can describe $\P_{1,2,3}$ by three charts $U_i$, where each $U_i$ covers all of $\P_{1,2,3}$ except 
the point $z_i=0$. This time, however, the different scalings will force us to take some power of the
homogeneous coordinates in order to arrive at affine coordinates. First
consider $U_1=\{z_1\neq 0\}$. In this patch we may take the coordinates $(\hat{z}_2,\hat{z}_3)=(z_2z_1^{-2},z_3z_1^{-3})$.
In the patch $U_2=\{z_2\neq 0\}$, however, we can only choose 
\be\label{twotooneap}
(\hat{z}_1,\hat{z}_3)=(z_1^2z_2^{-1},z_3^2z_2^{-3})
\ee. 
A similar choice is forced on us in $U_3$. The map from $U_2=\{z_2\neq 0\}$ to $\C^2$ given by \eqref{twotooneap}
is clearly not one-to one, as e.g. the two distinct points $(z_1,z_2,z_3)=(\pm 1 ,1 ,1 )$ are mapped to a single
affine coordinate.

This behavior signals the appearance of singularities. If we, instead of using affine coordinates, 
fix $z_2=1$ by choosing $\lambda$ appropriately, we note that we still can mod out $(z_1,1,z_3)\sim (- z_1,1,-z_3)$. Hence
$\P_{1,2,3}$ looks like $\C^2/\Z_2$ in the chart $U_2$ and there is a singularity at $z_1=z_3=0$. This space cannot be mapped
to someting that looks like $\C^2$ by using affine coordinates. The affine coordinates we have chosen above do not see this
singularity because we have ``squared'' $\C^2/\Z_2$, identifying the points $(z_1,z_3)$ and $(-z_1,z_3)$.
Again, we find a similar situation in the patch $U_3=\{z_3\neq 0\}$. As the charge of $z_3$ is three, we find that
$U_3$ is $\C^2/Z_3$.

Coming back to the corresponding fan, Figure~\ref{fan3}, we note that the fact that $v_1$ extends
so far has led to the high scalings, which ultimately have lead to the singularities. This logic is actually true
in general: if the lattice vectors $v_i$ that span a single cone do not generate the whole lattice, there is a {\bf singularity} \cite{Fulton}.
Coming back to the fan shown in Figure~\ref{fan3}, we observe that both $v_2$ and $v_1$, as well as $v_3$ and $v_1$,
span a common cone, but fail to generate the whole lattice. This gives rise to the two singularities we have observed above.
These singularities can be resolved by introducing further one-dimensional cones, i.e. subdividing the fan. Desingularizations
of this type are called blow-ups and are discussed this with the help of further tools below.

\subsubsection{Compactness}

Another property of toric varieties that can be read off their fan is compactness. Whereas the fan shown in
Figure~\ref{fan3} yields a compact toric variety, $\P_{1,2,3}$, the fan given in Figure~\ref{fan1} gives rise
to a non-compact toric variety: $\P^1\times \C$. The crucial difference between these two fans 
is that the two-dimensional cones of the fan of $\P_{1,2,3}$ fills out $\R^2$, whereas the fan of $\P^1\times \C$ fails to 
do so. This has a clear effect in the case of $\P^1\times \C$, where one homogeneous coordinate simply has charge zero. Hence the
coordinate $z_3$ does not take part in any scaling, yielding a factor of $\C$. If we were to add a one-dimensional
cone that is opposing $v_3$ (and the two corresponding two-dimensional cones) we compactify $\P^1\times \C$ to $\P^1\times \P^1$.
The general requirement for compactness is a straightforward generalization of what we have just discussed: a toric variety is
{\bf compact} if an only if the cones of the highest dimension fill out the whole vector space the fan is embedded in \cite{Fulton}.

\subsubsection{Fibrations}
\begin{figure}
\begin{center}
\includegraphics[height=3cm]{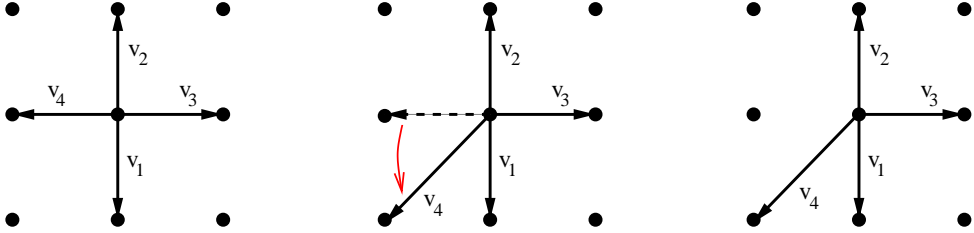}  
\caption{\textsl{The fan on the left hand side, which describes $\P^1\times \P^1$, also the structure of a
product. Tilting the lattice vector $v_4$, we obtain a $\P^1$ bundle over $\P^1$.}} \label{fanhirz} 
\end{center}
\end{figure}

A further property of toric varieties that can be easily seen from its fan (or the charges) are fibration structures.
Let us start this discussion by introducing another example, it is shown in Figure~\ref{fanhirz}. 
The fan on the left hand side can be quickly recognized to describe $\P^1\times \P^1$. The relations
among the $v_i$ are $v_1+v_2=0$ and $v_3+v_4=0$. 
This leads to the charges
\begin{center}
\begin{tabular}[h]{ccccc}
$z_1$ & $z_2$& $z_3$ & $z_4$ & \\  
0 & 0 & 1& 1 & \\
1 & 1& 0 & 0 & \ . 
\end{tabular}
\end{center} 
The table above is a shorthand for the equivalence relation $(z_1,z_2,z_3,z_4)\sim (\lambda z_1,\lambda z_2,\mu z_3, \mu z_4)$,
where as usual $\lambda, \mu \in \C^*$. If we now tilt $v_4$, yielding the fan at the right hand side of
Figure~\ref{fanhirz}, we still find the relation $v_1+v_2=0$. The second relation, however, has changed to $v_4+v_2+v_3=0$.
Hence we find the charges
\begin{center}
\begin{tabular}[h]{ccccc}
$z_1$ & $z_2$& $z_3$ & $z_4$ & \\  
0 & 1 & 1& 1 &\\
1 & 1& 0 & 0 & \ .
\end{tabular} 
\end{center} 
If we forget about $z_1$ and $z_2$, the remaining two coordinates $z_3$ and 
$z_4$ describe a $\P^1$. On the other hand, $z_1$ and $z_2$ describe a $\P^1$ if we fix $z_3$ and $z_4$. 
These two $\P^1$s are not independent, as one of the homogeneous coordinates of one of the $\P^1$s, $z_2$, takes part
in the scaling of the other $P^1$. Hence the fan on the right hand side of Figure~\ref{fanhirz} describes a bundle of $\P^1$ over 
$\P^1$. In the present case, this space is known as the first Hirzebruch surface, $\Hirz[1]$. More details about the Hirzebruch
surfaces can be found in Section~\ref{Hirzebruchsurfaces}.

%\subsubsection{Why ``toric'' ?}
%Before we proceed, let us pause and explain what is so toric about toric varieties, as our approach does
%not seem to justify this name so far\footnote{A simple two-torus is not a toric variety, so this cannot be the
%reason for this name. To see this, note that there are only two types of one-dimensional fans which yield the toric
%varieties $\P^1$, $\C$ or a quotient of one of the two.}. The name toric variety stems from the following definition: \\
%A toric variety $X$ is a complex algebraic variety that contains an algebraic torus $T=(C^*)^r$ as a dense
%open subset (hence the name). Furthermore there is a natural action of $T$ on $X$, whose
%restriction to $T$ is nothing but coordinatewise multiplication. We have constructed toric varieties from a fan
%$\Sigma$ by taking
%\be
%X(\Sigma)=\left(\C^n-Z(\Sigma)\right)/G \ .
%\ee
%As the exceptional set only contains points at which some of the homogeneous coordinates vanish, $X(\Sigma)$ always 
%contains $(C^*)^n/G$. This quotient gives another algebraic torus, $(C^*)^m$, where $m$ is the complex dimension of the 
%toric variety. This has a natural action on itself which extends to all of $X$, see \cite{Fulton, Hori:2003ic} for the details.

%\subsubsection{Polytopes}
%explain $T^n$ fibration

\section{Line bundles and divisors} \label{aplbundles}

In this section we discuss line bundles on complex manifolds and the close ties they have to divisors. 
See \cite{Griffiths:1978} and \cite{Huybrechts} for the general theory.

Let us start by considering line bundles. A {\bf line bundle} $L$, $\pi:L\rightarrow B$ 
over some base space $B$ is specified by local trivializations $\psi_i:\pi^{-1}(U_i) \simeq U_i\times \C$ and
holomorphic transition functions $g_{ij}=\psi_i \cdot \psi_j^{-1}$. The transition
function should be such that they do not vanish anywhere in $U_i\cap U_j$ and satisfy
\begin{equation}\label{cocylcecond}
 g_{ij}g_{ji}=1 , \hspace{2cm}  g_{ij}g_{jk}g_{ki}=1 \ .
\end{equation}

Sections $\sigma$ are patched together by the relation
\be
\sigma_j(b)=\sigma_i(b)g_{ij}(b) \ .
\ee
Here $\sigma_i(b)$ denotes the section as a function on the base in a trivializing neighborhood $U_i$.

A simple example is given by line bundles on $\P^1$. Let us take the two charts $U_1=\{z_1\neq 0\}$ and
$U_2=\{z_2\neq 0\}$ as the local trivializations and consider the transition functions: $g_{12}=\hat{z}_2^{-n}, n \in \Z$.
These line bundles carry the name ${\cal O}_{\P^1}(n)$. Any section of this line bundle must satisfy
\be
\sigma_1(\hat{z}_2)\hat{z}_2^{-n}=\sigma_2(\hat{z}_1) \ .
\ee
In $U_1\cap U_2$ we furthermore have the relation $\hat{z}_1=\hat{z}_2^{-1}$. Let us try to construct
a section of ${\cal O}_{\P^1}(n)$ by starting from the monomial $\hat{z}_2^k$ in $U_1$.
In the chart $U_2$ it continues as $\hat{z}_1^{-k}\hat{z}_1^n=\hat{z}_1^{n-k}$. If $k$ is equal to
$n$ or smaller, this section has a zero of order $k$ at $\hat{z}_2=0$ and a zero of order $n-k$ at
$\hat{z}_1=0$. If the $k$ is bigger than $n$, we find a zero of order $k$ at $\hat{z}_2=0$ and a 
pole of order $k-n$ at $\hat{z}_1=0$. Note that the difference of zeros and poles
(counted with multiplicity) is always equal to $n$. It is clear that this pattern remains if we consider 
polynomials instead of monomials: poles and zeros will merely become distributed over the whole $\P^1$, but
the difference between their numbers (again counted with multiplicities) will stay the same. This is the reason 
for the number $n$ appearing in the name in the bundle. 

We can only construct a holomorphic (as opposed to a meromorphic) section if we start in with a monomial of 
$\hat{z}_2^k$ in $U_1$ that has $k \leq n$. By the linearity of the transition function it is clear that any polynomial 
of $\hat{z}_2$ in $U_1$ that has degree less or equal to $n$ will extend to a holomorphic section in 
${\cal O}_{\P^1}(n)$. Note that the expressions for $\sigma_1(\hat{z}_2)$ and $\sigma_2(\hat{z}_1)$ that describe 
these holomorphic section are precisely the local descriptions of a homogeneous polynomial of degree $n$ in the
homogeneous coordinates. This only makes sense if $n \geq 0$. But as we have seen before, these are the only cases
in which there are holomorphic sections. Let us make this more explicit. We can write a homogeneous polynomial of 
degree $n$ in the homogeneous coordinates $z_1$ and $z_2$ as
\be
a_0 z_1^n + a_1 z_1^{n-1}z_2+...+a_n z_2^n \ .
\ee
In the chart $U_1$ we have $z_1\neq 0$, so that we can divide by $z_1^n$ to obtain
\be
\sigma_1(\hat{z}_2)=a_0 + a_1 \hat{z}_2+...+a_n \hat{z}_2^n \ .
\ee
In the chart $U_2$ we can likewise divide by $z_2^n$ and arrive at 
\be
\sigma_2(\hat{z}_1)=a_0 \hat{z}_1^n + a_1 \hat{z}_1^{n-1}+...+a_n \ .
\ee
Note that $\sigma_1(\hat{z}_2)\hat{z}_2^{-n}=\sigma_2(\hat{z}_1)$, as it should be. Hence
homogeneous polynomials of degree $n$ give a global description of holomorphic sections of the
bundle ${\cal O}_{\P^1}(n)$. 

We have seen that all sections of the line bundle ${\cal O}_{\P^1}(n)$ have a common number of zeros
minus poles. Using this information to characterize the bundle is at the heart of the correspondence
between line bundles and divisors. A {\bf divisor} is nothing but a collection of irreducible analytic hypersurfaces $V_i$
which carry multiplicities $d_i \in \Z$:
\be
D=\sum_i d_i D_i \ .
\ee
In the case of ${\cal O}_{\P^1}(n)$, these hypersurfaces are just points. For a trivial bundle,
whose sections are meromorphic functions $f$ on $B$, the number of poles is equal to the number of zeros. Such a
divisor is called a {\bf principal divisor} and is denoted by $(f)$ for a meromorphic function $f$. Two divisors $D$ and $D'$ are
said to be {\bf linearly equivalent}, $D\sim D'$, if they differ only by a principal divisor. 

Let us come back to the example of ${\cal O}_{\P^1}(n)$. Meromorphic functions on $\P^1$ have an equal number of zeros
and poles. Hence any divisor on $\P^1$ for which $\sum_i d_i=0$ is a principle divisor. From this is follows that any 
two points on $\P^1$ are linearly equivalent divisors. Hence all divisors associated to the sections of 
${\cal O}_{\P^1}(n)$ are the same up to linear equivalence:
\be
D({\cal O}_{\P^1}(n))\sim n\cdot \mbox{pt} \ ,
\ee
where pt denotes a point in $\P^1$. Note that linear equivalence gives the same relations as homology.

Through addition, Divisor classes naturally carry the structure of an abelian group. Line bundles carry the same
structure: the association, usually written as $L\otimes L'$ acts on the transition functions by ordinary multiplication. 
In our example, this means that ${\cal O}_{\P^1}(n)\otimes {\cal O}_{\P^1}(m)={\cal O}_{\P^1}(n+m)$. The identity is given 
by the trivial bundle and the inverse is called the dual line bundle, denoted by $L^*$. It is clear that we can get sections of 
$L\otimes L'$ by multiplying sections of $L$ with sections of $L'$, but that not all sections of $L\otimes L'$
can be constructed that way. This is familiar, if one thinks of sections as homogeneous polynomials. The group of all line bundles on a base space 
$B$ is called the {\bf Picard group} of $B$ and denoted by $\Pic(B)$.

As all divisor classes on $\P^1$ are given by $n$ points, one may wonder if this also exhausts the line bundles
that can be constructed. This is indeed true: Up to isomorphism, line bundles are classified by their first Chern class, 
$c_1(L)\in \Pic(B)=H^2(B,\Z)\cap H^{1,1}(B)$. The correspondence between the first Chern class and the Divisor class is
actually very simple. Let us stick to our example for a bit more. The first Chern class of a line bundle, integrated over
the base, counts the number of zeros any holomorphic section is forced to have. As we have seen that a holomorphic section 
of ${\cal O}_{\P^1}(n)$ has $n$ zeros, its first Chern class is $n\cdot H$, where $H$ is the volume form of $\P^1$.
This is nothing but the Poincar\'e dual of the homology class of the associated divisor. As first cohomology group of $\P^1$ is $\Z$, 
all line bundles on $\P^1$ are actually isomorphic to one of the bundles ${\cal O}_{\P^1}(n)$.

All we have said so far holds in great generality. Given any line bundle $L$ on a space $B$, we may consider the zero locus
and the poles of one of its sections $\sigma$. Let us denote the zero locus of $\sigma$ by $V$ and the locus of the poles of
$\sigma$ by $V'$. The algebraic
hypersurfaces $V$ and $V'$ can be uniquely written as the union of irreducible analytic hypersurfaces:
\be
V=V_1\cup...\cup V_n \ ,\hspace{1cm} V'=V'_1\cup...\cup V'_n \ .
\ee
The associated divisor is then given by
\be \label{weildiv}
D(\sigma)=\sum_i \ord(V_1)V_i +\ord(V'_1)V'_i \ .
\ee
Here $\ord(V_i)$ is the order of the sections over $V_i$ and $\ord(V'_i)$ counts the order
of the poles, which is always negative. Divisors that are defined in the form \eqref{weildiv} are called
{\bf Weil divisors}. A divisor is called {\bf effective} if only positive coefficients appear in \eqref{weildiv},
i.e. $V'=\{\ \}$. Hence holomorphic sections correspond to effective divisors.

Given a meromorphic section $\sigma$ of a line bundle, we can construct another meromorphic section $\tilde{\sigma}$ by
multiplying it with a meromorphic function $f$ on $B$: $\tilde{\sigma}=f\sigma$. The divisor $D(\tilde{\sigma})$ is
then given by 
\be
D(\tilde{\sigma})=D(\sigma)+(f) \ .
\ee
Hence $D(\tilde{\sigma})$ and $D(\sigma)$ are linearly equivalent divisors. As any quotient of two meromorphic sections
of $L$ defines a meromorphic function on $B$, we can write $\sigma_a=f_{ab}\sigma_{b}$ for any two sections. Hence all
sections of a line bundle give rise to linearly equivalent divisors.

We may associate a homology class to any divisor $D(\sigma)$ by using \eqref{weildiv}. As $V$ and $V'$ are a formal
sum of hypersurfaces, the Poincar\'e dual of $D(\sigma)$, $\PD(D(\sigma))$, is an integral $(1,1)$ form. This $(1,1)$ form is 
equal to the first Chern class of the line bundle $L$, see \cite{Griffiths:1978} for a proof of this important relation. 
As different sections of the same bundle must give rise to the same Chern class, it follows that divisors which are linearly 
equivalent must also be in the same homology class of $B$.

So far, we have discussed how divisors arise from line bundles. We have found that the proper objects to identify
are line bundles and equivalence classes of divisors. We have shown how to find the divisor associated to a line bundle.
It is also possible to construct a line bundle from a divisor. For a divisor $D$, this bundle is commonly denoted by $[D]$. 
The construction of $[D]$ starts from a description of divisors which differs slightly from \eqref{weildiv}. Suppose we can find an 
open cover $U_i$ of $B$ such that in each $U_i$ the divisor $D$, i.e. the hypersurfaces $V$ and $V'$ with multiplicities, is described 
by a local meromorphic function $f_i$, which we can always write as $f_i=\frac{g_i}{h_i}$, using holomorphic
functions $g_i$ and $h_i$. The functions $f_i$ are called the local defining functions of the divisor. Divisors 
defined in terms of local defining functions are called {\bf Cartier divisors}. If $B$ is smooth, there is no distinction between Weil and 
Cartier divisors, so that we can always switch between a description in terms of a collection of hypersurfaces and the local 
defining functions.

To construct the line bundle $[D]$, consider the functions $g_{ij}$ defined in $U_i\cap U_j$ by
\be
g_{ij}=\frac{f_i}{f_j} \ .
\ee
As $f_i$ and $f_j$ are local defining functions of the same divisor, $g_{ij}$ is a holomorphic function
which vanishes nowhere inside $U_i\cap U_j$. The conditions \eqref{cocylcecond} are also easily checked,
so that we have just defined a line bundle in terms of its transition functions. This line bundle is built
such that it patches together the local defining functions $f_i$ to a section of $[D]$. Hence the sections of the
bundle $[D]$ are such that they reproduce the divisor $D$.

The correspondence we have just outlined provides us with very powerful tools to discuss line bundles. To specify
a line bundle, all we have to do is give a section. This defines a divisor $D$ which completely determines the bundle.
Furthermore, this description immediately gives us access to topological data of the bundle. We only have to
compute the Poincar\'e dual of the divisor $D$ to find the Chern class of the line bundle. The
divisors of a smooth manifold are dual to the space of integral $(1,1)$-forms, $\Pic(B)=H^{1,1}(B)\cap H^2(B,\Z)$.
Hence we can compute $h^{1,1}(B)$ by inspecting the divisors that exist in $B$, or, equivalently, the line
bundles that can be constructed on $B$.

\subsection{Toric divisors}

Toric varieties are naturally equipped with line bundles and the associated divisors. Note that we have been able to 
construct all line bundles on $\P^1$ admitting holomorphic sections by considering homogeneous
polynomials, yielding the bundles ${\cal O}_{\P^1}(n), n\geq 0$. In the description of toric
varieties we have given in Section~\ref{toricvarsect}, this strategy has a natural generalization. We have 
constructed toric varieties from homogeneous coordinates $z_i$ by subtracting the exceptional set $Z(\Sigma)$
and modding out the group $G$. $G$ acts on the homogeneous coordinates by 
\be
z_i \mapsto z_i\prod_{K}\lambda_{K}^{Q_{K i}} \ , 
\ee
for $\lambda_{K} \in \C ^*$. Setting $z_i=0$ yields a divisor called $D_i$ on the toric variety $X$. Divisors of this type are
known as {\bf toric divisors}. A section of the corresponding bundle can be globally described by $z_i$. Coming back to the 
example of $\P^1$, we find that $[D_1]=[D_2]={\cal O}_{\P^1}(1)$. It is clear that the bundles $[D_1]$ and 
$[D_2]$ must be isomorphic as the corresponding divisors are linearly equivalent: they are both given by a point on $\P^1$. 

This fact holds quite generally: the bundle $[D_i]$ just depends on the charges of $z_i$. This can also be
expressed in the following way: given a fan $\Sigma$ and the lattice vectors spanning the one-dimensional cones,
the divisors $D_i$ satisfy the relation
\be \label{toricdivlinequiv}
\sum_i D_i \ v_{i}\cdot l \sim 0 \ .
\ee
for any lattice vector $l$. Here $v\cdot l$ is the scalar product between the two lattice vectors.
If the fan $\Sigma$ is $n$-dimensional, we find $n$ independent such relations.

Let us make another example and consider the toric divisors of the space $\Hirz[1]$. Its homogeneous 
coordinates have the charges 
\begin{center}
\begin{tabular}[h]{ccccc}
$z_1$ & $z_2$& $z_3$ & $z_4$ & \\  
0 & 1 & 1& 1 & \\
1 & 1& 0 & 0 & \ . 
\end{tabular}
\end{center} 
The corresponding lattice vectors are 
\begin{equation*}
v_1 = \left(\begin{aligned} 0 \\ -1 \end{aligned}\right)\ , \quad v_2 =
\left(\begin{aligned} 0 \\ 1 \end{aligned}\right)\ , \quad v_3 =
\left(\begin{aligned} 1 \\ 0\end{aligned}\right)\ , \quad v_4 = \left(\begin{aligned}
-1 \\ -1 \end{aligned}\right) \ .
\end{equation*}
From the charges, we can determine the relations $D_3=D_4$ and $D_1+D_4=D_2$. Note that the same relations
are obtained from equation \eqref{toricdivlinequiv} using the lattice vectors given above.

We have found that $\Hirz[1]$ has only two independent toric divisors, so that $h^{1,1}(\Hirz[1])=2$.
Let us choose $D_4$ and $D_1$ as a basis. Homogeneous polynomials of degree $(n,m)$ are hence 
sections of the line bundles $[nD_4+mD_1]$.

There is a simple formula for the total Chern class of the tangent bundle of a toric variety. We will not
derive it here, but merely sketch how it comes about. The trick is that the differentials with respect to the 
homogeneous coordinates, which are sections of the holomorphic tangent bundle, are themselves sections of line 
bundles on $X$. The holomorphic tangent bundle is hence a projective version of the sum of the line bundles
coming from the homogeneous coordinates. Using the properties of the the Chern classes one can then show that:
\be \label{cherntoricap}
c(X)=\prod_i (1+c_1([D_i]))=\prod_i (1+\PD(D_i)) \ .
\ee
Here $\PD(D_i)$ denotes the Poincar\'e-dual $(1,1)$-form of the divisor $D_i$.

\subsection{Homology, intersections and fans}\label{homdivtoric}

On a complex $n$-dimensional manifold $B$, one can define a topological {\bf intersection product} between differential forms 
by integration. On the homology cycles which are dual to these forms, this product counts intersections.
For $(1,1)$-forms $\omega_1,...,\omega_n$ it reads
\be
\omega_1\cdot...\cdot\omega_n =\int_B \omega_1 \wedge ...\wedge \omega_n\ .
\ee
The homology cycle dual to a $(1,1)$-form $\omega_i$, $\PD(\omega_i)$, has complex codimension one\footnote{We 
will mostly identify forms with their dual cycle in an abuse of notation, but here we make the distinction explicit.}. 
In a complex $n$-dimensional space, $n$ of these generically intersect in a number of points satisfying
\be
\#\left( \PD(\omega_1)\cap...\cap \PD(\omega_i)\right)=\omega_1\cdot...\cdot\omega_n \ .
\ee
As the intersection number only depends on the (co)-homology classes, it is a topological invariant.
It is clear that we can form similar intersection products between forms of other degrees as well. As in the
case discussed above, they count the number of points in which the Poincar\'e dual homology cycles intersect.
The intersection product has an obvious generalization to forms of higher degrees: a set of
cycles in a space $B$ has a topological intersection if the sum of the degree of the Poincar\'e dual forms matches
the dimension of $B$.

If we let $n-1$ divisors intersect in an $n$-dimensional space, this gives rise to a complex
curve. If these divisors are effective, the intersection produces a so-called {\bf effective curve}. 
In homology, these curves span the so-called {\bf Mori cone}. Let us denote the elements of the Mori cone 
by $C_i$. The integrals of the K\"ahler form $J$,
\begin{equation}
j_i=\int_{C_i} J \ ,
\end{equation}
gives the volumes of $C_i$. Varying the K\"ahler form hence changes the sizes of effective curves. 
The {\bf K\"ahler cone} is the cone of all K\"ahler forms for which all the $j_i$ are positive. When
we approach a boundary of this cone, one of the effective curves shrinks to zero size.

We can compute the intersection numbers between a basis of $(1,1)$-forms of toric varieties by using the
Poincar\'e dual toric divisors. Any one of the toric divisors is given by an equation of the form $z_i=0$. 
The number of intersections between $n$ divisors, e.g. $D_1\cap... \cap D_n$, is hence given by the 
number of solutions to the equations $z_1=..=z_n=0$. Remember that we can only simultaneously solve $n$ of these equations 
if the corresponding one-dimensional fans span a cone, for the solution to $z_i=0, i=1...n$ is otherwise part 
of the exceptional set $Z$ and is hence excluded from the toric variety $B$. If a toric variety is smooth, $n$
toric divisors have the intersection number $1$ if the span a common cone and zero otherwise\footnote{If a toric 
variety has quotient singularities one can still consistently assign intersection number, but finds that some need to
be fractional. Fractional intersection numbers occur if intersections happen at a quotient singularity.}.
Thus we can actually use the fan to read off intersection numbers. 

Let us demonstrate what we have discussed in the last paragraph for the example of $\Hirz[1]$. From the fan of 
$\Hirz[1]$ , Figure~\ref{fanhirz}, we find the relations
\begin{align}
D_1\cdot D_4=1 \hspace{1cm} D_3\cdot D_4=0 \hspace{1cm} D_1\cdot D_2=0 \ ,
\end{align}
as well as some further redundant ones. Using $D_3=D_4$ and $D_2=D_1+D_4$ we find
the following intersections for the basis $D_1, D_4$:
\begin{align}
D_1\cdot D_4=1  \hspace{1cm}  D_4\cdot D_4=0  \hspace{1cm}  D_1\cdot D_1=-1 \ .
\end{align}

\begin{figure}
\begin{center}
\includegraphics[height=7cm]{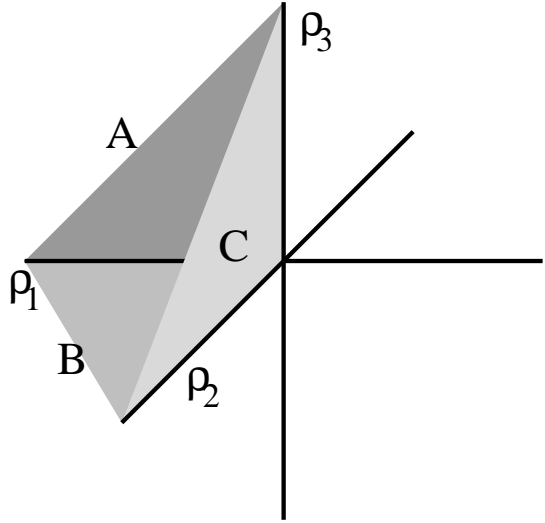}  
\caption{\textsl{A three-dimensional fan. We have drawn all of the one-dimensional and only three of the two-dimensional
cones. The corresponding toric variety is $\P^1\times \P^1\times \P^1$.}} \label{3dfan} 
\end{center}
\end{figure}

A $n$-dimensional toric variety $X$ whose fan $\Sigma$ contains $|\Sigma(1)|$ one-dimensional cones has $|\Sigma(1)|$ 
toric divisors which are subject to $n$ linear equivalence relations, see \eqref{toricdivlinequiv}. Hence we expect that 
\be \label{conjbettitoric}
b_{2n-2}=|\Sigma(1)|-n \ ,
\ee
where $b_m$ is the $m$-th betti number of $X$. If we consider a two-dimensional cone, we can identify 
it with the intersection of two toric divisors, i.e. if a cone $A\in \Sigma(2)$ is spanned by $D_1$
and $D_2\in \Sigma(1)$, the corresponding cycle is given by the equations $z_1=z_2=0$.
This cycle is nothing but the intersection of the two codimension-one cycles $D_1$ and $D_2$. 
Hence the cycles corresponding to a two-dimensional face have complex codimension two in $X$. 

All of this fits very well with the intersection product. Consider the fan given in Figure~\ref{3dfan}.
As the three one-dimensional cones $\rho_i$ span a common three-dimensional cone, we find that 
\be
D_1\cdot D_2\cdot D_3=1 .
\ee
Hence the codimension two cycle, i.e. curve, given by $z_1=z_2=0$ intersects the complex
surface $z_3=0$ in a point. Expanding our dictionairy between cones and cycles we 
identify the elements of $\Sigma(k)$, i.e. the k-dimensional cones of $\Sigma$, with cycles of complex 
codimension $k$ in $X$. Once we have found all toric divisors and computed their mutual intersections,
the homology ring of (even) cycles is just combinatorics.

One may wonder if toric varieties can have cycles that are not represented by some cone. Furthermore,
the only cycles we have discussed resulted from algebraic equations, hence they always have an even (real) 
dimension. For smooth toric varieties, one can show that all odd homology groups vanish, whereas the even betti
numbers are given by \cite{Fulton}
\be\label{bettitoric}
b_{2k}=\sum_{i=k}^{n}(-1)^{i-k}\left(\begin{aligned} i \\ k \end{aligned}\right)\ |\Sigma(n-i)|
\ee
The number of $m$-dimensional fans is denoted by $|\Sigma(m)|$ and 
$n$ is the (real) dimension of the fan, which is equal to the (complex) dimension of the
toric variety $X(\Sigma)$. 
We can use the above equation with $k=n-1$ to show that \eqref{conjbettitoric} actually holds:
\be
b_{2n-2}=|\Sigma(1)|-n|\Sigma(0)|=|\Sigma(1)|-n \ .
\ee

% final sentence on how nice that all is...

\subsection{The canonical bundle}

For any complex manifold $M$, there is a very important bundle called the {\bf canonical bundle}, $[K_M]$. It is the bundle
of holomorphic top-forms, i.e.
\be\label{defanticanoncialbundle}
[K_M]=\det T^*M \ ,
\ee
where $T^*M$ denotes the holomorphic cotangent bundle. The divisor associated to its dual bundle, $\det TM$ is
called {\bf anticanonical divisor}. It is Poincar\'e-dual to the first Chern class of $M$: 
\be\label{canbundlevsc1}
-K_M=\PD(c_1(M)) \ .
\ee
There is a particularly simple expression for the anticanonical divisor of toric varieties which is implicit in
\eqref{cherntoricap}, see also \cite{Fulton}. In terms of the toric divisors $D_i$, the canonical divisor of a toric variety $X$ is
\be \label{canbundletoric}
-K_M=\sum_i D_i \ .
\ee

\begin{figure}
\begin{center}
\includegraphics[height=6cm]{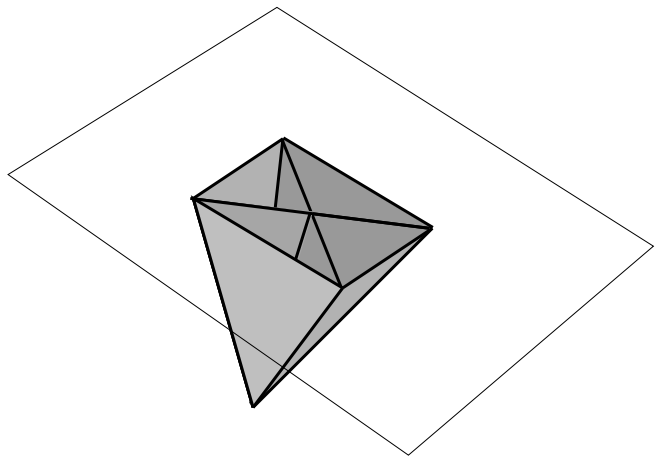}  
\caption{\textsl{This fan describes a toric variety which is Calabi-Yau because all generators of the one-dimensional
cones lie on a plane. As the three-dimensional cones do not fill out the whole space, it is not compact. Note that
one cannot expand the fan to turn it into a compact toric variety without violating the Calabi-Yau condition.}} \label{cyfan} 
\end{center}
\end{figure}

Note that this prevents the construction of compact toric Calabi-Yau manifolds. As $K_M=0$ for
Calabi-Yau manifolds, see Appendix~\ref{cyapp}, we find that $\sum_i D_i=0$ for toric Calabi-Yau manifolds. 
For this equation to be true, all of the lattice vectors $v_i$ must lie on a hyperplane normal to a
lattice vector $l$ so that $\sum_i  l\cdot v_i \ D_i=\sum_i D_i =0$. As all cones need
to be strongly convex, this means that the top-dimensional cones cannot span the entire vector space
the fan is embedded in, so that the toric variety is non-compact. We have illustrated this in Figure~\ref{cyfan}.

%fano + nakain moishezon

\subsection{Toric blow-ups}\label{blowupt}

We now introduce blow-ups from the perspective of toric varieties. We give a more general 
description of blow-ups in Section~\ref{blowup}. 

Blow-ups are the most natural way to desingularize toric varieties. Remember that singularities of toric 
varieties occur whenever there is a cone that is generated by lattice vectors that do not generate the whole
toric lattice. A natural way to remedy this is to add in further cones to the fan until we end up with a smooth
toric variety. It is a result of combinatorics that this can always be achieved, i.e. one
always finds a refinement of the fan such that the resulting toric variety is smooth \cite{Fulton}.

\begin{figure}
\begin{center}
\includegraphics[height=4cm]{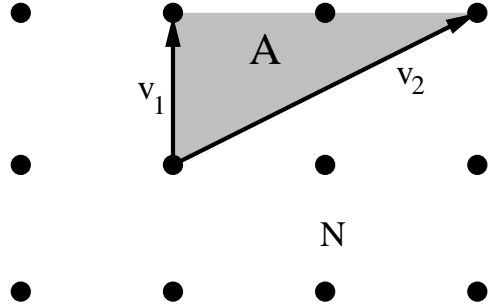}  
\caption{\textsl{The fan that describes the toric variety $\C^2/Z_2$. As the two lattice vectors $v_1$ and $v_2$ span
the cone $A$ but do not generate the whole lattice $N$, there is a singularity.}} \label{fanc2z2} 
\end{center}
\end{figure}

\begin{figure}
\begin{center}
\includegraphics[height=4cm]{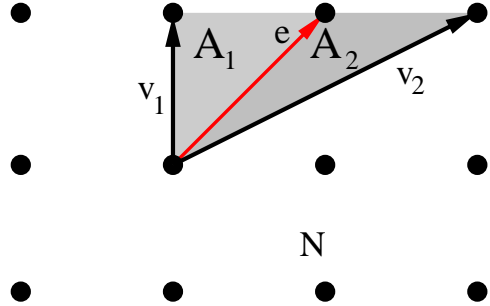}  
\caption{\textsl{The fan that describes the resolution of the toric variety $\C^2/Z_2$, $\widetilde{\C^2/Z_2}$.
Note that $v_1$ and $e$ generate the whole lattice $N$, as do $v_2$ and $e$.}} \label{fanc2z2res} 
\end{center}
\end{figure}

The prime example is $\C^2/Z_2$, see Figure~\ref{fanc2z2} for its fan. Let us first check that this
fan describes $\C^2/Z_2$. As both lattice vectors lie in a common cone, the exceptional set is empty.
Furthermore, there is no linear combination of $v_1=(0,1)$ and $v_2=(2,1)$ which is zero. Hence $G$ can only be
a finite group. Let us go back to the map $\phi$, \eqref{phimap}. In this case, it is given by:
\be
(t_1, t_2)\mapsto (t_2^2, t_1t_2) \ .
\ee
The kernel of $\phi$ is $(t_1,t_2)=(1,1)$ and $(t_1,t_2)=(-1,-1)$, so that the action of $G=\Z_2$ on the
homogeneous coordinates is
\be
(z_1,z_2)\mapsto -(z_1,z_2) \ .
\ee
Hence the fan shown in Figure~\ref{fanc2z2} yields $\C^2/Z_2$. The singularity at the origin is
also known as an $A_1$ singularity. The singularity is expected, because
$v_1$ and $v_2$ span the cone $A$, but fail to generate the entire lattice $N$. 
Let us now resolve this singularity by refining the fan. This involves adding another one-dimensional cone, 
spanned by the lattice vector $e$, and splitting $A$ into two cones: $A_1$ and $A_2$. We have depicted the
fan of the blow-up of $\C^2/Z_2$ in Figure~\ref{fanc2z2res}.

\begin{figure}
\begin{center}
\includegraphics[height=4cm]{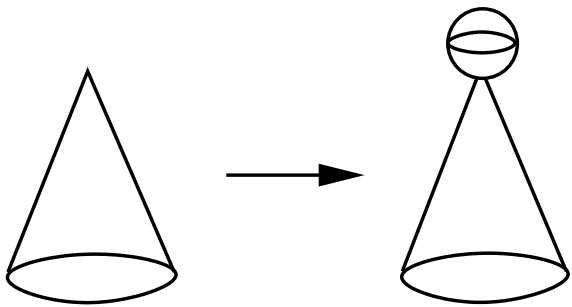}  
\caption{\textsl{$\C^2/\Z_2$, which looks like the complex version of a cone, has a singularity at the origin.
If we blow up the origin, the singularity is replaced by a $\P^1$.}} \label{resc2z2} 
\end{center}
\end{figure}

Let us now describe the associated toric variety, which we call $\widetilde{\C^2/Z_2}$. First note 
that the exceptional set is no longer empty. As $v_1$ and $v_2$ do not lie in any common cone, the exceptional set is now
given by $\{z_1=z_2=0\}$. Hence the singular point is now excluded.
As the three lattice vectors are $v_1=(0,1)$, $v_2=(2,1)$ and $e=(1,1)$, the map $\phi$ is now given by 
\be
(t_1,t_e,t_2)\mapsto(t_et_2^2,t_1t_2t_e) \ .
\ee
Hence the action of the group $G$ on the homogeneous coordinates is:
\be \label{Gresc2z2}
(z_1,z_e,z_2)\mapsto(\lambda z_1, \lambda^2 z_e, \lambda z_2) \ .
\ee
If we use $\lambda$ to fix $z_e=1$, we still need to mod out $\lambda=\pm 1$. Hence the patch
$z_e=1$ is 
\be
\left(\C^2-(0,0)\right)/\Z_2 \ .
\ee
Except for the singularity, which has been removed, 
this is space we started from.
Setting $z_e=0$ we are left with 
\be
(z_1,z_e,z_2)\mapsto(\lambda z_1, 0, \lambda z_2) \ .
\ee
Hence the divisor associated with the lattice vector $e$ we have introduced to resolve
the singularity is a $\P^1$. Divisors of this type are called {\bf exceptional divisors}. 
Let us follow a path inside $\widetilde{\C^2/Z_2}$ that approaches the former singularity:
$(z_1,z_e,z_2)=(\alpha z_1,z_e,\alpha z_2)$, $\alpha\rightarrow 0$. We can set $\lambda=\alpha^{-1}$ in
\eqref{Gresc2z2} to rewrite this as $(z_1,z_e,z_2)=( z_1,\alpha^2 z_e, z_2)$. Hence we approach
the exceptional divisor at $z_e=0$ instead of the former singularity. What has happend is
characteristic of the way blow-ups resolve singularities: the singularity is traded for a
new cycle, the exceptional divisor, see Figure~\ref{resc2z2}. 
One may also look at it the other way: the singularity of $\C^2/Z_2$ arises by shrinking the $\P^1$ sitting at
$\widetilde{\C^2/Z_2}$ to a point. As the sizes of effective submanifolds are given by integrating appropriate
powers of the K\"ahler form, blow-ups and blow-downs can also be described as variations of the K\"ahler form.

Blow-ups are essentially local operations, which only introduce
changes at the locus that is blown up, in this case the singular point at the origin of $\C^2/Z_2$. 

Although it may require much more work to work out all the details, toric blow-ups of singularities work very similar 
in more general cases. What is very beautiful about all this is that one can directly see the exceptional divisors appearing
through the extra cones that are introduced to resolve the singularities. If we place the lattice vector $v_2$ of the
example discussed above at the point $(n,1)$ instead of $(2,1)$, the resulting space is $\C^2/\Z_n$. It has a so-called
$A_{n-1}$ singularity at its origin\footnote{These singularities are introduced as singularities of hypersurfaces in Section~\ref{ellsurf}}. 
It should be intuitively clear that a resolution will result in $n-1$ exceptional $\P^1$.  They are arranged such that each one only 
intersects its two neighbors.

As blow-ups introduce new toric divisors, the canonical bundle, \eqref{canbundletoric}, will in general be changed. 
Denoting the exceptional divisor by $E$, the canonical divisor of the blow-up is
\be
K_{\tilde{X}}=K^*_X-E \ .
\ee
The $*$ denotes map that is induced under the blow-up. 
In the present case it is clear that the canonical bundle is trivial both before and after the blow-up,
because all lattice vectors spanning the one-dimensional cones lie on a plane in each case. Hence the blow-up we
have performed did not change the canonical bundle. Resolutions of this type are known as {\bf crepant resolutions}.
% this is actually a Calabi-Yau: ${\cal O}_{\P^1}(-2)$
Note that there is only one crepant resolution in the case of $\C^2/\Z_n$. In fans of dimension higher than two, 
there will in general be many crepant resolutions that lead to different smooth manifolds, see \cite{Reffert:2007im} 
for a collection of examples of this type.

Apart from blowing up singularities, one may also blow up toric varieties that are smooth. The template is the same:
one just enlarges the fan by new cones. The standard examples are the del Pezzo surfaces, discussed in Section~\ref{delPezzo}.
They are given by blow-ups of $\P^2$. Note that only the first $3$ blow-ups can be described torically.

\section{Hypersurfaces in toric varieties}\label{CYhyper}

In this section we consider algebraic hypersurfaces and complete intersections in toric varieties. 
% say more, give little intro...

We can construct an algebraic hypersurface $V$ in a toric variety $X$ by considering the vanishing locus of a homogeneous polynomial $P$.
As we have seen, homogeneous polynomials on a toric variety are sections of a line bundles. Let $P$ be a section of a line bundle called $L$. 
The vanishing locus of $P$ defines a divisor which is dual to the first Chern class of the line bundle $L$. At the level of
cohomology we have $\PD(\{P=0\})=\PD(V)=c_1([V])=c_1(L)$. 
The latter equations are of course to taken as equations of cohomology classes. Hence we can freely jump between the language of line bundles and
that of divisors.

\subsection{Smoothness}

Let us first discuss smoothness of hypersurfaces. A divisor $V$ in a smooth space $X$ is {\bf smooth} iff the differential of its 
defining functions, $df_i$, is non-vanishing on $V$. 
% singular space for emb..+example
Let us clarify this by making an example: consider a homogeneous polynomial $P_3(z_1, z_2, z_3)$ of degree $3$ in $\P^2$:
\be
P_3(z_1, z_2, z_3)=c_1 z_1^3 + c_2 z_2^3 + c_3 z_3^3 + c_4 z_1 z_2^2 + ...
\ee
Here $z_i$ are the homogeneous coodinates of $\P^2$ and $c_i$ are complex coefficients. Let us go to a patch 
in which we can use the affine coordinates $\hat{z}_1$ and $\hat{z}_2$. In this patch, the differential of 
$P_3(\hat{z}_1,\hat{z}_2)$ is given by 
\be\label{p3inp2}
\frac{\partial}{\partial \hat{z}_1}P_3 d\hat{z}_1+\frac{\partial}{\partial \hat{z}_2}P_3 d\hat{z}_2 \ .
\ee
We thus need to check if there is a solution of
\be\label{singp2}
\frac{\partial}{\partial \hat{z}_1}P_3(\hat{z}_1,\hat{z}_2)=\frac{\partial}{\partial \hat{z}_2}P_3(\hat{z}_1,\hat{z}_2)
=P_3(\hat{z}_1,\hat{z}_2)=0 \ .
\ee
These are three algebraic equations in a two-dimensional space. Of course we can choose the coefficients $c_i$ such
that there is a solution of \eqref{singp2}. For most choices of the $c_i$, however, there are no
solutions to \eqref{singp2}. The standard way to express such a situation is to say that the three equations 
\eqref{singp2} {\bf generically} have no common solutions. Note also that this fits with the topological intersections 
between the corresponding divisor classes. Hence, we have found that a generic hypersurface of degree $3$ in $\P^2$
is a smooth hypersurface.

The procedure outlined above can be performed more efficiently by using {\bf Bertini's theorem}\cite{Griffiths:1978, Hubsch:1992nu}. 
It states that a hypersurface $V$ in $X$ given by a generic polynomial $P$ is smooth away from the so-called {\bf base locus}.
The base locus consists of those points of $X$ that are in $V$ for \emph{any} choice of the coefficients of the
monomials in $P$. Given a line bundle $L$ and a generic section $P$ of this bundle, we
only have to check $V$ for smoothness at the base locus. As there the base locus is empty for \eqref{p3inp2}, it follows
directly that $P_3=0$ generically describes a smooth hypersurface. One can immediately see that the base locus is empty
in many examples, so that one is guaranteed to generically find a smooth hypersurface.

Let us make an example in which the base locus is non-empty. Consider the space $\Hirz[2]$. Its fan is
\begin{center}
\includegraphics[height=5cm]{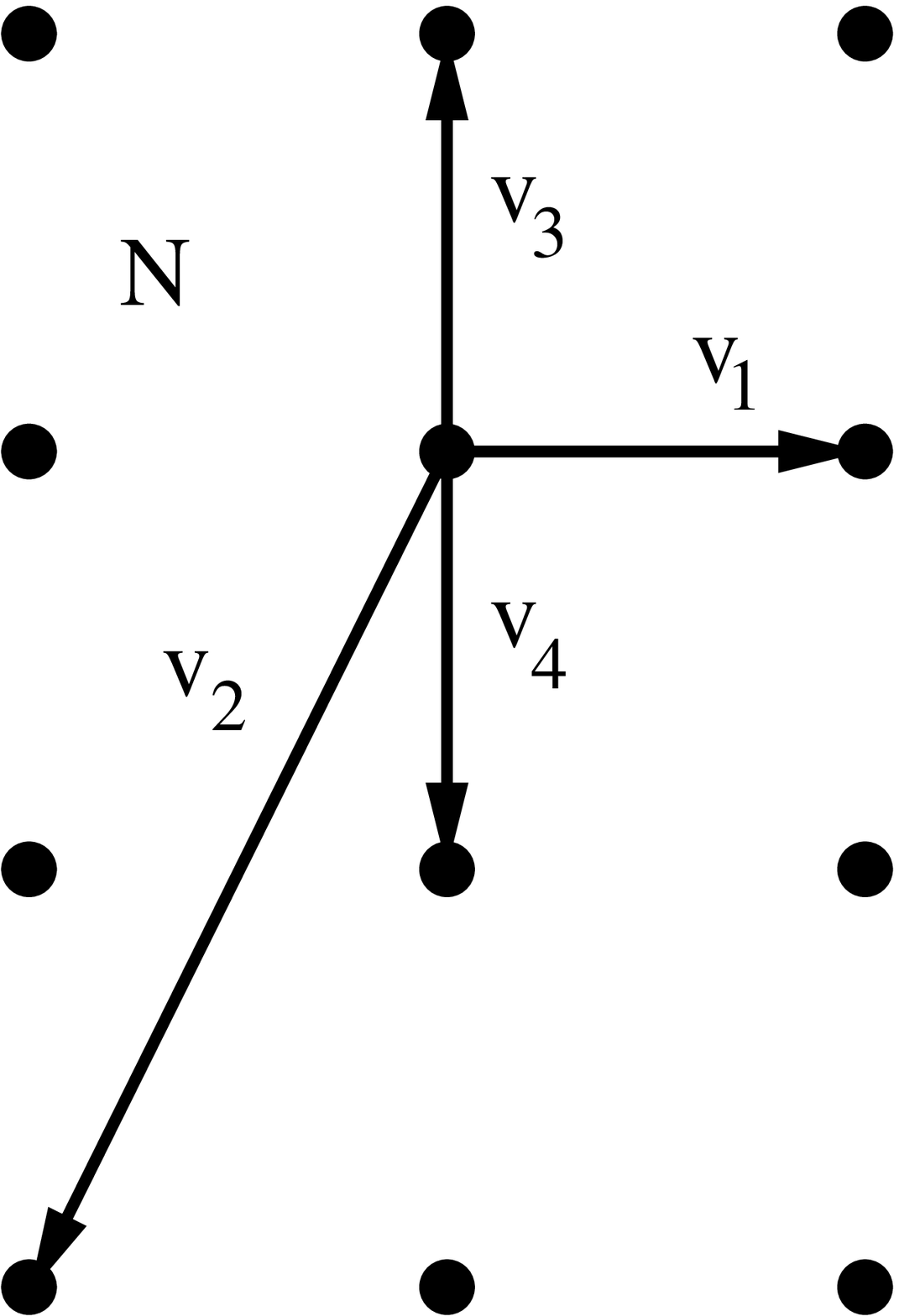} \ ,
 \end{center}
so that its homogeneous coordinates have the weights
\begin{center}
\begin{tabular}[h]{ccccc}
$z_1$ & $z_2$& $z_3$ & $z_4$ & \\  
1 & 1 & 2& 0 &\\
0 &0  & 1 & 1 & \ .
\end{tabular} 
\end{center} 
A homogeneous polynomial of degree $(2,2)$ is now forced to have the form
\be \label{eqhirz2}\alpha z_3 z_4 + z_4^2 P_2(z_1,z_2) \ .
\ee
Here $\alpha$ is a complex constant and $P_2(z_1,z_2)$ is a homogeneous polynomial of degree $2$ in $z_1$ and $z_2$.
The base locus is given by $z_4=0$, as this always solves \eqref{eqhirz2}. One may check that singularities can only
arise if $z_4=z_3=0$. As $v_3$ and $v_4$ do not lie in a common cone, this point is part of the exceptional set, i.e. it
is not a point of $\Hirz[2]$. Hence there is generically no singularity despite of the base locus being non-empty.

\subsection{Adjunction}\label{adjunkapp}
The main tool for determining topological properties of smooth hypersurfaces are the two adjunction formulae \cite{Griffiths:1978}.
An algebraic hypersurface $V$ of a space $X$ is a codimension one object. Hence its normal bundle inside $X$ is a line bundle. 
The {\bf first adjunction formula} states that this line bundle is the same as the line bundle $[V]$, restricted to $V$:
\be\label{adj1}
N_H=[V]|_V \ .
\ee
This can be seen as follows: consider local defining functions $f_i$ of $V$. By definition, they vanish
on $V$. The differentials, $df_i$, give a section of $N_V^*\times [V]$, where $N_V^*$ denotes the conormal bundle of $V$. 
If the hypersurface is smooth, these differentials do not vanish anywhere on $V$, so that they are sections of a trivial bundle
on $V$. Hence $N_V^*\times [V]|_V$ must be a trivial bundle, yielding \eqref{adj1}.

There is another intuitive way to understand \eqref{adj1}: sections in the normal bundle give infinitesimal deformations
of the hypersurface $V$ in $X$. Given that the hypersurface $V$ is defined through the vanishing locus of a section $\sigma(V)$ 
of $[V]$, we can achieve the same by infinitesimally changing $\sigma(V)$.

The {\bf second adjunction formula} arises from the decomposition of the tangent bundle of $X$, restricted to $V$:
\be
TX|_V=TV\oplus NV \ . 
\ee
Using the first adjunction formula we find that 
\be
TX|_V=TV\oplus [V]|_V \ ,
\ee
which means that
\be \label{adj2bundles}
TX|_V\oplus [-V]|_V =TV \ .
\ee
Taking the Chern classes of both sides and using the property \eqref{sumchernclass} we find that
\be\label{adj2gen}
c(TV)=\left(\frac{c(TX)}{c(V)}\right)|_V \ .
\ee 
If we restrict this equation to 2-forms, keeping in mind 
that $c_1(TM)=-K_M$, we find
\be
K_V=\left( K_X+V\right)|_V \ ,
\ee
which is equivalent to 
\be
[K_V]=\left([K_X]\otimes[V]\right)|_V \ .
\ee

Let us study an example: a hypersurface $V_n$ in $\P^2$ given by a homogeneous polynomial of
degree $n$. The three toric divisors of $\P^2$ are all equivalent. They are usually refereed to as
the hyperplane divisor and denoted by $H$. Hence we are studying a hypersurface which is linearly
equivalent to $nH$. The total Chern class of $\P^2$ is $(1+\PD(H))^3$, so that we find
\be
c(V_n)=\frac{(1+\PD(H))^3}{1+n\PD(H)}|_{V_n} \ .
\ee
Hence the first Chern class of $V_n$ is given by $(n-3)\PD(H)|_{V_n}$. We can use it to compute the Euler characteristic
of $V_n$ by integrating it over $V_n$:
\be \label{intvnp2}
\chi(V_n)=\int_{V_n}(3-n)\PD(H)|_{V_n}=\int_{V_n}(3-n)\PD(H) \ .
\ee
We have dropped the restriction to $V_n$ in the above equation, because it clearly does not
matter for the value of the integral. The reason that allows us to extend the Chern classes of hypersurfaces 
to the space they are embedded in is that the bundle on the left hand side of \eqref{adj2bundles} has an obvious 
extension to all of $X$, so that we can use this to extend the right hand side of \eqref{adj2bundles} to all of $X$.

As $V_n$ is homologous to $nH$ as a 2-cycle in $\P^2$, we may now use Poincar\'e duality to lift the
integral \eqref{intvnp2} to all of $\P^2$:
\be \label{intvnp2glob}
\chi(V_n)=\int_{\P^2}(3-n)\PD(H)\wedge n \PD(H)=3n-n^2 \ .
\ee
We have used $H\cap H=1$ in the last line. As it is a complex one-dimensional manifold, the hypersurface $V_n$ 
is a Riemann surface. Its genus is given by
\be \label{gofnp2}
g=\frac{1}{2}\chi-1=\frac{1}{2}(3n-n^2+2)=\frac{1}{2}(n-1)(n-2) \ .
\ee
Note that the first Chern class vanishes for $n=3$, in which case $V_n$ is a one-dimensional Calabi-Yau manifold:
a torus.

The strategy we have used above to compute the Euler characteristic can be used more generally to compute
all kinds of topological indices of hypersurfaces in toric varieties. After expressing all bundles in terms
of toric divisors, one can easily compute any characteristic class. By using Poincar\'e duality, integrals
of characteristic classes over $V$ can be transformed into integrals over the ambient space $X$. The
intersection numbers between the toric divisors are then sufficient to determine the value of this integral.

Note that we can also iterate the computation of characteristic classes, to find the Chern class of a 
complete intersection\footnote{Here one must of course take care that the complete intersection is transversal,
see \cite{Hubsch:1992nu} for some more details on this.}.
The Chern class of such a complete intersection given by $V=V_1\cap...\cap V_n$ is given by:
\be
c(V)=\frac{c(X)}{(1+\PD(V_1))\cdot...\cdot (1+\PD(V_n))} \ .
\ee
In a similar fashion we can iterate the trick leading from \eqref{intvnp2} to \eqref{intvnp2glob} and write
all integrals as integrals over the ambient space $X$.

\subsection{The Lefschetz hyperplane theorem}\label{lefschetz}

The adjunction formulae allow to compute the characteristic classes of hypersurfaces (or complete intersections),
which in turn can be used to compute topological numbers like the Euler characteristic or the arithmetic genera.
There is another relation that can be used to determine the lower homology groups of a hypersurface very quickly:
the {\bf Lefschetz hyperplane theorem}. In our context, it states that the cohomology groups $H^q(V)$ of a hypersurface 
$V$ embedded in a variety $X$ are isomorphic to $H^q(X)$ below some threshold if the bundle $[V]$ is positive.
A bundle $V$ on $X$ is called {\bf positive} if its divisor has a positive intersection number with all effective curves of
$X$.
Let $X$ be a complex manifold of (complex) dimension $d$, so that the (complex) dimension of $V$ is $d-1$. Then the
map 
\be
H^q(X)\rightarrow H^q(V)
\ee
induced by the inclusion of $V$ in $X$ is an isomorphism for $q \leq d-2$ and injective for $q=d-1$. Let us
discuss what this means for hypersurfaces embedded in toric fourfolds. In this case we have that $d=4$, so
that the first two cohomology groups of $X$ and $V$ are isomorphic. Hence the hypersurface all 2-cycles of $V$ descend
from the ambient space $X$. This is very useful for computing the hodge numbers of Calabi-Yau threefold hypersurfaces.

\subsection{Singularities and blow-ups}\label{blowup}

In this section we show how to describe blow-ups more generally. Let us
start from a simple example: the blow-up of $\C^2/\Z_2$. We have already discussed
this blow-up from the perspective of toric varieties in Section~\ref{blowupt}.
We first describe $\C^2/\Z_2$ as a hypersurface. Let $\C^2/\Z_2$ be given by modding
out $(z_1, z_2)\sim (-z_1, -z_2)$ from $\C^2$. Let us choose three coordinates invariant 
under $\Z_2$:
\be
x_1=z_1z_2, \hspace{1cm}x_2=z_1^2, \hspace{1cm}x_3=z_2^2 \ .
\ee
Note that given $z_1$ and $z_2$, the $x_i$ are uniquely determined.
The $x_i$ have to obey the equation
\be
x_1^2=x_2 x_3 \ .
\ee
in $\C^3$. By defining 
\be
\hat{x}_1=ix_1, \hspace{1cm} \hat{x}_2=x_2+ix_3, \hspace{1cm} \hat{x}_2=x_2-ix_3 \ ,
\ee
we arrive at the canonical form of an $A_1$ singularity, see Table~\ref{ADEtable}:
\be\label{a1sing}
\hat{x}_1^2+\hat{x}_2^2+\hat{x}_3^2=0 \ .
\ee
By computing the gradient it is easy to see that this equation has one singularity which is located at the origin.

To blow up the singularity one introduces a $\P^2$ parameterized by coordinates $\xi_i$, which are subject to
the equations:
\begin{equation}\label{blowp2eq}
\hat{x}_1\xi_2=\hat{x}_2\xi_1, \hspace{1cm}\hat{x}_1\xi_3=\hat{x}_3\xi_1, \hspace{1cm}\hat{x}_2\xi_3=\hat{x}_3\xi_2 \ .
\end{equation}
The $\xi_i$ are uniquely determined away from the origin $\hat{x}_i=0$. At the point $x_i=0, i=1,2,3$, however,
they are left completely undetermined. Let us choose a path approaching the origin to find the exceptional
divisor: $\hat{x}_i=c_it$, $t\rightarrow 0$. The $c_i$ are constants which parameterize different paths. To satisfy
\eqref{a1sing} they have to obey $c_1^2+c_2^2+c_3^2=0$. The $\xi_i$ along this path are given by $\xi_i=c_i$.
At $t=0$ we hence end up at the point 
\be  \label{excp2blowup}
\hat{x_i}=0, \hspace{1cm}\xi_1^2+\xi_2^2+\xi_3^2=0 \ .
\ee
This is a homogeneous equation of degree $2$ in $\P^2$. Note that it does not contain any singularities,
as the point $\xi_i=0, i=1,2,3$ is not contained in $\P^2$. By looking at \eqref{gofnp2} we find that the complex
curve described by \eqref{excp2blowup} has genus $0$, so it is a $\P^1$. Hence we find the same result as previously 
obtained: blowing up $\C^2/\Z_2$ at the origin introduces an exceptional $\P^1$.

This example shows in a nice way that blow-ups are local operations, as already mentioned in Section~\ref{blowupt}. 
The surface we have blown up is not changed away from the former singularity, which is now the location
of the exceptional divisor. 
%by introducing new coordinates $\xi_i$ parameterizing $\P^2$ and choose new 
%coordinates in the patch $\xi_z\neq 0$. They obey the equation 
%\be
%y^2\alpha+x^2\beta+z^2\gamma=0.
%\ee
%The resolution is not complete and there remains an $A_1$ singularity. 
%The closure of the point set within the extra $\P^2$ becomes

What we have done is easily generalized to complex manifolds of dimension $d$ \cite{Griffiths:1978}. To
blow up a complex surface at a point $p$ (this point may be a at singularity or smooth), we introduce
a $\P^d$ parameterized by the homogeneous coordinate $\xi_i$ subject to the equations:
\be
z_i\xi_j=z_j\xi_i \ .
\ee
The rest of the analysis proceeds in a fashion similar to what we have done before. 

Note that one can also blow up along some algebraic surface of codimension $m$ defined by 
a set of equations $P_i=0, i=1...m$ by introducing a $\P^m$ whose coordinates must satisfy
\be
P_i\xi_j=P_j\xi_i \ .
\ee

\subsection{Moduli}\label{modsections}

Given a hypersurface in a toric variety there is an easy way to determine the number 
of its deformations. Deformations of a hypersurface $V$ in $B$ are given by sections of the
normal bundle $NB|_V$. By the first adjunction formula we have
\begin{equation}
 N_{H/B} = [H]|_H \ .
\end{equation}
Hence we can count the number of deformations by counting the number of sections in the bundle $[V]$. 
If the bundle $[V]$ has a toric variety as its base, the number of sections can be determined by counting
the number of monomials in the homogeneous polynomial describing a generic section $\sigma$. If the base allows automorphisms, 
i.e. redefinitions of the homogeneous coordinates, sections that differ only by such a redefinition are to be considered 
identical. Furthermore, for a constant $a$, the divisors $\sigma=0$ and $a\sigma=0$ clearly define the same
hypersurface. Hence we should subtract a further complex degree of freedom related to a rescaling of the section.

Let us discuss this for some simple examples. First consider divisors $nH$ on $\P^1$, where $H$ is the hyperplane
divisor. Note that this divisor just corresponds to $n$ points on $\P^1$. Divisors of this type arise through 
sections which are homogeneous polynomials of degree $n$. In terms of homogeneous coordinates $z_1,z_2$ they can 
hence be written as
\be
\sigma=c_1z_1^n+c_2z_1^{n-1}z_2+...+c_{n+1}z_2^n \ ,
\ee
so that there are $n+1$ coefficients. 

We can redefine the homogeneous coordinates $z_1,z_2$ by an $SL(2,\C)$ matrix:
\begin{equation}
 \left(\begin{array}{c}z_1\\z_{2}\end{array}\right)\rightarrow
\left(\begin{array}{cc} 
	a & b \\
	c & d 
	\end{array}\right)\
\left(\begin{array}{c}z_{1}\\z_{2}\end{array}\right)\ ,
\end{equation}
which contains 3 complex numbers. Hence the automorphism group of $\P^1$ is 3-dimensional. As we can set one
of the coefficients $c_i$ to unity by rescaling $\sigma$, we obtain the result that $n$ points moving on $\P^1=S^2$
have $n-3$ complex degrees of freedom. 

Let us continue with $\P^2$. A homogeneous polynomial of degree $n$ on $\P^2$ has the form
\be
c_1z_1^n+c_2z_1^{n-1}z_2+c_3z_1^{n-1}z_3+... \ .
\ee
The number of terms which contain $z_1^m$ is $n-m+1$, because $z_1^m$ must be multiplied by a homogeneous
polynomial of degree $n-m$ in the two remaining coordinates $z_2$ and $z_3$. As $m$ ranges between $0$ and $m$
we find 
\be\label{modp2}
\sum_{m=0}^n n-m+1 = \sum_{m=1}^{n+1} m=\frac{(n+1)(n+2)}{2} \ .
\ee
Furthermore, $\P^2$ has $8$ automorphisms, so that we find
\be
\frac{(n+1)(n+2)}{2}-9 \ .
\ee
deformations for a hypersurface defined by a polynomial of degree $n$ on $\P^2$.

The examples only illustrate the well-known fact that a homogeneous polynomial of degree $n$ 
in $m$ variables has
\be
\frac{(n+m-1)!}{n!(m-1)!}
\ee
coefficients.

The number of deformations of hypersurfaces in other toric varieties can be studied
in a similar fashion. It is important to note that counting coefficients only
determines the number of deformations of the embedding.

\subsubsection{The Lie algebra of automorphisms of toric varieties} \label{toricAut}

When talking about possible deformations of a hypersurface $H$ embedded in a
complex manifold $Y$, we always encounter the question which deformations can be
undone by applying an automorphism of the embedding space $Y$. An easy example
can be given by the complex line in $Y=\C P^2$, that is, let $H$ be given by the
zero locus of a homogeneous polynomial of degree one. Fixing the overall
scale factor by setting one coefficient equal to one, $H$ reads
\begin{eqnarray}
H: && z_1 + \alpha z_2 + \beta z_3 = 0\ .
\end{eqnarray}
Naively one might think that the moduli space of $H$ is two-dimensional due to the
two coefficients $\alpha$ and $\beta$. However, this is not the case as the topology of $\C P^2$
stays unchanged if an element of the automorphism group $PGl(3,\C)$ is applied on the
homogeneous coordinates. In other words, we can use automorphisms of the
embedding space to set coefficients to zero, i.e.\ these
coefficients do not represent degrees of freedom. The Lie algebra of
$PGl(3,\C)$ is eight-dimensional. Thus, the moduli space of the
complex line is zero-dimensional and $H$ is unique in $\C P^2$.

It can be difficult to determine the automorphism group for more complicated
manifolds. However, for toric varieties Demazure analyzed the automorphism group in detail and determined the dimension of its algebra \cite{Demazure:1970}.
In this appendix we explain how to determine the dimension of the Lie
algebra of automorphisms, $\dim \aut_T$, in the case of a general
toric variety $T$. 

\begin{figure} \hspace{1.7cm}
\begin{tabular}{c@{\hspace{3cm}}c}
\includegraphics[height=4cm]{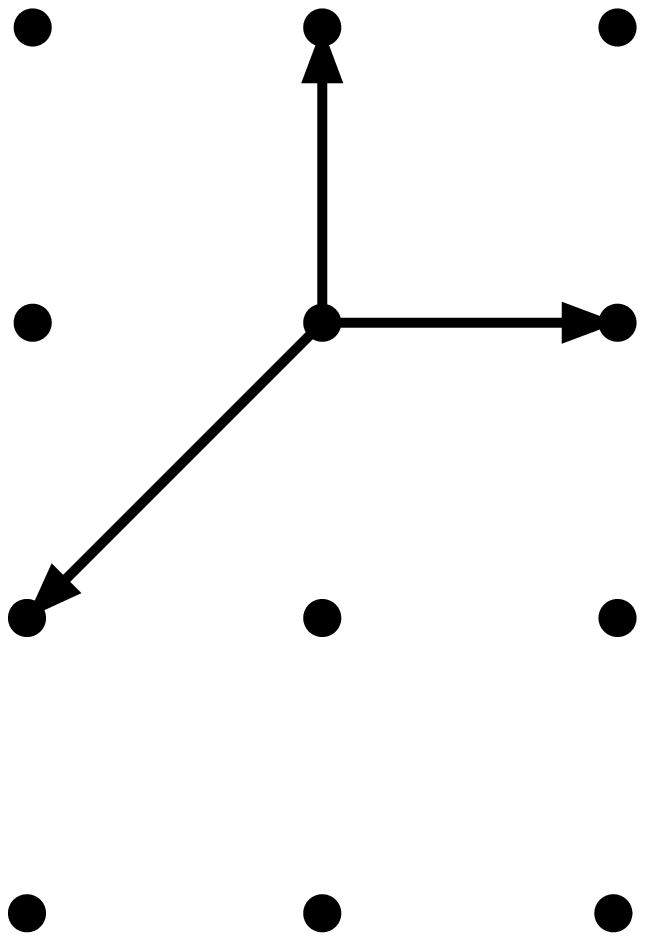}  &
\includegraphics[height=4cm]{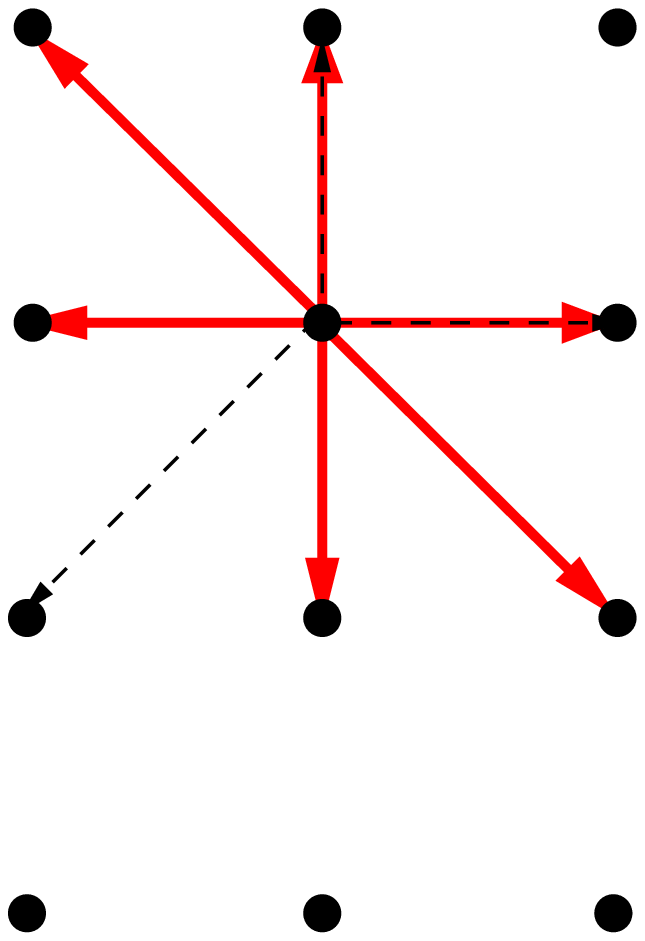} \\
the fan of $\C P^2$ & the same fan with its root system
\end{tabular}
\caption{\textsl{The toric diagram and root system of $\C P^2$}} 
\label{rootsystemCP2}
\end{figure}

Given the fan $\Sigma$ of $T$, we denote the set of $j$-dimensional
cones in $\Sigma$ by $\Sigma(j)$. For any one-dimensional cone $\rho \in
\Sigma(1)$ there is a primitive vector $n(\rho)$ generating $\rho$. A
primitive vector to a cone $\rho \in \Sigma(1)$ is the lattice vector $n(\rho)$
that spans $\rho$ such that there is no other lattice vector $m \neq n(\rho)$
for which $n = a m$ with $a \in \mathbb{N}$. Mapping any cone $\rho \in \Sigma(1)$ to its primitive vector $n(\rho)$ defines an embedding of $\Sigma(1)$ into a
lattice $N$. Similarly, the complete fan $\Sigma$ can be embedded into $N$. We denote the dual lattice of $N$ by $M$ and the natural Cartesian scalar product on $M \otimes N$ by $(\cdot,\cdot)$.
Furthermore, we introduce the root system $R(N,\Sigma)$ of a toric variety.
Abstractly, this is defined as the set of all elements $\alpha$
of the dual lattice $M$ for which exactly one cone $\rho_\alpha \in \Sigma(1)$
with $(\alpha,n(\rho_\alpha))=1$ exists and $(\alpha,n(\rho)) \leq 0$ holds for
all other cones $\rho \in \Sigma(1)$ \cite{Oda:1988}. 
This can be nicely illustrated for toric varieties of complex dimension
two. For example, take the fan of $\C P^2$ as given in the left diagram of Figure
\ref{rootsystemCP2}. The corresponding root system is given by the six vectors drawn in the right diagram of Figure \ref{rootsystemCP2}.

Now the theorem due to Demazure
\cite{Demazure:1970} states that for a compact, non-singular toric variety
the dimension of the automorphism algebra is given by
\begin{equation}
\dim \aut_T = \text{rank } N + \# R(N,\Sigma)\ ,
\end{equation}
where $\# R(N,\Sigma)$ is the number of elements of the root system $R(N,\Sigma)$. Applying this theorem to the
case of compact non-singular complex surfaces, we can use that its toric fan
can be embedded into a two-dimensional lattice $N$. Therefore we have $\text{rank
} N = 2$ so that we obtain
\begin{equation} \label{dimautomorphismgroup}
\dim\aut_T = 2 + \# R(N,\Sigma)\ .
\end{equation}
Thus we can deduce the dimension of the automorphism algebra by finding the number of roots. In the case
of $\C P^2$ we obtain six root vectors and hence
\begin{equation}
\dim \aut_{\C P^2} = 2 + 6 = 8\ ,
\end{equation}
which is in agreement with the fact that the automorphism group of $\C P^2$ is $PGl_3(\C)$. However, \eqref{dimautomorphismgroup} holds in general and thus can be applied to any other complex surface. In particular, for Hirzebruch surfaces it yields
\begin{equation}
\dim \aut_{\Hirz[n]} = \Big\{ \begin{array}{cl} 6 & \text{for $n=0$} \ ,
\\ 5+n & \text{for $n>0$} \ . \end{array}   \label{autF}
\end{equation}

\subsection{Calabi-Yau manifolds}\label{cyapp}

{\bf Calabi-Yau manifolds} are complex $n$-dimensional K\"ahler manifolds which can be characterized by one of the following 
equivalent properties \cite{Candelas:1987is, Bouchard:2007ik}:
\begin{itemize}
\item There exists a K\"ahler metric which is Ricci-flat.
\item The first Chern class of its holomorphic tangent bundle vanishes.
\item It has a trivial canonical bundle.
\item There exists a Ricci-flat metric.
\item There exists a covariantly constant spinor
\item Its holonomy group is contained in $SU(n)$
\item It admits a globally defined holomorphic top-form which vanishes nowhere.
\end{itemize}
Note that the vanishing of the first Chern class is identical to the triviality of the canonical bundle by
\eqref{canbundlevsc1}. Furthermore, the holomorphic top-forms are sections of the canonical bundle by definition, see 
\eqref{defanticanoncialbundle}, so that we can find a global non-vanishing section iff the canonical bundle is trivial. 
It was conjectured by Calabi that the vanishing of the first Chern class implies the existence of a Ricci-flat K\"ahler 
metric \cite{Calabi}. This was later proven by Yau \cite{Yau}. See \cite{Candelas:1987is, Bouchard:2007ik, Hori:2003ic} 
for a proof of the equivalence of the other conditions.

We will call Calabi-Yau manifolds that have the full $SU(n)$ holonomy group {\bf proper Calabi-Yau manifolds}.
This excludes tori and Calabi-Yau manifolds that are products of lower-dimensional Calabi-Yau manifolds.
For proper Calabi-Yau $n$-folds, all hodge numbers $h^{0,i}, i < n$ vanish. We have depicted the Hodge
diamond of the only proper Calabi-Yau two-fold, $K3$, in Section~\ref{k3mod}. The Hodge diamond of
Calabi-Yau threefolds is hence given by two numbers, $h^{1,1}$ and $h^{2,1}$:
\be
\label{CY3 hodge structure}
  {\arraycolsep=2pt
  \begin{array}{*{9}{c}}
   & &&&1&&&&\\&&&0&&0&&& \\& &0&&h^{1,1}&&0&& \\ 1&&&h^{2,1}&&h^{1,2}&&&1 . \\
    &&0&&h^{1,1}&&0&& \\&&&0&&0&&& \\ &&&&1&&&&
  \end{array}}
\ee
Hence the Euler characteristic of Calabi-Yau threefolds is $\chi=2(h^{1,1}-h^{1,2})$.
As $c_1=0$ for Calabi-Yau threefolds, the integral of the Todd class always gives zero, so that
$\chi_0=0$, as expected.

Calabi-Yau manifolds can easily be constructed as hypersurfaces in toric varieties by using \eqref{canbundletoric}
and the second adjunction formula, \eqref{adj2gen}.
Let us make a simple example and consider a Calabi-Yau hypersurface in $X=\P^2\times\P^2$. If we denote the
hyperplane divisors of the two $\P^2$ factors by $H_1$ and $H_2$, we find that theonly intersection among four elements 
of the basis $H_1, H_2$ that is non-vanishing is $H_1\cdot H_1 \cdot H_2 \cdot H_2 =1$. 
The canonical divisor is given by
\be
K_X=-3H_1-3H_2 \ ,
\ee
so that we can construct a Calabi-Yau manifold $M$ in $X$ by considering a homogeneous polynomial of degree $3,3$.
We can easily compute its total Chern class:
\begin{align}
c(M)=\frac{c(X)}{1+3H_1+3H_2}=&1+3(H_1^2+3H_1\cdot H_2+H_2^2)\\
    &-(8H_1^3+27H_1^2\cdot H_2+27H_1\cdot H_2^2+8H_2^3), \nn
\end{align}
where we have abbreviated $H\cdot ...\cdot H=H^n$. Its Euler characteristic is hence given by
\begin{align}
 &\int_M -(8H_1^3+27H_1^2\cdot H_2+27H_1\cdot H_2^2+8H_2^3) \nn \\
 =&\int_X-(8H_1^3+27H_1^2\cdot H_2+27H_1\cdot H_2^2+8H_2^3)\cdot(3H_1+3H_2) \nn\\
=&-162 \ .
\end{align}
Furthermore, the Lefschetz hyperplane gives $h^{1,1}(M)=h^{1,1}(X)=2$. Hence
we find that 
\be\label{h21p2tp2}
h^{2,1}(M)=\frac{2h^{1,1}-\chi(M)}{2}=83 \ .
\ee
Computations of this type can also be performed with the use of computer algebra. A this context, the 
package PALP \cite{Kreuzer:2002uu,palp} is very useful.

The moduli of Calabi-Yau manifolds are those variations of the metric which do not violate the condition of Ricci
flatness. By varying the Ricci form with respect to the metric and setting the result to zero, one finds that each such
variation either corresponds to an harmonic form in $H^{1,1}$ or $H^{2,1}$ for the case of proper
Calabi-Yau threefolds \cite{Candelas:1990pi}. The moduli that stem from $H^{1,1}$ are called
K\"ahler moduli, whereas the moduli coming from $H^{2,1}$ are called complex structure moduli.

Calabi-Yau manifolds have two special differential forms. As they are K\"ahler, they have a closed $(1,1)$ form $J$. 
Furthermore the Calabi-Yau condition is equivalent to the existence of a closed $3,0$ form $\Omega^{3,0}$. We can 
integrate these forms over the elements of the second and third homology groups to obtain the so-called {\bf periods}.
As we have already discussed, the periods of the K\"ahler form measure the size of effective curves, so that 
variations of the K\"ahler form correspond to changing the volumes of these curves. 

The periods of the
holomorphic 3-form are best discussed by introducing a symplectic basis of 3-cycles $A_i, B^i$ such that 
$A_i\cdot B^j=\delta_i^j$. In this basis the periods
\be
z_i=\int_{A_i}\Omega^{3,0}
\ee
can be chosen to be independent while the periods
\be
\pi^i(z_j)=\int_{B^i}\Omega^{3,0}
\ee
are functions of the periods of the $A_i$, as indicated. The $z_i$ are projective coordinates on the
moduli space of complex structures, as $\Omega^{3,0}$ $\lambda\Omega^{3,0}$, $\lambda\in \C^*$ define
the same complex structures. As $h^{3,0}=h^{0,3}$ this means that the dimension of the moduli space of
complex structures in $h^{2,1}$. 

For Calabi-Yau fourfolds, the story is essentially the same \cite{Klemm:1996ts}. K\"ahler deformations come from
$(1,1)$-forms, whereas the number of complex structure deformations is given by $h^{3,1}$.

For $K3$ surfaces, the forms $J$ and $\Omega$ both are elements of the second cohomology group. This
makes their moduli space somewhat special, see Section~\ref{k3mod} for the details.

As we have discussed in Section~\ref{modsections}, the deformations of hypersurfaces can be determined
by counting the number of deformations of the defining section. For Calabi-Yau manifolds, polynomial
deformations correspond to complex structure moduli. As the deformations of the equation which determines
the Calabi-Yau manifold as a hypersurface depends on the embedding, the number of polynomial deformations
does not equal the number of complex structure deformations in general. It is however an ``experimental'' fact
that the two numbers agree in many examples \cite{Green:1987rw}. Let us demonstrate this for the Calabi-Yau hypersurface embedded
in $\P^2\times \P^2$ discussed above. As a homogeneous polynomial of degree three on $\P^2$ has $4\times 5/2=10$
coefficients, see \eqref{modp2}, a homogeneous polynomial of degree $(3,3)$ on $\P^2\times \P^2$ has $10\times 10=100$
coefficients. As the automorphism group of $\P^2\times \P^2$ is $8+8=16$ dimensional, we obtain, after further
subtracting the freedom of rescaling, $83$ degrees of freedom. Note that this is precisely the same number as the number of 
complex structure deformations of the hypersurface Calabi-Yau manifold \eqref{h21p2tp2}.

\section{Rational surfaces} \label{complexSurfaces}
We know from the classification of Nikulin, reviewed in Appendix \ref{nikulinClassification}, that rational surfaces naturally appear as base spaces for $K3$ orientifolds in type IIB.
Rational surfaces can be
obtained by blowing up $\C P^2$ or Hirzebruch surfaces $\Hirz[n]$ and play an important role in our considerations.\footnote{$\C P^2$ and
$\Hirz[n]$ themselves are called minimal rational surfaces. A surface is called minimal
if it does not contain a curve with self-intersection $(-1)$. If a surface actually does
contain such curves, these curves can be blown down in order to obtain a minimal surface.}
Their main properties are
summarized in this appendix.
In Appendix \ref{delPezzo} we briefly discuss del Pezzo surfaces, which are (except $ \C P^1 \times \C
P^1$) blow-ups of $\C P^2$. In Appendix \ref{Hirzebruchsurfaces} we then give a short review on Hirzebruch surfaces.

\subsection{Del Pezzo surfaces} \label{delPezzo}
A complex two dimensional manifold $Y$ is called a del Pezzo surface if the
anticanonical bundle is positive definite, i.e.\ it has positive intersection
number with every curve in $Y$.

\begin{figure}
\begin{tabular}{c@{\hspace{3cm}}c@{\hspace{3cm}}c}
\includegraphics[height=4cm]{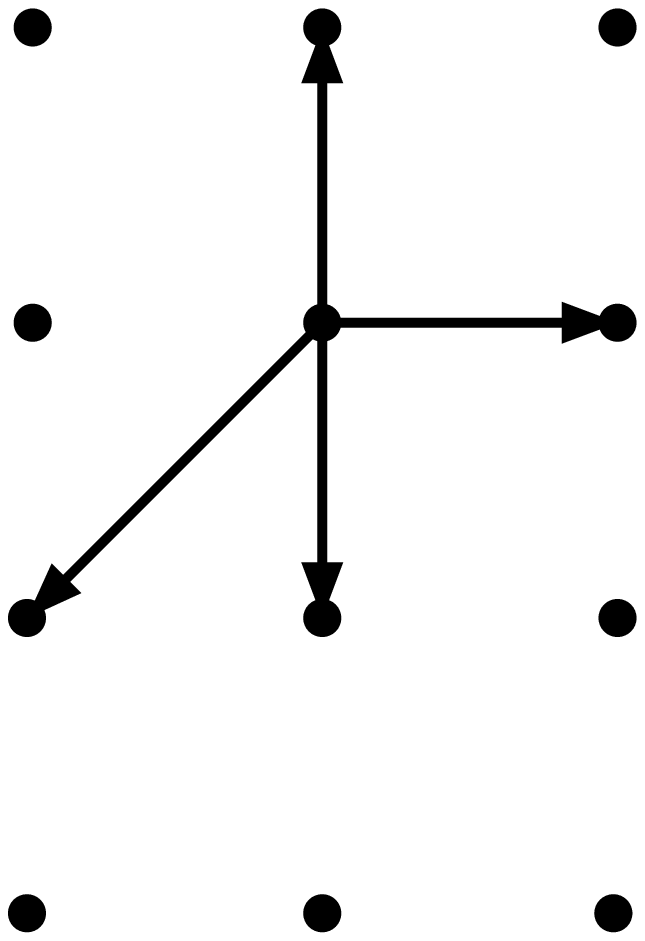}  &
\includegraphics[height=4cm]{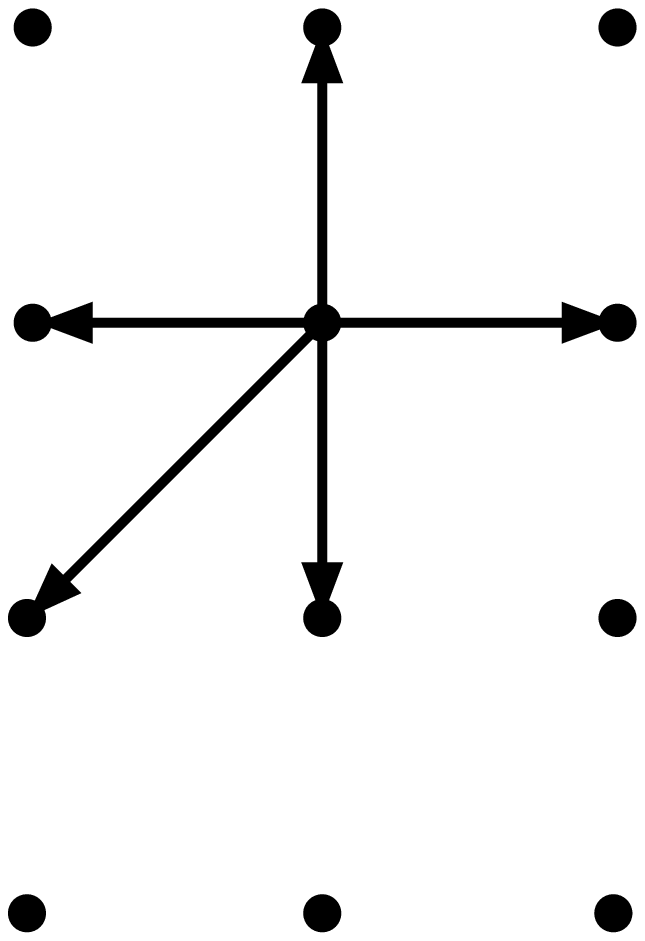}  &
\includegraphics[height=4cm]{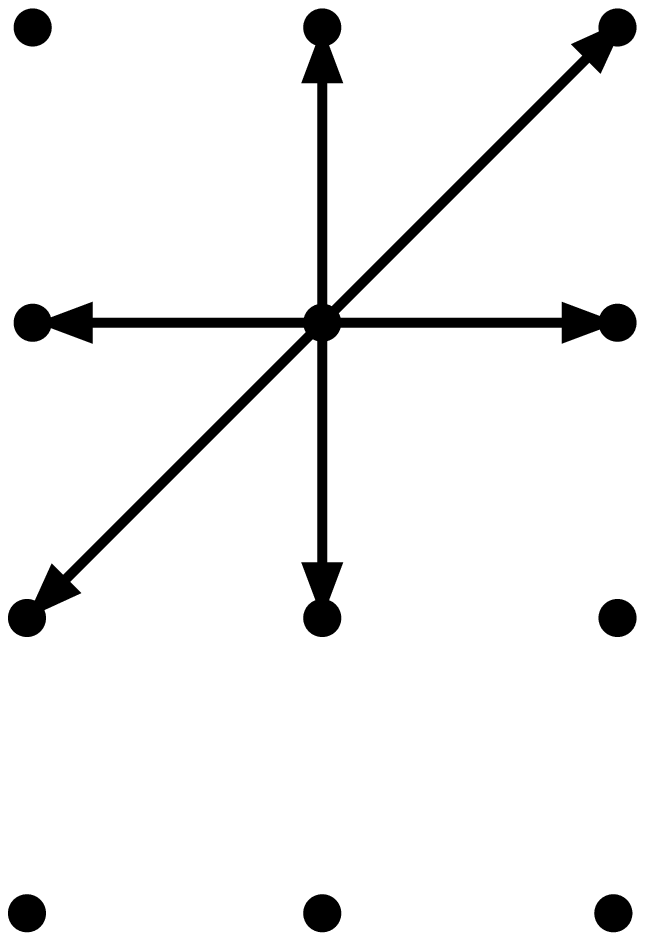} \\
$dP_1=\Hirz[1]$ & $dP_2$ &$dP_3$
\end{tabular}
\caption{\textsl{The fans of del Pezzo surfaces $dP_n$ with
$n=1,2,3$}} \label{delpezzo}
\end{figure}

There are ten topologically different del Pezzo surfaces. Nine of them are
blow-ups of $\C P^2$ at $n=0,...,8$ points. These surfaces are denoted by
$dP_n$. Additionally, there is $\Hirz[0]=\C P^1 \times \C
P^1$ which is also a Hirzebruch surface. 
Del Pezzo surfaces are completely classified by their Euler characteristic, except for the case $\chi = 4$, where there are the two del Pezzo surfaces $dP_1$ and $\C P^1 \times \C
P^1$ \cite{Hubsch:1992nu}.
Note that for $n \le 3$, $dP_n$ is a toric variety. Their toric
diagrams are given in Figure \ref{delpezzo} and \ref{rootsystemCP2}. For the remainder of this appendix, we focus on the del Pezzo surfaces $dP_n$. The surface $\Hirz[0]=\C P^1 \times \C P^1$ is discussed together with the other Hirzebruch surfaces in Appendix \ref{Hirzebruchsurfaces}.

The hyperplane divisor $H$ of $\C P^2$ has self-intersection $H^2=1$. Each
blow-up introduces one further exceptional divisor $E_i$. Thus, the dimension of the middle homology of a del Pezzo surface $dP_n$ is given by
\begin{equation}
\dim H_2(dP_n,\Z) = n+1\ .
\end{equation}
It is a well-known
fact that exceptional divisors at smooth points of a surface have
self-intersection number $E_i \cdotp E_j = - \delta_{ij}$. Furthermore, a
hypersurface in $\C P^2$ generically does not meet the blown-up points. As a result,
the hyperplane divisor $H$ does not intersect exceptional divisors, $H \cdotp
E_i = 0$. Therefore, we find the following intersection pattern of del Pezzo
surface $dP_n$:
\begin{equation}
H \cdotp H = 1\ , \hspace{2cm} E_i \cdotp H = 0\ , \hspace{2cm} E_i\cdot E_j =
-\delta_{ij}\ ,
\label{dPintersectionpattern}
\end{equation}
with $i=1,...,n$.

We can obtain the canonical divisor $K_{dP_n}$ of a del Pezzo
surface if we take the blown-up cycles into account. It is well-known that the canonical line bundle of a 
manifold $\tilde{Y}$ blown up at a smooth point is given by $[K_{\tilde{Y}}] = \sigma^*[K_Y]\otimes [(\dim Y - 1)E]$ where $\sigma: \tilde{Y} \rightarrow Y$ is the blow-up
and $[E]$ denotes the line bundle corresponding to the exceptional divisor
$E$ \cite{Hubsch:1992nu}.
Iteratively blowing up the exceptional divisors, we find
\begin{equation*}
[K_{dP_n}] = \sigma_n^*[K_{\C P^2}] \otimes \bigotimes_{i=1}^{n} [E_i]\ ,
\end{equation*}
where $\sigma_n: dP_n \rightarrow \C P^2$ is the blow-up at $n$ points.
For the corresponding divisors this reads
\begin{equation*}
K_{dP_n} = -3H + \sum_{i=1}^{n} E_i \ .
\end{equation*}
Here we used $\sigma_n^*[K_{\C P^2}] = [-3H]$.

\subsection{Hirzebruch surfaces} \label{Hirzebruchsurfaces}

\begin{figure}
\begin{tabular}{c@{\hspace{3cm}}c@{\hspace{3cm}}c}
\includegraphics[height=4cm]{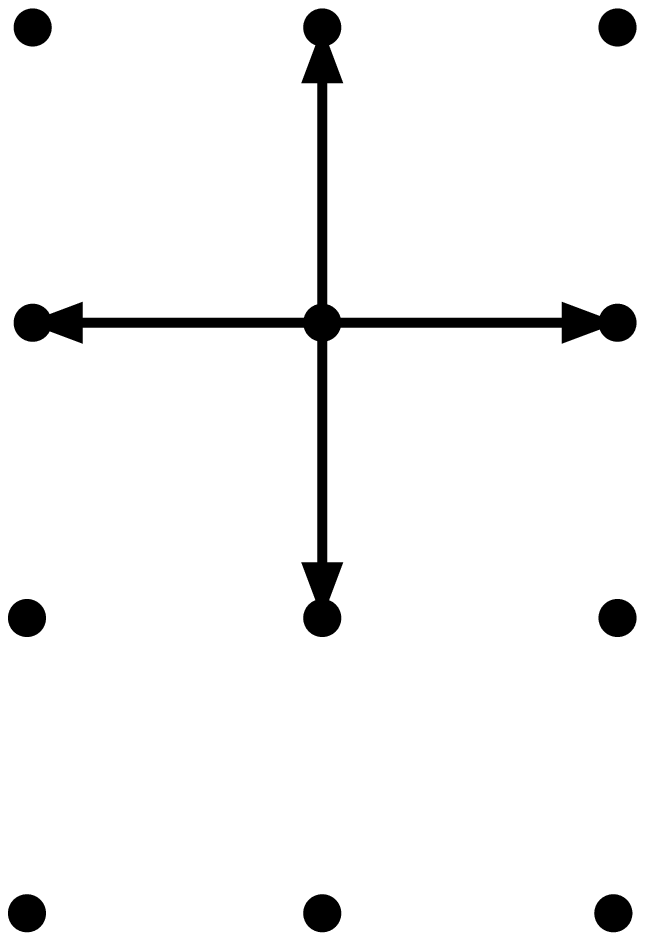}  &
\includegraphics[height=4cm]{F1.eps}  &
\includegraphics[height=4cm]{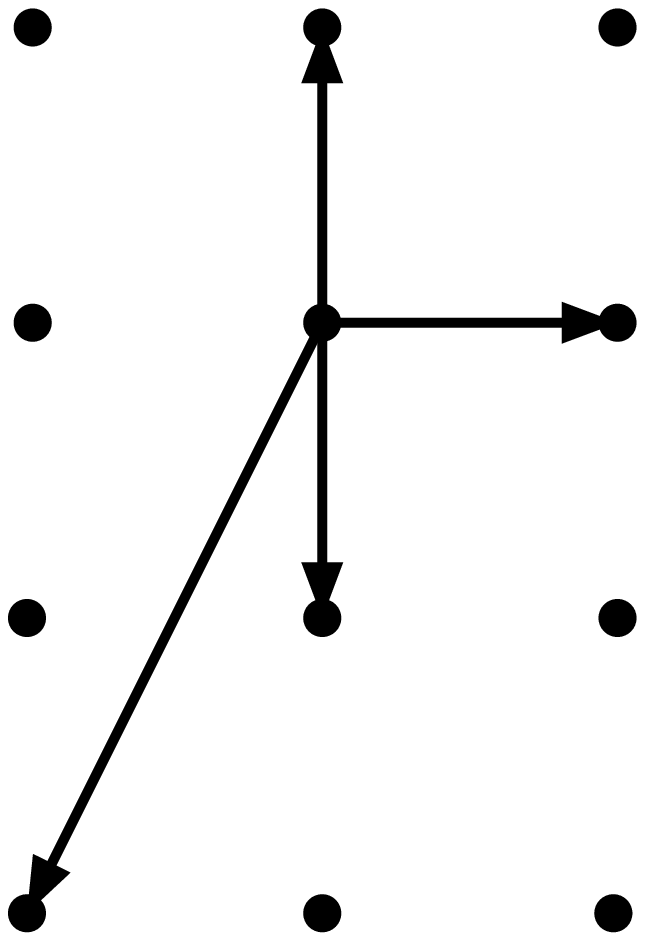} \\
$\Hirz[0]$ & $\Hirz[1]=dP_1$ &$\Hirz[2]$
\end{tabular}
\caption{\textsl{The toric diagrams of Hirzebruch surfaces in the case of
$n=0,1,2$}} \label{hirzebruch}
\end{figure}

Now we turn to Hirzebruch surfaces
$\Hirz[n]$. They are $\C P^1$ fibrations over $\C P^1$. All of them are toric varieties, and we display the fans of the first three Hirzebruch surfaces in Figure \ref{hirzebruch}.
In general, the fan of the $n$-th Hirzebruch surface $\Hirz[n]$ is given
by the cones corresponding to the vectors \cite{Hori:2003ic}
\begin{equation*}
v_1 = \left(\begin{aligned} 1 \\ 0 \end{aligned}\right)\ , \quad v_2 =
\left(\begin{aligned} -1 \\ -n \end{aligned}\right)\ , \quad v_3 =
\left(\begin{aligned} 0 \\ 1 \end{aligned}\right)\ , \quad v_4 = \left(\begin{aligned}
0 \\ -1 \end{aligned}\right) \ ,
\end{equation*}
with $n \in \mathbb{N}$.
Thus the $\C^*$-actions are
\begin{eqnarray*}
\C^*_\lambda: && (z_1,z_2,z_3,z_4) \rightarrow
(\lambda z_1,\lambda z_2,\lambda^n z_3,z_4) \ , \\
\C^*_\mu: && (z_1,z_2,z_3,z_4) \rightarrow (z_1,z_2,\mu z_3,\mu z_4) \ ,
\end{eqnarray*}
and the exceptional set is
\begin{equation}
Z_{\Hirz[n]} = \left\{ (z_1,z_2,z_3,z_4)\in \C^4 | (z_1,z_2) = 0 \text{ or }
(z_3, z_4) = 0\right\} \ .
\end{equation}

Let us now turn to the toric divisors of Hirzebruch
surfaces. The linear equivalences between the toric divisors $T_i$ with
$i=1,...,4$ of $\Hirz[n]$ are given by
\begin{equation}
T_1 = T_2 \quad \textrm{and} \quad T_3=nT_1+T_4 \ .
\end{equation}
and the middle homology group of Hirzebruch surfaces
$H_2(\Hirz[n])$ is generated by two 2-cycles satisfying
\begin{equation}
T_1^2=0\ , \quad T_1T_4 = 1\ , \quad T_4^2=-n \ . \label{HirzebruchDivisors}
\end{equation}
One can calculate the first Chern class
\begin{equation} \label{c1Hirz}
c_1(\Hirz[n])=\sum_{i=1}^4T_i=(2+n)T_1+2T_4 \ ,
\end{equation}
and the Euler characteristic
\begin{equation}
\chi(\Hirz[n]) = \int_{\Hirz[n]} c_2(\Hirz[n]) 
= \sum_{i<j} T_iT_j = 4 \ .
\end{equation}
Note that the first Chern class is positive definite if $n=0,1$. This means
that $\Hirz[0]$ and $\Hirz[1]$ are also del Pezzo surfaces. Indeed, we already
discussed this issue in the previous subsection. In the case of $n=2$, the first Chern class is
positive semi-definite.

Next, we turn to the investigation of curves in Hirzebruch surfaces. Let
$\cal H$ be such a curve of degree $(a,b)$, i.e.\ given by the zero
locus of a polynomial $p$ that is homogeneous of degree $a$ with respect to
$\C^*_\lambda$ and homogeneous of degree $b$ w.r.t.\ $\C^*_\mu$. Then $\cal H$ is
linearly equivalent to $aT_1+bT_4$. Thus we find
\begin{equation}
c_1({\cal H}) = c_1(\Hirz[n])-{\cal H}=(2+n-a)T_1+(2-b)T_4\ .
\end{equation}
Using the intersection numbers \eqref{HirzebruchDivisors}, the Euler
characteristic of $\cal H$ then turns out to be
\begin{equation}
\chi({\cal H}) = \left[2(1-a)-n(1-b)\right]b+2a\ .
\end{equation}
Hence the genus of $\cal H$ is
\begin{equation}
g_{(a,b)}=1-\frac{\chi}{2}=1+b\left[(a-1)+\frac{n}{2}(1-b)\right]-a\ .
\end{equation}
We can apply this result to the case of an O-plane $\cal H$ in $\Hirz[n]$.
In the weak coupling limit of F-Theory on $\Hirz[n]$, it is described
by a curve of degree $(2n+4,4)$ \cite{Morrison:1996na}. Plugging this into
the above equation we obtain $g_{(2n+4,4)}=9$, independent of $n$. 
We use this result in Table \ref{summExam} in Section~\ref{DonO}.

\section[Classification of non-symplectic involutions of $K3$ surfaces]{Classification of non-symplectic involutions of \boldmath$K3$ surfaces} \label{nikulinClassification}
Here we give a short review on the classification of $K3$ surfaces $X$ equipped with a non-symplectic involution $\iota: X \rightarrow X$, obtained by Nikulin in \cite{Nikulin:1979,Nikulin:1983,Nikulin:1986}.

Such pairs $(X, \iota)$ are classified in terms of a characteristic triplet $(r,a,\delta)$, where $r$ and $a$ are non-negative integers and $\delta$ is zero or one. We explain now how these numbers are defined. For that, we define $S_X$ to be
the Picard lattice of $X$ and $S$ to be the sublattice spanned by
$(1,1)$-forms that are even under the pullback $\iota^*$:
\begin{equation}
S := \left\{ \omega \in H^{(1,1)}(X) | \iota^* \omega = \omega\right\} \ .
\end{equation}
The numbers in the characteristic triplet can be obtained from the structure of
$S$. The first is $r:= \text{rank } S$. Furthermore, for each lattice there exists a dual
lattice $S^*=\text{Hom}(S,\Z)$. Each element $\alpha$ of $S^*$ can be
represented by an element $c$ of $S \otimes \R$ by identifying
$\alpha(\cdot)=(c,\cdot)$, where $(\cdot,\cdot)$ denotes the scalar product on
$S$. Since $(S,S) \subset \Z$ it is obvious that $S \subset S^*$. Indeed, it can
be shown that $S^*/S = (\Z_2)^a$ for an integer $a$ which is the
second entry of the characteristic triplet.
The third entry $\delta$ is defined as follows: Identify every linear form 
$(x, \cdot) \in S^*$ with the corresponding element $x \in S \otimes \R$ and determine $(x,x)$.
If for all such linear forms there is $(x,x) \in \Z$, then $\delta = 0$.
Otherwise, we set $\delta=1$. 
These three numbers $(r,a,\delta)$ determine the pair
$(X,\iota)$ is up to isomorphisms.

There is only a limited set of surfaces $Y$ that arise from $K3$ surfaces $X$
by modding out a non-symplectic involution $\iota: X \rightarrow X$. 
The surface $Y$ is either an Enriques surface (which is the only case with
no fixed point locus), or a rational surface, cf.\ Appendix \ref{complexSurfaces}.
Such manifolds $Y$ naturally admit a $K3$ double cover. 
It was shown by Nikulin that these surfaces correspond to so-called
non-singular DPN-pairs $(Y,C)$. By definition, $Y$ is a non-singular projective
algebraic surface with $b^1(Y)=0$ and $C$ is a non-singular effective divisor
$C=-2K_Y$ in $Y$.\footnote{The definition may be extended to singular effective
divisors $C$. See \cite{alexeev:2004} for a treatment of such cases.}
In this correspondence, $C$ denotes the fixed point locus $X^\iota$ of $\iota$ in $X$. In orientifold models, $C$ is the position of the O-plane.

An important result of Nikulin is that the numbers $(r,a,\delta)$ immediately
give certain properties of the fixed point locus $C\equiv X^\iota$. If the characteristic triplet is not $(10,10,0)$ or $(10,8,0)$,
the O-plane curve $C$ is of the form
\begin{equation}
C=C_g+\sum_{i=1}^k E_i\ ,
\end{equation}
where $C_g$ is a curve of genus $g$ and the $E_i$ are rational curves, that is,
of genus 0. The quantities $g$ and $k$ can be expressed in terms of the
characteristic triplet:
\begin{equation}\label{Nikulin}
\begin{aligned}
g&=\tfrac{1}{2} (22-r-a) \ , \\
k&=\tfrac{1}{2}(r-a)\ . 
\end{aligned}
\end{equation}
It is an important fact that $C_g$ and all $E_i$ do not intersect each other.

We analyze the case $Y=\Hirz[4]$ in Appendix \ref{reducible}. It is known \cite{alexeev:2004} that the characteristic
triplet is $(2,0,0)$ in this case. In the light of Section~\ref{DonO}, the O-plane is the disjoint union of a curve of 
genus 10 and a single rational curve in this case.

\section[An example: The weak coupling limit for base space
${\Hirz[4]}$]{An example: The weak coupling limit for base space
$\bf \Hirz[4]$} \label{reducible}

We have seen in Section~\ref{moreDoffO} that the suggested counting of F-theory 3-cycles holds in the case
of smooth Calabi-Yau threefolds. In this appendix, we want to discuss the weak
coupling limit of F-theory over a base $\Hirz[4]$. This surface is rational and
has a non-singular effective divisor $C = -2K_{\Hirz[4]}$. Thus it can be
obtained as the quotient space of a $K3$ surface $X$, where the quotient map is
a non-symplectic involution (see Appendix~\ref{nikulinClassification} for a
short review on the theory of such surfaces). However, the elliptically fibred
Calabi-Yau threefold is singular. 

Let us start with the F-theory perspective. We recall the Weierstrass equation 
\begin{equation}
y^2=x^3+fx+g \ ,
\end{equation}
where $f$ and $g$ are sections of $L^{\otimes 4}$ and $L^{\otimes 6}$,
respectively. Here $L$ is a line bundle defined by the Calabi-Yau condition
$c_1(L) = c_1(\Hirz[4])$. As discussed in Appendix~\ref{complexSurfaces}, the
first Chern class is given by
\begin{equation}
c_1(\Hirz[4])=6T_1+2T_4 \ ,
\end{equation}
and therefore we find
\begin{equation}
L=[T_1]^{\otimes 6}\otimes[T_4]^{\otimes 2}\ .
\end{equation}
This means that $f$ and $g$ are polynomials of degree $(24,8)$ and $(30,12)$.
We can write the general form of $f$ and $g$:
\begin{eqnarray}
f &=&
\left(\sum_{\alpha=0}^{6}\sum_{\beta=0}^{24-4\alpha}F_{\alpha\beta}z_1^\beta
z_2^{24-4\alpha-\beta}z_3^\alpha z_4^{6-\alpha}\right)z_4^2 \ , \\
g &=&
\left(\sum_{\alpha=0}^{7}\sum_{\beta=0}^{30-4\alpha}G_{\alpha\beta}z_1^\beta
z_2^{30-4\alpha-\beta}z_3^\alpha z_4^{7-\alpha}\right)z_4.
\end{eqnarray}
It was observed in \cite{Morrison:1996pp} that this elliptically fibred
Calabi-Yau threefold has generically a $D_4$ singularity. By Bertini's Theorem
\cite{Griffiths:1978}, this singularity is located at the base locus of the
threefold, which is $x=y=z_4=0$. Note that the base locus describes a
hyperplane in the base $\Hirz[4]$ given by $z_4=0$. Therefore we expect a
brane at $T_4$ that carries an $SO(8)$ gauge group. With this in mind we now turn
to the weak coupling limit.

In the weak coupling limit, the O-plane $O$ is described by the zero locus
of a polynomial $h$ of degree $(12,4)$. Its general form is
\begin{equation}
h=\left(\sum_{\alpha=0}^{3}\sum_{\beta=0}^{12-4\alpha}H_{\alpha\beta}z_1^\beta
z_2^{12-4\alpha-\beta}z_3^\alpha z_4^{4-\alpha}\right)z_4 \ .
\end{equation}
Note that $O$ is reducible. In particular, we find $O = O'+ T_4$.
This means that one O-plane splits from $O$ and coincide with $T_4$. Doing the
same analysis with the D-brane $\cal D$ which is given by the zero-locus of
$\eta^2+12h\chi$ we obtain that $\cal D$ is reducible as well. In particular, we
find ${\cal D} = {\cal D}'+4T_4$. We see that one O-plane and four D-branes
coincide with $T_4$ producing an $SO(8)$ gauge group on $T_4$ in
agreement with the results obtained in the F-theory picture.

By Nikulin's classification \cite{Nikulin:1986} we know that $\Hirz[4]$ has the
characteristic triplet $(2,0,0)$ and by the results in Appendix~\ref{nikulinClassification} this means that the fixed point locus, i.e.\ the
O-plane, is of the form
\begin{equation}
O = C_{10} + E_1 \ ,
\end{equation}
where $C_{10}$ is a curve of genus 10 and $E_1$ is a rational curve. This fits
with our results above by identifying $O'=C_{10}$ and $T_4=E_1$. Note
that $O$ is non-singular meaning that $C_{10}$ and $E_1$ do not intersect.

As we have seen, although the Calabi-Yau threefold is singular, F-theory on
$\Hirz[4]$ has a weak coupling limit. For $B=\Hirz[n]$ with
$n>4$ the corresponding elliptically fibred Calabi-Yau threefold has singularities of $E$-type which cannot appear in an orientifold model in perturbative type IIB . Indeed, $\Hirz[n\geq 5]$ do not have a $K3$ double
cover and thus there cannot exist a dual type IIB orientifold model. The base
$B=\Hirz[3]$ generically has an $A_3$-singularity, which might be obtained
in the perturbative type IIB orientifold model. However, $\Hirz[3]$ apparently does not admit a $K3$
double cover since the $O$ hypersurface one obtains in the weak
coupling limit is always singular \cite{Nikulin:1986}.

\section{Linear Algebra on Spaces with Indefinite Metric\label{app:linalg}}

Since some of the usual theorems about eigenvalues and eigenvectors of self-adjoint operators do not
carry over to the case of an indefinite scalar product, we collect some useful facts in this
appendix (see also \cite{indefMetric}). We consider a real vector space $\Vt$ equipped with a non-degenerate scalar product
$\left(\vt\cdot \wt\right)$ of signature $(n,m)$, where $n<m$ and $n$ refers to positive norm. In the
case we are interested in, $\Vt=H^2\!\left(\Kt\right)$ and the signature is $(3,19)$. Let $A$ be an
endomorphism of $\Vt$ which is selfadjoint with respect to this scalar product. We denote the set of
eigenvalues of $A$ by $\left\{\lambda_i\right\}$. Since the eigenvalues are the roots
of the real characteristic polynomial, they are either real or come in complex conjugate pairs. We
consider the complexification $\Vt_\mathbb{C}$ of $\Vt$, such that the scalar product involves complex
conjugation of the first entry.

In $\Vt_\mathbb{C}$, $A$ has $n+m$ eigenvalues. Note that a self-adjoint operator $A$ is not
necessarily diagonalizable in a space with indefinite metric.
However, this problem only occurs if there exists a zero-norm eigenvector
relative to a degenerate eigenvalue \cite{Pandit}. We will not consider this
non-generic case. Then $A$ is diagonalizable in $\Vt_\mathbb{C}$ with eigenvectors given by $\{e_i\}$.  From the selfadjointness, we have
\begin{align}\label{D1} 
  \left(\bar{\lambda}_i-\lambda_j\right)\left(e_i\cdot e_j\right)=0\,.
\end{align}
Since the metric is indefinite, $\left(e_i\cdot e_i\right)=0$
does not imply $e_i=0$, so that not all eigenvalues need to be real.

If there exist one non-real eigenvalue $\lambda$ with eigenvector $e$, then $\bar{\lambda}$ is also an eigenvalue.The corresponding eigenvector is $\bar{e}$. Equation \eqref{D1} tells us that $e$ and $\bar{e}$ are null. In the case we are considering, $\lambda$ is non-degenerate. Then, the non-degeneracy of the inner product implies $(\bar{e},e)\not =0$.
With these vector we can construct two real vectors
\begin{align}
  \vt^+&= e+ \bar{e}\,, & \vt^-&= -\I\left(e - \bar{e}\right)
\end{align}
that have opposite norm. Then, $\vt^\pm$ generate a subspace of the original real space $\Vt$, such that the scalar product on
this subspace is of signature $(1,1)$. One can define the orthogonal complement of this subspace in $\Vt$ and look for the next complex eigenvalue and the corresponding $2\times 2$ block. There can be at most $n$ of these $2\times 2$ blocks. Then there are at least $m-n$ real eigenvalues.

We conclude that the canonical form of a generic matrix $A$ selfadjoint with respect to a indefinite inner product with signature $(n,m)$ is block diagonal, with $n$ $2\times 2$ block relative to subspaces of signature $(1,1)$ and a positive definite $(m-n)$-diagonal block\footnote{
A matrix selfadjoint with respect to a definite metric is positive definite}. Vectors belonging to different blocks are orthogonal to each other.

Let us concentrate on a $2\times 2$ block. We choose a basis such that the metric has the matrix form
\begin{equation}
\Mt=\begin{pmatrix}
  0 &1\\  1& 0
\end{pmatrix}\:.
\end{equation}
The selfadjointness condition on $A$ is $A\,\Mt=\Mt\,A^T$, implying that
\begin{equation}
A=\begin{pmatrix}
  a &b\\  c& a
\end{pmatrix}\:.
\end{equation}
With a transformation that leaves $\Mt$ invariant, A can be brought to the canonical form\footnote{If $b,c$ are either both zero or both non-zero. Otherwise, the matrix is of the form we said before: It has a degenerate real eigenvalue relative to a zero norm eigenvector.}
\begin{equation}
A'=\begin{pmatrix}
  a &b\\  b& a
\end{pmatrix} \mbox{ or }
A'=\begin{pmatrix}
  a &-b\\  b& a
\end{pmatrix}\:.
\end{equation}
If we now change basis with the matrix $P=\frac{1}{\sqrt{2}}\begin{pmatrix} 1&1 \\ 1&-1  \end{pmatrix}$, then $M$ and $A$ go to:
\begin{equation}
\Mt=\begin{pmatrix}\label{Mt2x2}
  1 &0\\  0& -1
\end{pmatrix}\:, \qquad
A'=\begin{pmatrix}
   \lambda_1 &0\\  0& \lambda_2
\end{pmatrix} \mbox{ or }
A'=\begin{pmatrix}
  a &b\\  -b& a
\end{pmatrix}\:.
\end{equation}

\

Let us now specialize to the case of $A=G^a G$, i.e.~$V$ is another vector space, equipped with
a scalar product of the same signature, and $G$ is a map from $\Vt$ to $V$. $G^a$ denotes its adjoint
with respect to these scalar products, i.e.\ $\left(v,G \vt\right)=\left(G^a v,\vt\right)$ (where
$\vt\in \Vt$ and $v\in V$). Clearly, the composition $G^a G$ is a selfadjoint map from $\Vt$ to itself.

We want to determine the canonical form for $G$. It will be of the same structure of $A$, with $n$ $2\times 2$ blocks of signature $(1,1)$ and a diagonal part relative to a metric in the form $-{\mathbf 1}_{m-n}$. The diagonal part will be simply given by the square root of the diagonal block of $A$. 
Regarding the $2\times 2$ blocks, we find that both canonical forms can be written as $A'=g^a g$ with a ``square root'' matrix $g$.
%We take also for the corresponding vectors in $V$ a metric $M$ of the form \eqref{Mt2x2}. 
Since $A$ is of the form $G^aG$, the eigenvalues $\lambda_1,\lambda_2$ in \eqref{Mt2x2} must be either both positive or both negative. We consider these two cases separately. The canonical forms for $g$ are
\begin{align}
      g=\begin{pmatrix}
        \sqrt{\lambda_1} &0\\  0& \sqrt{\lambda_2}
      \end{pmatrix}\:,\qquad
      g=\begin{pmatrix}
        0& \sqrt{\left|\lambda_2\right|}\\ \sqrt{\left|\lambda_1\right|}& 0
      \end{pmatrix}\:,\quad
      g=\begin{pmatrix}
        \gamma & \delta\\-\delta & \gamma\\
      \end{pmatrix}\:,
\end{align}
where in the last matrix we have defined $\gamma$ and $\delta$ such that $\alpha=\gamma^2-\delta^2$ and $\beta=2\gamma\delta$.

Then, the matrix of $G$ can be brought with a change of basis into the form:
\begin{align}
  G_d&=
  \begin{pmatrix}
    g_1 &&&&&\\
    & \ddots&&&&\\
    & & g_n &&& \\
    &&& \sqrt{\lambda_1} &&\\
    &&&&\ddots  &\\
    &&&&&  \sqrt{\lambda_{n-m}}
  \end{pmatrix}\:.
\end{align}

If we call the matrix of the change of basis $\Pt$, then we can summarize our results as:
\begin{equation}
 \Pt^{-1} G^a G \Pt = G_d^aG_d \qquad,\qquad \Pt^T \Mt \Pt = \mathbb{M}
\end{equation}
where $\mathbb{M}$ is the diagonal matrix given by $n$ $2\times 2$ blocks $(+1,-1)$ and an $m-n$ block $(-1,...,-1)$.

\

We now show that there exists a change of basis in the space $V$ such that the matrix of $G$ can be brought to the form $G_d$, i.e. there exists a matrix $P$ such that 
\begin{equation}
P^{-1}G\Pt = G_d \:. 
\end{equation}
This matrix is given by $ P \equiv {G^a}^{-1} \Pt G_d^a$. Let us check that:
\begin{align}
 P^{-1} G \Pt = {G_d^a}^{-1}\Pt^{-1}G^aG\Pt = {G_d^a}^{-1}G_d^aG_d =G_d
\end{align}
Moreover, we obtain the relations:
\begin{equation}
 \Pt^{-1}G^a P = G_d^a \qquad\qquad P^{-1} GG^a P = G_dG_d^a \qquad\qquad P^T M P = \mathbb{M} \:.
\end{equation}

Only in the case of all eigenvalues being positive do we get a fully diagonal form for $G$, otherwise we
have non-diagonal $2\times2$ blocks.

Returning to the potential (and to the $K3$ case where $n=3$ and $m=19$), we see that if $G^aG$ is diagonalizable with non-negative eigenvalues, then $G$ and $G^a$ can be brought to the same diagonal form $G_d$ with respect to bases made up of three positive norm and nineteen negative norm vectors. This means that the minimum condition \eqref{eq:planetoplane} is satisfied. The converse is also true: If the condition \eqref{eq:planetoplane} is satisfied, then $G$ and $G^a$ can be brought to a diagonal form by changes of bases and so $G^aG$ becomes diagonal with non-negative entries.

\section[Explicit expressions for the $\sigma_{ij}^k$]{Explicit expressions for the \boldmath$\sigma_{ij}^k$}\label{appa}
In this appendix, we give explicit expression for the integral cycles $\sigma_{ij}^k$
appearing in Chapter~\ref{chapterk3}. 
{\small \begin{align}
\sigma_{14}^1  &=\frac{1}{2}\left(\pi_{14}-C_3-C_8-C_{12}-C_{15}\right)=e_3+E_4+E_8\nonumber \\
\sigma_{14}^2  &=\frac{1}{2}\left(\pi_{14}-C_4-C_7-C_{11}-C_{16}\right)=e_2+E_{12}+E_{16}\nonumber \\
\sigma_{14}^3  &=\frac{1}{2}\left(\pi_{14}-C_2-C_5-C_{9}-C_{14}\right) =e_1+e_3+e^2+e^3 \nonumber \\
&+\frac{1}{2}\left(E_1-E_2-E_3+E_4+E_5-E_6-E_7+E_8\right)\nonumber \\
&+\frac{1}{2}\left(-E_9-E_{10}-E_{11}+E_{12}-E_{13}-E_{14}-E_{15}+E_{16}\right) \nonumber \\
\sigma_{14}^4  &=\frac{1}{2}\left(\pi_{14}-C_1-C_6-C_{10}-C_{13}\right) =-e_1+e_2+e^3+e^2 \nonumber \\
&+\frac{1}{2}\left(-E_1-E_2-E_3+E_4-E_5-E_6-E_7+E_8\right)\nonumber \\
&+\frac{1}{2}\left(E_9-E_{10}-E_{11}+E_{12}+E_{13}-E_{14}-E_{15}+E_{16}\right).\\
\nonumber \\
\sigma_{23}^1 & =\frac{1}{2}\left(\pi_{23}-C_5-C_6-C_{7}-C_{8}\right)=E_7-E_6\nonumber \\
\sigma_{23}^2 & =\frac{1}{2}\left(\pi_{23}-C_1-C_2-C_{3}-C_{4}\right)=E_3-E_2\nonumber \\
\sigma_{23}^3 & =\frac{1}{2}\left(\pi_{23}-C_9-C_{10}-C_{11}-C_{12}\right)=E_{11}-E_{10}\nonumber \\
\sigma_{23}^4 & =\frac{1}{2}\left(\pi_{23}-C_{13}-C_{14}-C_{15}-C_{16}\right)=E_{15}-E_{14}\\
\nonumber \\
\sigma_{12}^1 & =\frac{1}{2}\left(\pi_{12}-C_1-C_3-C_{6}-C_{8}\right) \nonumber \\
&=\frac{1}{2}\left( -E_1-E_2+E_3+E_4-E_5-E_6+E_7+E_8\right)\nonumber \\
\sigma_{12}^2 & =\frac{1}{2}\left(\pi_{12}-C_2-C_4-C_{5}-C_{7}\right) \nonumber\\
&=e_1-e^2-e^3+\frac{1}{2}\left( E_1-E_2+E_3-E_4+E_5-E_6+E_7-E_8 \right)\nonumber \\
\sigma_{12}^3 & =\frac{1}{2}\left(\pi_{12}-C_9-C_{11}-C_{14}-C_{16}\right)\nonumber\\
&=e_1+\frac{1}{2}\left( -E_9-E_{10}+E_{11}+E_{12}-E_{13}-E_{14}+E_{15}+E_{16}\right)\nonumber \\
\sigma_{12}^4 & =\frac{1}{2}\left(\pi_{12}-C_{10}-C_{12}-C_{13}-C_{15}\right)\nonumber \\
&=-e^2-e^3+\frac{1}{2}\left( E_9-E_{10}+E_{11}-E_{12}+E_{13}-E_{14}+E_{15}-E_{16}\right)
\end{align}
\begin{align}
\sigma_{34}^1 & =\frac{1}{2}\left(\pi_{34}-C_5-C_6-C_{13}-C_{14}\right)=e_1+e^1+E_1-E_6-E_9-E_{14}\nonumber \\
\sigma_{34}^2 & =\frac{1}{2}\left(\pi_{34}-C_1-C_2-C_{9}-C_{10}\right)=e_1+e^1+E_1-E_2-E_{9}-E_{10}\nonumber \\
\sigma_{34}^3 & =\frac{1}{2}\left(\pi_{34}-C_3-C_{4}-C_{11}-C_{12}\right)=e_1+e^1-e^2-e^3+E_1+E_3-E_{9}+E_{11}\nonumber \\
\sigma_{34}^4 & =\frac{1}{2}\left(\pi_{34}-C_{7}-C_{8}-C_{15}-C_{16}\right)=e_1+e^1-e^2-e^3+E_1+E_7-E_{9}+E_{15}\\
\nonumber \\
\sigma_{13}^1 & =\frac{1}{2}\left(\pi_{13}-C_1-C_2-C_{5}-C_{6}\right)=e_2-e_3-E_2-E_6\nonumber \\
\sigma_{13}^2 & =\frac{1}{2}\left(\pi_{13}-C_9-C_{10}-C_{13}-C_{14}\right)=-E_{10}-E_{14}\nonumber \\
\sigma_{13}^3 & =\frac{1}{2}\left(\pi_{13}-C_{11}-C_{12}-C_{15}-C_{16}\right)=-e^2-e^3+e_2-e_3+E_{11}+E_{15}\nonumber \\
\sigma_{13}^4 & =\frac{1}{2}\left(\pi_{13}-C_{3}-C_{4}-C_{7}-C_{8}\right)=-e^2-e^3+E_3+E_7\\
\nonumber \\
\sigma_{42}^1 & =\frac{1}{2}\left(\pi_{42}-C_6-C_8-C_{13}-C_{15}\right)=e_2-e_3-e^3-E_5-E_6+E_{13}+E_{15}\nonumber \\
\sigma_{42}^2 & =\frac{1}{2}\left(\pi_{42}-C_5-C_7-C_{14}-C_{16}\right)=-e^3+e_2-e_3-E_6-E_8+E_{15}+E_{16}\nonumber \\
\sigma_{42}^3 & =\frac{1}{2}\left(\pi_{42}-C_2-C_4-C_{9}-C_{11}\right) = e_1+e_2-e_3-e^3 \nonumber \\
&+\frac{1}{2}\left(E_1-E_2+E_3-E_4-E_5-E_6-E_7-E_8\right)\nonumber \\
&+\frac{1}{2}\left(-E_9-E_{10}+E_{11}+E_{12}+E_{13}+E_{14}+E_{15}+E_{16}\right) \nonumber \\
\sigma_{42}^4 & =\frac{1}{2}\left(\pi_{42}-C_1-C_3-C_{10}-C_{12}\right) =-e_1+e_2-e_3-e^3 \nonumber \\
&+\frac{1}{2}\left(-E_1-E_2+E_3+E_4-E_5-E_6-E_7-E_8\right)\nonumber \\
&+\frac{1}{2}\left(E_9-E_{10}+E_{11}-E_{12}+E_{13}+E_{14}+E_{15}+E_{16}\right).
\end{align}}

\end{appendix}


\begin{thebibliography}{99}


%\cite{Polchinski:1998rq}
\bibitem{Polchinski:1998rq}
  J.~Polchinski,
  ``String theory. Vol. 1: An introduction to the bosonic string,''
%\href{http://www.slac.stanford.edu/spires/find/hep/www?irn=4634799}{SPIRES entry}
{\it  Cambridge, UK: Univ. Pr. (1998) 402 p}
  ``String theory. Vol. 2: Superstring theory and beyond,''
%\href{http://www.slac.stanford.edu/spires/find/hep/www?irn=4634802}{SPIRES entry}
{\it  Cambridge, UK: Univ. Pr. (1998) 531 p}



%\cite{Maldacena:1997re}
\bibitem{Maldacena:1997re}
  J.~M.~Maldacena,
  ``The large N limit of superconformal field theories and supergravity,''
  Adv.\ Theor.\ Math.\ Phys.\  {\bf 2} (1998) 231
  [Int.\ J.\ Theor.\ Phys.\  {\bf 38} (1999) 1113]
  [arXiv:hep-th/9711200].
  %%CITATION = IJTPB,38,1113;%%

 \bibitem{kontsevich-1994}
M.~Kontsevich,``Homological Algebra of Mirror Symmetry'',
[arXiv.org:alg-geom/9411018]



%\cite{Johnson:2000ch}
\bibitem{Johnson:2000ch}
  C.~V.~Johnson,
  ``D-brane primer,''
  arXiv:hep-th/0007170.
  %%CITATION = HEP-TH/0007170;%%



%\cite{Lust:2004ks}
\bibitem{Lust:2004ks}
  D.~Lust,
  ``Intersecting brane worlds: A path to the standard model?,''
  Class.\ Quant.\ Grav.\  {\bf 21} (2004) S1399
  [arXiv:hep-th/0401156].
  %%CITATION = CQGRD,21,S1399;%%



%\cite{Douglas:2006es}
\bibitem{Douglas:2006es}
  M.~R.~Douglas and S.~Kachru,
  ``Flux compactification,''
  Rev.\ Mod.\ Phys.\  {\bf 79} (2007) 733
  [arXiv:hep-th/0610102].
  %%CITATION = RMPHA,79,733;%%

%\cite{Grana:2005jc}
\bibitem{Grana:2005jc}
  M.~Grana,
  ``Flux compactifications in string theory: A comprehensive review,''
  Phys.\ Rept.\  {\bf 423}, 91 (2006)
  [arXiv:hep-th/0509003].
  %%CITATION = PRPLC,423,91;%%

%\cite{Blumenhagen:2006ci}
\bibitem{Blumenhagen:2006ci}
  R.~Blumenhagen, B.~Kors, D.~Lust and S.~Stieberger,
  ``Four-dimensional String Compactifications with D-Branes, Orientifolds   and
  Fluxes,''
  Phys.\ Rept.\  {\bf 445}, 1 (2007)
  [arXiv:hep-th/0610327].
  %%CITATION = PRPLC,445,1;%%


%\cite{Denef:2007pq}
\bibitem{Denef:2007pq}
  F.~Denef, M.~R.~Douglas and S.~Kachru,
  ``Physics of string flux compactifications,''
  Ann.\ Rev.\ Nucl.\ Part.\ Sci.\  {\bf 57} (2007) 119
  [arXiv:hep-th/0701050].
  %%CITATION = ARNUA,57,119;%%


%\cite{Denef:2004dm}
\bibitem{Denef:2004dm}
  F.~Denef, M.~R.~Douglas and B.~Florea,
  ``Building a better racetrack,''
  JHEP {\bf 0406} (2004) 034
  [arXiv:hep-th/0404257].
  %%CITATION = JHEPA,0406,034;%%


%\cite{Aspinwall:1993nu}
\bibitem{Aspinwall:1993nu}
  P.~S.~Aspinwall, B.~R.~Greene and D.~R.~Morrison,
  ``Calabi-Yau moduli space, mirror manifolds and spacetime topology  change in
  %string theory,''
  Nucl.\ Phys.\  B {\bf 416} (1994) 414
  [arXiv:hep-th/9309097].
  %%CITATION = NUPHA,B416,414;%%

%\cite{Strominger:1995cz}
\bibitem{Strominger:1995cz}
  A.~Strominger,
  ``Massless black holes and conifolds in string theory,''
  Nucl.\ Phys.\  B {\bf 451} (1995) 96
  [arXiv:hep-th/9504090].
  %%CITATION = NUPHA,B451,96;%%


\bibitem{Reid}
M.~Reid, ``The moduli space of 3-folds with K = 0 may nevertheless be irreducible'',
     Math. Ann. 278 (1987) 329–334.


%\cite{Avram:1995pu}
\bibitem{Avram:1995pu}
  A.~C.~Avram, P.~Candelas, D.~Jancic and M.~Mandelberg,
  ``On the Connectedness of Moduli Spaces of Calabi-Yau Manifolds,''
  Nucl.\ Phys.\  B {\bf 465} (1996) 458
  [arXiv:hep-th/9511230].
  %%CITATION = NUPHA,B465,458;%%
 
%\cite{Kreuzer:2000xy}
\bibitem{Kreuzer:2000xy}
  M.~Kreuzer and H.~Skarke,
  ``Complete classification of reflexive polyhedra in four dimensions,''
  Adv.\ Theor.\ Math.\ Phys.\  {\bf 4} (2002) 1209
  [arXiv:hep-th/0002240].
  %%CITATION = 00203,4,1209;%%

%\cite{Aspinwall:2005qw}
\bibitem{Aspinwall:2005qw}
  P.~S.~Aspinwall,
  ``An analysis of fluxes by duality,''
  arXiv:hep-th/0504036.
  %%CITATION = HEP-TH/0504036;%%

%\cite{Lerche:1986cx}
\bibitem{Lerche:1986cx}
  W.~Lerche, D.~Lust and A.~N.~Schellekens,
  ``Chiral Four-Dimensional Heterotic Strings from Selfdual Lattices,''
 Nucl.\ Phys.\  B {\bf 287} (1987) 477.
  %%CITATION = NUPHA,B287,477;%%



%\cite{Douglas:2006za}
\bibitem{Douglas:2006za}
  M.~R.~Douglas,
  ``Understanding the landscape,''
  arXiv:hep-th/0602266.
  %%CITATION = HEP-TH/0602266;%%


%\cite{Ashok:2003gk}
\bibitem{Ashok:2003gk}
  S.~Ashok and M.~R.~Douglas,
  ``Counting flux vacua,''
  JHEP {\bf 0401} (2004) 060
  [arXiv:hep-th/0307049].
  %%CITATION = JHEPA,0401,060;%%

%\cite{Douglas:2004kp}
\bibitem{Douglas:2004kp}
  M.~R.~Douglas,
  ``Statistics of string vacua,''
  arXiv:hep-ph/0401004.
  %%CITATION = HEP-PH/0401004;%%

%\cite{Denef:2004ze}
\bibitem{Denef:2004ze}
  F.~Denef and M.~R.~Douglas,
  ``Distributions of flux vacua,''
  JHEP {\bf 0405} (2004) 072
  [arXiv:hep-th/0404116].
  %%CITATION = JHEPA,0405,072;%%


%\cite{Blumenhagen:2004xx}
\bibitem{Blumenhagen:2004xx}
  R.~Blumenhagen, F.~Gmeiner, G.~Honecker, D.~Lust and T.~Weigand,
  ``The statistics of supersymmetric D-brane models,''
  Nucl.\ Phys.\  B {\bf 713} (2005) 83
  [arXiv:hep-th/0411173].
  %%CITATION = NUPHA,B713,83;%%


%\cite{Gmeiner:2005vz}
\bibitem{Gmeiner:2005vz}
  F.~Gmeiner, R.~Blumenhagen, G.~Honecker, D.~Lust and T.~Weigand,
  ``One in a billion: MSSM-like D-brane statistics,''
  JHEP {\bf 0601} (2006) 004
  [arXiv:hep-th/0510170].
  %%CITATION = JHEPA,0601,004;%%


%\cite{Denef:2008wq}
\bibitem{Denef:2008wq}
  F.~Denef,
  ``Les Houches Lectures on Constructing String Vacua,''
  arXiv:0803.1194 [hep-th].
  %%CITATION = ARXIV:0803.1194;%%
%\cite{Gasperini:2007zz}
\bibitem{Gasperini:2007zz}
  M.~Gasperini,
  ``Elements of string cosmology,''
%\href{http://www.slac.stanford.edu/spires/find/hep/www?irn=8402965}{SPIRES entry}
{\it  Cambridge, UK: Cambridge Univ. Pr. (2007) 552 p}

%\cite{Erdmenger:1900zz}
\bibitem{Erdmenger}
J.~Erdmenger (ed.)
``String Cosmology'',
Wiley-VCH, Berlin (2009)

%\cite{Bousso:2000xa}
\bibitem{Bousso:2000xa}
  R.~Bousso and J.~Polchinski,
  ``Quantization of four-form fluxes and dynamical neutralization of the
  cosmological constant,''
  JHEP {\bf 0006} (2000) 006
  [arXiv:hep-th/0004134].
  %%CITATION = JHEPA,0006,006;%%

%\cite{Kachru:2003aw}
\bibitem{Kachru:2003aw}
  S.~Kachru, R.~Kallosh, A.~D.~Linde and S.~P.~Trivedi,
  ``De Sitter vacua in string theory,''
  Phys.\ Rev.\  D {\bf 68} (2003) 046005
  [arXiv:hep-th/0301240].
  %%CITATION = PHRVA,D68,046005;%%

%\cite{Becker:2007zj}
\bibitem{Becker:2007zj}
  K.~Becker, M.~Becker and J.~H.~Schwarz,
  ``String theory and M-theory: A modern introduction,''
%\href{http://www.slac.stanford.edu/spires/find/hep/www?irn=7073143}{SPIRES entry}
{\it  Cambridge, UK: Cambridge Univ. Pr. (2007) 739 p}
%\cite{Baumann:2009ni}
\bibitem{Baumann:2009ni}
  D.~Baumann and L.~McAllister,
  ``Advances in Inflation in String Theory,''
  arXiv:0901.0265 [hep-th].
  %%CITATION = ARXIV:0901.0265;%%
%\cite{Acharya:2006zw}
\bibitem{Acharya:2006zw}
  B.~S.~Acharya and M.~R.~Douglas,
  ``A finite landscape?,''
  arXiv:hep-th/0606212.
  %%CITATION = HEP-TH/0606212;%%
%\cite{Klebanov:2000hb}
\bibitem{Klebanov:2000hb}
  I.~R.~Klebanov and M.~J.~Strassler,
  ``Supergravity and a confining gauge theory: Duality cascades and
  chiSB-resolution of naked singularities,''
  JHEP {\bf 0008} (2000) 052
  [arXiv:hep-th/0007191].
  %%CITATION = JHEPA,0008,052;%%
\bibitem{gkp01}
  S.~B.~Giddings, S.~Kachru and J.~Polchinski,
  ``Hierarchies from fluxes in string compactifications,''
  Phys.\ Rev.\  D {\bf 66} (2002) 106006
  {\ttfamily [arXiv:hep-th/0105097]}
  %%CITATION = PHRVA,D66,106006;%%
%\cite{Verlinde:1999fy}
\bibitem{Verlinde:1999fy}
  H.~L.~Verlinde,
  ``Holography and compactification,''
  Nucl.\ Phys.\  B {\bf 580} (2000) 264
  [arXiv:hep-th/9906182].
  %%CITATION = NUPHA,B580,264;%%

%\cite{Chan:2000ms}
\bibitem{Chan:2000ms}
  C.~S.~Chan, P.~L.~Paul and H.~L.~Verlinde,
  ``A note on warped string compactification,''
  Nucl.\ Phys.\  B {\bf 581} (2000) 156
  [arXiv:hep-th/0003236].
  %%CITATION = NUPHA,B581,156;%%


%\cite{Brummer:2005sh}
\bibitem{Brummer:2005sh}
  F.~Brummer, A.~Hebecker and E.~Trincherini,
  ``The throat as a Randall-Sundrum model with Goldberger-Wise
  stabilization,''
  Nucl.\ Phys.\  B {\bf 738} (2006) 283
  [arXiv:hep-th/0510113].
  %%CITATION = NUPHA,B738,283;%%
%\cite{Randall:1999ee}
\bibitem{Randall:1999ee}
  L.~Randall and R.~Sundrum,
  ``A large mass hierarchy from a small extra dimension,''
  Phys.\ Rev.\ Lett.\  {\bf 83} (1999) 3370
  [arXiv:hep-ph/9905221].
  %%CITATION = PRLTA,83,3370;%%
%\cite{Hebecker:2006bn}
\bibitem{Hebecker:2006bn}
  A.~Hebecker and J.~March-Russell,
  ``The ubiquitous throat,''
  Nucl.\ Phys.\  B {\bf 781} (2007) 99
  [arXiv:hep-th/0607120].
  %%CITATION = NUPHA,B781,99;%%
%\cite{Grimm:2004uq}
\bibitem{Grimm:2004uq}
  T.~W.~Grimm and J.~Louis,
  ``The effective action of N = 1 Calabi-Yau orientifolds,''
  Nucl.\ Phys.\  B {\bf 699} (2004) 387
  [arXiv:hep-th/0403067].
  %%CITATION = NUPHA,B699,387;%%
%\cite{Jockers:2004yj}
\bibitem{Jockers:2004yj}
  H.~Jockers and J.~Louis,
  ``The effective action of D7-branes in N = 1 Calabi-Yau orientifolds,''
  Nucl.\ Phys.\  B {\bf 705} (2005) 167
  [arXiv:hep-th/0409098].
  %%CITATION = NUPHA,B705,167;%%
%\cite{Collinucci:2008pf}
\bibitem{Collinucci:2008pf}
  A.~Collinucci, F.~Denef and M.~Esole,
  ``D-brane Deconstructions in IIB Orientifolds,''
  JHEP {\bf 0902} (2009) 005
  [arXiv:0805.1573 [hep-th]].
  %%CITATION = JHEPA,0902,005;%%
%\cite{Gomis:2005wc}
\bibitem{Gomis:2005wc}
  J.~Gomis, F.~Marchesano and D.~Mateos,
  ``An open string landscape,''
  JHEP {\bf 0511}, 021 (2005)
  [arXiv:hep-th/0506179].
  %%CITATION = JHEPA,0511,021;%%

%\cite{Douglas:2006xy}
\bibitem{Douglas:2006xy}
  M.~R.~Douglas and W.~Taylor,
  ``The landscape of intersecting brane models,''
  JHEP {\bf 0701}, 031 (2007)
  [arXiv:hep-th/0606109].
  %%CITATION = JHEPA,0701,031;%%
\bibitem{Dbranestorus1}
R.~Blumenhagen, L.~Goerlich, B.~K\"ors and D.~L\"ust,
  ``Noncommutative compactifications of type I strings on tori with  magnetic
  background flux,''
  JHEP {\bf 0010} (2000) 006
  [arXiv:hep-th/0007024];\\
\bibitem{Dbranestorus2}
C.~Angelantonj, I.~Antoniadis, E.~Dudas and A.~Sagnotti,
  ``Type-I strings on magnetised orbifolds and brane transmutation,''
  Phys.\ Lett.\ B {\bf 489} (2000) 223
  [arXiv:hep-th/0007090];\\
\bibitem{Dbranestorus3}
G.~Aldazabal, S.~Franco, L.~E.~Ibanez, R.~Rabadan and A.~M.~Uranga,
  ``D = 4 chiral string compactifications from intersecting branes,''
  J.\ Math.\ Phys.\  {\bf 42} (2001) 3103
  [arXiv:hep-th/0011073].
%\cite{Greene:1989ya}
\bibitem{Greene:1989ya}
  B.~R.~Greene, A.~D.~Shapere, C.~Vafa and S.~T.~Yau,
  ``Stringy Cosmic Strings And Noncompact Calabi-Yau Manifolds,''
  Nucl.\ Phys.\  B {\bf 337} (1990) 1.
  %%CITATION = NUPHA,B337,1;%%
%\cite{Vafa:1996xn}
\bibitem{Vafa:1996xn}
  C.~Vafa,
  ``Evidence for F-Theory,''
  Nucl.\ Phys.\  B {\bf 469} (1996) 403
  [arXiv:hep-th/9602022].
  %%CITATION = NUPHA,B469,403;%%
\bibitem{riemannsurfaces}
 R.~Narasimhan, ``Compact Riemann Surfaces'', Birkh\"auser Basel, Berlin (1996)

\bibitem{Chandrasekharan:1985}
  K.~Chandrasekharan,
  ``Elliptic Functions,''
{\it Berlin, Germany: Springer Verlag (1985) 189 p}

\bibitem{Gibbons:1995vg}
  G.~W.~Gibbons, M.~B.~Green and M.~J.~Perry,
  ``Instantons and Seven-Branes in Type IIB Superstring Theory,''
  Phys.\ Lett.\  B {\bf 370}, 37 (1996)
  [arXiv:hep-th/9511080].
%%CITATION = PHLTA,B370,37;%%
%\cite{Grimm:2009ef}


\bibitem{openperiods}

  M.~Alim, M.~Hecht, H.~Jockers, P.~Mayr, A.~Mertens and M.~Soroush,
  ``Hints for Off-Shell Mirror Symmetry in type II/F-theory
  Compactifications,''
  arXiv:0909.1842 [hep-th].
  %%CITATION = ARXIV:0909.1842;%%

  M.~Aganagic and C.~Beem,
  ``The Geometry of D-Brane Superpotentials,''
  arXiv:0909.2245 [hep-th].
  %%CITATION = ARXIV:0909.2245;%%

    S.~Li, B.~H.~Lian, and S.-T.~Yau,
    ``Picard-Fuchs Equations for Relative Periods and Abel-Jacobi Map for Calabi-Yau Hypersurfaces'',
    arXiv:0910.4215.
%%CITATION = 0910.4215;%%"

%\cite{Grimm:2009sy}
  T.~W.~Grimm, T.~W.~Ha, A.~Klemm and D.~Klevers,
  {\it Five-Brane Superpotentials and Heterotic/F-theory Duality},
  arXiv:0912.3250 [Unknown].
  %%CITATION = ARXIV:0912.3250;%%

%\cite{Jockers:2009ti}
  H.~Jockers, P.~Mayr and J.~Walcher,
  {\it On N=1 4d Effective Couplings for F-theory and Heterotic Vacua},
  arXiv:0912.3265 [Unknown].
  %%CITATION = ARXIV:0912.3265;%%

\bibitem{Witten:1995ex}
  E.~Witten,
  ``String theory dynamics in various dimensions,''
  Nucl.\ Phys.\  B {\bf 443}, 85 (1995)
  [arXiv:hep-th/9503124].
%%CITATION = NUPHA,B443,85;%%

\bibitem{bb96}
  K.~Becker and M.~Becker,
  ``M-Theory on Eight-Manifolds,''
  Nucl.\ Phys.\  B {\bf 477} (1996)~155
  {\ttfamily [arXiv:hep-th/9605053]}
  %%CITATION = NUPHA,B477,155;%%

%%%%%%%%%%%%% STATEMENT THAT m with flux is warped CY:
%\cite{Becker:2001pm}
\bibitem{Becker:2001pm}
  K.~Becker and M.~Becker,
  ``Supersymmetry breaking, M-theory and fluxes,''
  JHEP {\bf 0107} (2001) 038
  [arXiv:hep-th/0107044].
  %%CITATION = JHEPA,0107,038;%%

\bibitem{hl01}
  M.~Haack and J.~Louis,
  ``M-theory compactified on Calabi-Yau fourfolds with background flux,''
  Phys.\ Lett.\  B {\bf 507} (2001) 296
  {\ttfamily [arXiv:hep-th/0103068]}
  %%CITATION = PHLTA,B507,296;%%

\bibitem{gvw99}
  S.~Gukov, C.~Vafa and E.~Witten,
  ``CFT's from Calabi-Yau four-folds,''
  Nucl.\ Phys.\  B {\bf 584}, 69 (2000)
  [Erratum-ibid.\  B {\bf 608}, 477 (2001)]
  {\ttfamily [arXiv:hep-th/9906070]}
  %%CITATION = NUPHA,B584,69;%%

\bibitem{Gorlich:2004qm}
  L.~Gorlich, S.~Kachru, P.~K.~Tripathy and S.~P.~Trivedi,
  ``Gaugino condensation and nonperturbative superpotentials in flux
  compactifications,''
  JHEP {\bf 0412}, 074 (2004)
  [arXiv:hep-th/0407130].
%%CITATION = JHEPA,0412,074;%%

\bibitem{drs99}
  K.~Dasgupta, G.~Rajesh and S.~Sethi,
  ``M theory, orientifolds and G-flux,''
  JHEP {\bf 9908} (1999) 023
  {\ttfamily [arXiv:hep-th/9908088]}
  %%CITATION = JHEPA,9908,023;%%
%\cite{Donagi:2008ca}
\bibitem{Donagi:2008ca}
  R.~Donagi and M.~Wijnholt,
  ``Model Building with F-Theory,''
  arXiv:0802.2969 [hep-th].
  %%CITATION = ARXIV:0802.2969;%%

\bibitem{Beasley:2008dc}
  C.~Beasley, J.~J.~Heckman and C.~Vafa,
  ``GUTs and Exceptional Branes in F-theory - I,''
  arXiv:0802.3391 [hep-th].
  %%CITATION = ARXIV:0802.3391;%%

\bibitem{Beasley:2008kw}
  C.~Beasley, J.~J.~Heckman and C.~Vafa,
  ``GUTs and Exceptional Branes in F-theory - II: Experimental Predictions,''
  arXiv:0806.0102 [hep-th].
  %%CITATION = ARXIV:0806.0102;%%

%\cite{Donagi:2008kj}
\bibitem{Donagi:2008kj}
  R.~Donagi and M.~Wijnholt,
  ``Breaking GUT Groups in F-Theory,''
  arXiv:0808.2223 [hep-th].
  %%CITATION = ARXIV:0808.2223;%%

%\cite{Donagi:2009ra}
\bibitem{Donagi:2009ra}
  R.~Donagi and M.~Wijnholt,
  ``Higgs Bundles and UV Completion in F-Theory,''
  arXiv:0904.1218 [hep-th].
  %%CITATION = ARXIV:0904.1218;%%

%\cite{Blumenhagen:2007zk}
\bibitem{Blumenhagen:2007zk}
  R.~Blumenhagen, M.~Cvetic, D.~Lust, R.~2.~Richter and T.~Weigand,
  ``Non-perturbative Yukawa Couplings from String Instantons,''
  Phys.\ Rev.\ Lett.\  {\bf 100} (2008) 061602
  [arXiv:0707.1871 [hep-th]].
  %%CITATION = PRLTA,100,061602;%%

%\cite{Heckman:2008qt}
\bibitem{Heckman:2008qt}
  J.~J.~Heckman and C.~Vafa,
  ``F-theory, GUTs, and the Weak Scale,''
  JHEP {\bf 0909} (2009) 079
  [arXiv:0809.1098 [hep-th]].
  %%CITATION = JHEPA,0909,079;%%

%\cite{Heckman:2008es}
\bibitem{Heckman:2008es}
  J.~J.~Heckman, J.~Marsano, N.~Saulina, S.~Schafer-Nameki and C.~Vafa,
  ``Instantons and SUSY breaking in F-theory,''
  arXiv:0808.1286 [hep-th].
  %%CITATION = ARXIV:0808.1286;%%

%\cite{Marsano:2008jq}
\bibitem{Marsano:2008jq}
  J.~Marsano, N.~Saulina and S.~Schafer-Nameki,
  ``Gauge Mediation in F-Theory GUT Models,''
  Phys.\ Rev.\  D {\bf 80} (2009) 046006
  [arXiv:0808.1571 [hep-th]].
  %%CITATION = PHRVA,D80,046006;%%

%\cite{Marsano:2009ym}
\bibitem{Marsano:2009ym}
  J.~Marsano, N.~Saulina and S.~Schafer-Nameki,
  ``F-theory Compactifications for Supersymmetric GUTs,''
  JHEP {\bf 0908} (2009) 030
  [arXiv:0904.3932 [hep-th]].
  %%CITATION = JHEPA,0908,030;%%

%\cite{Marsano:2009gv}
\bibitem{Marsano:2009gv}
  J.~Marsano, N.~Saulina and S.~Schafer-Nameki,
  ``Monodromies, Fluxes, and Compact Three-Generation F-theory GUTs,''
  JHEP {\bf 0908} (2009) 046
  [arXiv:0906.4672 [hep-th]].
  %%CITATION = JHEPA,0908,046;%%
%\cite{Blumenhagen:2009yv}
\bibitem{Blumenhagen:2009yv}
  R.~Blumenhagen, T.~W.~Grimm, B.~Jurke and T.~Weigand,
  ``Global F-theory GUTs,''
  arXiv:0908.1784 [hep-th].
  %%CITATION = ARXIV:0908.1784;%%

\bibitem{Blumenhagen:2009up}
  R.~Blumenhagen, T.~W.~Grimm, B.~Jurke and T.~Weigand,
  ``F-theory uplifts and GUTs,''
  arXiv:0906.0013 [hep-th].
  %%CITATION = ARXIV:0906.0013;%%
%\cite{Blumenhagen:2008zz}
\bibitem{Blumenhagen:2008zz}
  R.~Blumenhagen, V.~Braun, T.~W.~Grimm and T.~Weigand,
  ``GUTs in Type IIB Orientifold Compactifications,''
  Nucl.\ Phys.\  B {\bf 815} (2009) 1
  [arXiv:0811.2936 [hep-th]].
  %%CITATION = NUPHA,B815,1;%%
%\cite{Braun:2008ua}
\bibitem{Braun:2008ua}
  A.~P.~Braun, A.~Hebecker and H.~Triendl,
  ``D7-Brane Motion from M-Theory Cycles and Obstructions in the Weak Coupling
  Limit,''
  Nucl.\ Phys.\  B {\bf 800} (2008) 298
  [arXiv:0801.2163 [hep-th]].
  %%CITATION = NUPHA,B800,298;%%
%\cite{Braun:2009wh}
\bibitem{Braun:2009wh}
  A.~P.~Braun, R.~Ebert, A.~Hebecker and R.~Valandro,
  ``Weierstrass meets Enriques,''
  arXiv:0907.2691 [hep-th].
  %%CITATION = ARXIV:0907.2691;%%
%\cite{Braun:2008pz}
\bibitem{c3fold}
  A.~P.~Braun, S.~Gerigk, A.~Hebecker and H.~Triendl,
  ``D7-Brane Moduli vs. F-Theory Cycles in Elliptically Fibred Threefolds,''
  arXiv:0912.1596 [hep-th].
  %%CITATION = ARXIV:0912.1596;%%


\bibitem{Braun:2008pz}
  A.~P.~Braun, A.~Hebecker, C.~Ludeling and R.~Valandro,
  ``Fixing D7 Brane Positions by F-Theory Fluxes,''
  Nucl.\ Phys.\  B {\bf 815} (2009) 256
  [arXiv:0811.2416 [hep-th]].
  %%CITATION = NUPHA,B815,256;%%

%\cite{Blumenhagen:2010at}
\bibitem{Blumenhagen:2010at}
  R.~Blumenhagen,
  ``Basics of F-theory from the Type IIB Perspective,''
  arXiv:1002.2836 [Unknown].
  %%CITATION = ARXIV:1002.2836;%%



%\cite{Sen:1997gv}
\bibitem{Sen:1997gv}
  A.~Sen,
  ``Orientifold limit of F-theory vacua,''
  Phys.\ Rev.\  D {\bf 55} (1997) 7345
  [arXiv:hep-th/9702165].
  %%CITATION = PHRVA,D55,7345;%%
  ``Orientifold limit of F-theory vacua,''
  Nucl.\ Phys.\ Proc.\ Suppl.\  {\bf 68} (1998) 92
  [Nucl.\ Phys.\ Proc.\ Suppl.\  {\bf 67} (1998) 81]
  [arXiv:hep-th/9709159].
  %%CITATION = NUPHZ,67,81;%%


\bibitem{Klemm:1996ts}
  A.~Klemm, B.~Lian, S.~S.~Roan and S.~T.~Yau,
  ``Calabi-Yau fourfolds for M- and F-theory compactifications,''
  Nucl.\ Phys.\  B {\bf 518}, 515 (1998)
  [arXiv:hep-th/9701023].
  %%CITATION = NUPHA,B518,515;%%

\bibitem{Berglund:1998va}
  P.~Berglund, A.~Klemm, P.~Mayr and S.~Theisen,
  ``On type IIB vacua with varying coupling constant,''
  Nucl.\ Phys.\  B {\bf 558} (1999) 178
  [arXiv:hep-th/9805189].
  %%CITATION = NUPHA,B558,178;%%

%\cite{Bershadsky:1998vn}
\bibitem{Bershadsky:1998vn}
  M.~Bershadsky, T.~Pantev and V.~Sadov,
  ``F-theory with quantized fluxes,''
  Adv.\ Theor.\ Math.\ Phys.\  {\bf 3} (1999) 727
  [arXiv:hep-th/9805056].
  %%CITATION = 00203,3,727;%%
%\cite{Collinucci:2008zs}
\bibitem{Collinucci:2008zs}
  A.~Collinucci,
  ``New F-theory lifts,''
  arXiv:0812.0175 [hep-th].\\
  %%CITATION = ARXIV:0812.0175;%%
  ``New F-theory lifts II: Permutation orientifolds and enhanced
  singularities,''
  arXiv:0906.0003 [hep-th].
  %%CITATION = ARXIV:0906.0003;%%
%\cite{Hubsch:1992nu}
\bibitem{Hubsch:1992nu}
  T.~Hubsch,
  ``Calabi-Yau manifolds: A Bestiary for physicists,''
%\href{http://www.slac.stanford.edu/spires/find/hep/www?irn=2618117}{SPIRES entry}
{\it  Singapore, Singapore: World Scientific (1992) 362 p}

\bibitem{fano}
V. A. Iskovskih,
``Fano threefolds. I'', Math. USSR-Izv. 11 (3) (1977)\ ,
``Fano threefolds II'', Math. USSR-Izv. 12 (3) (1978)\ ,
``Anticanonical models of three-dimensional algebraic varieties'', \
Current problems in mathematics, Vol. 12 (Russian), VINITI, Moscow (1979)
\\
S. Mori,S.  Mukai, 
``Classification of Fano 3-folds with B2≥2'', 
Manuscripta Mathematica 36 (2) (1981)
%\cite{Hanany:2009vx}
\bibitem{Hanany:2009vx}
  A.~Hanany and Y.~H.~He,
  ``Chern-Simons: Fano and Calabi-Yau,''
  arXiv:0904.1847 [hep-th].
  %%CITATION = ARXIV:0904.1847;%%

%\cite{Sethi:1996es}
\bibitem{Sethi:1996es}
  S.~Sethi, C.~Vafa and E.~Witten,
  ``Constraints on low-dimensional string compactifications,''
  Nucl.\ Phys.\  B {\bf 480} (1996) 213
  [arXiv:hep-th/9606122].
  %%CITATION = NUPHA,B480,213;%%

\bibitem{peters}
W. Barth, C. Peters and A. Van de Ven, Compact complex surfaces, 
Ergeb. Math. Grenzgeb. (3) 4, Springer-Verlag, Berlin, 1984.
\bibitem{Dimca}
 A. Dimca, ``Singularities and Topology of Hypersurfaces,'' 
(1992) New York: Springer.
\bibitem{Kodaira2}
 K.Kodaira
``On Compact Analytic Surfaces: II,''
Annals of Mathematics, Vol. 77, No.3 (1963)

\bibitem{Aspinwall:1996mn}
P.~S.~Aspinwall,
  ``K3 surfaces and string duality,''
  [arXiv:hep-th/9611137]
%%CITATION = HEP-TH/9611137;%%


%\cite{Bergshoeff:2006jj}
\bibitem{Bergshoeff:2006jj}
  E.~A.~Bergshoeff, J.~Hartong, T.~Ortin and D.~Roest,
  ``Seven-branes and supersymmetry,''
  JHEP {\bf 0702} (2007) 003
  [arXiv:hep-th/0612072].
  %%CITATION = JHEPA,0702,003;%%


\bibitem{Candelas:1989ug}
  P.~Candelas, P.~S.~Green and T.~Hubsch,
  ``Rolling Among Calabi-Yau Vacua,''
  Nucl.\ Phys.\  B {\bf 330}, 49 (1990).
%%CITATION = NUPHA,B330,49;%%

%\cite{Aspinwall:1996nk}
\bibitem{Aspinwall:1996nk}
  P.~S.~Aspinwall and M.~Gross,
  ``The SO(32) Heterotic String on a K3 Surface,''
  Phys.\ Lett.\  B {\bf 387} (1996) 735
  [arXiv:hep-th/9605131].
  %%CITATION = PHLTA,B387,735;%%

%\cite{Perevalov:1997vw}
\bibitem{Perevalov:1997vw}
  E.~Perevalov and H.~Skarke,
  ``Enhanced gauge symmetry in type II and F-theory compactifications:  Dynkin
  diagrams from polyhedra,''
  Nucl.\ Phys.\  B {\bf 505} (1997) 679
  [arXiv:hep-th/9704129].
  %%CITATION = NUPHA,B505,679;%%
%\cite{Aspinwall:2000kf}
\bibitem{Aspinwall:2000kf}
  P.~S.~Aspinwall, S.~H.~Katz and D.~R.~Morrison,
  ``Lie groups, Calabi-Yau threefolds, and F-theory,''
  Adv.\ Theor.\ Math.\ Phys.\  {\bf 4} (2000) 95
  [arXiv:hep-th/0002012].
  %%CITATION = 00203,4,95;%%

%\cite{Szendroi:2002rf}
\bibitem{Szendroi:2002rf}
  B.~Szendroi,
  ``Enhanced gauge symmetry and braid group actions,''
  Commun.\ Math.\ Phys.\  {\bf 238} (2003) 35
  [arXiv:math/0210122].
  %%CITATION = CMPHA,238,35;%%
\bibitem{Tate}
J. Tate, Algorithm for determining the type of a singular fiber in an elliptic pencil, in: 
Modular Functions of One Variable IV, Lecture Notes in Mathematics, Vol. 476, Springer, Berlin, 1975

%\cite{Bershadsky:1996nh}
\bibitem{Bershadsky:1996nh}
  M.~Bershadsky, K.~A.~Intriligator, S.~Kachru, D.~R.~Morrison, V.~Sadov and C.~Vafa,
  ``Geometric singularities and enhanced gauge symmetries,''
  Nucl.\ Phys.\  B {\bf 481} (1996) 215
  [arXiv:hep-th/9605200].
  %%CITATION = NUPHA,B481,215;%%

\bibitem{Sen:1996vd}
  A.~Sen,
  ``F-theory and Orientifolds,''
  Nucl.\ Phys.\  B {\bf 475}, 562 (1996)
  [arXiv:hep-th/9605150].
%%CITATION = NUPHA,B475,562;%%

\bibitem{Sen:1997kw}
  A.~Sen,
  ``F-theory and the Gimon-Polchinski orientifold,''
  Nucl.\ Phys.\  B {\bf 498}, 135 (1997)
  [arXiv:hep-th/9702061].
%%CITATION = NUPHA,B498,135;%%
%\cite{Aluffi:2009tm}
\bibitem{Aluffi:2009tm}
  P.~Aluffi and M.~Esole,
  ``New Orientifold Weak Coupling Limits in F-theory,''
  arXiv:0908.1572 [hep-th].
  %%CITATION = ARXIV:0908.1572;%%

\bibitem{Bianchi:1989du}
  M.~Bianchi and A.~Sagnotti,
  ``Open Strings and the Relative Modular Group,''
  Phys.\ Lett.\  B {\bf 231}, 389 (1989).
%%CITATION = PHLTA,B231,389;%%

%\cite{Gimon:1996rq}
\bibitem{Gimon:1996rq}
  E.~G.~Gimon and J.~Polchinski,
  ``Consistency Conditions for Orientifolds and D-Manifolds,''
  Phys.\ Rev.\  D {\bf 54} (1996) 1667
  [arXiv:hep-th/9601038].
  %%CITATION = PHRVA,D54,1667;%%

%\cite{Sen:1996xx}
\bibitem{Sen:1996xx}
  A.~Sen,
  ``A non-perturbative description of the Gimon-Polchinski orientifold,''
  Nucl.\ Phys.\  B {\bf 489} (1997) 139
  [arXiv:hep-th/9611186].
  %%CITATION = NUPHA,B489,139;%%

%\cite{Aspinwall:1995zi}
\bibitem{Aspinwall:1995zi}
  P.~S.~Aspinwall,
  ``Enhanced gauge symmetries and K3 surfaces,''
  Phys.\ Lett.\  B {\bf 357}, 329 (1995)
  [arXiv:hep-th/9507012].
  %%CITATION = PHLTA,B357,329;%%
%\cite{Katz:1996ht}
\bibitem{Katz:1996ht}
  S.~H.~Katz, D.~R.~Morrison and M.~Ronen Plesser,
  ``Enhanced Gauge Symmetry in Type II String Theory,''
  Nucl.\ Phys.\  B {\bf 477} (1996) 105
  [arXiv:hep-th/9601108].
  %%CITATION = NUPHA,B477,105;%%


%\cite{Johansen:1996am}
\bibitem{Johansen:1996am}
  A.~Johansen,
  ``A comment on BPS states in F-theory in 8 dimensions,''
  Phys.\ Lett.\  B {\bf 395} (1997) 36
  [arXiv:hep-th/9608186].
  %%CITATION = PHLTA,B395,36;%%

%\cite{Gaberdiel:1997ud}
\bibitem{Gaberdiel:1997ud}
  M.~R.~Gaberdiel and B.~Zwiebach,
  ``Exceptional groups from open strings,''
  Nucl.\ Phys.\  B {\bf 518} (1998) 151
  [arXiv:hep-th/9709013].
  %%CITATION = NUPHA,B518,151;%%

%\cite{DeWolfe:1998zf}
\bibitem{DeWolfe:1998zf}
  O.~DeWolfe and B.~Zwiebach,
  ``String junctions for arbitrary Lie algebra representations,''
  Nucl.\ Phys.\  B {\bf 541} (1999) 509
  [arXiv:hep-th/9804210].
  %%CITATION = NUPHA,B541,509;%%

\bibitem{Lerche:1999de}
  W.~Lerche,
  ``On the heterotic/F-theory duality in eight dimensions,''
  arXiv:hep-th/9910207.
  %%CITATION = HEP-TH/9910207;%%
\bibitem{Vafa:1997pm}
  C.~Vafa,
  ``Lectures on strings and dualities,''
  arXiv:hep-th/9702201.
 %%CITATION = HEP-TH/9702201;%%

\bibitem{LopesCardoso:1996hq}
  G.~Lopes Cardoso, G.~Curio, D.~Lust and T.~Mohaupt,
  ``On the duality between the heterotic string and F-theory in 8
  dimensions,''
  Phys.\ Lett.\  B {\bf 389} (1996) 479
  [arXiv:hep-th/9609111].
  %%CITATION = PHLTA,B389,479;%%

\bibitem{Lerche:1998nx}
  W.~Lerche and S.~Stieberger,
  ``Prepotential, mirror map and F-theory on K3,''
  Adv.\ Theor.\ Math.\ Phys.\  {\bf 2}, 1105 (1998)
  [Erratum-ibid.\  {\bf 3}, 1199 (1999)]
  [arXiv:hep-th/9804176].
  %%CITATION = 00203,2,1105;%%

%\cite{Friedman:1997yq}
\bibitem{Friedman:1997yq}
  R.~Friedman, J.~Morgan and E.~Witten,
  ``Vector bundles and F theory,''
  Commun.\ Math.\ Phys.\  {\bf 187} (1997) 679
  [arXiv:hep-th/9701162].
  %%CITATION = CMPHA,187,679;%%

%\cite{Donagi}
\bibitem{Donagi}
R.Donagi,
``Principal bundles on elliptic fibrations'' Asian \ J.\ Math. {\bf1}, 214 (1997)
[arXiv:alg-geom/9702002]

%\cite{Tatar:2009jk}
\bibitem{Tatar:2009jk}
  R.~Tatar, Y.~Tsuchiya and T.~Watari,
  ``Right-handed Neutrinos in F-theory Compactifications,''
  Nucl.\ Phys.\  B {\bf 823} (2009) 1
  [arXiv:0905.2289 [hep-th]].
  %%CITATION = NUPHA,B823,1;%%

\bibitem{Griffiths:1978} P.~Griffiths and J.~Harris, ''Principles of 
Algebraic Geometry''; John Wiley \& Sons, Inc. (1978) (USA).

\bibitem{Candelas:1990pi}
  P.~Candelas and X.~de la Ossa,
  ``MODULI SPACE OF CALABI-YAU MANIFOLDS,''
  Nucl.\ Phys.\  B {\bf 355} (1991) 455.
  %%CITATION = NUPHA,B355,455;%%


\bibitem{Green:1987rw}
  P.~Green and T.~Hubsch,
  ``POLYNOMIAL DEFORMATIONS AND COHOMOLOGY OF CALABI-YAU MANIFOLDS,''
  Commun.\ Math.\ Phys.\  {\bf 113} (1987) 505.
  %%CITATION = CMPHA,113,505;%%

\bibitem{Morrison:1996na}
  D.~R.~Morrison and C.~Vafa,
  ``Compactifications of F-Theory on Calabi--Yau Threefolds -- I,''
  Nucl.\ Phys.\  B {\bf 473} (1996) 74
  [arXiv:hep-th/9602114].
  %%CITATION = NUPHA,B473,74;%%

\bibitem{Morrison:1996pp}
  D.~R.~Morrison and C.~Vafa,
  ``Compactifications of F-Theory on Calabi--Yau Threefolds -- II,''
  Nucl.\ Phys.\  B {\bf 476} (1996) 437
  [arXiv:hep-th/9603161].
  %%CITATION = NUPHA,B476,437;%%
\bibitem{conwaysloane}J. H. Conway and N. J. A. Sloane, \textit{Sphere
Packings, Lattices and Groups}, Berlin, Heidelberg and New York, 1988.

\bibitem{helgason}
S.Helgason,
Differential Geometry and Symmetric Spaces, Oxford University Press; (2001)   .


\bibitem{key-51}
  K.~S.~Choi, K.~Hwang and J.~E.~Kim,
  ``Dynkin diagram strategy for orbifolding with Wilson lines,''
  Nucl.\ Phys.\  B {\bf 662}, 476 (2003)
  [arXiv:hep-th/0304243];
  %%CITATION = NUPHA,B662,476;%%

\bibitem{key-51bis}
  A.~Hebecker and M.~Ratz,
  ``Group-theoretical aspects of orbifold and conifold GUTs,''
  Nucl.\ Phys.\  B {\bf 670}, 3 (2003)
  [arXiv:hep-ph/0306049].
  %%CITATION = NUPHA,B670,3;%%

\bibitem{key-55}
  E.~B.~Dynkin,
  ``Semisimple subalgebras of semisimple Lie algebras,''
  Trans.\ Am.\ Math.\ Soc.\  {\bf 6}, 111 (1957).
  %%CITATION = TAMTA,6,111;%%

\bibitem{key-54}
  R.~Slansky,
  ``Group Theory For Unified Model Building,''
  Phys.\ Rept.\  {\bf 79}, 1 (1981).
  %%CITATION = PRPLC,79,1;%%

\bibitem{key-65}
  R.~Gopakumar and S.~Mukhi,
  ``Orbifold and orientifold compactifications of F-theory and M-theory  to six
  and four dimensions,''
  Nucl.\ Phys.\  B {\bf 479}, 260 (1996)
  [arXiv:hep-th/9607057].
  %%CITATION = NUPHA,B479,260;%%

%key dings = dieses Paper hier:

\bibitem{alexeev:2004}
  V.~Alexeev, V.~V.~Nikulin,
  ``Classification of log del Pezzo surfaces of index $\leq 2$,''
  arXiv:math/0406536

\bibitem{Nikulin:1986}
  V.~Nikulin, ``Discrete reflection groups in Lobachevsky spaces and
  algebraic surfaces,'' 
  Proceedings of the International Congress of Mathematicians,
  Vol. 1, 2 (Berkeley, Calif., 1986) (Providence, RI), Amer. Math. Soc.,
  1987, pp. 654–671.

%\cite{Lust:2006zh}
\bibitem{Lust:2006zh}
  D.~Lust, S.~Reffert, E.~Scheidegger and S.~Stieberger,
  ``Resolved toroidal orbifolds and their orientifolds,''
  Adv.\ Theor.\ Math.\ Phys.\  {\bf 12} (2008) 67
  [arXiv:hep-th/0609014].
  %%CITATION = 00203,12,67;%%

\bibitem{Kumar:2009zc}
  V.~Kumar and W.~Taylor,
  ``Freedom and Constraints in the K3 Landscape,''
  arXiv:0903.0386 [hep-th].
  %%CITATION = ARXIV:0903.0386;%%
%\cite{Nibbelink:2008tv}
\bibitem{Nibbelink:2008tv}
  S.~G.~Nibbelink, D.~Klevers, F.~Ploger, M.~Trapletti and P.~K.~S.~Vaudrevange,
  ``Compact heterotic orbifolds in blow-up,''
  JHEP {\bf 0804} (2008) 060
  [arXiv:0802.2809 [hep-th]].
  %%CITATION = JHEPA,0804,060;%%

\bibitem{Heter1}
  G.~Honecker and M.~Trapletti,
  ``Merging heterotic orbifolds and K3 compactifications with line bundles,''
  JHEP {\bf 0701} (2007) 051
  [arXiv:hep-th/0612030];
  %%CITATION = JHEPA,0701,051;%%

\bibitem{Heter2}
  S.~G.~Nibbelink, M.~Trapletti and M.~Walter,
  ``Resolutions of $C^n/Z_n$ Orbifolds, their U(1) Bundles, and Applications to
  String Model Building,''
  JHEP {\bf 0703} (2007) 035
  [arXiv:hep-th/0701227];
  %%CITATION = JHEPA,0703,035;%%

\bibitem{Heter3}
  S.~Groot Nibbelink, H.~P.~Nilles and M.~Trapletti,
  ``Multiple anomalous U(1)s in heterotic blow-ups,''
  Phys.\ Lett.\  B {\bf 652} (2007) 124
  [arXiv:hep-th/0703211];
  %%CITATION = PHLTA,B652,124;%%

\bibitem{Heter4}
  S.~G.~Nibbelink, T.~W.~Ha and M.~Trapletti,
  ``Toric Resolutions of Heterotic Orbifolds,''
  Phys.\ Rev.\  D {\bf 77} (2008) 026002
  [arXiv:0707.1597 [hep-th]].
  %%CITATION = PHRVA,D77,026002;%%

%\cite{Valandro:2008zg}
\bibitem{Valandro:2008zg}
  R.~Valandro,
  ``Type IIB Flux Vacua from M-theory via F-theory,''
  JHEP {\bf 0903} (2009) 122
  [arXiv:0811.2873 [hep-th]].
  %%CITATION = JHEPA,0903,122;%%
\bibitem{Nikulin:1979}
  V.~Nikulin, 
  ``On factor groups of the automorphism group of hyperbolic
  forms modulo subgroups generated by 2-reﬂections,''
  Soviet.\ Math.\ Dokl.{\bf 20} (1979), 1156–1158

\bibitem{Nikulin:1983}
  V.~Nikulin, ``Quotient-groups of groups of automorphisms of hyperbolic
  forms by subgroups generated by 2-reﬂections. Algebro-geometric
  applications,''
  J.Soviet Math.{\bf 22} (1983), 1401–1476

\bibitem{Brunner:2003zm}
  I.~Brunner and K.~Hori,
  ``Orientifolds and mirror symmetry,''
  JHEP {\bf 0411} (2004) 005
  [arXiv:hep-th/0303135].
  %%CITATION = JHEPA,0411,005;%%

\bibitem{Hori:2003ic}
  K.~Hori {\it et al.},
  ``Mirror symmetry,''
%\href{http://www.slac.stanford.edu/spires/find/hep/www?irn=6746918}
{\it  Providence, USA: AMS (2003) 929 p}

\bibitem{Narasimhan}
Raghavan~Narasimhan, ``Compact Riemann Surfaces'', Birkh\"auser Basel (1996) 


\bibitem{Friedman-Morgan(1994)}
R. Friedman and J. W. Morgan, ``Smooth four-manifolds and complex surfaces'',
Ergebnisse der Mathematik und ihrer Grenzgebiete, Vol. 27, Springer (1994)

\bibitem{Safarevic}
Igor'~R.~Safarevic, ``Basic algebraic geometry'', Springer (1977)

\bibitem{Vick}
James~W.~Vick, ``Homology Theory'', Springer Verlag, New York, Berlin (1994) 
\bibitem{Hatcher}
Allen~Hatcher, ``Algebraic topology'', Cambridge Univ. Press, Cambridge (2003)

%\cite{Cordova:2009fg}
\bibitem{Cordova:2009fg}
  C.~Cordova,
  ``Decoupling Gravity in F-Theory,''
  arXiv:0910.2955 [Unknown].
  %%CITATION = ARXIV:0910.2955;%%

\bibitem{ak05} 
  P.~S.~Aspinwall and R.~Kallosh,
  ``Fixing all moduli for M-theory on K3 x K3,''
  JHEP {\bf 0510}, 001 (2005)
  {\ttfamily [arXiv:hep-th/0506014]}
  %%CITATION = JHEPA,0510,001;%%
\bibitem{hkl06}
  M.~Haack, D.~Krefl, D.~Lust, A.~Van Proeyen and M.~Zagermann,
  ``Gaugino condensates and D-terms from D7-branes,''
  JHEP {\bf 0701} (2007) 078
  {\ttfamily [arXiv:hep-th/0609211]}
  %%CITATION = JHEPA,0701,078;%%

\bibitem{aaf03}
  L.~Andrianopoli, R.~D'Auria, S.~Ferrara and M.~A.~Lledo,
  ``4-D gauged supergravity analysis of type IIB vacua on K3 x T**2/Z(2),''
  JHEP {\bf 0303} (2003) 044
  {\ttfamily [arXiv:hep-th/0302174]}
  %%CITATION = JHEPA,0303,044;%%


\bibitem{aaft03}
  C.~Angelantonj, R.~D'Auria, S.~Ferrara and M.~Trigiante,
  ``K3 x T**2/Z(2) orientifolds with fluxes, open string moduli and  critical
  points,''
  Phys.\ Lett.\  B {\bf 583} (2004) 331
  {\ttfamily [arXiv:hep-th/0312019]}
  %%CITATION = PHLTA,B583,331;%%

\bibitem{cjs78}
  E.~Cremmer, B.~Julia and J.~Scherk,
  ``Supergravity theory in 11 dimensions,''
  Phys.\ Lett.\  B {\bf 76} (1978) 409
  %%CITATION = PHLTA,B76,409;%%
\bibitem{dlm95}
  M.~J.~Duff, J.~T.~Liu and R.~Minasian,
  ``Eleven-dimensional origin of string / string duality: A one-loop test,''
  Nucl.\ Phys.\  B {\bf 452} (1995) 261
  {\ttfamily [arXiv:hep-th/9506126]}
  %%CITATION = NUPHA,B452,261;%%


\bibitem{t00}
  A.~A.~Tseytlin,
  ``R**4 terms in 11 dimensions and conformal anomaly of (2,0) theory,''
  Nucl.\ Phys.\  B {\bf 584} (2000) 233
  {\ttfamily [arXiv:hep-th/0005072]}
  %%CITATION = NUPHA,B584,233;%%


\bibitem{w96}
  E.~Witten,
  ``On flux quantization in M-theory and the effective action,''
  J.\ Geom.\ Phys.\  {\bf 22} (1997) 1
  {\ttfamily [arXiv:hep-th/9609122]}
  %%CITATION = JGPHE,22,1;%%

\bibitem{lmr05}
  D.~Lust, P.~Mayr, S.~Reffert and S.~Stieberger,
  ``F-theory flux, destabilization of orientifolds and soft terms on
  D7-branes,''
  Nucl.\ Phys.\  B {\bf 732}, 243 (2006)
  {\ttfamily [arXiv:hep-th/0501139]}
  %%CITATION = NUPHA,B732,243;%%
\bibitem{a00ag04}
  B.~S.~Acharya,
  ``On realising N = 1 super Yang-Mills in M theory,''
  {\ttfamily [arXiv:hep-th/0011089]},
  %%CITATION = HEP-TH/0011089;%%
  %{\it For a review, see also} 
  B.~S.~Acharya and S.~Gukov,
  ``M theory and Singularities of Exceptional Holonomy Manifolds,''
  Phys.\ Rept.\  {\bf 392}, 121 (2004)
  {\ttfamily [arXiv:hep-th/0409191]}
  %%CITATION = PRPLC,392,121;%%

%\cite{Bouchard:2007ik}
\bibitem{Bouchard:2007ik}
  V.~Bouchard,
  ``Lectures on complex geometry, Calabi-Yau manifolds and toric geometry,''
  arXiv:hep-th/0702063.
  %%CITATION = HEP-TH/0702063;%%

%\cite{Nakahara:2003nw}
\bibitem{Nakahara:2003nw}
  M.~Nakahara,
  ``Geometry, topology and physics,''
%\href{http://www.slac.stanford.edu/spires/find/hep/www?irn=7208855}{SPIRES entry}
Boca Raton, USA: Taylor and Francis (2003) 

%\cite{Candelas:1987is}
\bibitem{Candelas:1987is}
  P.~Candelas,
  ``LECTURES ON COMPLEX MANIFOLDS,''
%\href{http://www.slac.stanford.edu/spires/find/hep/www?irn=2116014}{SPIRES entry}
{\it  IN *TRIESTE 1987, PROCEEDINGS, SUPERSTRINGS '87* 1-88. }

\bibitem{BottandTu}
R.~Bott and L.~W.~Tu, ``Differential forms in algebraic topology'',
Springer, New York, Heidelberg, Berlin, (1982)

\bibitem{Huybrechts}
D. Huybrechts, ``Complex geometry'' Universitext, Springer-Verlag, Berlin (2005)

\bibitem{milnorstash}
J.~W.~Milnor and J.~D.~Stasheff, ``Characteristic Classes'',
Annals of Mathematics Studies 76, Princeton University Press (1974)

\bibitem{bertlmann}
R.~A.~Bertlmann, ``Anomalies in Quantum Field Theory'',International series of monographs on physics, vol. 91, Clarendon Press, Oxford (1996)

\bibitem{GS}
M.~G\"ockeler, T.~Sch\"ucker, ``Differential Geometry, Gauge Theories, and Gravity'', Cambridge University Press (1987)

\bibitem{Roe}
J.~Roe, ``Elliptic operators, topology and asymptotic methods'', Pitman Research Notes in Mathematics Series, 395. Longman, Harlow (1998)

\bibitem{Gaume}
L.~Alvarez-Gaume, ``Supersymmetry and the Atiyah-Singer index theorem'', Commun. Math. Phys.,. 90,. 161-173 (1983)

\bibitem{Oda:1988}
  T.~Oda,
  ``Convex Bodies and Algebraic Geometry,''
  {\it Springer-Verlag\ (1988)}

\bibitem{Fulton}
W.~Fulton, ``Introduction to toric varieties'', Annals of mathematics studies, Princeton
University Press, Princeton (1993)

%\cite{Skarke:1998yk}
\bibitem{Skarke:1998yk}
  H.~Skarke,
  ``String dualities and toric geometry: An introduction,''
  arXiv:hep-th/9806059.
  %%CITATION = HEP-TH/9806059;%%

%\cite{Reffert:2007im}
\bibitem{Reffert:2007im}
  S.~Reffert,
  ``The Geometer's Toolkit to String Compactifications,''
  arXiv:0706.1310 [hep-th].
  %%CITATION = ARXIV:0706.1310;%%

\bibitem{Calabi}
E.~Calabi, ``The space of K\"ahler metrics'', in {\it Proceedings of the International
Congress of Mathematics, Amsterdam, 1954}, North-Holland, Amsterdam (1956) \\
``On K\"ahler manifolds with vanishing first Chern class'', in {\it  geometry
and topology, a symposium in honour of S.Lefschetz}, Princeton University Press, Princeton (1957)

\bibitem{Yau}
S.~-T.~Yau, ``On Calabi's conjecture and some new results in algebraic geometry'', {\it Proceedings
of the National Academy of Sciences of the U.S.A.} {\bf 74} (1977) \\
``On the Ricci curvature of a compact K\"ahler manifold and the complex Monge-Amp\`ere equations.I.'',
{\it Communications on pure and applied mathematics} {\bf 31} (1978)

%\cite{Kreuzer:2002uu}
\bibitem{Kreuzer:2002uu}
  M.~Kreuzer and H.~Skarke,
  ``PALP: A Package for analyzing lattice polytopes with applications to toric
  geometry,''
  Comput.\ Phys.\ Commun.\  {\bf 157} (2004) 87
  [arXiv:math/0204356].
  %%CITATION = CPHCB,157,87;%%


\bibitem{palp}
PALP: A Package for Analyzing Lattice Polytopes, 
 http://hep.itp.tuwien.ac.at/~kreuzer/CY/CYpalp.html


\bibitem{Demazure:1970}
  M.~Demazure,
  ``Sous-groupes algebriques de rang maximum du groupe de Cremona,''
  Ann.\ Sci. \ Ecole\ Norm.\ Sup.\ (4) {\bf 3} (1970), 507-588


\bibitem{indefMetric}
  T.~Ya.~Azizov and I.~S.~Iokhvidov, 
  ``Linear Operators in Spaces with an Indefinite Metric,'' (Wiley, New York, 1989)

\bibitem{Pandit}
  L.~K.~Pandit,
  ``Linear Vector Spaces with Indefinite Metric,''
  Nuovo\ Cimento,\ 10\ (1959),\ 157








\end{thebibliography}
\end{document}